\documentclass[11pt,a4paper]{article}
\usepackage[utf8]{inputenc}
\usepackage[T1]{fontenc}
\usepackage{lmodern,textcomp,amsmath,amssymb,mathtools,braket,amsthm,csquotes,comment,bm,cancel}
\allowdisplaybreaks
\usepackage[english]{babel}
\usepackage[a4paper,margin=2.5cm]{geometry}
\usepackage{authblk}
\usepackage{xcolor}
\definecolor{cblue}{rgb}{0.16, 0.32, 0.75}
\definecolor{cred}{rgb}{0.7, 0.11, 0.11}
\usepackage{array,longtable,booktabs}
\usepackage{makeidx}
\makeindex
\usepackage[numbers,sort&compress,square]{natbib}
\usepackage[unicode=true,bookmarks=true,colorlinks=true,linkcolor=cred,citecolor=cblue,urlcolor=cblue]{hyperref}
\newcommand{\doi}[1]{\href{https://doi.org/#1}{\nolinkurl{doi:#1}}}
\hypersetup{pdftitle={Mathematical foundations of second quantization: an introduction through spin–boson models}, pdfauthor={Davide Lonigro}}
\usepackage{orcidlink}

\numberwithin{equation}{section}

\theoremstyle{plain}
\newtheorem{theorem}{Theorem}[section]
\newtheorem{proposition}[theorem]{Proposition}
\newtheorem{lemma}[theorem]{Lemma}
\newtheorem{corollary}[theorem]{Corollary}
\theoremstyle{definition}
\newtheorem{definition}[theorem]{Definition}
\newtheorem{example}[theorem]{Example}
\newtheorem{exercise}[theorem]{Exercise}
\newtheorem{axiom}{Axiom}
\theoremstyle{definition}
\newtheorem{remark}[theorem]{Remark}

\newcommand{\hilb}{\mathcal{H}}

\newcommand{\hfrak}{\mathfrak{H}}
\newcommand{\sph}{\mathfrak{h}}
\newcommand{\e}{\mathrm{e}}
\newcommand{\g}{\mathrm{g}}
\renewcommand{\Im}{\operatorname{Im}}
\renewcommand{\Re}{\operatorname{Re}}
\renewcommand{\a}[1]{a\!\left(#1\right)}
\newcommand{\adag}[1]{a^\dag\!\left(#1\right)}
\newcommand{\ii}{\mathrm{i}}

\newcommand{\N}{\mathcal{N}}
\newcommand{\Dom}{\operatorname{Dom}}
\newcommand{\Ran}{\operatorname{Ran}}
\newcommand{\Ker}{\operatorname{Ker}}

\newcommand{\sgn}{\operatorname{sgn}}
\newcommand{\seesol}[1]{\par\smallskip\noindent\emph{Solution}: see Appendix~\ref{sol:#1}.}
\newcommand{\solsec}[2]{\subsubsection{\texorpdfstring{Exercise~\ref{exerc:#1}}{Exercise}: #2}\label{sol:#1}}

\begin{document}
	
	\title{\textbf{Mathematical foundations of second quantization\\{\large An introduction through spin--boson models}}}
	\author[$1,\star$]{Davide Lonigro\hspace{3pt}\orcidlink{0000-0002-0792-8122}\hspace{1pt}}
	\affil[$1$]{\small Department of Physics, Friedrich-Alexander-Universit\"at Erlangen-N\"urnberg, Staudtstra\ss e 7, 91058 Erlangen, Germany}
	\affil[$\star$]{\small \texttt{davide.lonigro@fau.de}}
	\date{\today}
	\maketitle
	
	\begin{abstract}
		These lecture notes originate from a one-semester course on the mathematical foundations of second quantization, first given at the Department of Physics of the Friedrich-Alexander-Universit\"at Erlangen-N\"urnberg in the Winter Semester 2025/26 to a mixed audience of physics and mathematics students. Their guiding thread is a class of physical models: spin--boson models, describing a two-level system coupled to a boson field.
		
		After summarizing the necessary background on unbounded operators, up to the Kato--Rellich and KLMN theorems, and developing direct sums and tensor products of Hilbert spaces and of unbounded operators, we treat single-mode boson fields, constructing the ladder and number operators and characterizing their domains, before moving to Fock spaces, second quantization, bosonic and fermionic annihilation and creation operators, and the canonical (anti)commutation relations. Throughout, spin--boson models provide both the motivation and the test bench for the formalism: we rigorously establish their self-adjointness, solve the pure dephasing and Jaynes--Cummings models explicitly, deriving their spectra and the qubit channels they induce, show how their multimode counterparts display irreversible decoherence and decay, and, in an appendix, prove the rotating-wave approximation for the quantum Rabi model with an explicit error bound. Finally, we use the KLMN theorem to construct spin--boson models with non-normalizable form factors, and briefly discuss stronger ultraviolet divergences and the need for renormalization. Exercises with full solutions are included.
	\end{abstract}
	
	\tableofcontents
	
	\section{Introduction}\label{sec:intro}
	
	These notes are based on a course that I designed and first gave in the Winter Semester 2025/26 at the Department of Physics of the Friedrich-Alexander-Universit\"at Erlangen-N\"urnberg (FAU), under the title \emph{Mathematical Foundations of Second Quantization}, with two hours of lectures and two hours of exercise classes per week. The course was conceived as a follow-up to the course \emph{Mathematical Foundations of Quantum Theory}, which covers most of the basics of unbounded operators on Hilbert spaces. The notes expand and revise the material presented in the lectures. 
	
	With respect to the course, the material has been rearranged as follows. The basics of unbounded operators and the axioms of quantum mechanics, which the course recalls from \emph{Mathematical Foundations of Quantum Theory}, are summarized in Section~\ref{sec:recap}. The perturbation theory of self-adjoint operators is instead the first new topic of the course, and is presented there with proofs and exercises; being standard textbook material, it has been included in the same summary. All the remaining material, which forms the core of the notes---second quantization and its application to spin--boson models---is presented with proofs and exercises.
	
	\subsection{Why these notes}
	
	Second quantization lies at the crossroads of two communities. Physics students typically meet it early, as a set of rules: one introduces annihilation and creation operators, imposes canonical (anti)commutation relations, and manipulates expressions such as
	\begin{equation}
		H=\int\omega(k)a^\dag(k)a(k)\;\mathrm{d}k
	\end{equation}
	but often with little idea of which mathematical objects are hiding behind such expressions: on which domain they are defined, whether a given Hamiltonian is self-adjoint (and thus generates a dynamics at all), or in which sense an approximation is justified. Mathematics students, on the other hand, are perfectly at ease with the formalism, but rarely get to see what physicists actually do with Fock spaces: which models they study, why, and which questions they care about. Both communities are served by excellent literature: second quantization is an everyday tool in textbooks on many-body physics and quantum optics~\cite{fetterwalecka,negeleorland,altlandsimons,gerryknight}, and Fock spaces and quantum fields are treated with full mathematical rigor in several monographs~\cite{reedsimon2,brattelirobinson2,spohn,derezinskigerard,arai}.
	
	These notes are an attempt to address a mixed audience of physicists and mathematicians, making each half of the story accessible to the readers who are more familiar with the other one. To this end, they are built around a single class of models: \emph{spin--boson models}, describing a two-level system---a spin, or qubit---coupled to a boson field. Models of this kind are ubiquitous in physics, from quantum optics and cavity or circuit quantum electrodynamics to the theory of open quantum systems, where they are the standard models of decoherence and dissipation; already in their simplest instances---the pure dephasing, Jaynes--Cummings and quantum Rabi models---they display decoherence, Rabi oscillations, and the subtleties of the rotating-wave approximation. At the same time, they are arguably the clearest example of a class of models whose mathematics is largely under control; and yet, they are typically handled only formally in the physics literature, while rigorous treatments of field--matter interaction usually address physically complete but technically demanding models, such as the Pauli--Fierz model\index{Pauli--Fierz model} of non-relativistic quantum electrodynamics (see e.g.~\cite{spohn}), at research level. These notes aim to fill the gap in between. The rigorous treatment of spin--boson models requires the whole machinery of second quantization---Fock spaces, annihilation and creation operators, second quantized operators, and tensor products of unbounded operators---and, conversely, these models provide the motivation and the test bench for each piece of this machinery: everything in these notes is developed with them in mind.
	
	This goal has shaped the presentation in several ways. First, the notes aim to be as self-contained as possible: the background on unbounded operators that we need is summarized in Section~\ref{sec:recap}, while the constructions on which second quantization is based---direct sums and tensor products of Hilbert spaces and of unbounded operators---are developed, with proofs, in Section~\ref{sec:directsum}. Since a Fock space is a direct sum of tensor products, and every light--matter Hamiltonian is a sum of tensor products of operators, we spend somewhat more time than usual on these constructions, especially for unbounded operators. Second, rather than introducing the general Fock space formalism and recovering single-mode fields as a particular instance, we treat the single-mode case first (Section~\ref{sec:singlemode}), and only afterward the general one (Sections~\ref{sec:sq} and~\ref{sec:bf}). For a single mode, the annihilation, creation and number operators and their maximal domains can be described explicitly in terms of the Hermite functions, every step can be checked by hand, and the essential mathematical features of the general theory already appear in their simplest form. The general formalism is then built on top of this, so that each new ingredient (Fock spaces, annihilation and creation operators $a(f),a^\dag(f)$, (anti)symmetrization) can be compared with its single-mode counterpart.
	
	Accordingly, spin--boson models are discussed twice: first in the single-mode case (Section~\ref{sec:singlemodesb}), and then for general boson fields (Section~\ref{sec:multimodesb}), including form factors which are not square-integrable and a glimpse of ultraviolet divergences and renormalization (Section~\ref{sec:singular}); a rigorous proof of the rotating-wave approximation, with an explicit error bound, is given in Appendix~\ref{app:rwa}. In both cases, rigor never comes at the expense of explicit computations---perhaps the most distinctive feature of these notes. The pure dephasing and Jaynes--Cummings models, for instance, are not only proven to be self-adjoint, but solved explicitly, with their spectra, eigenvectors and the qubit channels they induce derived step by step (Sections~\ref{subsec:dephasing} and~\ref{subsec:jc}); their multimode counterparts, which display irreversible decoherence and decay, are then treated in the same spirit (Sections~\ref{subsec:multidephasing} and~\ref{subsec:multijc}). Along the way, we point out the connections with current research, e.g. on the self-adjointness of nonlinear light--matter Hamiltonians (Remark~\ref{rem:quadratic}). Ultimately, the notes aim to bring the reader to the point where the research literature on field--matter models becomes accessible.
	
	Let me stress that these notes make no claim to completeness, nor to originality. Most of their content is standard: Section~\ref{sec:recap} summarizes textbook results without proofs, and the material of Sections~\ref{sec:directsum}, \ref{sec:singlemode} and~\ref{sec:sq} can be found, in various forms, in the monographs cited above, where the reader will also find much more than is covered here. Spin--boson models, on the other hand, are mostly treated in the research literature, including some of my own works, on which our discussion draws and which are cited in the corresponding sections. Nor are these notes meant to compete with the existing introductions to Fock spaces and second quantization written at the level of lecture notes or textbook chapters, such as~\cite{derezinskiccr}, devoted to representations of the canonical (anti)commutation relations, or the chapters on second quantization and on non-relativistic matter coupled to radiation in~\cite{gustafsonsigal}: what distinguishes them is the choice of objects, namely simple models of field--matter interaction, from a single field mode to infinitely many, treated rigorously and at a textbook level.
	
	\subsection{Prerequisites, notation and conventions}\label{subsec:notation}
	
	The notes assume familiarity with linear algebra, basic measure theory and $L^2$ spaces (the notions we need are recalled in Appendix~\ref{app:measure}), and elementary quantum mechanics, as well as a first exposure to unbounded operators on Hilbert spaces, at the level of a first course on the mathematical foundations of quantum mechanics: adjoints and closures, (essential) self-adjointness, the spectral theorem. All results of this kind that we need, together with perturbation theory up to the Kato--Rellich and KLMN theorems, are summarized without proofs in Section~\ref{sec:recap}, mostly following~\cite{teschl}; readers meeting this material for the first time may want to read that section alongside the first chapters of~\cite{teschl}.
	
	Throughout the notes, $\mathbb{N}=\{0,1,2,\ldots\}$ includes zero, and $\overline{z}$ denotes the complex conjugate of $z\in\mathbb{C}$; accordingly, $\overline{f}$ is the pointwise complex conjugate of a function $f$. The same overline also denotes the closure of sets and of closable operators, the meaning being always clear from the context. Hilbert spaces are complex and, unless otherwise stated, separable; the scalar product $\braket{\cdot,\cdot}$ is antilinear in the first entry and linear in the second one, and orthonormal bases are always understood in the Hilbert space sense, i.e. as complete orthonormal sets. When several Hilbert spaces are involved, we indicate by a subscript the space to which a norm or scalar product refers, as in $\|\cdot\|_{\hilb_1}$ or $\braket{\cdot,\cdot}_{\hilb_1\otimes\hilb_2}$, unless this is clear from the context. All measure spaces are $\sigma$-finite (Appendix~\ref{app:measure}). Operators are linear, and always come with their domain: we write $A:\Dom A\subseteq\hilb\rightarrow\hilb$, and $\Ran A$ and $\Ker A$ for the range and kernel of $A$; $A\subseteq B$ means that $B$ is an extension of $A$, $A\restriction\mathcal{D}$ denotes the restriction of $A$ to a subspace $\mathcal{D}$, $\overline{A}$ is the closure of a closable operator, and $A^*$ its adjoint. $\mathcal{B}(\hilb)$ denotes the space of bounded operators on $\hilb$, and $I$ the identity on any space. Given symmetric operators $A,B$, we write $A\geq0$ if $\braket{\psi,A\psi}\geq0$ for all $\psi\in\Dom A$, and $A\leq B$ if $B-A\geq0$. Commutators and anticommutators are denoted by $[A,B]:=AB-BA$ and $\{A,B\}:=AB+BA$, always understood as acting on vectors for which both products are defined. We write $M_v$ for the multiplication operator associated with a measurable function $v$, $\mathcal{S}(\mathbb{R})$ for the Schwartz space (Definition~\ref{def:schwartz}), and $\mathrm{H}^k$ for Sobolev spaces. 
	
	The symbol $\odot$ denotes the \emph{algebraic} tensor product of spaces or operators and $\otimes$ its \emph{Hilbert space} counterpart, i.e. its completion or closure (Section~\ref{sec:tensor}); $\cong$ denotes unitary equivalence\index{unitary equivalence}, which we often use to identify spaces and operators. The single-particle Hilbert space underlying a Fock space is denoted by $\sph$ (Section~\ref{sec:sq}), while $\hilb$ denotes a generic Hilbert space. Number operators are denoted by $\N$, and the vectors $\phi_n$, $n\in\mathbb{N}$, always denote the corresponding number states (in particular, the Hermite functions when $\hilb=L^2(\mathbb{R})$). For two-level systems, $\sigma_x,\sigma_y,\sigma_z$ are the Pauli matrices, $\sigma_\pm$ the raising and lowering matrices, and $e_\pm$ the eigenvectors of $\sigma_z$ with eigenvalues $\pm1$. Boldface letters denote vectors in $\mathbb{R}^3$, such as $\bm{x}$ or $\bm{g}=(g_x,g_y,g_z)$, or vector fields with values in $\mathbb{R}^3$, such as the fields $\bm{E}$, $\bm{B}$ of Section~\ref{subsec:prelude}; accordingly, $\bm{\sigma}=(\sigma_x,\sigma_y,\sigma_z)$. Physicists' bra-ket notation is used occasionally, when it makes expressions more transparent: given $v\in\hilb$, $\ket{v}$ denotes the bounded operator $\mathbb{C}\ni z\mapsto zv\in\hilb$, and $\bra{v}=\braket{v,\cdot}:\hilb\rightarrow\mathbb{C}$ its adjoint, i.e. the continuous linear functional associated with $v$. Accordingly, $\ket{v}\!\bra{w}$ is the rank-one operator $u\mapsto\braket{w,u}v$ on $\hilb$, which, for $v=w$ normalized, is the orthogonal projection onto the span of $v$. Vectors are often labeled by their quantum numbers, as in $\ket{n}\equiv\phi_n$ for the number states, or $\ket{e_+,\phi_n}\equiv e_+\otimes\phi_n$. We work in units with $\hbar=1$ throughout. Finally, the imaginary unit, Euler's number and differentials are written in roman type: $\ii$, $\e$, $\mathrm{d}x$. 
	
	A list of the main symbols used in the notes, with pointers to the place where each of them is introduced, is given at the end of the notes (page~\pageref{sec:symbols}).
	
	\subsection{Outline}
	
	Section~\ref{sec:recap} summarizes the necessary background on unbounded operators: self-adjointness and its criteria, the spectral and Stone theorems, the mathematical axioms of quantum theory and mixed states, the Kato--Rellich theorem (Theorem~\ref{thm:katorellich})---our main tool to establish the self-adjointness of the models in Sections~\ref{sec:singlemodesb} and~\ref{sec:multimodesb}---and quadratic forms up to the KLMN theorem (Theorem~\ref{thm:klmn}). Section~\ref{sec:directsum} introduces direct sums and tensor products of Hilbert spaces and operators, including tensor products of $L^2$ spaces and of multiplication operators.
	
	Section~\ref{sec:singlemode} deals with single-mode boson fields: starting from the normal mode expansion of the electromagnetic field, we revisit the quantum harmonic oscillator in its ladder representation, characterize the maximal domains of the ladder and number operators, and abstract the construction to an arbitrary separable Hilbert space. Section~\ref{sec:singlemodesb} introduces single-mode spin--boson models: we prove their self-adjointness, solve the pure dephasing and Jaynes--Cummings models exactly, and discuss the quantum Rabi model and the rotating-wave approximation. Section~\ref{sec:sq} develops the general theory: Fock spaces, second quantization of single-particle operators, and annihilation and creation operators. Section~\ref{sec:bf} implements the symmetry constraints for indistinguishable particles: after collecting the necessary facts on permutations, it introduces symmetric and antisymmetric tensors, bosonic and fermionic Fock spaces and the corresponding annihilation and creation operators, the canonical commutation and anticommutation relations, and the expression of second quantized operators in terms of annihilation and creation operators. Section~\ref{sec:multimodesb} applies this machinery to general spin--boson models, including the multimode counterparts of the pure dephasing and Jaynes--Cummings models, which display irreversible decoherence and decay. Section~\ref{sec:singular} shows how the KLMN theorem allows one to define spin--boson models whose form factor is not square-integrable, and briefly discusses stronger ultraviolet divergences, for which renormalization is needed. 
	
	Appendix~\ref{app:measure} recalls the basic notions of measure theory used throughout the notes, and Appendix~\ref{app:rwa} contains a rigorous proof of the rotating-wave approximation for the quantum Rabi model, with an explicit error bound. Exercises are scattered throughout the text; many of them complete arguments of the main text, and their solutions are collected in Appendix~\ref{app:solutions}, grouped by section.

	\section{Preliminaries}\label{sec:recap}
	
	This section collects the mathematical background on unbounded operators that will be needed in the rest of the notes. We recall the basic notions on Hilbert spaces and linear operators (Section~\ref{subsec:hilbertops}), closed operators and the criteria for (essential) self-adjointness (Section~\ref{subsec:esa}), the spectral theorem, Stone's theorem and resolvent convergence (Section~\ref{subsec:spectrum}), the axioms of quantum mechanics (Section~\ref{subsec:qmfoundations}) and their extension to mixed states (Section~\ref{subsec:mixed}); we then discuss perturbations of self-adjoint operators, up to the Kato--Rellich theorem (Section~\ref{sec:perturbation}), and sesquilinear forms, up to the KLMN theorem (Section~\ref{subsec:forms}). Since this material is standard, results are stated without proofs. Sections~\ref{subsec:hilbertops}--\ref{subsec:spectrum}, \ref{sec:perturbation} and~\ref{subsec:forms} closely follow the first chapters (Chapters 0--6) of Teschl's textbook~\cite{teschl}: all results stated there without a reference can be found, with proofs, in~\cite{teschl}, and citations are only given for material taken from other sources or from later chapters of~\cite{teschl}, and for pointers to topics not discussed here. Other excellent references are~\cite{reedsimon1,reedsimon2,hall,deoliveira,schmuedgen,katobook}. Readers already familiar with some of this material may skip it, and come back to it when needed: Table~\ref{tab:recap} shows where the main results of each subsection are used in the rest of the notes.
	
	\begin{table}[ht]
		\centering
		\small
		\renewcommand{\arraystretch}{1.25}
		\begin{tabular}{@{}>{\raggedright\arraybackslash}p{3.3cm}>{\raggedright\arraybackslash}p{6.3cm}>{\raggedright\arraybackslash}p{5.5cm}@{}}
			\toprule
			Section & Main results & Mainly used in\\
			\midrule
			\ref{subsec:hilbertops} Hilbert spaces and linear operators & $L^2$ spaces, adjoints, multiplication operators (Example~\ref{examp:multiplication}), momentum operator (Example~\ref{examp:momentum}) & Throughout; $L^2$ spaces and multiplication operators in Sections~\ref{sec:directsum} and~\ref{sec:sq}--\ref{sec:singular}; momentum in Section~\ref{subsec:dephasing}\\
			\ref{subsec:esa} Closed operators and (essential) self-adjointness & Closures, criteria of (essential) self-adjointness (Theorems~\ref{thm:criterion-sa} and~\ref{thm:criterion-esssa}), bases of eigenvectors (Proposition~\ref{prop:eigenvectors}), harmonic oscillator (Example~\ref{examp:qho}) & Direct sums of operators (Section~\ref{subsec:directsumops}); harmonic oscillator (Section~\ref{sec:singlemode})\\
			\ref{subsec:spectrum} Spectrum, spectral theorem and Stone's theorem & Spectral theorem (Theorems~\ref{thm:spectraltheorem1} and~\ref{thm:spectraltheorem2}), unitary groups and Stone's theorem (Proposition~\ref{prop:unitarygroup}, Theorem~\ref{thm:stone}), resolvent convergence & Spectral theorem: Sections~\ref{sec:directsum}--\ref{sec:singular}; unitary groups: Sections~\ref{subsec:gammau} and~\ref{subsec:multidephasing}, Appendix~\ref{app:rwa}; resolvent convergence: Sections~\ref{subsec:multidephasing} and~\ref{subsec:formsum}\\
			\ref{subsec:qmfoundations} Axioms of quantum mechanics & States, observables, evolution, composite systems & Physical interpretation throughout; composite systems in Section~\ref{sec:tensor}\\
			\ref{subsec:mixed} Mixed states & Density operators, Liouville--von Neumann equation & Partial trace (Proposition~\ref{prop:partialtrace}); qubit channels (Sections~\ref{subsec:dephasing}, \ref{subsec:jc}, \ref{subsec:multidephasing} and~\ref{subsec:multijc})\\
			\ref{sec:perturbation} Kato--Rellich theorem & Relative boundedness, Kato--Rellich theorem (Theorem~\ref{thm:katorellich}), Example~\ref{examp:wrtsquare} & Self-adjointness of spin--boson models (Sections~\ref{subsec:singlemodesb} and~\ref{subsec:multimodesb})\\
			\ref{subsec:forms} Sesquilinear forms and the KLMN theorem & Closed forms, representation theorem, KLMN theorem (Theorem~\ref{thm:klmn}) & Only Section~\ref{sec:singular}\\
			\bottomrule
		\end{tabular}
		\caption{Where the results of Section~\ref{sec:recap} are used in the rest of the notes.}\label{tab:recap}
	\end{table}
	
	\subsection{Hilbert spaces and linear operators}\label{subsec:hilbertops}
	
	We assume familiarity with the basic theory of Hilbert spaces, and only fix some terminology.
	
	\begin{definition}[Hilbert space]\label{def:scalarproduct}\index{Hilbert space}\index{scalar product}
		A \emph{scalar product} on a complex vector space $\hilb$ is a positive definite, Hermitian, sesquilinear form $\hilb\times\hilb\ni(u,v)\mapsto\braket{u,v}\in\mathbb{C}$, antilinear in its first entry and linear in its second one. Equipped with a scalar product, $\hilb$ is a \emph{pre-Hilbert space}\index{pre-Hilbert space}, and a \emph{Hilbert space} if it is complete with respect to the induced norm $\|u\|:=\sqrt{\braket{u,u}}$.
	\end{definition}
	
	All scalar products satisfy the \emph{Cauchy--Schwarz inequality}\index{Cauchy--Schwarz inequality} $|\braket{u,v}|\leq\|u\|\|v\|$. A countable orthonormal set $(e_n)_{n\in\mathbb{N}}\subset\hilb$ is \emph{complete}, or an \emph{orthonormal basis}\index{orthonormal basis}, if $u=\sum_{n\in\mathbb{N}}\braket{e_n,u}e_n$ for all $u\in\hilb$, the series converging in norm; $\hilb$ is \emph{separable}\index{Hilbert space!separable} if it admits a countable orthonormal basis. All Hilbert spaces of interest in quantum mechanics are separable.
	
	We will freely use the basic topological notions associated with the norm. A sequence $(u_n)_{n\in\mathbb{N}}\subset\hilb$ \emph{converges} to $u\in\hilb$ if $\|u_n-u\|\to0$, and is a \emph{Cauchy sequence} if $\|u_n-u_m\|\to0$ as $n,m\to\infty$; completeness means that every Cauchy sequence converges. A subset $\mathcal{M}\subseteq\hilb$ is \emph{closed} if it contains the limits of all its convergent sequences; its \emph{closure} $\overline{\mathcal{M}}$ is the smallest closed set containing $\mathcal{M}$, i.e. the set of all limits of convergent sequences in $\mathcal{M}$; and $\mathcal{M}$ is \emph{dense}\index{dense subspace} if $\overline{\mathcal{M}}=\hilb$, i.e. if every $u\in\hilb$ can be approximated arbitrarily well by elements of $\mathcal{M}$. For instance, the finite linear combinations of the elements of an orthonormal basis form a dense subspace of $\hilb$. Every closed subspace $\mathcal{M}$ has an \emph{orthogonal complement} $\mathcal{M}^\perp:=\{u\in\hilb:\braket{v,u}=0\text{ for all }v\in\mathcal{M}\}$, and every $u\in\hilb$ can be uniquely decomposed as $u=v+w$ with $v\in\mathcal{M}$ and $w\in\mathcal{M}^\perp$ (projection theorem)\index{projection theorem}; the map $u\mapsto v$ is the \emph{orthogonal projection}\index{orthogonal projection} onto $\mathcal{M}$.
	
	\begin{example}[The $\ell^2$ space]\label{examp:ell2}\index{l2@$\ell^2$ space}
		Let $\ell^2$ be the space of square-summable complex sequences, equipped with the scalar product $\braket{\cdot,\cdot}$:
		\begin{equation}
			\ell^2:=\left\{x=(x_n)_{n\in\mathbb{N}}:\sum_{n\in\mathbb{N}}|x_n|^2<\infty\right\},\qquad\braket{x,y}:=\sum_{n\in\mathbb{N}}\overline{x_n}y_n.
		\end{equation}
		This is an infinite-dimensional separable Hilbert space. An orthonormal basis $(e_n)_{n\in\mathbb{N}}$ can be obtained by setting $e_n=(\delta_{nm})_{m\in\mathbb{N}}$, i.e. $e_0=(1,0,0,\ldots)$, $e_1=(0,1,0,\ldots)$, and so on.
		In fact, all infinite-dimensional separable Hilbert spaces are unitarily equivalent to $\ell^2$, via $v\mapsto(\braket{e_n,v})_{n\in\mathbb{N}}$ for any orthonormal basis $(e_n)_{n\in\mathbb{N}}$.
	\end{example}
	
	\begin{example}[$L^2$ spaces]\label{examp:l2}\index{L2@$L^2$ space}
		Let $(\mathcal{M},\Sigma,\mu)$ be a $\sigma$-finite measure space (i.e. $\mathcal{M}=\cup_{j\in\mathbb{N}}\mathcal{M}_j$ with $\mu(\mathcal{M}_j)<\infty$), and consider the space of functions\footnote{More precisely, the elements of $L^2(\mathcal{M},\Sigma,\mu)$ are equivalence classes of functions, where one defines two functions $f,g$ to be equivalent if $f(x)=g(x)$ $\mu$-almost everywhere (i.e. up to the values on a set $E$ with $\mu(E)=0$); see Appendix~\ref{app:measure}.}
		\begin{equation}
			L^2(\mathcal{M},\Sigma,\mu):=\left\{f:\mathcal{M}\rightarrow\mathbb{C}:\int_\mathcal{M}|f(x)|^2\;\mathrm{d}\mu(x)<\infty\right\},
		\end{equation}
		the integral above being a Lebesgue integral, equipped with the scalar product
		\begin{equation}
			\braket{f,g}:=\int_\mathcal{M}\overline{f(x)}g(x)\;\mathrm{d}\mu(x).
		\end{equation}
		This is a Hilbert space; for all measure spaces of interest in these notes (e.g. $\mathcal{M}=\Omega\subseteq\mathbb{R}^d$ with the Lebesgue measure, see below) it is separable, and it is infinite-dimensional except in degenerate cases (e.g. $\mathcal{M}$ a finite set).
		
		When $\mathcal{M}=\Omega\subseteq\mathbb{R}^d$, $\Sigma$ is the Borel $\sigma$-algebra, and $\mu$ is the $d$-dimensional Lebesgue measure on it, we simply write $L^2(\Omega)$ and $\mathrm{d}\mu(x)\equiv\mathrm{d}x$.
		
	\end{example}
	
	\begin{example}[Vector-valued $L^2$ spaces]\label{examp:l2vec}
		The same definitions apply to functions with values in a Hilbert space: given a Hilbert space $\hfrak$ with scalar product $\braket{\cdot,\cdot}_\hfrak$, one can define the space
		\begin{equation}
			L^2(\mathcal{M},\Sigma,\mu;\hfrak):=\left\{f:\mathcal{M}\rightarrow\hfrak:\int_\mathcal{M}\|f(x)\|_\hfrak^2\;\mathrm{d}\mu(x)<\infty\right\},
		\end{equation}
		equipped with the scalar product
		\begin{equation}
			\braket{f,g}:=\int_\mathcal{M}\braket{f(x),g(x)}_\hfrak\;\mathrm{d}\mu(x).
		\end{equation}
	\end{example}
	
	From now on, unless otherwise stated, $\hilb$ will denote a complex separable Hilbert space, typically infinite-dimensional.
	
	\begin{definition}[Linear operators]\label{def:linop}\index{operator!linear}\index{operator!bounded}\index{operator!densely defined}
		A \emph{linear operator} on $\hilb$ is a function $A:\Dom A\subseteq\hilb\rightarrow\hilb$, where $\Dom A$ is a vector subspace of $\hilb$, such that $A(u+v)=Au+Av$ and $A(\alpha u)=\alpha Au$ for all $u,v\in\Dom A,\alpha\in\mathbb{C}$. 		
		$A$ is said to be \emph{bounded} if $\|A\|:=\sup_{0\neq u\in\Dom A}\|Au\|/\|u\|<\infty$, and \emph{unbounded} otherwise. 		
		$A$ is said to be \emph{densely defined} if $\Dom A$ is dense in $\hilb$.
	\end{definition}
	
	A linear operator is bounded if and only if it is continuous: if $A$ is bounded, then $\|Au-Av\|\leq\|A\|\|u-v\|$ for all $u,v\in\Dom A$; conversely, if $A$ is continuous at $0$, there is $\delta>0$ such that $\|Au\|\leq1$ whenever $\|u\|\leq\delta$, whence $\|A\|\leq1/\delta$ by linearity. Unbounded operators are thus discontinuous: $u_n\to u$ does not imply $Au_n\to Au$, which is the root of most technical subtleties discussed in these notes. By virtue of the continuous linear extension theorem\index{continuous linear extension theorem}, all densely defined bounded operators can be uniquely extended to an operator with $\Dom A=\hilb$. Accordingly, we will always regard bounded operators as defined on the whole space; they form a $C^*$-algebra, denoted by $\mathcal{B}(\hilb)$.
	
	More generally, given two Hilbert spaces $\hilb_1,\hilb_2$, a map $U:\hilb_1\rightarrow\hilb_2$ is \emph{unitary}\index{unitary operator} if it is a bijective isometry, i.e. $\Ran U=\hilb_2$ and $\|Uv\|_{\hilb_2}=\|v\|_{\hilb_1}$ for all $v\in\hilb_1$; by the polarization identity, unitary operators also preserve scalar products. They are characterized by their action on orthonormal bases:
	
	\begin{lemma}\label{lem:unitary}
		Let $\hilb_1,\hilb_2$ be two separable Hilbert spaces.
		\begin{itemize}
			\item[(i)] If $U:\hilb_1\rightarrow\hilb_2$ is unitary, then $(Ue_n)_{n\in\mathbb{N}}$ is an orthonormal basis of $\hilb_2$ for every orthonormal basis $(e_n)_{n\in\mathbb{N}}$ of $\hilb_1$.
			\item[(ii)] Conversely, let $(e_n)_{n\in\mathbb{N}}$ be an orthonormal basis of $\hilb_1$, and $U_0:\operatorname{Span}(e_n)_{n\in\mathbb{N}}\rightarrow\hilb_2$ a linear map such that $(U_0e_n)_{n\in\mathbb{N}}$ is an orthonormal basis of $\hilb_2$. Then $U_0$ extends uniquely to a bounded operator $U:\hilb_1\rightarrow\hilb_2$, which is unitary and given by
			\begin{equation}
				U\sum_{n\in\mathbb{N}}c_ne_n=\sum_{n\in\mathbb{N}}c_nU_0e_n\qquad\text{for all }(c_n)_{n\in\mathbb{N}}\in\ell^2.
			\end{equation}
		\end{itemize}
		In particular, a unitary operator is uniquely determined by its action on a single orthonormal basis.
	\end{lemma}

	The domain $\Dom A$ is an integral part of the definition of an operator: two operators acting in the same way on different domains generally describe different observables. This also affects elementary operations between operators: given $A:\Dom A\subseteq\hilb\rightarrow\hilb$ and $B:\Dom B\subseteq\hilb\rightarrow\hilb$, their sum $A+B$ is defined on $\Dom(A+B):=\Dom A\cap\Dom B$, while their composition $AB$ (resp. $BA$) is defined on $\Dom AB:=\left\{u\in\Dom B:Bu\in\Dom A\right\}$ (resp. $\Dom BA:=\left\{u\in\Dom A:Au\in\Dom B\right\}$). In particular, $A+B$, $AB$, $BA$ might not be densely defined even though $A$ and $B$ are.
	
	\begin{definition}[Extension/restriction of operators]\label{def:extension}\index{operator!extension}\index{operator!restriction}
		Let $A:\Dom A\subseteq\hilb\rightarrow\hilb$, $B:\Dom B\subseteq\hilb\rightarrow\hilb$. Then $A\subseteq B$ (equivalently, $B\supseteq A$) if and only if $\Dom A\subseteq\Dom B$ and $Au=Bu$ for all $u\in\Dom A$. In this case, $B$ is an \emph{extension} of $A$, and $A$ is a \emph{restriction} of $B$.
	\end{definition}
	
	\begin{definition}[Adjoint operator]\label{def:adjoint}\index{adjoint}
		Let $A:\Dom A\subseteq\hilb\rightarrow\hilb$ be densely defined. The \emph{adjoint} of $A$ is the operator $A^*:\Dom A^*\subseteq\hilb\rightarrow\hilb$ uniquely defined by
		\begin{align}
			\Dom A^*&:=\left\{v\in\hilb:\exists\tilde{v}\in\hilb\text{ s.t. }\braket{v,Au}=\braket{\tilde{v},u}\;\text{ for all }u\in\Dom A\right\};\\
			A^*v&:=\tilde{v}.
		\end{align}
	\end{definition}
	This definition is well-posed, since the density of $\Dom A$ ensures that $\tilde{v}$ is unique. Note that, if $A\in\mathcal{B}(\hilb)$, then $A^*\in\mathcal{B}(\hilb)$ as well.
	
	\begin{definition}[Symmetric and self-adjoint operators]\index{operator!symmetric}\index{operator!self-adjoint}
		A densely defined operator $A$ on $\hilb$ is said to be
		\begin{itemize}
			\item \emph{symmetric}, if $A\subseteq A^*$, i.e. $\braket{v,Au}=\braket{Av,u}$ for all $u,v\in\Dom A$;
			\item \emph{self-adjoint}, if $A=A^*$.
		\end{itemize}
	\end{definition}
	
	\begin{example}[Multiplication operators]\label{examp:multiplication}\index{multiplication operator}
		Let $\Omega\subseteq\mathbb{R}^d$, $v:\Omega\rightarrow\mathbb{C}$ a measurable function (Appendix~\ref{app:measure}; for our purposes it suffices to note that all continuous functions are measurable). Then the operator $M_v:\Dom M_v\subseteq L^2(\Omega)\rightarrow L^2(\Omega)$ defined by
		\begin{align}
			\Dom M_v&:=\left\{\varphi\in L^2(\Omega):\int_\Omega|v(x)\varphi(x)|^2\;\mathrm{d}x<\infty\right\};\label{eq:multdomain}\\
			(M_v\varphi)(x)&:=v(x)\varphi(x)
		\end{align}
		satisfies $M_v^*=M_{\overline{v}}$; in particular, $M_v$ is self-adjoint if and only if $v(x)$ is real-valued.
	\end{example}
	
	One can prove that $M_v$ is bounded if and only if $v$ is an (essentially) bounded function on $\Omega$. For instance, the \emph{position operator}\index{position operator} $q$ (multiplication by $x$) is bounded if and only if $\Omega$ is a bounded region.
	
	All this can be generalized to multiplication operators on a general $\sigma$-finite measure space $(\mathcal{M},\Sigma,\mu)$. 
	
	\begin{example}[Momentum operator on the real line]\label{examp:momentum}\index{momentum operator}
		Let $\mathrm{H}^1(\mathbb{R})$ be the \emph{first Sobolev space} on $\mathbb{R}$, i.e. the space of all functions $f\in L^2(\mathbb{R})$ whose (weak) derivative exists and satisfies $f'\in L^2(\mathbb{R})$. Then the \emph{momentum operator} $p:\Dom p\subseteq L^2(\mathbb{R})\rightarrow L^2(\mathbb{R})$ defined by
		\begin{align}
			\Dom p&:=\mathrm{H}^1(\mathbb{R});\\
			(p\varphi)(x)&:=-\ii\varphi'(x)
		\end{align}
		is self-adjoint: via the Fourier transform, it is unitarily equivalent to the position operator on $L^2(\mathbb{R})$ \cite[Sections 7.2 and 8.1]{teschl}.
	\end{example}
	
	\begin{example}[Momentum operator on the interval]\label{examp:momentuminterval}
		Let $\mathrm{H}^1(0,1)$ be the first Sobolev space on $(0,1)$. Then the momentum operator $p:\Dom p\subseteq L^2(0,1)\rightarrow L^2(0,1)$ defined by
		\begin{align}
			\Dom p&:=\mathrm{H}^1(0,1);\\
			(p\varphi)(x)&:=-\ii\varphi'(x)
		\end{align}
		is not self-adjoint---in fact, not even symmetric: given $\varphi\in\Dom p$, an integration by parts shows
		\begin{equation}
			\braket{\varphi,p\varphi}-\braket{p\varphi,\varphi}=-\ii\left(|\varphi(1)|^2-|\varphi(0)|^2\right),
		\end{equation}
		which is generally nonzero. To obtain a self-adjoint operator, one must restrict $p$ by imposing suitable boundary conditions in its domain.
		
		It can be shown that $p$ admits a one-parameter family of self-adjoint restrictions $p_\alpha$, $\alpha\in[0,2\pi)$, defined by
		\begin{align}
			\Dom p_\alpha&:=\left\{\varphi\in\mathrm{H}^1(0,1):\varphi(1)=\e^{\ii\alpha}\varphi(0)\right\};\\
			(p_\alpha\varphi)(x)&:=-\ii\varphi'(x).
		\end{align}
		All these operators have the same formal expression but different domains, corresponding to different (quasi-periodic) boundary conditions, and thus describe different physical observables.
	\end{example}
	
	\begin{example}[Laplace operator on the real line]\label{examp:laplace}\index{Laplace operator}
		Let $\mathrm{H}^2(\mathbb{R})$ be the \emph{second Sobolev space} on $\mathbb{R}$, i.e. the space of all functions $f\in L^2(\mathbb{R})$ whose (weak) first and second derivatives exist and satisfy $f',f''\in L^2(\mathbb{R})$. Then the Laplace operator $-\Delta:\Dom(-\Delta)\subseteq L^2(\mathbb{R})\rightarrow L^2(\mathbb{R})$ defined by
		\begin{align}
			\Dom (-\Delta)&:=\mathrm{H}^2(\mathbb{R});\\
			(-\Delta\varphi)(x)&:=-\varphi''(x)
		\end{align}
		is self-adjoint and nonnegative \cite[Theorem 7.17]{teschl}. Besides, $-\Delta=p^2$. 	This operator represents, up to multiplicative constants, the kinetic energy of a massive quantum particle on the real line.		
	\end{example}

	\begin{example}[Laplace operator on the segment]\label{examp:laplacesegment}
		Let $\mathrm{H}^2(0,1)$ be the second Sobolev space on $(0,1)$. Then, just like the momentum operator, the Laplace operator $-\Delta:\Dom(-\Delta)\subseteq L^2(0,1)\rightarrow L^2(0,1)$ defined by
		\begin{align}
			\Dom(-\Delta)&:=\mathrm{H}^2(0,1);\\
			(-\Delta\varphi)(x)&:=-\varphi''(x)
		\end{align}
		is not even symmetric, since, given a generic $\varphi\in\mathrm{H}^2(0,1)$,
		\begin{equation}\label{eq:boundarylaplacian}
			\braket{\varphi,(-\Delta)\varphi}-\braket{(-\Delta)\varphi,\varphi}=-2\ii\left[\operatorname{Im}\left(\overline{\varphi(1)}\varphi'(1)\right)-\operatorname{Im}\left(\overline{\varphi(0)}\varphi'(0)\right)\right],
		\end{equation}
		which is generally nonzero.
		Again, one must restrict the domain to wavefunctions satisfying a boundary condition making the boundary term above vanish.		
		It can be shown~\cite{bonneau2001,asorey2005} that all such boundary conditions are parametrized by $2\times2$ unitary matrices $U$, as follows:
		\begin{align}
			\Dom(-\Delta_U)&:=\left\{\varphi\in\mathrm{H}^2(0,1):\ii(I+U)\begin{bmatrix}
				\varphi(0)\\\varphi(1)
			\end{bmatrix}=(I-U)\begin{bmatrix}
				\varphi'(0)\\-\varphi'(1)
			\end{bmatrix}\right\};\\
			(-\Delta_U\varphi)(x)&:=-\varphi''(x).
		\end{align}
		These include Dirichlet ($U=I$) and Neumann ($U=-I$) boundary conditions, and more generally all \emph{separated} boundary conditions, each involving only one endpoint (diagonal $U$), as well as (quasi-)periodic ones, which couple the two endpoints; see also \cite[Section 9.2]{teschl}.
	\end{example}
	
	\begin{remark}[Momentum, Laplacian, and boundary conditions]\label{rem:momentumlaplacian}\index{boundary conditions}
		Examples~\ref{examp:momentuminterval} and~\ref{examp:laplacesegment} reveal a striking asymmetry: the Laplacian admits a four-parameter family of self-adjoint realizations, including the \emph{local} Dirichlet and Neumann ones, which impose a condition at each endpoint separately, whereas every self-adjoint momentum $p_\alpha$ ties $\varphi(1)$ to $\varphi(0)$. The boundary terms explain why. For the Laplacian, the boundary term~\eqref{eq:boundarylaplacian} equals $-2\ii\left(j(1)-j(0)\right)$, with $j=\operatorname{Im}(\overline{\varphi}\varphi')$ the probability current, and thus vanishes whenever no probability flows through either wall, as local conditions ensure. For the momentum, the boundary term only involves $\varphi(0)$ and $\varphi(1)$, and vanishes if and only if $|\varphi(0)|=|\varphi(1)|$. Neumann conditions are thus invisible to it, and the only local choice is $\varphi(0)=\varphi(1)=0$, which yields a symmetric but not self-adjoint operator $p_{\rm min}$, its adjoint being $p$ on the whole $\mathrm{H}^1(0,1)$.
		
		Physically, a classical particle bouncing elastically between two walls has a conserved energy, but not a conserved momentum: at each bounce, $p\mapsto-p$. Local boundary conditions are the quantum counterpart of such walls, while a self-adjoint momentum generates a unitary group of translations (Theorem~\ref{thm:stone} below), which can only be unitary on a segment if whatever leaves through one end re-enters through the other: $p_\alpha$ describes a particle on a ring threaded by a magnetic flux, not a particle in a box. Mathematically, this also shows that the formal identity $-\Delta=p^2$ only holds in the operator sense for specific boundary conditions: $p_\alpha^2$ is the Laplacian with the nonlocal, quasi-periodic conditions $\varphi(1)=\e^{\ii\alpha}\varphi(0)$, $\varphi'(1)=\e^{\ii\alpha}\varphi'(0)$, and neither the Dirichlet nor the Neumann Laplacian is the square of a self-adjoint momentum---they are rather $p_{\rm min}^*p_{\rm min}$ and $p_{\rm min}p_{\rm min}^*$, respectively. For more on these operators, and on some surprising properties of their dynamics, see~\cite{bonneau2001,asorey2005}.
	\end{remark}

	\subsection{Closed operators and (essential) self-adjointness}\label{subsec:esa}
	
	Recall that, given two Hilbert spaces $\hilb_j$ ($j=1,2$) with scalar products $\braket{\cdot,\cdot}_{\hilb_j}$, $\hilb_1\oplus\hilb_2$ is the Hilbert space obtained by equipping the Cartesian product $\hilb_1\times\hilb_2$ with the scalar product
	\begin{equation}\label{eq:directsumsp}
		\Braket{(u_1,u_2),(v_1,v_2)}_{\hilb_1\oplus\hilb_2}:=\braket{u_1,v_1}_{\hilb_1}+\braket{u_2,v_2}_{\hilb_2}.
	\end{equation}
	We will come back to this in Section~\ref{sec:directsum}.
	
	\begin{definition}[Closable operator, closure, closed operator]\label{def:closure}\index{operator!closed}\index{operator!closable}\index{closure}
		Let $A:\Dom A\subseteq\hilb\rightarrow\hilb$. $A$ is said to be
		\begin{itemize}
			\item \emph{closed}, if its \emph{graph}\index{graph of an operator} $\operatorname{Graph}A:=\left\{(u,Au):u\in\Dom A\right\}\subseteq\hilb\oplus\hilb$ is a closed subspace of $\hilb\oplus\hilb$, that is: given $(u_n)_{n\in\mathbb{N}}\subset\Dom A$, if $u_n\to u\in\hilb$ and $Au_n\to v\in\hilb$, then $u\in\Dom A$ and $v=Au$;
			\item \emph{closable}, if there exists an operator $\overline{A}:\Dom\overline{A}\subseteq\hilb\rightarrow\hilb$ (the \emph{closure} of $A$) such that $\operatorname{Graph}\overline{A}=\overline{\operatorname{Graph}A}$, that is: given $(u_n)_{n\in\mathbb{N}}\subset\Dom A$, if $u_n\to u\in\hilb$ and $Au_n\to v\in\hilb$, then $u\in\Dom \overline{A}$ and $v=\overline{A}u$.
		\end{itemize}
	\end{definition}
	
	If $A$ is closable, a subspace $\mathcal{D}\subseteq\Dom\overline{A}$ is a \emph{core}\index{core} for $\overline{A}$ if the closure of the restriction $\overline{A}\restriction\mathcal{D}$ is $\overline{A}$ itself; equivalently, if $\mathcal{D}$ is dense in $\Dom\overline{A}$ with respect to the \emph{graph norm}\index{graph norm} $\|u\|_{\overline{A}}:=(\|u\|^2+\|\overline{A}u\|^2)^{1/2}$. In particular, $\Dom A$ is a core for $\overline{A}$.
	
	The following important characterization holds:
	
	\begin{proposition}\label{prop:closableiffadjointdense}
		Let $A:\Dom A\subseteq\hilb\rightarrow\hilb$ be densely defined. Then $A$ is closable if and only if its adjoint $A^*$ is densely defined, in which case $\overline{A}=A^{**}$.
	\end{proposition}
	
	In particular, all densely defined symmetric operators are closable, and all self-adjoint operators are closed.
	
	\begin{definition}[Essentially self-adjoint operator]\index{operator!essentially self-adjoint}\index{core}
		A symmetric operator $A:\Dom A\subseteq\hilb\rightarrow\hilb$ is \emph{essentially self-adjoint} if its closure $\overline{A}$ is self-adjoint; in this case, $\Dom A$ is a core for the self-adjoint operator $\overline{A}$.
	\end{definition}
	
	The name is motivated by the following fact: if (and only if) $\overline{A}$ is self-adjoint, then $\overline{A}$ is the only self-adjoint extension of $A$, so that $A$ unambiguously defines an observable.
	
	\begin{example}[Multiplication operators]
		We have seen (Example~\ref{examp:multiplication}) that, given a real-valued measurable function $v:\Omega\subseteq\mathbb{R}^d\rightarrow\mathbb{R}$, the corresponding multiplication operator $M_v$ is self-adjoint on its maximal domain~\eqref{eq:multdomain}. If $v$ is in addition locally bounded (e.g. continuous), it can be shown that the restriction $M_v^0$ of this operator to the ($v$-independent) domain
		\begin{equation}
			\Dom M_v^0:=\left\{\varphi\in L^2(\Omega):\varphi\text{ has compact support}\right\}
		\end{equation}
		satisfies $\overline{M_v^0}=M_v$. Therefore, $M_v^0$ is essentially self-adjoint and its domain is a core for all multiplication operators with locally bounded functions. (Local boundedness is needed: for $v(x)=|x|^{-1}$ on $\Omega=\mathbb{R}$, the compactly supported function $\chi_{[-1,1]}$ is not even in $\Dom M_v$.)
	\end{example}
	
	\begin{example}[Momentum and Laplacian]\label{examp:momentumlaplacian}
		As stated in Examples~\ref{examp:momentum} and~\ref{examp:laplace}, the momentum operator $p$ and the Laplace operator $-\Delta$ on $L^2(\mathbb{R})$, defined on the Sobolev spaces\index{Sobolev space} $\mathrm{H}^1(\mathbb{R})$ and $\mathrm{H}^2(\mathbb{R})$ respectively, are self-adjoint. However, the (smaller) spaces of compactly supported, $k$-times continuously differentiable functions:
		\begin{equation}
			\mathrm{C}_{\rm c}^1(\mathbb{R})\text{ (for $p$ only)},\quad\mathrm{C}_{\rm c}^2(\mathbb{R}),\quad\ldots,\quad\mathrm{C}_{\rm c}^\infty(\mathbb{R}),
		\end{equation}
		are all cores for these operators (for $-\Delta$, see \cite[Lemma 7.19]{teschl}; the same cutoff argument applies to $p$).
	\end{example}
	
	Another very common---and useful---choice of domain is the following one.
	
	\begin{definition}[Schwartz space]\label{def:schwartz}\index{Schwartz space}
		The \emph{Schwartz space} is the space of all smooth, rapidly decreasing functions on $\mathbb{R}$:
		\begin{equation}
			\mathcal{S}(\mathbb{R}):=\left\{f\in\mathrm{C}^\infty(\mathbb{R}):\sup_{x\in\mathbb{R}}|x^\alpha f^{(\beta)}(x)|<\infty\;\forall\alpha,\beta\in\mathbb{N}\right\}.
		\end{equation}
	\end{definition}
	
	That is, $f$ and all its derivatives $f^{(\beta)}$ remain bounded (equivalently, square-integrable) even when multiplied by any power $x^\alpha$, i.e. they decay as $|x|\to\infty$ faster than any inverse power of $|x|$ (e.g. exponentially). Clearly $\mathrm{C}^\infty_{\rm c}(\mathbb{R})\subseteq\mathcal{S}(\mathbb{R})$, since functions in $\mathrm{C}^\infty_{\rm c}(\mathbb{R})$ are identically zero near infinity; thus $\mathcal{S}(\mathbb{R})$ is a dense subspace of $L^2(\mathbb{R})$. By construction, $\mathcal{S}(\mathbb{R})$ is invariant under differentiation and under multiplication by polynomials, and thus under the action of arbitrary polynomials in the position and momentum operators $q$ and $p$. One can prove that $\mathcal{S}(\mathbb{R})$ is again a core for both $p$ and $-\Delta$ (see \cite[Problem 7.10 and Section 8.1]{teschl}).
	
	Since computing the adjoint explicitly is generally impractical, the following criteria are often more practical:
	
	\begin{theorem}[Criterion of self-adjointness]\label{thm:criterion-sa}\index{criterion of self-adjointness}
		Let $A:\Dom A\subseteq\hilb\rightarrow\hilb$ be symmetric. The following are equivalent:
		\begin{itemize}
			\item[(i)] $A$ is self-adjoint;
			\item[(ii)] $\Ker(A^*\mp\ii)=\{0\}$ and $A$ is closed;
			\item[(iii)] $\Ran(A\pm\ii)=\hilb$.
		\end{itemize}
	\end{theorem}
	
	\begin{theorem}[Criterion of essential self-adjointness]\label{thm:criterion-esssa}\index{criterion of essential self-adjointness}
		Let $A:\Dom A\subseteq\hilb\rightarrow\hilb$ be symmetric. The following are equivalent:
		\begin{itemize}
			\item[(i)] $A$ is essentially self-adjoint;
			\item[(ii)] $\Ker(A^*\mp\ii)=\{0\}$;
			\item[(iii)] $\overline{\Ran(A\pm\ii)}=\hilb$.
		\end{itemize}
	\end{theorem}
	
	\begin{remark}
		In both statements, $\pm\ii$ can be replaced with any pair $z_0,\overline{z_0}\in\mathbb{C}\setminus\mathbb{R}$. Furthermore, if $A\geq E_0$ (i.e. $\braket{\psi,A\psi}\geq E_0\|\psi\|^2$ for all $\psi\in\Dom A$), then for every $E<E_0$ the following are equivalent:
		\begin{itemize}
			\item[(i)] $A$ is self-adjoint;
			\item[(ii)] $\Ker(A^*-E)=\{0\}$ and $A$ is closed;
			\item[(iii)] $\Ran(A-E)=\hilb$,
		\end{itemize}
		and similarly for essential self-adjointness.
	\end{remark}
	
	Concretely, $\Ran(A\pm\ii)=\hilb$ (resp. $\overline{\Ran(A\pm\ii)}=\hilb$) means the following: for every $\psi\in\hilb$ (resp. for every $\psi$ in a given dense subspace of $\hilb$), the equations $(A\pm\ii)\varphi_\pm=\psi$
	admit a solution $\varphi_\pm\in\Dom A$.
	
	In finite dimensions, an operator is Hermitian if and only if it admits an orthonormal basis of eigenvectors with real eigenvalues. This is no longer true in infinite dimensions, where a self-adjoint operator may have no eigenvectors at all; nevertheless, the following holds:
	
	\begin{proposition}\label{prop:eigenvectors}
		Let $A:\Dom A\subseteq\hilb\rightarrow\hilb$ be symmetric. The following properties hold:
		\begin{itemize}
			\item Let $0\neq u\in\Dom A$ be an eigenvector of $A$, i.e. $Au=\lambda u$ for some $\lambda\in\mathbb{C}$. Then $\lambda\in\mathbb{R}$.
			\item Assume $u,u'\in\Dom A$ are eigenvectors of $A$ corresponding to distinct eigenvalues $\lambda\neq\lambda'$. Then $\braket{u,u'}=0$.
			\item Assume there exists an orthonormal basis $(u_n)_{n\in\mathbb{N}}\subset\Dom A$ of eigenvectors of $A$. Then $A$ is essentially self-adjoint, and its restriction to $\operatorname{Span}(u_n)_{n\in\mathbb{N}}$ is also essentially self-adjoint.
		\end{itemize}
	\end{proposition}
	
	\begin{example}[The quantum harmonic oscillator]\label{examp:qho}\index{harmonic oscillator}
		Let $\hilb=L^2(\mathbb{R})$, $\omega>0$, and $H_0:\Dom H_0\subseteq\hilb\rightarrow\hilb$ be defined by $\Dom H_0:=\mathcal{S}(\mathbb{R})$ (Definition~\ref{def:schwartz}) and
		\begin{equation}\label{eq:qho}
			H_0=\frac{1}{2}\left(-\frac{\mathrm{d}^2}{\mathrm{d}x^2}+\omega^2x^2\right)=\frac{1}{2}\left(p^2+\omega^2q^2\right).
		\end{equation}
		This is the well-known \emph{quantum harmonic oscillator} with frequency $\omega$ (in units in which the mass is $1$). It is well-defined on $\mathcal{S}(\mathbb{R})$, since $\mathcal{S}(\mathbb{R})$ is invariant under differentiation and under multiplication by polynomials, and an integration by parts, as for the free Laplacian, shows that it is symmetric.		
		Besides, $H_0$ admits an orthonormal basis of eigenvectors. These are the \emph{Hermite functions}\index{Hermite functions}
		\begin{equation}\label{eq:hermite}
			\phi_n(x):=\left(\frac{\omega}{\pi}\right)^{1/4}\frac{1}{\sqrt{2^nn!}}\e^{-\omega x^2/2}H_n(\sqrt{\omega}x),\quad\text{where }H_n(y):=(-1)^n\e^{y^2}\frac{\mathrm{d}^n}{\mathrm{d}y^n}\e^{-y^2},
		\end{equation}
		which are in $\mathcal{S}(\mathbb{R})$ and satisfy the following properties:
		\begin{itemize}
			\item[(i)] For all $n\in\mathbb{N}$, $H_0\phi_n=\omega\left(n+\frac{1}{2}\right)\phi_n$;
			\item[(ii)] They are orthonormal: $\braket{\phi_n,\phi_m}=\delta_{nm}$;
			\item[(iii)] They are a basis of $L^2(\mathbb{R})$.
		\end{itemize}
		Hence $H_0$ is essentially self-adjoint by Proposition~\ref{prop:eigenvectors}, and its restriction to $\operatorname{Span}(\phi_n)_{n\in\mathbb{N}}$ is also essentially self-adjoint; besides, its eigenvalues are all simple.
		
		Property (i) can be checked by a direct computation in the variable $y=\sqrt{\omega}x$, using the identities $H_n'(y)=2yH_n(y)-H_{n+1}(y)$ (which follows by differentiating the definition of $H_n$) and $H_n'(y)=2nH_{n-1}(y)$; together, they yield Hermite's equation $H_n''-2yH_n'+2nH_n=0$, which is equivalent to (i). See also \cite[Section 8.3]{teschl}, where the Hermite functions are obtained via ladder operators, as we will do in Section~\ref{sec:singlemode}.
		The orthogonality of the functions $\phi_n$ follows directly from the fact that eigenvectors of a symmetric operator corresponding to distinct eigenvalues are orthogonal (Proposition~\ref{prop:eigenvectors}), so that proving (ii) only requires showing $\|\phi_n\|=1$. Property (iii) requires more sophisticated arguments; see e.g. \cite[Lemma 8.3 and Section 8.3]{teschl}.
		
		Note that, even in this simple case, the domain of the self-adjoint operator $\overline{H_0}$ is not easy to characterize: we will see in Example~\ref{examp:qho-counter} that, unlike for the free Laplacian, $\Dom\overline{H_0}\neq\mathrm{H}^2(\mathbb{R})$.
		
	\end{example}
	
	\subsection{Spectrum, spectral theorem and Stone's theorem}\label{subsec:spectrum}
	
	We shall only consider closed operators here. This is not a significant restriction, since we are ultimately interested in symmetric and self-adjoint operators, which are respectively closable and closed.
	
	\begin{definition}[Resolvent set, resolvent operator, spectrum]\label{def:resolvent}\index{resolvent}\index{spectrum}
		Let $A:\Dom A\subseteq\hilb\rightarrow\hilb$ be a densely defined, closed operator. The \emph{resolvent set} $\rho(A)$ and the \emph{spectrum} $\sigma(A)$ of $A$ are the subsets of $\mathbb{C}$ defined as follows:
		\begin{align}
			\rho(A)&:=\left\{z\in\mathbb{C}:A-z\text{ is invertible, and }(A-z)^{-1}\in\mathcal{B}(\hilb)\right\};\\
			\sigma(A)&:=\mathbb{C}\setminus\rho(A).
		\end{align}
		Furthermore, for every $z\in\rho(A)$, the \emph{resolvent operator} of $A$ at $z$ is the operator $R_A(z)\in\mathcal{B}(\hilb)$ defined by $R_A(z):=(A-z)^{-1}$.
	\end{definition}
	Note that, for a non-closed operator, the resolvent set is empty and thus the spectrum is the whole $\mathbb{C}$.
	
	A key tool in this context is the closed graph theorem:
	
	\begin{theorem}[Closed graph theorem]\label{thm:closedgraph}\index{closed graph theorem}
		Let $A$ be an operator on $\hilb$ with domain $\Dom A=\hilb$. Then the following are equivalent:
		\begin{itemize}
			\item[(i)] $A$ is bounded;
			\item[(ii)] $A$ is closed;
			\item[(iii)] $A$ is closable.
		\end{itemize}	
	\end{theorem}
	Therefore, closable operators defined on the whole Hilbert space are necessarily bounded; equivalently, unbounded closed operators (in particular, unbounded self-adjoint operators) are never defined on the whole of $\hilb$.	
	
	By virtue of the closed graph theorem, $z\in\rho(A)$ if and only if the operator $A-z$ is bijective, i.e. if the equation $(A-z)\varphi_z=\psi$
	admits a unique solution $\varphi_z\in\Dom A$ for every $\psi\in\hilb$.	
	Hence, $z\in\sigma(A)$ if and only if either $A-z$ is not injective, i.e. $\Ker(A-z)\neq\{0\}$, which means that $z$ is an eigenvalue of $A$; or $A-z$ is injective but not surjective, i.e. $\Ran(A-z)\neq\hilb$.
	As a direct consequence of the criterion of self-adjointness (Theorem~\ref{thm:criterion-sa}), a closed symmetric operator is self-adjoint if and only if $\sigma(A)\subseteq\mathbb{R}$. A closed symmetric non-self-adjoint operator has spectrum either equal to $\mathbb{C},\mathbb{C}^+\cup\mathbb{R}$, or $\mathbb{C}^-\cup\mathbb{R}$ \cite[Theorem X.1]{reedsimon2}. Conversely, non-symmetric operators may well have a real spectrum, as in $\mathcal{PT}$-symmetric quantum mechanics~\cite{benderboettcher1998}.
	
	The resolvent formalism allows us to generalize the spectral theorem from linear algebra to unbounded self-adjoint operators, and thus to introduce a functional calculus that makes sense of expressions like $\e^{-\ii tA}$ without resorting to (possibly problematic) series expansions.
	
	Recall the spectral theorem in finite dimensions: all self-adjoint matrices $A\in\mathbb{C}^{p\times p}$ can be diagonalized, i.e. written as $A=\sum_{j=1}^{p'}\lambda_jP_j$,
	where $p'\leq p$, $\lambda_1,\dots,\lambda_{p'}$ are the eigenvalues of $A$, and $P_1,\dots,P_{p'}$ are the orthogonal projectors onto the corresponding eigenspaces, satisfying $P_iP_j=\delta_{ij}P_i$ and $P_i^*=P_i$.	
	To generalize this to infinite dimensions, we need the language of \emph{projection-valued measures} (PVMs):
	
	\begin{definition}[Projection-valued measure]\label{def:pvm}\index{projection-valued measure}
		A \emph{projection-valued measure} is a function $P:\mathfrak{B}(\mathbb{R})\rightarrow\mathcal{B}(\hilb)$ satisfying the following properties:
		\begin{itemize}
			\item[(i)] For all $B\in\mathfrak{B}(\mathbb{R})$, $P(B)$ is a self-adjoint projector: $P(B)^2=P(B)=P(B)^*$.
			\item[(ii)] $P(\mathbb{R})=I$.
			\item[(iii)] \emph{Strong $\sigma$-additivity}: given any countable family $(B_n)_{n\in\mathbb{N}}\subset\mathfrak{B}(\mathbb{R})$ of disjoint sets,
			\begin{equation}
				P\left(\bigcup_{n\in\mathbb{N}}B_n\right)\psi=\sum_{n\in\mathbb{N}}P(B_n)\psi
			\end{equation}
			for all $\psi\in\hilb$.
		\end{itemize}
	\end{definition}
	
	Here $\mathfrak{B}(\mathbb{R})$ is the Borel $\sigma$-algebra, i.e. the smallest collection of subsets of $\mathbb{R}$ that contains all intervals and is closed under complements and countable unions (Appendix~\ref{app:measure}).
	
	\begin{theorem}[Spectral theorem, PVM form]\label{thm:spectraltheorem1}\index{spectral theorem}
		For every self-adjoint operator $A:\Dom A\subseteq\hilb\rightarrow\hilb$ there is a unique PVM $P_A:\mathfrak{B}(\mathbb{R})\rightarrow\mathcal{B}(\hilb)$ such that
		\begin{equation}\label{eq:spectraldomain}
			A=\int_{\sigma(A)}\lambda\,\mathrm{d}P_A(\lambda),\qquad
			\Dom A=\left\{\psi\in\hilb:\int_{\sigma(A)}\lambda^2\,\mathrm{d}\mu_{\psi,A}(\lambda)<\infty\right\},
		\end{equation}
		where, for every $\psi\in\hilb$, $\mu_{\psi,A}$ is the \emph{spectral measure}\index{spectral measure} of $\psi$ with respect to $A$, i.e. the finite Borel measure defined by
		\begin{equation}\label{eq:spectralmeasure}
			\mu_{\psi,A}(B):=\braket{\psi,P_A(B)\psi},\qquad B\in\mathfrak{B}(\mathbb{R}),
		\end{equation}
		with total mass $\mu_{\psi,A}(\mathbb{R})=\|\psi\|^2$.
	\end{theorem}
	The integral with respect to a PVM is defined by a procedure similar to the construction of the Lebesgue integral with respect to a measure.
	
	An equivalent formulation of the spectral theorem states that every self-adjoint operator is unitarily equivalent to a position operator, i.e. to the multiplication by the variable, on some measure space:
	
	\begin{theorem}[Spectral theorem, multiplicative form]\label{thm:spectraltheorem2}\index{spectral theorem!multiplicative form}
		For every self-adjoint operator $A:\Dom A\subseteq\hilb\rightarrow\hilb$ there exist a $\sigma$-finite Borel measure $\mu$ on $\mathcal{M}:=\mathbb{R}\times\mathbb{N}$ (i.e. on countably many copies of $\mathbb{R}$) and a unitary operator $U:\hilb\rightarrow L^2(\mathcal{M},\mu)$ such that $UAU^*$ is the multiplication operator associated with the function $(\lambda,j)\mapsto\lambda$, i.e.
		\begin{align}
			\Dom\left(	UAU^*\right)&=\left\{\psi\in L^2(\mathcal{M},\mu):\int_\mathcal{M}\lambda^2|\psi(\lambda,j)|^2\;\mathrm{d}\mu(\lambda,j)<\infty\right\};\\
			(UAU^*\psi)(\lambda,j)&=\lambda\psi(\lambda,j).
		\end{align}		
	\end{theorem}
	
	\begin{remark}
		The equivalence between these two seemingly distinct formulations is discussed in detail e.g. in \cite[Chapter 10.1]{hall} or \cite[Chapter 8.3]{deoliveira}. In~\cite{hall} the theorem states that $A$ is equivalent to multiplication by some real measurable function $h$ on some space $L^2(\mathcal{M},\Sigma,\mu)$. The version above is obtained by decomposing $\hilb$ into countably many mutually orthogonal \emph{cyclic subspaces} of $A$: on each of them $A$ is equivalent to multiplication by $\lambda$ on some $L^2(\mathbb{R},\mu_j)$. A single copy of $\mathbb{R}$ does not suffice in general because of spectral multiplicity (e.g. $A=I$ on $\mathbb{C}^2$). With a slight abuse of notation, we will often suppress the index $j$ and write $(UAU^*\psi)(\lambda)=\lambda\psi(\lambda)$.
	\end{remark}
	
	The spectral theorem enables us to define \emph{functions of a self-adjoint operator}\index{functional calculus}. Given $f:\sigma(A)\subseteq\mathbb{R}\rightarrow\mathbb{C}$ measurable, we can define $f(A):\Dom f(A)\subseteq\hilb\rightarrow\hilb$ via
	\begin{align}
		\Dom f(A)&:=\left\{\psi\in\hilb:\int_{\sigma(A)}|f(\lambda)|^2\;\mathrm{d}\mu_{\psi,A}(\lambda)<\infty\right\};\\
		f(A)&:=\int_{\sigma(A)}f(\lambda)\,\mathrm{d}P_A(\lambda).
	\end{align}
	Moreover, $\|f(A)\psi\|^2=\int_{\sigma(A)}|f(\lambda)|^2\,\mathrm{d}\mu_{\psi,A}(\lambda)$ for all $\psi\in\Dom f(A)$, which explains the definition of $\Dom f(A)$; besides, $f(A)$ is bounded if and only if $f$ is $P_A$-essentially bounded, and is self-adjoint if and only if $f$ is real-valued $P_A$-almost everywhere.
	
	In particular, each self-adjoint operator generates an evolution group:
	\begin{proposition}\label{prop:unitarygroup}
		Let $A:\Dom A\subseteq\hilb\rightarrow\hilb$ be self-adjoint. The family of bounded operators $(U_A(t))_{t\in\mathbb{R}}$, with
		\begin{equation}
			U_A(t):=	\e^{-\ii tA}=\int_{\sigma(A)}\e^{-\ii\lambda t}\;\mathrm{d}P_A(\lambda),
		\end{equation}
		is a \emph{strongly continuous one-parameter unitary group}\index{unitary group, strongly continuous}, i.e.:
		\begin{itemize}
			\item[(i)] For all $t\in\mathbb{R}$, $U_A(t)$ is unitary;
			\item[(ii)] $U_A(0)=I$ and, for all $t,s\in\mathbb{R}$, $U_A(t+s)=U_A(t)U_A(s)$;
			\item[(iii)] For all $\psi\in\hilb$, the function $t\mapsto U_A(t)\psi\in\hilb$ is continuous.
		\end{itemize}
		Furthermore, $A$ can be uniquely reconstructed as the \emph{generator}\index{unitary group!generator} of $U_A(t)$:
		\begin{equation}\label{eq:generator}
			\Dom A=\left\{\psi\in\hilb:\exists\frac{\mathrm{d}}{\mathrm{d}t}U_A(t)\psi\big|_{t=0}\in\hilb\right\},\quad
			A\psi=\ii\frac{\mathrm{d}}{\mathrm{d}t}U_A(t)\psi\big|_{t=0}.
		\end{equation}
	\end{proposition}
	
	The converse also holds:
	
	\begin{theorem}[Stone's theorem]\label{thm:stone}\index{Stone's theorem}
		Let $(U(t))_{t\in\mathbb{R}}$ be a strongly continuous one-parameter unitary group. Then its generator $A:\Dom A\subseteq\hilb\rightarrow\hilb$, defined as in Eq.~\eqref{eq:generator} with $U_A(t)$ replaced by $U(t)$, is self-adjoint and $U(t)=\e^{-\ii tA}$.
	\end{theorem}
	
	Self-adjoint operators and strongly continuous one-parameter unitary groups are thus in one-to-one correspondence: self-adjointness cannot be relaxed (e.g. to mere symmetry) if the evolution is to be unitary, strongly continuous, and to satisfy the group property.
	
	Finally, we will need a notion of convergence for sequences of unbounded self-adjoint operators. Let $A$ and $A_n$, $n\in\mathbb{N}$, be self-adjoint. We say that $A_n\to A$ in the \emph{norm resolvent sense}\index{resolvent convergence} if $\|R_{A_n}(z)-R_A(z)\|\to0$, and in the \emph{strong resolvent sense} if $R_{A_n}(z)\psi\to R_A(z)\psi$ for all $\psi\in\hilb$, for some $z\in\mathbb{C}\setminus\mathbb{R}$; in both cases, the convergence then holds for all $z\in\mathbb{C}\setminus\mathbb{R}$ and, for operators uniformly bounded from below, for all real $z$ sufficiently far below their spectra. Norm resolvent convergence implies strong resolvent convergence, and the latter is equivalent to the strong convergence of the corresponding unitary groups, $\e^{-\ii tA_n}\psi\to\e^{-\ii tA}\psi$ for all $\psi\in\hilb$ and $t\in\mathbb{R}$ (Trotter's theorem\index{Trotter's theorem}).
	
	\subsection{The axioms of quantum mechanics}\label{subsec:qmfoundations}
	
	We can now summarize (compare \cite[Section 2.1]{teschl}) the description of a closed quantum system in the nonrelativistic regime in terms of (1) states, (2) observables, (3) evolution, and (4) composite systems. Here we only consider pure states; mixed states are discussed in Section~\ref{subsec:mixed}.
	
	\begin{axiom}[States]\index{axioms of quantum mechanics}
		Each pure \emph{state} of a quantum system is described by a unit vector $\psi$ in a complex separable Hilbert space $\hilb$, unique up to a global phase.
	\end{axiom}
	
	\begin{axiom}[Observables]\label{ax:observables}
		Each \emph{observable} of a quantum system is described by a self-adjoint operator $A:\Dom A\subseteq\hilb\rightarrow\hilb$ such that the spectral measure $\mu_{\psi,A}$ defined in Eq.~\eqref{eq:spectralmeasure}, with $P_A$ being the PVM uniquely associated with $A$, is the probability distribution of the outcomes of a measurement of this observable on the system prepared in the state described by $\psi$.
	\end{axiom}
	
	In this language, the physical meaning of domains becomes transparent: by Eq.~\eqref{eq:spectraldomain}, a state $\psi$ belongs to $\Dom A$ if and only if the probability distribution of the outcomes of a measurement of $A$ in the state $\psi$ has finite second moment (equivalently, finite mean and variance), and analogously for higher moments and powers of $A$.
	
	\begin{axiom}[Evolution]
		The evolution of the system is described by a strongly continuous one-parameter unitary group $(U(t))_{t\in\mathbb{R}}$, whose generator $H:\Dom H\subseteq\hilb\rightarrow\hilb$ (the \emph{Hamiltonian})\index{Hamiltonian} is the self-adjoint operator associated with the energy of the system.
	\end{axiom}
	
	Therefore $U(t)=\e^{-\ii tH}$. For all $\psi\in\hilb$, the evolution function $\mathbb{R}\ni t\mapsto U(t)\psi\in\hilb$
	is always well-defined and continuous, but possibly not differentiable. 	
	In addition, if $\psi\in\Dom H$---i.e. the energy distribution of $\psi$ has finite second moment---the evolution is differentiable and satisfies the Schr\"odinger equation\index{Schr\"odinger equation} (see \cite[Theorem 5.1 and Lemma 5.2]{teschl}):
	\begin{equation}
		\ii\frac{\mathrm{d}}{\mathrm{d}t}U(t)\psi=HU(t)\psi.
	\end{equation}
	
	Finally, composite systems are described as follows. 
	
	\begin{axiom}[Composite systems]\label{ax:composite}
		Given two quantum systems S$_{1}$, S$_{2}$ respectively associated with Hilbert spaces $\hilb_{\rm 1}$ and $\hilb_{\rm 2}$, the \emph{composite system} S$_{12}$ is associated with the Hilbert space tensor product of the two spaces: $\hilb_{\rm 12}=\hilb_{1}\otimes\hilb_{2}$.
		Furthermore, in the absence of interaction, the evolution of S$_{12}$ is given by $U_{12}(t)=U_1(t)\otimes U_2(t)$,
		where $U_1(t)$ and $U_2(t)$ are the unitary groups describing the free evolution of S$_{1}$ and S$_{2}$ respectively.
	\end{axiom}
	
	The concept of tensor product of operators has not been introduced yet; we will come to that in Section~\ref{subsec:tensoroperators}.
	
	\subsection{Mixed states}\label{subsec:mixed}
	
	The axioms above describe the states of a quantum system by unit vectors, i.e. as \emph{pure states}\index{pure state}. In many situations, however---a system prepared by a noisy source, or a subsystem of a larger system---our knowledge of the state is only statistical, and the appropriate description is in terms of \emph{density operators}; see e.g.~\cite[Section 2.4]{nielsenchuang} for a physics-oriented introduction, and~\cite[Section VI.6]{reedsimon1} for the mathematical background on trace-class operators.
	
	A bounded operator $T\in\mathcal{B}(\hilb)$ is \emph{trace class}\index{trace class} if $\sum_n\braket{e_n,|T|e_n}<\infty$ for some, and then every, orthonormal basis $(e_n)_{n\in\mathbb{N}}$ of $\hilb$, where $|T|:=(T^*T)^{1/2}$. In this case, the \emph{trace} $\operatorname{tr}T:=\sum_n\braket{e_n,Te_n}$ converges absolutely and does not depend on the choice of the basis. Trace-class operators are \emph{compact}\index{compact operator}, i.e. they map bounded sets into sets with compact closure; equivalently, they are norm limits of finite-rank operators~\cite[Section VI.5]{reedsimon1}.
	Besides, if $T$ is trace class and $A\in\mathcal{B}(\hilb)$, then $AT$ and $TA$ are trace class, with $\operatorname{tr}(AT)=\operatorname{tr}(TA)$.
	
	\begin{definition}[Density operators]\label{def:density}\index{density operator}\index{mixed state}
		A \emph{density operator} on $\hilb$ is a nonnegative trace-class operator $\rho\in\mathcal{B}(\hilb)$ with $\operatorname{tr}\rho=1$.
	\end{definition}
	
	Being compact and nonnegative, every density operator admits a spectral decomposition
	\begin{equation}\label{eq:density}
		\rho=\sum_kp_k\ket{\psi_k}\!\bra{\psi_k},\qquad p_k\geq0,\qquad\sum_kp_k=1,
	\end{equation}
	with $(\psi_k)_k$ an orthonormal family, where $\ket{\psi}\!\bra{\psi}$ denotes the orthogonal projection onto the span of a unit vector $\psi$. Every unit vector $\psi$ thus defines a density operator $\ket{\psi}\!\bra{\psi}$, which does not depend on the global phase of $\psi$: pure states are precisely the density operators which are rank-one projections, and they are characterized by $\operatorname{tr}\rho^2=1$, while $\operatorname{tr}\rho^2<1$ for all other density operators, the \emph{mixed states}. By Eq.~\eqref{eq:density}, a mixed state $\rho$ describes a statistical mixture, in which the system is in the pure state $\psi_k$ with probability $p_k$.
	
	The axioms extend naturally to mixed states. The probability distribution of the outcomes of a measurement of an observable $A$ in the state $\rho$ is $\mu_{\rho,A}(B):=\operatorname{tr}(\rho P_A(B))$, $B\in\mathfrak{B}(\mathbb{R})$, which reduces to the spectral measure~\eqref{eq:spectralmeasure} for $\rho=\ket{\psi}\!\bra{\psi}$; in particular, if $A$ is bounded, its expectation value is $\operatorname{tr}(\rho A)$. The evolution generated by a Hamiltonian $H$ is $\rho(t)=\e^{-\ii tH}\rho\,\e^{\ii tH}$, which reduces to $\psi(t)=\e^{-\ii tH}\psi$ for pure states.
	
	The counterpart of the Schr\"odinger equation for density operators is the \emph{Liouville--von Neumann equation}\index{Liouville--von Neumann equation}, formally $\ii\frac{\mathrm{d}}{\mathrm{d}t}\rho(t)=[H,\rho(t)]$. For unbounded $H$, some care is needed: the commutator $H\rho-\rho H$ is only defined on vectors $\psi\in\Dom H$ such that $\rho\psi\in\Dom H$, and one must specify in which topology the derivative is taken. A convenient setting is the Hilbert space $\mathcal{B}_2(\hilb)$ of \emph{Hilbert--Schmidt operators}\index{Hilbert--Schmidt operator}, i.e. bounded operators $T$ with $\operatorname{tr}(T^*T)<\infty$, with scalar product $\braket{S,T}_2:=\operatorname{tr}(S^*T)$; every trace-class operator, and in particular every density operator, is Hilbert--Schmidt. The maps $\rho\mapsto\e^{-\ii tH}\rho\,\e^{\ii tH}$ form a strongly continuous one-parameter unitary group on $\mathcal{B}_2(\hilb)$, whose self-adjoint generator $\mathcal{L}$ (Theorem~\ref{thm:stone}), the \emph{Liouvillian}\index{Liouvillian}, plays the role of the Hamiltonian: for every $\rho\in\Dom\mathcal{L}$, $\rho(t)$ is the unique solution of $\ii\frac{\mathrm{d}}{\mathrm{d}t}\rho(t)=\mathcal{L}\rho(t)$, the derivative being taken in the Hilbert--Schmidt norm. One can show that $\rho\in\Dom\mathcal{L}$ if and only if $\rho\Dom H\subseteq\Dom H$ and the commutator $H\rho-\rho H$, defined on $\Dom H$, extends to a Hilbert--Schmidt operator, which is then $\mathcal{L}\rho$: the Liouvillian is the rigorous counterpart of the commutator with $H$. See~\cite{lonigro2024lvn} for details, including further characterizations of $\Dom\mathcal{L}$. One can equivalently formulate the Liouville--von Neumann equation on the space of trace-class operators, with the derivative taken in the trace norm; the two formulations lead to essentially the same results.
	
	Finally, mixed states arise naturally when describing a subsystem of a composite system: as we will see in Section~\ref{sec:tensor} (Proposition~\ref{prop:partialtrace} and Remark~\ref{rem:schmidt}), the state of a subsystem is described by a density operator, the reduced state, which is mixed whenever the global pure state is entangled.
	
	\begin{example}[Qubit density matrices]\label{examp:qubitdensity}
		For $\hilb=\mathbb{C}^2$, with $e_\pm$ the eigenvectors of $\sigma_z$, every density operator can be written as
		\begin{equation}
			\rho=\begin{bmatrix}
				\rho_{++}&\rho_{+-}\\\rho_{-+}&\rho_{--}
			\end{bmatrix},\qquad\rho_{++},\rho_{--}\geq0,\quad\rho_{++}+\rho_{--}=1,\quad\rho_{-+}=\overline{\rho_{+-}},\quad|\rho_{+-}|^2\leq\rho_{++}\rho_{--}.
		\end{equation}
		The diagonal elements $\rho_{\pm\pm}$ are the \emph{populations}\index{qubit!populations and coherences} of the two levels, while the off-diagonal elements $\rho_{\pm\mp}$ are called \emph{coherences}. Equivalently, $\rho=\frac{1}{2}(I+\bm{r}\cdot\bm{\sigma})$ for a unique $\bm{r}\in\mathbb{R}^3$ with $|\bm{r}|\leq1$ (the \emph{Bloch ball}\index{Bloch ball}); the eigenvalues of $\rho$ are $\frac{1}{2}(1\pm|\bm{r}|)$, so that $\rho$ is pure if and only if $|\bm{r}|=1$. For instance, $\rho=\frac{1}{2}I$ is the \emph{maximally mixed} state.
	\end{example}
	
	\subsection{Perturbations of self-adjoint operators: the Kato--Rellich theorem}\label{sec:perturbation}
	
	By Axiom~\ref{ax:observables}, quantum observables are described by self-adjoint (or at least essentially self-adjoint) operators. We will repeatedly face the following question: let $H_0:\Dom H_0\subseteq\hilb\rightarrow\hilb$ be a self-adjoint operator, and $V:\Dom V\subseteq\hilb\rightarrow\hilb$ (at least) symmetric; under what conditions is $H_0+V$ self-adjoint as well? Typically, $H_0$ is a free Hamiltonian---the kinetic energy of a particle, or the free energy of two noninteracting systems---and $V$ an external potential or an interaction; intuitively, a ``sufficiently small'' symmetric perturbation $V$ should preserve self-adjointness.
	
	For unbounded operators this is not a trivial question: $H_0+V$ must be densely defined, and in general $(H_0+V)^*\neq H_0^*+V^*$. If $V$ is bounded and symmetric, however, $H_0+V$ is self-adjoint on $\Dom H_0$; for instance, a \emph{Schr\"odinger operator}---a Laplacian plus a real potential $v$---is self-adjoint on the domain of the Laplacian whenever $v$ is bounded. In many applications, however, $V$ is unbounded, and a more refined criterion is needed.
	
	\begin{definition}[Relative boundedness]\label{def:relbound}\index{relative boundedness}
		Let $A:\Dom A\subseteq\hilb\rightarrow\hilb$, $B:\Dom B\subseteq\hilb\rightarrow\hilb$. $B$ is \emph{relatively bounded with respect to $A$} (or $A$-bounded) if the following conditions hold:
		\begin{itemize}
			\item[(i)] $\Dom B\supseteq\Dom A$;
			\item[(ii)] There exist $a,b\geq0$ such that, for all $u\in\Dom A$,
			\begin{equation}\label{eq:relbound}
				\|Bu\|\leq a\|Au\|+b\|u\|.
			\end{equation}
		\end{itemize}
		The infimum of all values of $a$ such that the inequality above holds for some $b\geq0$ is the \emph{$A$-bound}\index{relative boundedness!relative bound} of $B$.
	\end{definition}
	
	\begin{definition}[Infinitesimal relative boundedness]\label{def:infrelbound}\index{relative boundedness!infinitesimal}
		$B$ is said to be \emph{infinitesimally} relatively bounded with respect to $A$ (or infinitesimally $A$-bounded) if it is $A$-bounded with $A$-bound \emph{equal to zero}, that is: 
		\begin{itemize}
			\item[(i)] $\Dom B\supseteq\Dom A$;
			\item[(ii)] For all $\epsilon>0$ there exists $b_\epsilon\geq0$ such that, for all $u\in\Dom A$,
			\begin{equation}
				\|Bu\|\leq \epsilon\|Au\|+b_\epsilon\|u\|.
			\end{equation}
		\end{itemize}
	\end{definition}
	
	Clearly, a bounded operator $B\in\mathcal{B}(\hilb)$ is (infinitesimally) relatively bounded with respect to any operator $A$, since Eq.~\eqref{eq:relbound} holds with $a=0$ and $b=\|B\|$. However, the converse is not true: $B$ can be (even infinitesimally) $A$-bounded without being bounded.
	A first example is the following. Recall the \emph{Young inequality}\index{Young inequality}: given $a,b\geq0$, for every $\epsilon>0$,
	\begin{equation}\label{eq:young}
		ab\leq\frac{\epsilon}{2} a^2+\frac{1}{2\epsilon}b^2,
	\end{equation}
	which follows from $\left(\sqrt{\epsilon}\,a-b/\sqrt{\epsilon}\right)^2\geq0$.
	
	\begin{example}\label{examp:wrtsquare}
		Let $A:\Dom A\subseteq\hilb\rightarrow\hilb$ be a possibly unbounded symmetric operator. Then $A$ is infinitesimally relatively bounded with respect to $A^2$. Indeed, $\Dom A^2\subseteq\Dom A$ and, for all $u\in\Dom A^2$ and $\epsilon>0$, by the Cauchy--Schwarz and Young inequalities,
		\begin{equation}
			\|Au\|^2=\braket{u,A^2u}\leq\|u\|\|A^2u\|\leq\frac{\epsilon}{2}\|A^2u\|^2+\frac{1}{2\epsilon}\|u\|^2,
		\end{equation}
		whence $\|Au\|\leq\sqrt{\epsilon/2}\,\|A^2u\|+\sqrt{1/(2\epsilon)}\,\|u\|$.
	\end{example}
	
	For example, the momentum operator $p$ on the real line is infinitesimally bounded with respect to the Laplace operator $-\Delta=p^2$. Example~\ref{examp:wrtsquare} will be a key ingredient in the proof of self-adjointness of spin--boson models (Propositions~\ref{prop:visrelbound} and~\ref{prop:sb_sa}).
	
	On $L^2$ spaces it is often convenient to work with squared norms, which are integrals. The two formulations are equivalent, as one shows by an elementary application of the Young inequality:
	\begin{proposition}\label{prop:squared_relbound}
		Let $A:\Dom A\subseteq\hilb\rightarrow\hilb$ and $B:\Dom B\subseteq\hilb\rightarrow\hilb$, and assume $\Dom B\supseteq\Dom A$. Then the following are equivalent:
		\begin{itemize}
			\item[(i)] There exist $a,b\geq0$ such that Eq.~\eqref{eq:relbound} holds for all $u\in\Dom A$;
			\item[(ii)] There exist $\tilde{a},\tilde{b}\geq0$ such that, for all $u\in\Dom A$,
			\begin{equation}
				\|Bu\|^2\leq \tilde{a}^2\|Au\|^2+\tilde{b}^2\|u\|^2.
			\end{equation}
		\end{itemize}
		Besides, if (i) holds for some $a$, then (ii) holds for every $\tilde{a}>a$, and vice versa; in particular, the $A$-bound of $B$ can be computed from either formulation.
	\end{proposition}
	
	The fundamental result on the perturbation of self-adjoint operators is due to Rellich~\cite{rellich1939} and Kato~\cite{kato1951} (see also \cite[Theorem X.12]{reedsimon2} and~\cite{katobook}):
	
	\begin{theorem}[Kato--Rellich theorem]\label{thm:katorellich}\index{Kato--Rellich theorem}
		Let $A:\Dom A\subseteq\hilb\rightarrow\hilb$ be self-adjoint (resp. essentially self-adjoint), and $B:\Dom B\subseteq\hilb\rightarrow\hilb$ be symmetric and $A$-bounded with $A$-bound $a_\infty$ strictly smaller than $1$. Then:
		\begin{itemize}
			\item The operator $A+B:\Dom A\subseteq\hilb\rightarrow\hilb$ is self-adjoint (resp. essentially self-adjoint);
			\item Furthermore, if $A$ is bounded from below, $A+B$ is also bounded from below.
		\end{itemize}	
	\end{theorem}
	The idea of the proof is the following: writing $A+B\mp\ii\lambda=\left[1+B(A\mp\ii\lambda)^{-1}\right](A\mp\ii\lambda)$ for $\lambda>0$, one shows that $\|B(A\mp\ii\lambda)^{-1}\|$ tends to the $A$-bound of $B$ as $\lambda\to\infty$, and is thus smaller than $1$ for $\lambda$ large enough; hence the operator in square brackets is invertible, and $\Ran(A+B\mp\ii\lambda)=\hilb$, so that $A+B$ is self-adjoint by the criterion of self-adjointness (Theorem~\ref{thm:criterion-sa}), applied to $\lambda^{-1}(A+B)$.	
	Note that the Kato--Rellich theorem provides a sufficient---not necessary---criterion for self-adjointness: for $a_\infty\geq1$, $A+B$ might or might not be self-adjoint.	
	
	\begin{remark}
		The condition $a_\infty<1$ is optimal in the sense that the statement is not generally true when $a_\infty=1$. Let $A$ be self-adjoint and unbounded, and $B=-A$: the operator $A+B$ is a restriction of the self-adjoint null operator, therefore it is not self-adjoint---but still essentially self-adjoint.
		A slight generalization of the Kato--Rellich theorem is \emph{W\"ust's theorem}\index{W\"ust theorem}: if $A$ is (essentially) self-adjoint, $B$ is symmetric with $\Dom B\supseteq\Dom A$, and the relative bound holds with $a=1$, i.e. $\|Bu\|\leq\|Au\|+b\|u\|$ for some $b\geq0$ and all $u\in\Dom A$, then $A+B$ is essentially self-adjoint~\cite{wust1971} (see also \cite[Theorem X.14]{reedsimon2}). In this case, $A+B$ need not be self-adjoint, as the previous example shows.
	\end{remark}
	
	In physical models, the interaction term often comes with a coupling constant $g\in\mathbb{R}$. If $B$ is $A$-bounded with $A$-bound $a_\infty$, then $gB$ is $A$-bounded with $A$-bound $|g|a_\infty$, so that $A+gB$ is (essentially) self-adjoint on $\Dom A$ for all $|g|<1/a_\infty$ (with the convention $1/0=\infty$): a ``sufficiently small'' perturbation does not spoil self-adjointness.
	
	\begin{example}[Schr\"odinger operators on the segment]\label{examp:schrsegment}\index{Schr\"odinger operator}
		Let $\hilb=L^2(0,1)$ and $H_0:\Dom H_0\subseteq\hilb\rightarrow\hilb$ be the \emph{Dirichlet Laplacian}\index{Laplace operator!Dirichlet} on the segment, i.e.
		\begin{align}
			\Dom H_0&=\left\{\varphi\in\mathrm{H}^2(0,1):\varphi(0)=\varphi(1)=0\right\};\\
			(H_0\varphi)(x)&=-\varphi''(x),
		\end{align}
		which, as discussed in Example~\ref{examp:laplacesegment}, is self-adjoint. Let $v:(0,1)\rightarrow\mathbb{R}$ be a possibly unbounded, but square-integrable potential.\footnote{All bounded functions are in $L^2(0,1)$, but not vice versa---take e.g. $v(x)=x^{-\alpha}$ ($x\in(0,1)$) for some $0<\alpha<1/2$: this function is unbounded near $x=0$ but still square-integrable.} Then the corresponding multiplication operator $M_v$ is infinitesimally $H_0$-bounded, and thus, by the Kato--Rellich theorem (Theorem~\ref{thm:katorellich}), $H=H_0+M_v$ is self-adjoint on $\Dom H=\Dom H_0$. The key ingredient is the estimate
		\begin{equation}\label{eq:ineq_derivative}
			\|\varphi\|^2_\infty=\sup_{x\in(0,1)}|\varphi(x)|^2\leq\|\varphi\|^{3/2}\|\varphi''\|^{1/2}\qquad\text{for all }\varphi\in\Dom H_0,
		\end{equation}
		which follows by combining $\|\varphi\|_\infty^2\leq\|\varphi\|\|\varphi'\|$ (write $|\varphi(x)|^2$ as the integral of its derivative from either endpoint, and average the two resulting bounds, as in Example~\ref{examp:delta} below) and $\|\varphi'\|^2\leq\|\varphi\|\|\varphi''\|$ (integrate by parts). Applying the Young inequality~\eqref{eq:young} twice---first to the product of $\|\varphi\|$ and $(\|\varphi\|\|\varphi''\|)^{1/2}$, then to $\|\varphi\|\|\varphi''\|$---Eq.~\eqref{eq:ineq_derivative} implies that, for every $\epsilon>0$, there is $c_\epsilon\geq0$ such that $\|\varphi\|_\infty^2\leq\epsilon\|\varphi''\|^2+c_\epsilon\|\varphi\|^2$, whence
		\begin{equation}
			\|M_v\varphi\|^2=\int_0^1v(x)^2|\varphi(x)|^2\;\mathrm{d}x\leq\|v\|^2\|\varphi\|_\infty^2\leq\epsilon\|v\|^2\|H_0\varphi\|^2+c_\epsilon\|v\|^2\|\varphi\|^2.
		\end{equation}
	\end{example}
	
	\begin{example}[Schr\"odinger operators on the real line]
		An analogous result holds on the real line. Let $\hilb=L^2(\mathbb{R})$, $H_0:\Dom H_0\subseteq\hilb\rightarrow\hilb$ be the Laplace operator on the real line, and $v:\mathbb{R}\rightarrow\mathbb{R}$ a square-integrable potential.
		The estimate~\eqref{eq:ineq_derivative} holds for all $\varphi\in\Dom H_0=\mathrm{H}^2(\mathbb{R})$, with the supremum taken over $\mathbb{R}$: the same argument applies, the boundary values being replaced by the limits of $\varphi$ and $\varphi'$ at $\pm\infty$, which vanish (cf. Example~\ref{examp:delta} below).\footnote{Estimates of this kind also hold on $\mathbb{R}^2$ and $\mathbb{R}^3$; see e.g. \cite[Lemma 10.1]{teschl} or \cite[Proposition~8.7]{schmuedgen}.} Hence, as in the segment case (Example~\ref{examp:schrsegment}), $M_v$ is infinitesimally $H_0$-bounded, and $H=H_0+M_v$ is self-adjoint on $\Dom H=\Dom H_0$.
		
		The statement also holds (see \cite[Theorem 10.2]{teschl}) for $v\in L^\infty(\mathbb{R})+L^2(\mathbb{R})$, i.e. for potentials that are the sum of a bounded potential $v_1\in L^\infty(\mathbb{R})$, possibly not square-integrable (e.g. $v_1(x)=v_0\neq0$), and a square-integrable potential $v_2\in L^2(\mathbb{R})$, possibly not bounded (e.g. $v_2(x)=|x|^{-1/4}\e^{-x^2}$). Indeed, in this case $M_v=M_{v_1}+M_{v_2}$, with $M_{v_1}$ bounded and $M_{v_2}$ infinitesimally $H_0$-bounded.
	\end{example}
	
	\begin{example}[Counterexample: the quantum harmonic oscillator]\label{examp:qho-counter}
		As shown in Example~\ref{examp:qho}, the quantum harmonic oscillator $H_0=\frac{1}{2}(-\Delta+\omega^2x^2)$, with domain $\Dom H_0=\mathcal{S}(\mathbb{R})$, is essentially self-adjoint and thus admits a unique self-adjoint extension $\overline{H_0}$.		
		However, this cannot be deduced from the Kato--Rellich theorem applied to the free Laplacian $-\Delta$, since the multiplication operator $M_v$ corresponding to $v(x)=x^2$ is not even $(-\Delta)$-bounded: its domain does not contain $\Dom(-\Delta)=\mathrm{H}^2(\mathbb{R})$. For example, the function $\varphi(x)=(1+x^2)^{-1/2}$ is in $\mathrm{H}^2(\mathbb{R})$, but $x^2\varphi\notin L^2(\mathbb{R})$, since $x^4|\varphi(x)|^2\sim x^2$ for large $|x|$.
		
		The same function shows that, unlike for the Laplacian, $\Dom\overline{H_0}\neq\mathrm{H}^2(\mathbb{R})$. Indeed, for $\psi\in\mathcal{S}(\mathbb{R})$, two integrations by parts give $2\Re\braket{-\psi'',x^2\psi}=2\|x\psi'\|^2-2\|\psi\|^2$, whence
		\begin{equation}
			4\|H_0\psi\|^2=\|\psi''\|^2+\omega^4\|x^2\psi\|^2+2\omega^2\|x\psi'\|^2-2\omega^2\|\psi\|^2\geq\omega^4\|x^2\psi\|^2-2\omega^2\|\psi\|^2.
		\end{equation}
		Since $\mathcal{S}(\mathbb{R})$ is a core for $\overline{H_0}$ and $M_v$ is closed, this inequality extends to all $\psi\in\Dom\overline{H_0}$, which is therefore contained in $\Dom M_v$; but $\varphi\notin\Dom M_v$. (In fact, one can show that $\Dom\overline{H_0}=\mathrm{H}^2(\mathbb{R})\cap\Dom M_v$.)
	\end{example}
	
	\begin{example}[Quantum harmonic oscillator with linear potential]\label{examp:linearpot}
		Let $H_0$ be the quantum harmonic oscillator of Example~\ref{examp:qho}, defined on $\Dom H_0=\mathcal{S}(\mathbb{R})$ (thus essentially self-adjoint), and $v(x)=x$. Integrating by parts, $\braket{\varphi,H_0\varphi}=\frac{1}{2}\|\varphi'\|^2+\frac{\omega^2}{2}\|M_v\varphi\|^2$ for all $\varphi\in\mathcal{S}(\mathbb{R})$, whence, by the Cauchy--Schwarz and Young inequalities, for every $\epsilon>0$,
		\begin{equation}
			\|M_v\varphi\|^2\leq\frac{2}{\omega^2}\braket{\varphi,H_0\varphi}\leq\frac{2}{\omega^2}\|\varphi\|\|H_0\varphi\|\leq\epsilon\|H_0\varphi\|^2+\frac{1}{\epsilon\omega^4}\|\varphi\|^2.
		\end{equation}
		Thus $M_v$ is infinitesimally $H_0$-bounded (Proposition~\ref{prop:squared_relbound}) and, by the Kato--Rellich theorem, $H_0+M_v$ is essentially self-adjoint on $\mathcal{S}(\mathbb{R})$. Up to rescalings and constant shifts, this is the ``displaced'' oscillator that we will meet in the pure dephasing spin--boson model (Section~\ref{subsec:dephasing}).
	\end{example}
	
	\subsection{Sesquilinear forms and the KLMN theorem}\label{subsec:forms}
	
	Many operators of interest in quantum mechanics are more naturally described through their ``expectation values'' $\braket{\psi,A\psi}$ than through their action on vectors. This idea is made precise by the theory of sesquilinear forms; see e.g.~\cite{katobook,schmuedgen} for details and proofs.
	
	\begin{definition}[Sesquilinear and quadratic forms]\label{def:forms}\index{sesquilinear form}\index{quadratic form}\index{form domain}
		Let $\Dom q\subseteq\hilb$ be a dense subspace. A \emph{sesquilinear form} on $\hilb$ with \emph{form domain} $\Dom q$ is a map $q:\Dom q\times\Dom q\rightarrow\mathbb{C}$ which is antilinear in its first entry and linear in its second one; its \emph{quadratic form} is $q(u):=q(u,u)$. The form $q$ is said to be
		\begin{itemize}
			\item \emph{Hermitian}, if $q(u,v)=\overline{q(v,u)}$ for all $u,v\in\Dom q$;
			\item \emph{bounded from below}, if there exists $c\in\mathbb{R}$ such that $q(u)\geq-c\|u\|^2$ for all $u\in\Dom q$; we then write $q\geq-c$.
		\end{itemize}
	\end{definition}
	
	A sesquilinear form is uniquely determined by its quadratic form, via the polarization identity
	\begin{equation}
		q(u,v)=\frac{1}{4}\sum_{k=0}^3\ii^{-k}q\left(u+\ii^kv\right),
	\end{equation}
	and it is Hermitian if and only if $q(u)\in\mathbb{R}$ for all $u\in\Dom q$. The prototypical example is the following one.
	
	\begin{example}[The form of a self-adjoint operator bounded from below]\label{examp:operatorform}
		Let $A:\Dom A\subseteq\hilb\rightarrow\hilb$ be self-adjoint and bounded from below, $A\geq-c$. By the spectral theorem, $(A+c)^{1/2}$ is a well-defined nonnegative self-adjoint operator, and
		\begin{equation}
			\Dom q_A:=\Dom(A+c)^{1/2},\qquad q_A(u,v):=\Braket{(A+c)^{1/2}u,(A+c)^{1/2}v}-c\braket{u,v}
		\end{equation}
		defines a Hermitian form $q_A\geq-c$, the \emph{form associated with $A$}; one easily checks, again by the spectral theorem, that neither $\Dom q_A$ nor $q_A$ depends on the choice of $c$. We will denote by $\mathcal{Q}(A):=\Dom q_A$ the form domain of $A$. Clearly $\Dom A\subseteq\mathcal{Q}(A)$ and
		\begin{equation}
			q_A(u,v)=\braket{u,Av}\qquad\text{for all }u\in\mathcal{Q}(A),\;v\in\Dom A.
		\end{equation}
		For instance, if $A\geq0$, then $q_A(u,v)=\braket{A^{1/2}u,A^{1/2}v}$ on $\mathcal{Q}(A)=\Dom A^{1/2}$.
	\end{example}
	
	Not every Hermitian form bounded from below arises from an operator in this way. The forms that do are characterized by a completeness property.
	
	\begin{definition}[Closed forms, form cores]\label{def:closedform}\index{closed form}\index{form core}
		Let $q$ be a Hermitian form with $q\geq-c$. The \emph{form norm} on $\Dom q$ is $\|u\|_q:=\left(q(u)+(c+1)\|u\|^2\right)^{1/2}$.
		The form $q$ is \emph{closed} if $\Dom q$ is complete with respect to $\|\cdot\|_q$. A subspace $\mathcal{D}\subseteq\Dom q$ which is dense in $\Dom q$ with respect to $\|\cdot\|_q$ is a \emph{form core} for $q$.
	\end{definition}
	
	The form norm is induced by the scalar product $q(u,v)+(c+1)\braket{u,v}$, and different admissible choices of $c$ yield equivalent norms, so that neither closedness nor the notion of form core depends on $c$. For instance, the form $q_A$ of Example~\ref{examp:operatorform} is closed: $\|u\|_{q_A}^2=\|(A+c)^{1/2}u\|^2+\|u\|^2$ is the squared graph norm of the closed operator $(A+c)^{1/2}$. The converse is the content of the following fundamental result.
	
	\begin{theorem}[Representation theorem]\label{thm:representation}\index{representation theorem}
		Let $q$ be a densely defined, Hermitian, closed form on $\hilb$, with $q\geq-c$. Then there exists a unique self-adjoint operator $A$ such that $\Dom A\subseteq\Dom q$ and
		\begin{equation}
			q(u,v)=\braket{u,Av}\qquad\text{for all }u\in\Dom q,\;v\in\Dom A.
		\end{equation}
		Moreover, $A\geq-c$, $q=q_A$ (in particular, $\Dom q=\mathcal{Q}(A)$), and $\Dom A$ is a form core for $q$. Explicitly,
		\begin{align}
			\Dom A&=\left\{v\in\Dom q:\;\exists w\in\hilb\text{ s.t. }q(u,v)=\braket{u,w}\text{ for all }u\in\Dom q\right\},\\
			Av&=w.
		\end{align}
	\end{theorem}
	
	In other words, closed Hermitian forms bounded from below and self-adjoint operators bounded from below are in one-to-one correspondence.
	
	\begin{example}[Laplace operators on the segment, revisited]\label{examp:formslaplacian}
		Let $\mathrm{H}^1(0,1)$ be the first Sobolev space on $(0,1)$, i.e. the space of functions $u\in L^2(0,1)$ whose (weak) derivative exists and satisfies $u'\in L^2(0,1)$; its elements are absolutely continuous, so their boundary values $u(0),u(1)$ are well-defined. The two forms
		\begin{equation}
			q_{\rm D}(u,v)=q_{\rm N}(u,v)=\int_0^1\overline{u'(x)}\,v'(x)\;\mathrm{d}x,
		\end{equation}
		with form domains $\Dom q_{\rm D}=\{u\in\mathrm{H}^1(0,1):u(0)=u(1)=0\}$ and $\Dom q_{\rm N}=\mathrm{H}^1(0,1)$, are nonnegative and closed, and one can show that the associated operators are the Laplace operators on the segment (Example~\ref{examp:laplacesegment}) with Dirichlet and Neumann boundary conditions, respectively. Note that the Dirichlet condition is imposed in the form domain, while the Neumann condition is not: it emerges automatically, when integrating by parts, from the requirement that $q_{\rm N}(u,v)=\braket{u,w}$ for all $u\in\mathrm{H}^1(0,1)$.
	\end{example}
	
	Next we discuss form perturbations. The Kato--Rellich theorem requires the perturbation $B$ to be an operator defined (at least) on $\Dom A$. In several situations of interest, however, the perturbation only makes sense as a form, i.e. through its ``expectation values''; a typical example is a delta potential (Example~\ref{examp:delta} below), and we will meet another one when dealing with spin--boson models with singular form factors (Section~\ref{sec:singular}). In such cases one can work at the level of forms, and use the representation theorem (Theorem~\ref{thm:representation}) to go back to operators. See e.g.~\cite{katobook,schmuedgen} for more details.
	
	\begin{definition}[Relative form boundedness]\label{def:formbounded}\index{relative form bound}\index{form-bounded operator}
		Let $A:\Dom A\subseteq\hilb\rightarrow\hilb$ be self-adjoint with $A\geq-c$, and $\mathcal{Q}(A)$ its form domain. A Hermitian form $b$ with $\Dom b\supseteq\mathcal{Q}(A)$ is \emph{$A$-form-bounded} (or relatively form-bounded with respect to $A$) if there exist $a,\beta\geq0$ such that
		\begin{equation}\label{eq:formbound}
			|b(u)|\leq a\left\|(A+c)^{1/2}u\right\|^2+\beta\|u\|^2\qquad\text{for all }u\in\mathcal{Q}(A).
		\end{equation}
		The infimum of all $a$ such that Eq.~\eqref{eq:formbound} holds for some $\beta$ is the \emph{$A$-form bound} of $b$. If it is zero, $b$ is said to be \emph{infinitesimally $A$-form-bounded}.
	\end{definition}
	
	Changing $c$ only changes the constant $\beta$ in Eq.~\eqref{eq:formbound}, so the definition does not depend on the choice of $c$. Relative form boundedness is weaker than relative boundedness: if $B$ is a symmetric operator which is $A$-bounded with $A$-bound $a_\infty$, then the form $u\mapsto\braket{u,Bu}$, initially defined on $\Dom A$, uniquely extends to an $A$-form-bounded form on $\mathcal{Q}(A)$ with $A$-form bound not larger than $a_\infty$.
	
	\begin{theorem}[KLMN theorem]\label{thm:klmn}\index{KLMN theorem}
		Let $A:\Dom A\subseteq\hilb\rightarrow\hilb$ be self-adjoint with $A\geq-c$, and let $b$ be an $A$-form-bounded Hermitian form with $A$-form bound strictly smaller than $1$. Then there exists a unique self-adjoint operator $H$ such that $\Dom H\subseteq\mathcal{Q}(A)$ and
		\begin{equation}\label{eq:formsum}
			\braket{u,Hv}=q_A(u,v)+b(u,v)\qquad\text{for all }u\in\mathcal{Q}(A),\;v\in\Dom H.
		\end{equation}
		Moreover, $\mathcal{Q}(H)=\mathcal{Q}(A)$ and $H$ is bounded from below: if Eq.~\eqref{eq:formbound} holds with $a<1$, then $H\geq-(c+\beta)$.
	\end{theorem}
	
	The theorem is named after Kato, Lions, Lax, Milgram and Nelson~\cite{kato1955,lions1961,laxmilgram1954,nelson1964}. The operator $H$ is called the \emph{form sum}\index{form sum} of $A$ and $b$; if $b$ is the form of an operator $B$, one often writes $H=A\dotplus B$.
	
	The proof is a direct consequence of the representation theorem (Theorem~\ref{thm:representation}): since the form bound is smaller than $1$, the form norm of $q_A+b$ is equivalent to that of $q_A$, so that $q_A+b$ is closed, and bounded from below by $-(c+\beta)$.
	
	\begin{remark}\label{rem:krklmn}
		If $B$ is a symmetric operator which is $A$-bounded with $A$-bound smaller than $1$, both the Kato--Rellich and the KLMN theorems apply, and they yield the same operator. Indeed, by the Kato--Rellich theorem $A+B$ is self-adjoint on $\Dom A\subseteq\mathcal{Q}(A)$; besides, for $v\in\Dom A$, the extended form $b$ satisfies $b(u,v)=\braket{u,Bv}$ for all $u\in\mathcal{Q}(A)$, so that $A+B$ satisfies Eq.~\eqref{eq:formsum}. By uniqueness, $H=A+B$. The KLMN theorem, however, gives less information on the domain: in general, it only yields $\Dom H\subseteq\mathcal{Q}(A)$, and $\Dom H$ need not coincide with $\Dom A$. An example of this phenomenon is discussed after Theorem~\ref{thm:singularsb}.
	\end{remark}
	
	\begin{example}[Delta potential on the real line]\label{examp:delta}\index{delta potential}
		Let $-\Delta$ be the Laplace operator on $L^2(\mathbb{R})$ (Example~\ref{examp:laplace}). Via the Fourier transform, one shows that $\mathcal{Q}(-\Delta)=\mathrm{H}^1(\mathbb{R})$, the space of functions $u\in L^2(\mathbb{R})$ with weak derivative $u'\in L^2(\mathbb{R})$, and $q_{-\Delta}(u,v)=\int\overline{u'(x)}v'(x)\,\mathrm{d}x$ \cite[Eq.~(7.37)]{teschl}. Functions in $\mathrm{H}^1(\mathbb{R})$ are continuous and vanish at infinity, so for $\alpha\in\mathbb{R}$ the Hermitian form
		\begin{equation}
			b(u,v):=\alpha\,\overline{u(0)}\,v(0),\qquad\Dom b=\mathrm{H}^1(\mathbb{R}),
		\end{equation}
		is well-defined. Formally, $b$ is the form of the ``operator'' $\alpha\,\delta(x)$; but there is no operator $B$ on $L^2(\mathbb{R})$ such that $b(u,v)=\braket{u,Bv}$, since $v\mapsto v(0)\delta$ does not map into $L^2(\mathbb{R})$. Nevertheless, $b$ is infinitesimally $(-\Delta)$-form-bounded. Indeed, for $u\in\mathrm{H}^1(\mathbb{R})$,
		\begin{equation}
			|u(0)|^2=-\int_0^\infty\frac{\mathrm{d}}{\mathrm{d}x}|u(x)|^2\;\mathrm{d}x\leq2\int_0^\infty|u(x)||u'(x)|\;\mathrm{d}x,
		\end{equation}
		and similarly on $(-\infty,0)$; averaging the two bounds and using the Cauchy--Schwarz inequality, $|u(0)|^2\leq\|u\|\|u'\|$, whence, for every $\epsilon>0$,
		\begin{equation}
			|b(u)|\leq|\alpha|\|u\|\|u'\|\leq\epsilon\|u'\|^2+\frac{\alpha^2}{4\epsilon}\|u\|^2.
		\end{equation}
		By the KLMN theorem, the form sum $-\Delta\dotplus\alpha\delta$ is a well-defined self-adjoint operator, bounded from below, with form domain $\mathrm{H}^1(\mathbb{R})$. Point interactions of this kind are discussed in detail in~\cite{albeverio2005}; the analogous point interaction on a segment is treated in \cite[Section 6.5]{teschl}.
	\end{example}
	
	\section{Direct sums and tensor products}\label{sec:directsum}
	
	A section on direct sums and tensor products, complete with proofs, may seem out of place in notes on second quantization. However, a Fock space is a direct sum of tensor products (Section~\ref{subsec:fock}), and light--matter Hamiltonians are sums of tensor products of operators (Sections~\ref{sec:singlemodesb} and~\ref{sec:multimodesb}). Many of the properties that matter most for our purposes---for instance, the self-adjointness of spin--boson models---thus become easy corollaries once direct sums and tensor products of unbounded operators are under control: dealing with the technicalities here allows the later sections to focus on the applications without sacrificing rigor.
	
	Consider two Hilbert spaces $\hilb_1$, $\hilb_2$. There are two natural ways to construct a new Hilbert space out of them, intuitively corresponding to two logical operations. The \emph{orthogonal direct sum} $\hilb_1\oplus\hilb_2$ corresponds to ``OR'': for example, the spin state of a spin-$\frac12$ particle lives in $\mathbb{C}\oplus\mathbb{C}$: spin up or spin down. The \emph{(Hilbert space) tensor product} $\hilb_1\otimes\hilb_2$ corresponds to ``AND'': for example, the state space of two quantum particles is the tensor product of the two single-particle spaces: particle $1$ and particle $2$. In the finite-dimensional case, this important difference is also reflected in the dimension of the resulting space: $\dim(\hilb_1\oplus\hilb_2)=\dim\hilb_1+\dim\hilb_2$, while $\dim(\hilb_1\otimes\hilb_2)=\dim\hilb_1\cdot\dim\hilb_2$.
	
	We first define direct sums of Hilbert spaces (Section~\ref{subsec:directsumspaces}) and of operators (Section~\ref{subsec:directsumops}), and then tensor products of Hilbert spaces (Section~\ref{sec:tensor}), with special attention to $L^2$ spaces (Section~\ref{subsec:tensorl2}). Tensor products of unbounded operators require more care: we first define their algebraic tensor product (Section~\ref{subsec:tensoroperators}), study the case of multiplication operators (Section~\ref{subsec:tensormulti}), and finally define their Hilbert space tensor product, proving in particular the self-adjointness of polynomials in tensor products of self-adjoint operators (Section~\ref{subsec:hstensorops}).
	
	\subsection{Direct sums of Hilbert spaces}\label{subsec:directsumspaces}
	
	We already encountered the direct sum $\hilb_1\oplus\hilb_2$ when discussing closed operators (Section~\ref{subsec:esa}). Let us define it formally:
	
	\begin{definition}[Direct sum of two Hilbert spaces]\label{def:directsum_two}\index{direct sum!of Hilbert spaces}
		Let $\hilb_1$, $\hilb_2$ be two Hilbert spaces, with scalar products $\braket{\cdot,\cdot}_{\hilb_1}$, $\braket{\cdot,\cdot}_{\hilb_2}$ respectively, and let $\hilb_1\times\hilb_2:=\left\{(u_1,u_2):u_1\in\hilb_1,\,u_2\in\hilb_2\right\}$ be their Cartesian product.
		The \emph{orthogonal direct sum} $\hilb_1\oplus\hilb_2$ is the vector space $\hilb_1\times\hilb_2$ endowed with the scalar product~\eqref{eq:directsumsp}.
	\end{definition}
	
	From now on, whenever we just refer to the direct sum of Hilbert spaces, we always mean the orthogonal direct sum. It is easy to show that this is a legitimate scalar product, inducing the norm
	\begin{equation}
		\|(u_1,u_2)\|^2_{\hilb_1\oplus\hilb_2}=\|u_1\|^2_{\hilb_1}+\|u_2\|^2_{\hilb_2},
	\end{equation}
	and that $\hilb_1\oplus\hilb_2$ is complete---thus, a Hilbert space itself.
	
	Sometimes, the notation $u_1\oplus u_2$ is used in place of $(u_1,u_2)$, especially if one wants to stress the choice of scalar product to avoid confusion with the tensor product. Also, as we will see, it is often convenient to represent $(u_1,u_2)$ as a formal column vector:
	\begin{equation}
		\begin{bmatrix}
			u_1\\u_2
		\end{bmatrix},\qquad u_1\in\hilb_1,\;u_2\in\hilb_2.
	\end{equation}
	
	One readily notices that the subspaces $\{(u_1,0):u_1\in\hilb_1\}$ and $\{(0,u_2):u_2\in\hilb_2\}$ of $\hilb_1\oplus\hilb_2$ are unitarily equivalent to $\hilb_1$ and $\hilb_2$ respectively, via the maps $u_1\mapsto(u_1,0)$ and $u_2\mapsto(0,u_2)$.
	With a slight abuse of notation, one often identifies $(u_1,0)$ with $u_1$ and $(0,u_2)$ with $u_2$.
	
	Separability is preserved by direct sums:
	
	\begin{proposition}\label{prop:directsumbases}
		Assume that $\hilb_1$ and $\hilb_2$ are separable. Then $\hilb_1\oplus\hilb_2$ is separable and, given two orthonormal bases $(e_{1}^{(k)})_{k\in\mathbb{N}}\subset\hilb_1$, $(e_{2}^{(k)})_{k\in\mathbb{N}}\subset\hilb_2$, the family of vectors
		\begin{equation}
			\left((e_{1}^{(k)},0)\right)_{k\in\mathbb{N}}\cup\left((0,e_2^{(k)})\right)_{k\in\mathbb{N}}\subset\hilb_1\oplus\hilb_2
		\end{equation}
		is an orthonormal basis of $\hilb_1\oplus\hilb_2$.
	\end{proposition}
	
	\begin{exercise}\label{exerc:directsumbases}
		Prove this. Hint: orthonormality follows directly from Eq.~\eqref{eq:directsumsp}; for completeness, show that a vector $(u_1,u_2)$ orthogonal to all these vectors must have $u_1=0$ and $u_2=0$.
		\seesol{directsumbases}
	\end{exercise}
	
	\begin{example}[Spin-$\frac12$ particle]\label{examp:spinonehalf}\index{spin-1/2 particle@spin-$\frac12$ particle}
		As discussed in Example~\ref{examp:l2}, the Hilbert space of a spinless quantum particle in a region of space $\Omega\subseteq\mathbb{R}^d$ is $L^2(\Omega)$. However, particles may also carry an internal degree of freedom, the spin. The states of a spin-$\frac12$ particle, e.g. an electron, are elements of the Hilbert space
		\begin{equation}
			\hilb=L^2(\Omega)\oplus L^2(\Omega)=\left\{\begin{bmatrix}
				\varphi_+\\\varphi_-
			\end{bmatrix}:\varphi_+,\varphi_-\in L^2(\Omega)\right\}.
		\end{equation}
		Equivalently, the pair $(\varphi_+,\varphi_-)$ can be written as a $\mathbb{C}^2$-valued function
		\begin{equation}
			\Omega\ni x\mapsto\Phi(x)=\begin{bmatrix}
				\varphi_+(x)\\\varphi_-(x)
			\end{bmatrix}\in\mathbb{C}^2
		\end{equation}
		which satisfies
		\begin{equation}
			\int_\Omega\|\Phi(x)\|_{\mathbb{C}^2}^2\;\mathrm{d}x=\int_\Omega|\varphi_+(x)|^2\;\mathrm{d}x+\int_\Omega|\varphi_-(x)|^2\;\mathrm{d}x<\infty.
		\end{equation}
		That is, $L^2(\Omega)\oplus L^2(\Omega)$ can be identified with $L^2(\Omega,\mathbb{C}^2)$, the space of square-integrable, $\mathbb{C}^2$-valued functions on $\Omega$ (cf. Example~\ref{examp:l2vec}).
	\end{example}	
	The same definitions and properties immediately generalize to the direct sum of finitely many Hilbert spaces:
	\begin{equation}
		\hilb_1\oplus\hilb_2\oplus\cdots\oplus\hilb_N\qquad\text{or}\qquad	\bigoplus_{j=1}^N\hilb_j,
	\end{equation}
	which is the space of $N$-tuples $(u_j)_{j=1}^N$, with $u_1\in\hilb_1,\ldots,u_N\in\hilb_N$, and scalar product
	\begin{equation}
		\Braket{(u_j)_{j=1}^N,(v_j)_{j=1}^N}:=\sum_{j=1}^N\braket{u_j,v_j}_{\hilb_j}.
	\end{equation}
	
	For our purposes, we also need to consider the direct sum of countably many Hilbert spaces:
	
	\begin{definition}[Direct sum of a countable family of Hilbert spaces]\label{def:directsum_countable}\index{direct sum!countable}
		Let $(\hilb_n)_{n\in\mathbb{N}}$ be a countable family of Hilbert spaces, each with scalar product $\braket{\cdot,\cdot}_{\hilb_n}$ and associated norm $\|\cdot\|_{\hilb_n}$. Their \emph{orthogonal direct sum} $\oplus_{n\in\mathbb{N}}\hilb_n$ is the space
		\begin{equation}
			\left\{(u_n)_{n\in\mathbb{N}}:\,u_n\in\hilb_n,\;\sum_{n\in\mathbb{N}}\|u_n\|_{\hilb_n}^2<\infty\right\}
		\end{equation}
		endowed with the scalar product
		\begin{equation}
			\Braket{(u_n)_{n\in\mathbb{N}},(v_n)_{n\in\mathbb{N}}}_{\oplus_n\hilb_n}:=\sum_{n\in\mathbb{N}}\braket{u_n,v_n}_{\hilb_n}.
		\end{equation}
	\end{definition}
	
	\begin{example}
		$\ell^2=\bigoplus_{n\in\mathbb{N}}\mathbb{C}$.
	\end{example}
	
	As we will see later on (Definition~\ref{def:fock}), countable direct sums enter the definition of Fock spaces, with each $\hilb_n$ representing the space of all $n$-particle states of a field.
	
	\begin{proposition}
		Let $(\hilb_n)_{n\in\mathbb{N}}$ be a countable family of Hilbert spaces. Their orthogonal direct sum $\oplus_{n\in\mathbb{N}}\hilb_n$ is a Hilbert space. Furthermore, if each $\hilb_n$ is separable, then $\oplus_{n\in\mathbb{N}}\hilb_n$ is separable.
	\end{proposition}
	
	\begin{proof}
		Clearly $\oplus_{n\in\mathbb{N}}\hilb_n$ is a vector space, and $\braket{\cdot,\cdot}_{\oplus_n\hilb_n}$ is a well-defined scalar product. Completeness is proven exactly as for $\ell^2$ (see e.g. \cite[Section 0.2]{teschl}): if $(u^{(k)})_{k\in\mathbb{N}}$ is a Cauchy sequence, then, for each $n$, $(u^{(k)}_n)_{k\in\mathbb{N}}$ is a Cauchy sequence in $\hilb_n$, and thus converges to some $u_n\in\hilb_n$; bounding the finite partial sums $\sum_{n=0}^N\|u_n-u_n^{(k)}\|^2$ uniformly in $N$, one shows that $u=(u_n)_{n\in\mathbb{N}}\in\oplus_{n\in\mathbb{N}}\hilb_n$ and $\|u^{(k)}-u\|\to0$. Separability follows by joining together orthonormal bases of all spaces $\hilb_n$, as in the finite case (Proposition~\ref{prop:directsumbases}).
	\end{proof}
	
	\begin{definition}\label{def:hfin}
		Given a countable family of Hilbert spaces $(\hilb_n)_{n\in\mathbb{N}}$, let $\hilb_{\rm fin}\subset\oplus_{n\in\mathbb{N}}\hilb_n$ be defined by
		\begin{equation}
			\hilb_{\rm fin}:=\bigcup_{N\in\mathbb{N}}\bigoplus_{n=0}^N\hilb_n,
		\end{equation}
		i.e., explicitly, the subspace of all sequences $(u_n)_{n\in\mathbb{N}}$ with \emph{finitely many nonzero} entries:
		\begin{equation}
			\hilb_{\rm fin}=\left\{(u_n)_{n\in\mathbb{N}}\in\bigoplus_{n\in\mathbb{N}}\hilb_n:\;\exists N\in\mathbb{N}\,\text{ s.t. }u_n=0\,\text{ for all }\,n> N\right\}.
		\end{equation}
	\end{definition}
	
	\begin{proposition}\label{prop:hfin}
		$\hilb_{\rm fin}$ is a dense subspace of $\oplus_{n\in\mathbb{N}}\hilb_n$.
	\end{proposition}
	
	\begin{exercise}\label{exerc:hfin}
		Prove this. Hint: given $u=(u_n)_{n\in\mathbb{N}}\in\oplus_n\hilb_n$, consider the truncated sequences $u^{(k)}\in\hilb_{\rm fin}$, with $u^{(k)}_n=u_n$ for $n\leq k$ and $u^{(k)}_n=0$ for $n>k$, and prove $\|u^{(k)}-u\|\to0$.
		\seesol{hfin}
	\end{exercise}
	
	When dealing with Fock spaces, this will mean that states with a finite---but arbitrarily large---number of particles form a dense subspace (Proposition~\ref{prop:finitefock_dense}).
	
	\subsection{Direct sums of operators}\label{subsec:directsumops}
	
	From now on we state our definitions and results directly for countably many spaces; everything applies a fortiori to the finite case.
	
	\begin{definition}[Orthogonal direct sum of operators]\label{def:directsumoperators}\index{direct sum!of operators}
		Let $(\hilb_n)_{n\in\mathbb{N}}$ be a countable family of Hilbert spaces, with direct sum $\hilb=\oplus_{n\in\mathbb{N}}\hilb_n$, and $(A_n)_{n\in\mathbb{N}}$ a countable family of operators on them, with $A_n:\Dom A_n\subseteq\hilb_n\rightarrow\hilb_n$. Their \emph{direct sum}, written as $A:=\bigoplus_{n\in\mathbb{N}}A_n$,
		is the operator on $\hilb$ defined by
		\begin{align}
			\Dom A&:=\left\{u=(u_n)_{n\in\mathbb{N}}\in\hilb:\,u_n\in\Dom A_n,\;\sum_{n\in\mathbb{N}}\|A_nu_n\|^2_{\hilb_n}<\infty\right\};\\
			(Au)_n&:=A_nu_n.
		\end{align}
	\end{definition}
	
	Note that the domain of $A$ consists of sequences $u\in\hilb$ satisfying two conditions: each component $u_n$ is in the domain of the corresponding operator $A_n$, and additionally (if the direct sum involves infinitely many spaces) the sum of all $\|A_nu_n\|^2_{\hilb_n}$ must converge. If so, clearly $\|Au\|^2_{\hilb}=\sum_{n}\|A_nu_n\|^2_{\hilb_n}$.
	
	We expect $A$ to inherit the properties of its components $A_n$. We begin with the following:
	
	\begin{proposition}\label{prop:densedomain}
		Let $(A_n)_{n\in\mathbb{N}}$ be a family of densely defined operators on $(\hilb_n)_{n\in\mathbb{N}}$, and $A=\oplus_{n\in\mathbb{N}}A_n$ their direct sum on $\hilb=\oplus_{n\in\mathbb{N}}\hilb_n$. Then:
		\begin{itemize}
			\item[(i)] $\Dom A$ is dense in $\hilb$;
			\item[(ii)] $\Dom A\cap\hilb_{\rm fin}$ (cf. Definition~\ref{def:hfin}) is also dense in $\hilb$.
		\end{itemize}
	\end{proposition}
	
	\begin{proof}
		Since $\Dom A\cap\hilb_{\rm fin}\subseteq\Dom A$, it suffices to prove (ii). Given $u\in\hilb$ and $\epsilon>0$, by Proposition~\ref{prop:hfin} we can first approximate $u$ within $\epsilon/2$ by some $u'\in\hilb_{\rm fin}$, with $u'_n=0$ for $n>N$; then, since each $\Dom A_n$ is dense in $\hilb_n$, we can approximate each of the finitely many components $u'_0,\dots,u'_N$ by $u''_n\in\Dom A_n$, with $\|u'_n-u''_n\|<\epsilon/(2\sqrt{N+1})$. The vector $u''=(u''_0,\dots,u''_N,0,0,\dots)$ belongs to $\Dom A\cap\hilb_{\rm fin}$ and satisfies $\|u-u''\|<\epsilon$.
	\end{proof}
	
	\begin{proposition}[Adjoint of direct sum]\label{prop:adjoint_directsum}
		Let $(A_n)_{n\in\mathbb{N}}$ be a family of densely defined operators on $(\hilb_n)_{n\in\mathbb{N}}$, and $A=\oplus_{n\in\mathbb{N}}A_n$ their direct sum on $\hilb=\oplus_{n\in\mathbb{N}}\hilb_n$. Then
		\begin{equation}
			A^*=\bigoplus_{n\in\mathbb{N}}A_n^*.
		\end{equation}
		In particular, $A$ is symmetric if and only if each $A_n$ is symmetric, and self-adjoint if and only if each $A_n$ is self-adjoint.
	\end{proposition}
	
	\begin{proof}
		We must show that the operators $\oplus_{n\in\mathbb{N}}A_n^*$, defined by
		\begin{align}
			\Dom\left(\oplus_nA_n^*\right)&=\left\{v=(v_n)_{n\in\mathbb{N}}:v_n\in\Dom A_n^*,\;\sum_n\|A_n^*v_n\|^2_{\hilb_n}<\infty\right\};\\
			(\oplus_nA_n^*v)_n&=A^*_nv_n,
		\end{align}
		and $A^*$, defined by
		\begin{align}
			\Dom A^*&=\left\{v=(v_n)_{n\in\mathbb{N}}:\exists\tilde{v}\in\hilb\,\text{ s.t. }\,\braket{\tilde{v},u}_\hilb=\braket{v,Au}_\hilb\;\forall u\in\Dom A\right\};\\
			A^*v&=\tilde{v},
		\end{align}
		coincide, i.e. have the same domain and the same action.
		
		\medskip 
		\emph{Step 1}: prove $\oplus_nA^*_n\subseteq A^*$. Let $v\in\Dom(\oplus_nA_n^*)$. This means that each component $v_n$ is in $\Dom A^*_n$, that is, there exists $\tilde{v}_n\in\hilb_n$ such that
		\begin{equation}
			\braket{\tilde{v}_n,u_n}_{\hilb_n}=\braket{v_n,A_nu_n}_{\hilb_n}\qquad\text{for all }u_n\in\Dom A_n,
		\end{equation}
		and moreover $\sum_n\|A^*_nv_n\|^2_{\hilb_n}<\infty$, i.e. $\sum_n\|\tilde{v}_n\|^2_{\hilb_n}<\infty$.
		Therefore $\tilde{v}=(\tilde{v}_n)_{n\in\mathbb{N}}$ is an element of $\hilb$, and, for all $u=(u_n)_{n\in\mathbb{N}}\in\Dom A$,	
		\begin{equation}
			\braket{\tilde{v},u}_\hilb=\sum_{n\in\mathbb{N}}\braket{\tilde{v}_n,u_n}_{\hilb_n}=\sum_{n\in\mathbb{N}}\braket{v_n,A_nu_n}_{\hilb_n}=\braket{v,Au}_\hilb,
		\end{equation}
		meaning that $v\in\Dom A^*$ and $A^*v=\tilde{v}=\oplus_nA_n^*v_n$.
		
		\medskip
		\emph{Step 2}: prove $A^*\subseteq\oplus_nA^*_n$.		
		Let $v\in\Dom A^*$, that is, there exists $\tilde{v}=A^*v\in\hilb$ such that, for all $u\in\Dom A$, $\braket{A^*v,u}_\hilb=\braket{v,Au}_{\hilb}$.
		In particular, choosing $u=u^{(k)}:=(0,\ldots,0,u_k,0,\ldots)$ with $u_k\in\Dom A_k$, so that $u^{(k)}\in\Dom A$ and $Au^{(k)}=(0,\ldots,0,A_ku_k,0,\ldots)$, this equation reduces to $\braket{(A^*v)_k,u_k}_{\hilb_k}=\braket{v_k,A_ku_k}_{\hilb_k}$.
		This means precisely that each component $v_k$ of $v$ is in $\Dom A_k^*$, with $(A^*v)_k=A^*_kv_k$. Moreover $\sum_k\|A_k^*v_k\|^2_{\hilb_k}=\|A^*v\|^2_\hilb<\infty$, so $v\in\Dom(\oplus_nA_n^*)$, thus proving the claim.
	\end{proof}
	
	In particular, the direct sum of self-adjoint operators is self-adjoint (cf. \cite[Theorem 2.24]{teschl}). The same holds for essential self-adjointness:
	
	\begin{proposition}\label{prop:directsum_esssa}
		Let $(A_n)_{n\in\mathbb{N}}$ be a family of (essentially) self-adjoint operators on $(\hilb_n)_{n\in\mathbb{N}}$, and $A=\oplus_{n\in\mathbb{N}}A_n$ their direct sum on $\hilb=\oplus_{n\in\mathbb{N}}\hilb_n$. Then $A$ is (essentially) self-adjoint. Besides, the space $\Dom A\cap\hilb_{\rm fin}$ (cf. Definition~\ref{def:hfin}) is a core for $A$, i.e. the restriction of $A$ to $\Dom A\cap\hilb_{\rm fin}$ is also essentially self-adjoint.
	\end{proposition}
	
	\begin{exercise}\label{exerc:directsumesssa}
		Prove this by using the criterion of (essential) self-adjointness (Theorems~\ref{thm:criterion-sa}--\ref{thm:criterion-esssa}).
		\seesol{directsumesssa}
	\end{exercise}
	
	\begin{proposition}\label{prop:closure_directsum}
		Let $(A_n)_{n\in\mathbb{N}}$ be a family of densely defined, closable operators on $(\hilb_n)_{n\in\mathbb{N}}$, and $A=\oplus_{n\in\mathbb{N}}A_n$. Then $A$ is closable and
		\begin{equation}
			\overline{A}=\bigoplus_{n\in\mathbb{N}}\overline{A_n}.
		\end{equation}
		In particular, the direct sum of closed operators is closed.
	\end{proposition}
	
	\begin{proof}
		Since each $A_n$ is closable, each $A_n^*$ is densely defined (Proposition~\ref{prop:closableiffadjointdense}). By Proposition~\ref{prop:adjoint_directsum}, $A^*=\oplus_nA_n^*$, which is densely defined by Proposition~\ref{prop:densedomain}; hence $A$ is closable. Applying Proposition~\ref{prop:adjoint_directsum} once more,
		\begin{equation}
			\overline{A}=A^{**}=\Bigl(\bigoplus_{n\in\mathbb{N}}A_n^*\Bigr)^*=\bigoplus_{n\in\mathbb{N}}A_n^{**}=\bigoplus_{n\in\mathbb{N}}\overline{A_n}.\qedhere
		\end{equation}
	\end{proof}
	
	One property that does not automatically translate to the direct sum is boundedness: the direct sum of countably many bounded operators may well be unbounded.
	
	\begin{example}\label{examp:protonumber}
		Let $(\hilb_n)_{n\in\mathbb{N}}$ be a family of Hilbert spaces, and $I_n$ the identity operator on each space. The operator $\N:=\oplus_{n\in\mathbb{N}}nI_n$ on $\hilb=\oplus_n\hilb_n$ is unbounded. To see this, take any sequence of vectors $(u^{(k)})_{k\in\mathbb{N}}\subset\hilb$ such that
		\begin{equation}
			u^{(k)}=(0,0,\ldots,0,\overbrace{u_k}^{k\text{th}},0,\ldots),\qquad u_k\in\hilb_k,\;\|u_k\|_{\hilb_k}=1.
		\end{equation}
		Then $\|u^{(k)}\|_\hilb=1$ but $\|\N u^{(k)}\|_\hilb=k\to\infty$ as $k\to\infty$. This is also reflected in the domain of $\N$, which by definition is
		\begin{equation}
			\Dom\N=\left\{u\in\hilb:\sum_{n\in\mathbb{N}}n^2\|u_n\|^2_{\hilb_n}<\infty\right\}\subsetneq\hilb.
		\end{equation}
		If each $\hilb_n$ is interpreted as the space of $n$-particle states of a quantum field, $\N$ ``counts'' the particles: as we will see in Example~\ref{examp:number_fock}, it is precisely the number operator on the Fock space.
	\end{example}
	
	\begin{exercise}\label{exerc:uniformly}
		Show that the direct sum of a family of uniformly bounded operators---that is, $\|A_n\|\leq M$ for all $n\in\mathbb{N}$---is a bounded operator on $\hilb=\oplus_{n\in\mathbb{N}}\hilb_n$, with $\|A\|\leq M$ (cf. \cite[Problem 2.27]{teschl}). Hint: estimate $\|Av\|^2=\sum_n\|A_nv_n\|^2$ for $v=(v_n)_{n\in\mathbb{N}}$.
		\seesol{uniformly}
	\end{exercise}
	
	\subsection{Tensor products of Hilbert spaces}\label{sec:tensor}
	
	As discussed in Axiom~\ref{ax:composite}, the tensor product implements the logical ``AND'' between two systems, written as $\ket{u}\otimes\ket{v}$, $\ket{u}\ket{v}$, or $\ket{u,v}$ in the bra-ket notation. In finite dimensions, the construction is straightforward; for instance, for complex tuples,
	
	\begin{equation}
		u=\begin{bmatrix}
			u_1\\u_2\\u_3
		\end{bmatrix}\in\mathbb{C}^{3},\quad	
		v=\begin{bmatrix}
			v_1\\v_2
		\end{bmatrix}\in\mathbb{C}^{2},\quad u\otimes v=\begin{bmatrix}
			u_1v_1\\u_1v_2\\u_2v_1\\u_2v_2\\u_3v_1\\u_3v_2\\
		\end{bmatrix}\in\mathbb{C}^{3\cdot 2}.
	\end{equation}
	The extension to generic finite-dimensional spaces---and its independence of the particular choice of bases on them---is then immediate.
	
	The tensor product is bilinear: $(\alpha u)\otimes v=u\otimes(\alpha v)=\alpha(u\otimes v)$ and
	\begin{equation}
		u\otimes(v+w)=u\otimes v+u\otimes w,\qquad (u+v)\otimes w=u\otimes w+v\otimes w.
	\end{equation}	
	In infinite dimensions, as usual, some additional care is required. The safest option is to rely on a basis-independent construction---one that avoids dealing directly with infinite sums.	
	Our starting point is to make sense of expressions like $u\otimes v$, with $u\in\hilb_1$ and $v\in\hilb_2$ elements of two Hilbert spaces. Among the various equivalent constructions, we choose one based on \emph{conjugate-bilinear forms}; here and in Section~\ref{subsec:tensoroperators}, our presentation closely follows~\cite[Section 7.5]{schmuedgen}, including the notation $\odot$ and $\otimes$ for algebraic tensor products and their completions or closures.
	
	Recall that, by Riesz's representation theorem\index{Riesz representation theorem}, every element $v$ of a Hilbert space $\hilb$ is uniquely associated with a bounded linear form on $\hilb$: $v\leftrightarrow \braket{v,\cdot}$,
	that is, the bounded linear form defined by $\hilb\ni u\mapsto\braket{v,u}\in\mathbb{C}$; or, equivalently, with a bounded conjugate-linear form: $v\leftrightarrow \braket{\cdot,v}$.
	With this second choice, the form depends linearly on $v$: $\braket{\cdot,\alpha v+\beta w}=\alpha\braket{\cdot,v}+\beta\braket{\cdot,w}$.
	
	The idea is the following: given $v_1\in\hilb_1$ and $v_2\in\hilb_2$, we define $v_1\otimes v_2$ as a bounded conjugate-bilinear form on $\hilb_1\times\hilb_2$.
	
	\begin{definition}[Algebraic tensor product]\label{def:algtensorhs}\index{tensor product!of Hilbert spaces, algebraic}
		Let $\hilb_1,\hilb_2$ be Hilbert spaces. Given $v_1\in\hilb_1$ and $v_2\in\hilb_2$, we define their \emph{tensor product} $v_1\otimes v_2$ as the bounded conjugate-bilinear form $v_1\otimes v_2:\hilb_1\times\hilb_2\rightarrow\mathbb{C}$ defined by
		\begin{equation}
			(u_1,u_2)\in\hilb_1\times\hilb_2\mapsto (v_1\otimes v_2)(u_1,u_2):=\braket{u_1,v_1}_{\hilb_1}\braket{u_2,v_2}_{\hilb_2}.
		\end{equation}
		The \emph{algebraic tensor product} of $\hilb_1$ and $\hilb_2$, denoted by $\hilb_1\odot\hilb_2$, is the space of all conjugate-bilinear forms that can be obtained as finite linear combinations of tensor products of elements of $\hilb_1$ and $\hilb_2$:
		\begin{equation}
			\hilb_1\odot\hilb_2=\left\{\Psi=\sum_{j=1}^nv_j\otimes w_j,\; n\in\mathbb{N},\;v_1,\dots,v_n\in\hilb_1,\;w_1,\dots,w_n\in\hilb_2\right\}.
		\end{equation}
	\end{definition}
	The reason for denoting this space by $\hilb_1\odot\hilb_2$ rather than $\hilb_1\otimes\hilb_2$ will become clear after Definition~\ref{def:hs_tensorproduct}.
	
	Clearly, $\hilb_1\odot\hilb_2$ is a vector space; we now want to equip it with a scalar product. The idea is to set
	\begin{equation}\label{eq:tensorsp}
		\Braket{u_1\otimes u_2,v_1\otimes v_2}_{\hilb_1\odot\hilb_2}:=\braket{u_1,v_1}_{\hilb_1}\braket{u_2,v_2}_{\hilb_2},
	\end{equation}
	and then extend it by linearity to all elements of $\hilb_1\odot\hilb_2$. 	
	An important caveat is in order: since elements of $\hilb_1\odot\hilb_2$ admit multiple representations as linear combinations of pure tensors,
	\begin{equation}
		\Psi=\sum_{j=1}^nv_j\otimes w_j=\sum_{j'=1}^{n'}v'_{j'}\otimes w'_{j'},
	\end{equation}
	it is not obvious a priori that, given $\Phi,\Psi\in\hilb_1\odot\hilb_2$, the quantity $\Braket{\Psi,\Phi}_{\hilb_1\odot\hilb_2}$ is independent of the particular decomposition; if it were not, the product above would not be well-defined. We now show that it is.
	
	\begin{proposition}\label{prop:welldef_scalarproduct}
		Let $\hilb_1,\hilb_2$ be two Hilbert spaces, and $\hilb_1\odot\hilb_2$ their algebraic tensor product. Then the product $\braket{\cdot,\cdot}_{\hilb_1\odot\hilb_2}$ on $\hilb_1\odot\hilb_2$ defined by Eq.~\eqref{eq:tensorsp}, and extended by linearity to all vectors in $\hilb_1\odot\hilb_2$, uniquely defines a scalar product on $\hilb_1\odot\hilb_2$.
	\end{proposition}
	
	\begin{proof}
		We first show that the product is independent of the representation, and thus well-defined. By sesquilinearity, this amounts to proving the following: $\braket{\Phi,\Psi}_{\hilb_1\odot\hilb_2}=0$ whenever either $\Psi$ or $\Phi$ is the null element of $\hilb_1\odot\hilb_2$, i.e. the null conjugate-bilinear form on $\hilb_1\times\hilb_2$.
		
		Let us write $\Psi=\sum_{j=1}^nu_j\otimes v_j$ and $\Phi=\sum_{k=1}^{n'}u'_k\otimes v'_k$, and assume that $\Phi$ is the null form: $\Phi=0$. In particular, $\Phi(u_j,v_j)=0$ for all $j=1,\dots,n$, whence
		\begin{align}
			0&=\sum_{j=1}^n\Phi(u_j,v_j)
			=\sum_{j=1}^n\left(\sum_{k=1}^{n'}u'_k\otimes v'_k\right)(u_j,v_j)\nonumber\\
			&=\sum_{j=1}^n\sum_{k=1}^{n'}\braket{u_j,u'_k}_{\hilb_1}\braket{v_j,v'_k}_{\hilb_2}=\overline{\sum_{j=1}^n\sum_{k=1}^{n'}\braket{u'_k,u_j}_{\hilb_1}\braket{v'_k,v_j}_{\hilb_2}},
		\end{align}
		where we expressed $\Phi$ as a linear combination of tensor products, and used the fact that, by definition, $(u'_k\otimes v'_k)(u_j,v_j)=\braket{u_j,u'_k}_{\hilb_1}\braket{v_j,v'_k}_{\hilb_2}$.		
		But the double sum under the complex conjugation is exactly $\Braket{\Phi,\Psi}_{\hilb_1\odot\hilb_2}$. Indeed, by linearity,
		\begin{align}
			\Braket{\Phi,\Psi}_{\hilb_1\odot\hilb_2}&=\Braket{\sum_{k=1}^{n'}u'_k\otimes v'_k,\sum_{j=1}^nu_j\otimes v_j}=\sum_{j=1}^n\sum_{k=1}^{n'}\braket{u'_k\otimes v'_k,u_j\otimes v_j}_{\hilb_1\odot\hilb_2}\nonumber\\
			&=\sum_{j=1}^n\sum_{k=1}^{n'}\braket{u'_k,u_j}_{\hilb_1}\braket{v'_k,v_j}_{\hilb_2}.
		\end{align}
		We have thus proved that $\Braket{\Phi,\Psi}_{\hilb_1\odot\hilb_2}=0$ whenever one of the two vectors is the zero of $\hilb_1\odot\hilb_2$. 		
		Hence $\Braket{\cdot,\cdot}_{\hilb_1\odot\hilb_2}$ is well-defined; it is clearly sesquilinear and Hermitian, and we are left with proving positive definiteness.
		
		Let $\Psi\in\hilb_1\odot\hilb_2$, thus $\Psi=\sum_{j=1}^nu_j\otimes v_j$ for some $u_1,\dots,u_n\in\hilb_1$ and $v_1,\dots,v_n\in\hilb_2$. Consider the vector spaces $\mathcal{U}:=\operatorname{Span}(u_1,\dots,u_n)$, $\mathcal{V}:=\operatorname{Span}(v_1,\dots,v_n)$, 
		with dimensions $d,d'\leq n$ respectively. Picking orthonormal bases $(e_1,\dots,e_{d})$ and $(f_1,\dots,f_{d'})$ of $\mathcal{U}$ and $\mathcal{V}$, we have 
		\begin{equation}
			u_j=\sum_{k=1}^d\braket{e_k,u_j}_{\hilb_1}e_k,\qquad v_j=\sum_{k'=1}^{d'}\braket{f_{k'},v_j}_{\hilb_2}f_{k'},
		\end{equation}
		thus
		\begin{align}
			\Psi&=\sum_{j=1}^n\sum_{k=1}^d\sum_{k'=1}^{d'}\braket{e_k,u_j}_{\hilb_1}\braket{f_{k'},v_j}_{\hilb_2}e_k\otimes f_{k'}\nonumber\\
			&=\sum_{k=1}^d\sum_{k'=1}^{d'}\left(\sum_{j=1}^n\braket{e_k,u_j}_{\hilb_1}\braket{f_{k'},v_j}_{\hilb_2}\right)e_k\otimes f_{k'}\equiv\sum_{k=1}^d\sum_{k'=1}^{d'}c_{kk'}\,e_k\otimes f_{k'},
		\end{align}
		and therefore
		\begin{align}
			\Braket{\Psi,\Psi}_{\hilb_1\odot\hilb_2}&=\sum_{k,h=1}^d\sum_{k',h'=1}^{d'}c_{kk'}\overline{c_{hh'}}\Braket{e_h\otimes f_{h'},e_k\otimes f_{k'}}_{\hilb_1\odot\hilb_2}\nonumber\\
			&=\sum_{k,h=1}^d\sum_{k',h'=1}^{d'}c_{kk'}\overline{c_{hh'}}\braket{e_h,e_k}_{\hilb_1}\braket{f_{h'},f_{k'}}_{\hilb_2}=\sum_{k=1}^d\sum_{k'=1}^{d'}|c_{kk'}|^2,
		\end{align}
		and $\Braket{\Psi,\Psi}_{\hilb_1\odot\hilb_2}\geq0$, and $\Braket{\Psi,\Psi}_{\hilb_1\odot\hilb_2}=0$ if and only if all coefficients $c_{kk'}$ vanish, in which case $\Psi=0$, proving positive definiteness.		
	\end{proof}
	
	Therefore, $\hilb_1\odot\hilb_2$ is a pre-Hilbert space. However, in general it is not complete.
	
	\begin{example}
		Let $\hilb$ be a separable infinite-dimensional Hilbert space with orthonormal basis $(e_k)_{k\in\mathbb{N}}$, and
		\begin{equation}
			v_n:=\sum_{k=1}^n\frac{1}{k}e_k\otimes e_k\in\hilb\odot\hilb,\qquad n\in\mathbb{N}.
		\end{equation}
		Since the vectors $e_k\otimes e_k$ are orthonormal, $\|v_n-v_m\|^2_{\hilb\odot\hilb}=\sum_{k=m+1}^n k^{-2}$ for $n\geq m$, so that $(v_n)_{n\in\mathbb{N}}$ is a Cauchy sequence. However, it does not converge in $\hilb\odot\hilb$. Indeed, to every $v=\sum_{r=1}^Nv_r\otimes v'_r\in\hilb\odot\hilb$ we can associate the (representation-independent) infinite matrix
		\begin{equation}
			c_{ij}(v):=\braket{e_i\otimes e_j,v}_{\hilb\odot\hilb}=\sum_{r=1}^N\braket{e_i,v_r}\braket{e_j,v'_r},\qquad i,j\in\mathbb{N},
		\end{equation}
		whose columns all belong to the span of the $N$ sequences $(\braket{e_i,v_r})_{i\in\mathbb{N}}$, $r=1,\dots,N$: that is, $(c_{ij}(v))_{i,j}$ has finite rank. If $v_n\to v\in\hilb\odot\hilb$, then $c_{ij}(v)=\lim_nc_{ij}(v_n)$ equals $\delta_{ij}/i$ for $i,j\geq1$ (and vanishes if $i=0$ or $j=0$), which is a matrix of infinite rank---a contradiction.
	\end{example}
	
	It is therefore necessary to take the completion, which brings vectors such as $\sum_{k=1}^\infty \frac{1}{k}e_k\otimes e_k$ into the picture:
	
	\begin{definition}[Hilbert space tensor product]\label{def:hs_tensorproduct}\index{tensor product!of Hilbert spaces}
		Let $\hilb_1,\hilb_2$ be two Hilbert spaces. Their \emph{Hilbert space tensor product} $\hilb_1\otimes\hilb_2$ is the \emph{completion} of the algebraic tensor product $\hilb_1\odot\hilb_2$ with respect to the norm $\|\cdot\|_{\hilb_1\odot\hilb_2}$:
		\begin{equation}
			\hilb_1\otimes\hilb_2:=\overline{\hilb_1\odot\hilb_2}^{\|\cdot\|_{\hilb_1\odot\hilb_2}}.
		\end{equation}
	\end{definition}
	From now on, the scalar product and norm of $\hilb_1\otimes\hilb_2$ are denoted by $\braket{\cdot,\cdot}_{\hilb_1\otimes\hilb_2}$ and $\|\cdot\|_{\hilb_1\otimes\hilb_2}$.	
	If at least one of the two spaces is finite-dimensional, no completion is needed:
	
	\begin{proposition}\label{prop:tensorproduct_finitedim}
		Let $\hilb_1,\hilb_2$ be two Hilbert spaces, and $\dim\hilb_1=:d<\infty$. Then $\hilb_1\odot\hilb_2$ is unitarily equivalent to the direct sum of $d$ copies of $\hilb_2$:
		\begin{equation}
			\hilb_1\odot\hilb_2\cong\bigoplus_{j=1}^d\hilb_2=:(\hilb_2)^{\oplus d}.
		\end{equation}
		In particular, $\hilb_1\odot\hilb_2$ is already complete, and thus $\hilb_1\otimes\hilb_2=\hilb_1\odot\hilb_2$.
	\end{proposition}
	
	\begin{exercise}\label{exerc:tensorfinitedim}
		Prove this. Hint: take an orthonormal basis $(e_j)_{j=1,\dots,d}$ of $\hilb_1$. First, prove that any element of $\hilb_1\odot\hilb_2$ can be written as $\sum_{j=1}^de_j\otimes u_j$
		for some $u_1,\dots,u_d\in\hilb_2$. Then define a map $U:\hilb_1\odot\hilb_2\rightarrow(\hilb_2)^{\oplus d}$ via $U(\sum_{j=1}^de_j\otimes u_j):=(u_1,\dots,u_d)$, and prove that this map is isometric and bijective (hence unitary).
		Since $(\hilb_2)^{\oplus d}$ is complete, so is $\hilb_1\odot\hilb_2$.
		\seesol{tensorfinitedim}
	\end{exercise}
	
	\begin{example}\label{examp:spinonehalf2}
		In Example~\ref{examp:spinonehalf} we saw that the Hilbert space of a spin-$\frac12$ particle in $\Omega\subseteq\mathbb{R}^d$ is $\hilb=L^2(\Omega)\oplus L^2(\Omega)\cong L^2(\Omega,\mathbb{C}^2)$. By Proposition~\ref{prop:tensorproduct_finitedim}, $\mathbb{C}^2\otimes L^2(\Omega)\cong L^2(\Omega)\oplus L^2(\Omega)$, whence the Hilbert space of such a particle can be equivalently represented as $\mathbb{C}^2\otimes L^2(\Omega)$. Notice that, since $\mathbb{C}^2$ is finite-dimensional, the algebraic and Hilbert space tensor products coincide.
	\end{example}
	
	Orthonormal bases behave as expected under tensor products (cf. \cite[Lemma 1.10]{teschl}):
	
	\begin{proposition}\label{prop:tensorproductbases}
		Let $\hilb_1,\hilb_2$ be two separable Hilbert spaces. Then $\hilb_1\otimes\hilb_2$ is also separable and, given two orthonormal bases $(e_{1k})_{k\in\mathbb{N}}\subset\hilb_1$ and $(e_{2k})_{k\in\mathbb{N}}\subset\hilb_2$, the set $(e_{1h}\otimes e_{2k})_{h,k\in\mathbb{N}}\subset\hilb_1\otimes\hilb_2$ is an orthonormal basis of $\hilb_1\otimes\hilb_2$.
	\end{proposition}
	
	\begin{exercise}\label{exerc:tensorbases}
		Prove this. Hint: since $\hilb_1\odot\hilb_2$ (which, recall, is the space of vectors of the form $\sum_{r=1}^Nv_r\otimes w_r$) is dense in $\hilb_1\otimes\hilb_2$ by construction, it suffices to show that $\operatorname{Span}(e_{1h}\otimes e_{2k})_{k,h\in\mathbb{N}}$ is dense in such a subspace.
		\seesol{tensorbases}
	\end{exercise}
	
	Tensor products also allow us to describe the state of a subsystem of a composite system, in terms of the density operators of Section~\ref{subsec:mixed}:
	
	\begin{proposition}[Partial trace]\label{prop:partialtrace}\index{partial trace}
		Let $\hilb_1,\hilb_2$ be two Hilbert spaces, and $T$ a trace-class operator on $\hilb_1\otimes\hilb_2$. Then there exists a unique trace-class operator $\operatorname{tr}_{\hilb_2}T$ on $\hilb_1$, the \emph{partial trace} of $T$ over $\hilb_2$, such that
		\begin{equation}\label{eq:partialtrace}
			\operatorname{tr}\left((A\otimes I)T\right)=\operatorname{tr}\left(A\operatorname{tr}_{\hilb_2}T\right)\qquad\text{for all }A\in\mathcal{B}(\hilb_1),
		\end{equation}
		where $A\otimes I$ is the bounded operator on $\hilb_1\otimes\hilb_2$ acting as $(A\otimes I)(u\otimes v)=Au\otimes v$ on pure tensors. Explicitly, given any orthonormal basis $(f_k)_{k\in\mathbb{N}}$ of $\hilb_2$, $\braket{u,(\operatorname{tr}_{\hilb_2}T)v}=\sum_k\braket{u\otimes f_k,T(v\otimes f_k)}$ for all $u,v\in\hilb_1$. Furthermore, if $T\geq0$, then $\operatorname{tr}_{\hilb_2}T\geq0$ and $\operatorname{tr}(\operatorname{tr}_{\hilb_2}T)=\operatorname{tr}T$; in particular, $\operatorname{tr}_{\hilb_2}$ maps density operators on $\hilb_1\otimes\hilb_2$ into density operators on $\hilb_1$.
	\end{proposition}
	
	\begin{proof}
		We first make sense of $A\otimes I$. Every $\Psi\in\hilb_1\odot\hilb_2$ can be written as $\Psi=\sum_{k=1}^mw_k\otimes g_k$ with $(g_k)_{k=1}^m$ orthonormal in $\hilb_2$ (expand the second factors on an orthonormal basis of their span, as in the proof of Proposition~\ref{prop:welldef_scalarproduct}), so that $\|\Psi\|^2=\sum_k\|w_k\|^2$ and $\|(A\odot I)\Psi\|^2=\sum_k\|Aw_k\|^2\leq\|A\|^2\|\Psi\|^2$. Hence $A\odot I$ extends by continuity to a bounded operator $A\otimes I$ on $\hilb_1\otimes\hilb_2$ (consistently with Definition~\ref{def:hs_tensorproduct_op} below; see also~\cite[Proposition 7.20(ii)]{schmuedgen}), and $(A\otimes I)^*=A^*\otimes I$, as one checks on pure tensors.
		
		Fix orthonormal bases $(e_n)_{n\in\mathbb{N}}$ of $\hilb_1$ and $(f_k)_{k\in\mathbb{N}}$ of $\hilb_2$, so that $(e_n\otimes f_k)_{n,k\in\mathbb{N}}$ is an orthonormal basis of $\hilb_1\otimes\hilb_2$ (Proposition~\ref{prop:tensorproductbases}). Recall that trace-class operators form a vector space, and that every trace-class operator is a linear combination of four nonnegative trace-class operators~\cite[Section VI.6]{reedsimon1}.
		
		Let first $T\geq0$. Given a unit vector $u\in\hilb_1$, the vectors $u\otimes f_k$ form an orthonormal family, which can be completed to an orthonormal basis of $\hilb_1\otimes\hilb_2$; since the trace does not depend on the basis and all diagonal matrix elements of $T$ are nonnegative, $\sum_k\braket{u\otimes f_k,T(u\otimes f_k)}\leq\operatorname{tr}T$, and thus $\sum_k\braket{u\otimes f_k,T(u\otimes f_k)}\leq\operatorname{tr}T\,\|u\|^2$ for all $u\in\hilb_1$. Applying the Cauchy--Schwarz inequality first to the nonnegative Hermitian form $(\Phi,\Psi)\mapsto\braket{\Phi,T\Psi}$, term by term, and then to the resulting sum, we see that the series
		\begin{equation}\label{eq:partialtraceform}
			\tau(u,v):=\sum_{k\in\mathbb{N}}\braket{u\otimes f_k,T(v\otimes f_k)}
		\end{equation}
		converges absolutely for all $u,v\in\hilb_1$, with $|\tau(u,v)|\leq\operatorname{tr}T\,\|u\|\|v\|$. Being sesquilinear and bounded, $\tau$ defines a bounded operator $S$ on $\hilb_1$ via $\braket{u,Sv}=\tau(u,v)$ (by Riesz's representation theorem, applied for fixed $v$). Clearly $S\geq0$, and, since nonnegative double series can be summed in any order,
		\begin{equation}
			\sum_{n\in\mathbb{N}}\braket{e_n,Se_n}=\sum_{n\in\mathbb{N}}\sum_{k\in\mathbb{N}}\braket{e_n\otimes f_k,T(e_n\otimes f_k)}=\operatorname{tr}T<\infty,
		\end{equation}
		so that $S=|S|$ is trace class, with $\operatorname{tr}S=\operatorname{tr}T$. For a general trace-class $T$, writing it as a linear combination of nonnegative trace-class operators, we see that the series~\eqref{eq:partialtraceform} still converges absolutely, and defines a trace-class operator $S$: the corresponding linear combination of the operators constructed above.
		
		We now show that $\operatorname{tr}_{\hilb_2}T:=S$ satisfies Eq.~\eqref{eq:partialtrace}. Let $A\in\mathcal{B}(\hilb_1)$. As $(A\otimes I)T$ is trace class, its trace, computed in the basis $(e_n\otimes f_k)_{n,k\in\mathbb{N}}$, converges absolutely, and can thus be summed in any order; using $(A\otimes I)^*=A^*\otimes I$,
		\begin{align}
			\operatorname{tr}\left((A\otimes I)T\right)&=\sum_{n\in\mathbb{N}}\sum_{k\in\mathbb{N}}\braket{A^*e_n\otimes f_k,T(e_n\otimes f_k)}\nonumber\\
			&=\sum_{n\in\mathbb{N}}\braket{A^*e_n,Se_n}=\sum_{n\in\mathbb{N}}\braket{e_n,ASe_n}=\operatorname{tr}(AS).
		\end{align}
		As for uniqueness, for the rank-one operator $A=\ket{v}\!\bra{u}$ we have $\operatorname{tr}(AS)=\sum_n\braket{e_n,v}\braket{u,Se_n}=\braket{u,Sv}$; hence Eq.~\eqref{eq:partialtrace} determines all matrix elements of $S$. In particular, $S$ does not depend on the choice of the bases, and the explicit expression in the statement is Eq.~\eqref{eq:partialtraceform}. Finally, if $T\geq0$, we have shown that $S\geq0$ and $\operatorname{tr}S=\operatorname{tr}T$.
	\end{proof}
	
	For instance, $\operatorname{tr}_{\hilb_2}(\rho_1\otimes\rho_2)=(\operatorname{tr}\rho_2)\rho_1$ and $\operatorname{tr}_{\hilb_2}\ket{u\otimes v}\!\bra{u'\otimes v'}=\braket{v',v}\ket{u}\!\bra{u'}$. If $\rho$ is the state of a composite system, Eq.~\eqref{eq:partialtrace} states that $\operatorname{tr}_{\hilb_2}\rho$ reproduces the expectation values of all bounded observables $A\otimes I$ pertaining to the first subsystem: it is the \emph{reduced state}\index{reduced state} of the first subsystem. The partial trace $\operatorname{tr}_{\hilb_1}$ over the first factor is defined analogously. See e.g.~\cite[Section 2.4.3]{nielsenchuang} for a physics-oriented discussion of partial traces and reduced states in finite dimensions.
	
	\begin{remark}[Entanglement and Schmidt decomposition]\label{rem:schmidt}\index{entanglement}\index{Schmidt decomposition}
		In the language of quantum information theory, a state of a bipartite system, described by a unit vector $\Psi\in\hilb_1\otimes\hilb_2$, is a \emph{product state} if $\Psi=u\otimes v$ for some $u\in\hilb_1$, $v\in\hilb_2$, and is \emph{entangled} otherwise. Every $\Psi\in\hilb_1\otimes\hilb_2$ admits a \emph{Schmidt decomposition}
		\begin{equation}\label{eq:schmidt}
			\Psi=\sum_{k}s_k\,e_k\otimes f_k,\qquad s_1\geq s_2\geq\ldots>0,\qquad\sum_ks_k^2=\|\Psi\|^2,
		\end{equation}
		with $(e_k)_k\subset\hilb_1$ and $(f_k)_k\subset\hilb_2$ orthonormal families, the sum being finite or countable. Indeed, by Proposition~\ref{prop:tensorproductbases}, $\Psi=\sum_{h,k}c_{hk}\,e_{1h}\otimes e_{2k}$ with $\sum_{h,k}|c_{hk}|^2=\|\Psi\|^2<\infty$; the matrix $(c_{hk})_{h,k}$ thus defines a Hilbert--Schmidt, and in particular compact, operator on $\ell^2$, and Eq.~\eqref{eq:schmidt} is nothing but its singular value decomposition (see e.g. \cite[Theorem VI.17]{reedsimon1}).
		The number of nonzero coefficients $s_k$, the \emph{Schmidt rank}, equals $1$ if and only if $\Psi$ is a product state; vectors of the algebraic tensor product $\hilb_1\odot\hilb_2$ are precisely those with finite Schmidt rank, which is another way to see why the vector $\sum_{k\geq1}k^{-1}e_k\otimes e_k$ does not belong to $\hilb\odot\hilb$. For $\|\Psi\|=1$, the reduced states of the two subsystems (Proposition~\ref{prop:partialtrace}) are $\operatorname{tr}_{\hilb_2}\ket{\Psi}\!\bra{\Psi}=\sum_ks_k^2\ket{e_k}\!\bra{e_k}$ and $\operatorname{tr}_{\hilb_1}\ket{\Psi}\!\bra{\Psi}=\sum_ks_k^2\ket{f_k}\!\bra{f_k}$: in particular, $\Psi$ is entangled if and only if its reduced states are mixed (Section~\ref{subsec:mixed}), and the \emph{entanglement entropy}\index{entanglement!entropy} $-\sum_ks_k^2\log s_k^2$---which, in infinite dimensions, may be infinite---quantifies how entangled $\Psi$ is.
		
		Spin--boson dynamics typically turns product states into entangled ones. In the pure dephasing model of Section~\ref{subsec:dephasing}, for instance, an initial product state $(c_+e_++c_-e_-)\otimes\Omega$ evolves into $c_+e_+\otimes\Omega_+(t)+c_-e_-\otimes\Omega_-(t)$, with $\Omega_\pm(t)=\e^{-\ii tH_\pm}\Omega$: for $c_\pm\neq0$, this vector has Schmidt rank $2$ unless $\Omega_+(t)$ and $\Omega_-(t)$ are parallel, and the loss of coherence of the qubit, measured by the overlap $\braket{\Omega_-(t),\Omega_+(t)}$, is precisely a manifestation of its entanglement with the field.
	\end{remark}
	
	\subsection{Tensor products of $L^2$ spaces}\label{subsec:tensorl2}
	
	We now turn to an important example: the tensor product of $L^2$ spaces (cf.~\cite[Example 7.9]{schmuedgen}). Recall (Appendix~\ref{app:measure}) that two $\sigma$-finite measure spaces $(\mathcal{M}_1,\Sigma_1,\mu_1)$ and $(\mathcal{M}_2,\Sigma_2,\mu_2)$ define a product measure space $(\mathcal{M}_1\times\mathcal{M}_2,\Sigma_1\otimes\Sigma_2,\mu_1\otimes\mu_2)$, on which integrals can be computed iteratively by Fubini's theorem; if $\mu_1$ and $\mu_2$ are the Lebesgue measures on $\Omega_1\subseteq\mathbb{R}^{d_1}$ and $\Omega_2\subseteq\mathbb{R}^{d_2}$, $\mu_1\otimes\mu_2$ is simply the Lebesgue measure on $\Omega_1\times \Omega_2\subseteq\mathbb{R}^{d_1+d_2}$.
	
	\begin{proposition}[Tensor product of $L^2$ spaces]\label{prop:tensorproduct_l2}\index{tensor product!of L2 spaces@of $L^2$ spaces}
		Let $(\mathcal{M}_1,\Sigma_1,\mu_1)$, $(\mathcal{M}_2,\Sigma_2,\mu_2)$ be two $\sigma$-finite measure spaces. Then $L^2(\mathcal{M}_1,\Sigma_1,\mu_1)\otimes L^2(\mathcal{M}_2,\Sigma_2,\mu_2)$ is unitarily equivalent to the $L^2$ space on $(\mathcal{M}_1\times\mathcal{M}_2,\Sigma_1\otimes\Sigma_2,\mu_1\otimes\mu_2)$:
		\begin{equation}
			L^2(\mathcal{M}_1,\Sigma_1,\mu_1)\otimes L^2(\mathcal{M}_2,\Sigma_2,\mu_2)\cong L^2(\mathcal{M}_1\times\mathcal{M}_2,\Sigma_1\otimes\Sigma_2,\mu_1\otimes\mu_2),
		\end{equation}
		via the unitary operator acting as
		\begin{equation}\label{eq:l2tol2}
			U(f_1\otimes f_2)(x,y)=f_1(x)f_2(y),\qquad x\in \mathcal{M}_1,\;y\in \mathcal{M}_2.
		\end{equation}
		Besides, the subspace of $L^2(\mathcal{M}_1\times\mathcal{M}_2,\Sigma_1\otimes\Sigma_2,\mu_1\otimes\mu_2)$ consisting of functions of the form
		\begin{equation}\label{eq:l2tol2dense}
			\Psi(x,y)=\sum_{j=1}^nf_{1,j}(x)f_{2,j}(y),\qquad f_{1,j}\in L^2(\mathcal{M}_1,\Sigma_1,\mu_1),\;f_{2,j}\in L^2(\mathcal{M}_2,\Sigma_2,\mu_2)
		\end{equation}
		is dense.
		
		The same statements hold for the tensor product of finitely many $L^2$ spaces.
	\end{proposition}

	\begin{proof}
		Recall that all Hilbert spaces are assumed to be separable (Section~\ref{subsec:notation}), and write $\hilb_j:=L^2(\mathcal{M}_j,\Sigma_j,\mu_j)$ and, omitting the $\sigma$-algebra and the measure, $L^2(\mathcal{M}_1\times\mathcal{M}_2)$ for the $L^2$ space on the product. Given $f_1\in\hilb_1$ and $f_2\in\hilb_2$, let $f_1\cdot f_2$ be the function $(x,y)\mapsto f_1(x)f_2(y)$. By Fubini's theorem (Appendix~\ref{app:measure}), $f_1\cdot f_2\in L^2(\mathcal{M}_1\times\mathcal{M}_2)$ and $\braket{f_1\cdot f_2,g_1\cdot g_2}=\braket{f_1,g_1}\braket{f_2,g_2}$, the integrand being integrable by the Cauchy--Schwarz inequality. In particular, $\|f_1\cdot f_2\|=\|f_1\|\|f_2\|$, so that the bilinear map $(f_1,f_2)\mapsto f_1\cdot f_2$ is continuous, since $\|f_1\cdot f_2-g_1\cdot g_2\|\leq\|f_1-g_1\|\|f_2\|+\|g_1\|\|f_2-g_2\|$.
		
		Now let $(e_n)_{n\in\mathbb{N}}$ and $(f_m)_{m\in\mathbb{N}}$ be orthonormal bases of $\hilb_1$ and $\hilb_2$. By the above, the functions $e_n\cdot f_m$ form an orthonormal set in $L^2(\mathcal{M}_1\times\mathcal{M}_2)$. This set is complete. Indeed, let $\Psi\in L^2(\mathcal{M}_1\times\mathcal{M}_2)$ be orthogonal to all of them. By Tonelli's theorem,
		\begin{equation}
			\int\left(\int|\Psi(x,y)|^2\,\mathrm{d}\mu_2(y)\right)\mathrm{d}\mu_1(x)=\|\Psi\|^2<\infty;
		\end{equation}
		thus, for almost all $x\in\mathcal{M}_1$, the function $\psi_x:=\Psi(x,\cdot)$ belongs to $\hilb_2$, and $x\mapsto\|\psi_x\|$ is square-integrable. For every $m$, the function $F_m(x):=\braket{f_m,\psi_x}$ is measurable (by Fubini's theorem) and satisfies $|F_m(x)|\leq\|\psi_x\|$, whence $F_m\in\hilb_1$. Since $|e_n(x)f_m(y)\Psi(x,y)|$ is integrable on the product space, Fubini's theorem yields
		\begin{equation}
			0=\braket{e_n\cdot f_m,\Psi}=\int\overline{e_n(x)}\left(\int\overline{f_m(y)}\Psi(x,y)\,\mathrm{d}\mu_2(y)\right)\mathrm{d}\mu_1(x)=\braket{e_n,F_m}
		\end{equation}
		for all $n$, whence $F_m=0$ almost everywhere. As there are countably many indices $m$, for almost all $x$ we have $\braket{f_m,\psi_x}=0$ for all $m$, i.e. $\psi_x=0$; hence $\|\Psi\|^2=\int\|\psi_x\|^2\,\mathrm{d}\mu_1(x)=0$.
		
		By Proposition~\ref{prop:tensorproductbases}, $(e_n\otimes f_m)_{n,m\in\mathbb{N}}$ is an orthonormal basis of $\hilb_1\otimes\hilb_2$; therefore, by Lemma~\ref{lem:unitary}(ii), the map $e_n\otimes f_m\mapsto e_n\cdot f_m$ extends to a unitary operator $U:\hilb_1\otimes\hilb_2\rightarrow L^2(\mathcal{M}_1\times\mathcal{M}_2)$. It acts as in Eq.~\eqref{eq:l2tol2}: writing $f_1=\sum_na_ne_n$ and $f_2=\sum_mb_mf_m$, by the continuity of $(f_1,f_2)\mapsto f_1\otimes f_2$ (recall $\|f_1\otimes f_2\|=\|f_1\|\|f_2\|$) and of $(f_1,f_2)\mapsto f_1\cdot f_2$,
		\begin{equation}
			U(f_1\otimes f_2)=U\sum_{n,m}a_nb_m\,e_n\otimes f_m=\sum_{n,m}a_nb_m\,e_n\cdot f_m=f_1\cdot f_2.
		\end{equation}
		Finally, the image under $U$ of the dense subspace $\hilb_1\odot\hilb_2$ is precisely the space of functions in Eq.~\eqref{eq:l2tol2dense}, which is therefore dense. The case of finitely many spaces is analogous.
	\end{proof}
	
	All properties discussed here immediately extend to the tensor product of finitely many Hilbert spaces: given $\hilb_1,\dots,\hilb_N$, one defines their algebraic tensor product $\bigodot_{j=1}^N\hilb_j$
	as the space of linear combinations of conjugate-multilinear forms on $\hilb_1\times\hilb_2\times\cdots\times\hilb_N$, with scalar product defined by
	\begin{equation}
		\Braket{\bigotimes_{j=1}^Nu_j,\bigotimes_{j=1}^Nv_j}_{\otimes_j\hilb_j}:=\prod_{j=1}^N\braket{u_j,v_j}_{\hilb_j}
	\end{equation}
	extended by linearity; their Hilbert space tensor product $\bigotimes_{j=1}^N\hilb_j$
	is again defined as the completion of $\odot_{j=1}^N\hilb_j$ with respect to the corresponding norm.	
	We will not need infinite tensor products.
	
	\subsection{The algebraic tensor product of operators}\label{subsec:tensoroperators}
	
	We now turn to the tensor product of operators. Again, we will need two distinct notions, denoted by $\odot$ and $\otimes$; as before, everything extends immediately to finitely many operators.
	
	\begin{definition}[Algebraic tensor product of two operators]\label{def:tensor_operators}\index{tensor product!of operators, algebraic}
		Let $\hilb_1,\hilb_2$ be two Hilbert spaces, and $A_1:\Dom A_1\subseteq\hilb_1\rightarrow\hilb_1$, $A_2:\Dom A_2\subseteq\hilb_2\rightarrow\hilb_2$ two operators on them. The \emph{algebraic tensor product} of $A_1$ and $A_2$ is the operator $A_1\odot A_2:\Dom(A_1\odot A_2)\subseteq\hilb_1\otimes\hilb_2\rightarrow\hilb_1\otimes\hilb_2$ with domain
		\begin{equation}
			\Dom(A_1\odot A_2):=\Dom A_1\odot\Dom A_2=\left\{\sum_{j=1}^nu_j\otimes v_j,\;u_j\in\Dom A_1,\,v_j\in\Dom A_2\right\},
		\end{equation}
		and whose action is defined by setting $(A_1\odot A_2)(u\otimes v):=A_1u\otimes A_2v$ for all pure tensors $u\otimes v$, with $u\in\Dom A_1$ and $v\in\Dom A_2$, and then extended to the whole domain by linearity.
	\end{definition}
	
	As for the scalar product, since vectors in $\Dom A_1\odot\Dom A_2$ admit multiple representations as linear combinations of pure tensors, we must check that $A_1\odot A_2$ is well-defined, i.e. that, if $\sum_{j=1}^nu_j\otimes v_j=\sum_{j'=1}^{n'}u'_{j'}\otimes v'_{j'}$, then $\sum_{j=1}^n A_1u_j\otimes A_2v_j=\sum_{j'=1}^{n'}A_1u'_{j'}\otimes A_2v'_{j'}$, thus making Definition~\ref{def:tensor_operators} meaningful. 
	
	\begin{proposition}\label{prop:welldef_tensorproduct}
		Let $A_1,A_2$ be two operators on $\hilb_1,\hilb_2$ respectively. Then $A_1\odot A_2$ is well-defined.
	\end{proposition}
	
	\begin{proof}
		As in Proposition~\ref{prop:welldef_scalarproduct}, by linearity it suffices to prove that, if $\Psi:=\sum_{j=1}^nu_j\otimes v_j=0$,
		then
		\begin{equation}\label{eq:giveszero}
			\sum_{j=1}^nA_1u_j\otimes A_2v_j=0.
		\end{equation}		
		We fix orthonormal bases $(e_1,\dots,e_d)$ and $(f_1,\dots,f_{d'})$ of $\operatorname{Span}(u_1,\dots,u_n)\subset\Dom A_1$ and $\operatorname{Span}(v_1,\dots,v_n)\subset\Dom A_2$ respectively, and write $\Psi=\sum_{k=1}^d\sum_{k'=1}^{d'}c_{kk'}e_k\otimes f_{k'}$ for some coefficients $(c_{kk'})_{k,k'}$. By assumption, this is the null element of $\hilb_1\otimes\hilb_2$; since the vectors $e_k\otimes f_{k'}$ are orthonormal, and thus linearly independent, all coefficients vanish: $c_{kk'}=0$.
		But, in this representation, the left-hand side of Eq.~\eqref{eq:giveszero} reads $\sum_{k,k'}c_{kk'}A_1e_k\otimes A_2f_{k'}$, which vanishes since $c_{kk'}=0$.
	\end{proof}
	
	\begin{proposition}\label{prop:tensorproductdense}
		Let $A_1:\Dom A_1\subseteq\hilb_1\rightarrow\hilb_1$ and $A_2:\Dom A_2\subseteq\hilb_2\rightarrow\hilb_2$ be densely defined operators on their respective spaces. Then $A_1\odot A_2$ is also densely defined on $\hilb_1\otimes\hilb_2$.
	\end{proposition}
	
	\begin{exercise}\label{exerc:denseindense}
		To prove this, prove the following, more general statement: if $\mathcal{V}_1\subseteq\hilb_1$ and $\mathcal{V}_2\subseteq\hilb_2$ are dense subspaces of $\hilb_1$ and $\hilb_2$, then their algebraic tensor product $\mathcal{V}_1\odot\mathcal{V}_2$, i.e. the space of all vectors
		\begin{equation}
			\sum_{j=1}^Nu_j\otimes v_j,\qquad N\in\mathbb{N},\;u_1,\dots,u_N\in\mathcal{V}_1,\;v_1,\dots,v_N\in\mathcal{V}_2,
		\end{equation}
		is dense in $\hilb_1\otimes\hilb_2$.
		
		Hint: any element of $\hilb_1\otimes\hilb_2$ can be approximated by an element of $\hilb_1\odot\hilb_2$, i.e. a finite linear combination of terms $u_j\otimes v_j$ with $u_j\in\hilb_1$ and $v_j\in\hilb_2$; but then each $u_j$ (resp. $v_j$) can be approximated by an element of $\mathcal{V}_1$ (resp. $\mathcal{V}_2$), and $\|u\otimes v-u'\otimes v'\|\leq\|u-u'\|\|v\|+\|u'\|\|v-v'\|$\dots
		\seesol{denseindense}
	\end{exercise}

	\begin{proposition}\label{prop:tensorproductclosable}
		Let $A_1:\Dom A_1\subseteq\hilb_1\rightarrow\hilb_1$, $A_2:\Dom A_2\subseteq\hilb_2\rightarrow\hilb_2$ be densely defined. Then
		\begin{equation}\label{eq:adjointtensor}
			(A_1\odot A_2)^*\supseteq A_1^*\odot A_2^*.
		\end{equation}
		Besides, if $A_1$ and $A_2$ are closable, then $A_1\odot A_2$ is also closable.
	\end{proposition}
	
	\begin{proof}
		Since $A_1$ and $A_2$ are densely defined, their adjoints $A_1^*$ and $A_2^*$ exist; furthermore, by Proposition~\ref{prop:tensorproductdense}, $A_1\odot A_2$ is also densely defined, whence $(A_1\odot A_2)^*$ exists as well.
		
		To check Eq.~\eqref{eq:adjointtensor}, let $\Psi\in\Dom(A_1^*\odot A_2^*)=\Dom A_1^*\odot\Dom A_2^*$; that is,
		\begin{equation}
			\Psi=\sum_{j=1}^nv_j\otimes v'_j,\qquad v_1,\dots,v_n\in\Dom A_1^*,\;v'_1,\dots,v'_n\in\Dom A_2^*.
		\end{equation}
		But then, for every $\Phi\in\Dom(A_1\odot A_2)$, i.e. any vector in the form
		\begin{equation}
			\Phi=\sum_{k=1}^mu_k\otimes u'_k,\qquad u_1,\dots,u_m\in\Dom A_1,\;u'_1,\dots,u'_m\in\Dom A_2,
		\end{equation}
		we have
		\begin{align}
			\Braket{(A_1^*\odot A_2^*)\Psi,\Phi}_{\hilb_1\otimes\hilb_2}&=\sum_{j=1}^n\sum_{k=1}^m\Braket{A_1^*v_j\otimes A_2^*v'_j,u_k\otimes u'_k}_{\hilb_1\otimes\hilb_2}\nonumber\\
			&=\sum_{j=1}^n\sum_{k=1}^m\Braket{A_1^*v_j,u_k}_{\hilb_1}\Braket{A_2^*v'_j,u'_k}_{\hilb_2}\nonumber\\
			&=\sum_{j=1}^n\sum_{k=1}^m\Braket{v_j,A_1u_k}_{\hilb_1}\Braket{v'_j,A_2u'_k}_{\hilb_2}\nonumber\\
			&=\sum_{j=1}^n\sum_{k=1}^m\Braket{v_j\otimes v'_j,(A_1\odot A_2)u_k\otimes u'_k}_{\hilb_1\otimes\hilb_2}=\Braket{\Psi,(A_1\odot A_2)\Phi}_{\hilb_1\otimes\hilb_2}.
		\end{align}
		This shows $\Dom(A_1\odot A_2)^*\supseteq\Dom A_1^*\odot\Dom A_2^*$ and $(A_1\odot A_2)^*\Psi=(A_1^*\odot A_2^*)\Psi$ for all $\Psi\in\Dom A_1^*\odot\Dom A_2^*$, i.e. $(A_1\odot A_2)^*\supseteq A_1^*\odot A_2^*$, as claimed.
		
		Finally, let $A_1$ and $A_2$ be closable. This happens if and only if $A_1^*$ and $A_2^*$ are densely defined (recall Proposition~\ref{prop:closableiffadjointdense}). But then, since $(A_1\odot A_2)^*\supseteq A_1^*\odot A_2^*$, $(A_1\odot A_2)^*$ is also densely defined and thus $A_1\odot A_2$ is closable.
	\end{proof}
	
	Unlike for direct sums, the inclusion $(A_1\odot A_2)^*\supseteq A_1^*\odot A_2^*$ is strict in general; in particular, if $A_1$ and $A_2$ are self-adjoint, we can only conclude $(A_1\odot A_2)^*\supseteq A_1\odot A_2$, but in general $A_1\odot A_2$ will not be self-adjoint. 
	
	Before continuing, it will be useful to explicitly consider the case of multiplication operators.
	
	\subsection{Tensor products of multiplication operators}\label{subsec:tensormulti}
	
	To keep the notation simple, we consider two multiplication operators on $L^2(\Omega_1)$ and $L^2(\Omega_2)$; everything generalizes to $N$ multiplication operators on arbitrary $L^2$ spaces. Let $\Omega_1\subseteq\mathbb{R}^{d_1},\Omega_2\subseteq\mathbb{R}^{d_2}$, and consider two measurable functions $v_1:\Omega_1\rightarrow\mathbb{C}$ and $v_2:\Omega_2\rightarrow\mathbb{C}$, and $M_{v_1}:\Dom M_{v_1}\subseteq L^2(\Omega_1)\rightarrow L^2(\Omega_1)$, $M_{v_2}:\Dom M_{v_2}\subseteq L^2(\Omega_2)\rightarrow L^2(\Omega_2)$ the corresponding multiplication operators. By Proposition~\ref{prop:tensorproduct_l2}, $L^2(\Omega_1)\otimes L^2(\Omega_2)\cong L^2(\Omega_1\times \Omega_2)$, the pure tensor $\varphi_1\otimes\varphi_2$ being identified with the function $(x,y)\mapsto\varphi_1(x)\varphi_2(y)$. Up to this identification, by Definition~\ref{def:tensor_operators}, the operator $M_{v_1}\odot M_{v_2}$ has domain
	\begin{equation}
		\Dom\left(M_{v_1}\odot M_{v_2}\right)=\left\{\Psi(x,y)=\sum_{j=1}^n\varphi_{1,j}(x)\varphi_{2,j}(y):\;\varphi_{1,j}\in\Dom M_{v_1},\;\varphi_{2,j}\in \Dom M_{v_2}\right\},
	\end{equation}
	on which it acts as the multiplication by the function $v_{12}(x,y):=v_1(x)v_2(y)$. Is it true that $M_{v_1}\odot M_{v_2}\cong M_{v_{12}}$? Recall that multiplication operators are naturally defined on their maximal domain (Example~\ref{examp:multiplication}); for multiplication by $v_{12}$, this is
	\begin{equation}
		\Dom M_{v_{12}}=\left\{\Psi\in L^2(\Omega_1\times \Omega_2):\int_{\Omega_1\times \Omega_2}|v_1(x)v_2(y)\Psi(x,y)|^2\;\mathrm{d}x\,\mathrm{d}y<\infty\right\}.
	\end{equation}
	On this domain, $M_{v_{12}}$ is closed, since $M_{v_{12}}^*=M_{\overline{v_{12}}}$, and in particular self-adjoint if $v_{12}(x,y)$ is real.
	
	\begin{proposition}\label{prop:tensorproduct_multi0}
		Up to the identification $L^2(\Omega_1)\otimes L^2(\Omega_2)\cong L^2(\Omega_1\times \Omega_2)$, we have
		\begin{equation}
			\overline{M_{v_1}\odot M_{v_2}}=M_{v_{12}},
		\end{equation}
		i.e., the space of functions $\sum_{j=1}^N\varphi_{1,j}\otimes\varphi_{2,j}$ with $\varphi_{1,j}\in\Dom M_{v_1},\varphi_{2,j}\in\Dom M_{v_2}$ is a core for $M_{v_{12}}$. 
	\end{proposition}
	
	\begin{proof}
		To keep the notation manageable, the identification $L^2(\Omega_1)\otimes L^2(\Omega_2)\cong L^2(\Omega_1\times \Omega_2)$ is understood throughout the proof.
		
		Clearly $M_{v_1}\odot M_{v_2}\subseteq M_{v_{12}}$ and thus, since $M_{v_{12}}$ is closed, $\overline{M_{v_1}\odot M_{v_2}}\subseteq M_{v_{12}}$.		
		We must prove the converse inclusion $M_{v_{12}}\subseteq\overline{M_{v_1}\odot M_{v_2}}$, i.e. $\Dom M_{v_{12}}\subseteq\Dom(\overline{M_{v_1}\odot M_{v_2}})$. Recalling the definition of operator closure, this amounts to proving that, for every $\Psi\in\Dom M_{v_{12}}$, there exists $(\tilde\Psi_n)_{n\in\mathbb{N}}\subset\Dom(M_{v_1}\odot M_{v_2})$ such that $\tilde\Psi_n\to\Psi$ and $v_{12}\tilde\Psi_n\to v_{12}\Psi$. We proceed by a two-step approximation: we first approximate $\Psi$ by suitable truncations $\Psi_n$, and then approximate each $\Psi_n$ by linear combinations of pure tensors.
		
		\medskip
		\emph{Step 1}: truncate $\Psi$.
		For the first step, define the rectangles
		\begin{equation}
			E_n:=A_n\times B_n,\qquad A_n:=\left\{x\in \Omega_1:|v_1(x)|\leq n\right\},\quad B_n:=\left\{y\in \Omega_2:|v_2(y)|\leq n\right\},
		\end{equation}
		and set $\Psi_n:=\Psi\chi_{E_n}$. Since $|v_{12}(x,y)|=|v_1(x)||v_2(y)|\leq n^2$ on $E_n$, we have $\Psi_n\in\Dom M_{v_{12}}$:
		\begin{equation}
			\|v_{12}\Psi_n\|^2=\int_{E_n}|v_{12}(x,y)|^2|\Psi(x,y)|^2\;\mathrm{d}x\,\mathrm{d}y\leq n^4\|\Psi\|^2<\infty.
		\end{equation}
		Besides, $E_n\uparrow \Omega_1\times \Omega_2$ (as $v_1,v_2$ are finite everywhere), so by dominated convergence (Appendix~\ref{app:measure}) $\Psi_n\to\Psi$ and $v_{12}\Psi_n\to v_{12}\Psi$.
		
		Fix $n\in\mathbb{N}$ for now.
		
		\medskip
		\emph{Step 2}: approximate each $\Psi_n$ with linear combinations of pure tensors. 		
		$\Psi_n$ vanishes outside the rectangle $E_n=A_n\times B_n$, i.e. it is an element of $L^2(A_n\times B_n)\cong L^2(A_n)\otimes L^2(B_n)$ (Proposition~\ref{prop:tensorproduct_l2}). Since the algebraic tensor product is dense in the Hilbert space tensor product, there are functions
		\begin{equation}
			\Psi_n^{(m)}=\sum_{r=1}^{r_{n,m}}\varphi^{(m,r)}_{1}\otimes\varphi^{(m,r)}_{2},\qquad \varphi^{(m,r)}_{1}\in L^2(\Omega_1),\;\varphi^{(m,r)}_{2}\in L^2(\Omega_2),
		\end{equation}
		with $\varphi^{(m,r)}_{1}$ and $\varphi^{(m,r)}_{2}$ vanishing outside $A_n$ and $B_n$ respectively,
		such that $\Psi_n^{(m)}\to\Psi_n$ as $m\to\infty$.
		
		Since $|v_1|\leq n$ on $A_n$ and $|v_2|\leq n$ on $B_n$, each $\varphi^{(m,r)}_1$ is in $\Dom M_{v_1}$ and each $\varphi^{(m,r)}_2$ is in $\Dom M_{v_2}$, whence $\Psi_n^{(m)}\in\Dom(M_{v_1}\odot M_{v_2})$. Besides, all functions involved vanish outside $E_n$, where $|v_{12}|\leq n^2$, so that $\|v_{12}\Psi_n^{(m)}-v_{12}\Psi_n\|\leq n^2\|\Psi_n^{(m)}-\Psi_n\|\to0$ as $m\to\infty$.
		
		Summing up: each $\Psi\in\Dom M_{v_{12}}$ can be approximated by a sequence $(\Psi_n)_{n\in\mathbb{N}}$ of functions supported in the rectangles $E_n$, on which $|v_{12}(x,y)|\leq n^2$, with $\Psi_n\to\Psi$ and $v_{12}\Psi_n\to v_{12}\Psi$; in turn, each $\Psi_n$ can be approximated by a sequence $(\Psi_n^{(m)})_{m\in\mathbb{N}}$ of functions in $\Dom(M_{v_1}\odot M_{v_2})$, with $\Psi_n^{(m)}\to\Psi_n$ and $v_{12}\Psi_n^{(m)}\to v_{12}\Psi_n$.
		
		\medskip
		\emph{Step 3}: combine the approximations. For each $n$, choose $m_n$ such that $\|\Psi_n^{(m_n)}-\Psi_n\|\leq1/(n+1)$ and $\|v_{12}\Psi_n^{(m_n)}-v_{12}\Psi_n\|\leq1/(n+1)$; then, by the triangle inequality, $\tilde{\Psi}_n:=\Psi_n^{(m_n)}\in\Dom(M_{v_1}\odot M_{v_2})$ satisfies $\tilde{\Psi}_n\to\Psi$ and $v_{12}\tilde{\Psi}_n\to v_{12}\Psi$, which was precisely our claim.		
	\end{proof}
	
	\subsection{The Hilbert space tensor product of operators}\label{subsec:hstensorops}
	
	The previous considerations motivate the following definition:
	
	\begin{definition}[Hilbert space tensor product of operators]\label{def:hs_tensorproduct_op}\index{tensor product!of operators}
		Let $A_1:\Dom A_1\subseteq\hilb_1\rightarrow\hilb_1$ and $A_2:\Dom A_2\subseteq\hilb_2\rightarrow\hilb_2$ be two closable operators on $\hilb_1$ and $\hilb_2$. Their \emph{Hilbert space tensor product} is the operator $A_1\otimes A_2$ defined as the closure of their algebraic tensor product:
		\begin{equation}
			A_1\otimes A_2:=\overline{A_1\odot A_2}.
		\end{equation}
	\end{definition}
	
	This definition is well-posed since, by Proposition~\ref{prop:tensorproductclosable}, $A_1\odot A_2$ is closable whenever $A_1$ and $A_2$ are. One can show that the equality $(A_1\otimes A_2)^*=A_1^*\otimes A_2^*$ holds for all closed, densely defined $A_1,A_2$; we only consider the self-adjoint case of interest for us here; see \cite[Proposition~7.26]{schmuedgen} for the general case. We begin by noting that Proposition~\ref{prop:tensorproduct_multi0} can be immediately rephrased as follows:
	
	\begin{proposition}\label{prop:tensorproduct_multi}
		Let $(\mathcal{M}_1,\Sigma_1,\mu_1)$, $(\mathcal{M}_2,\Sigma_2,\mu_2)$ be two $\sigma$-finite measure spaces, and let $v_1:\mathcal{M}_1\rightarrow\mathbb{C}$ and $v_2:\mathcal{M}_2\rightarrow\mathbb{C}$ be measurable. Then, up to the identification $L^2(\mathcal{M}_1,\Sigma_1,\mu_1)\otimes L^2(\mathcal{M}_2,\Sigma_2,\mu_2)\cong L^2(\mathcal{M}_1\times \mathcal{M}_2,\Sigma_1\otimes\Sigma_2,\mu_1\otimes\mu_2)$ (cf. Proposition~\ref{prop:tensorproduct_l2}), we have
		\begin{equation}
			M_{v_1}\otimes M_{v_2}=M_{v_{12}},
		\end{equation}
		where $v_{12}:\mathcal{M}_1\times \mathcal{M}_2\rightarrow\mathbb{C}$ is given by $v_{12}(x,y):=v_1(x)v_2(y)$. In particular, if $v_1$ and $v_2$ are real-valued, $M_{v_1}\otimes M_{v_2}$ is self-adjoint.
		
		The same holds for the tensor product of finitely many multiplication operators.
	\end{proposition}
	
	For general operators, we focus on the case of self-adjoint $A_1$ and $A_2$, where the spectral theorem (Theorem~\ref{thm:spectraltheorem2}) allows us to prove $(A_1\otimes A_2)^*=A_1\otimes A_2$. Since we will need it later, we directly consider a slightly more general case.
	
	\begin{proposition}\label{prop:polytensor}
		Let $\hilb_1,\dots,\hilb_n$ be Hilbert spaces, $\hilb=\otimes_{j=1}^n\hilb_j$ their Hilbert space tensor product, and $p:\mathbb{R}^n\rightarrow\mathbb{R}$ a real-valued polynomial of degree $d$, written in multi-index notation ($\alpha=(\alpha_1,\dots,\alpha_n)$, $|\alpha|=\alpha_1+\ldots+\alpha_n$) as
		\begin{equation}
			p(x_1,\dots,x_n)=\sum_{|\alpha|\leq d}c_{\alpha} x_1^{\alpha_1}x_2^{\alpha_2}\dots x_n^{\alpha_n}.
		\end{equation}
		Then, given self-adjoint operators $A_1,\dots,A_n$ on $\hilb_1,\dots,\hilb_n$ respectively, the operator defined by
		\begin{equation}
			p(A_1,\dots,A_n):=\sum_{|\alpha|\leq d}c_\alpha A_1^{\alpha_1}\otimes A_2^{\alpha_2}\otimes \cdots\otimes A_n^{\alpha_n}
		\end{equation}
		is essentially self-adjoint, and therefore $\overline{p(A_1,\dots,A_n)}$ is self-adjoint.
	\end{proposition}
	
	In particular, the following operators are self-adjoint whenever $A_1,\dots,A_n$ are self-adjoint:
	\begin{itemize}
		\item Their Hilbert space tensor product $A_1\otimes A_2\otimes\cdots\otimes A_n$;
		\item The operators acting as $A_j$ on $\hilb_j$ and trivially elsewhere:
		\begin{equation}
			I\otimes\cdots\otimes I\otimes\overbrace{A_j}^{j\text{th}}\otimes I\otimes\cdots\otimes I,\qquad j=1,\dots,n.
		\end{equation} 
		\item The closure of the sum of the operators above (or any subset of them):
		\begin{equation}
			\overline{\sum_{j=1}^nI\otimes\cdots\otimes I\otimes\overbrace{A_j}^{j\text{th}}\otimes I\otimes\cdots\otimes I}.
		\end{equation}
		
	\end{itemize}
	The latter operator will play a fundamental role in the remainder of these notes: for $n=2$ it yields the free part of spin--boson models (Sections~\ref{sec:singlemodesb} and~\ref{sec:multimodesb}), and for $A_1=\ldots=A_n$ it is the $n$-particle component of the second quantization of an operator (Definition~\ref{def:secondquantization}).
	
	\begin{proof}
		The proof relies on two unitary equivalences, which map $p(A_1,\dots,A_n)$ into a multiplication operator on a suitable $L^2$ space.
		
		First, since each $A_j$ is self-adjoint, the spectral theorem---in its multiplication operator formulation, Theorem~\ref{thm:spectraltheorem2}---ensures that $A_j$ is unitarily equivalent to a multiplication operator on some $L^2(\mathcal{M}_j,\Sigma_j,\mu_j)$ space, i.e. $(U_jA_jU_j^*\psi_j)(\lambda_j)=\lambda_j\psi_j(\lambda_j)$, and so are its powers: $(U_jA_j^{\alpha_j}U_j^*\psi_j)(\lambda_j)=\lambda_j^{\alpha_j}\psi_j(\lambda_j)$. Hence, via the unitary operator $U:=U_1\otimes U_2\otimes\cdots \otimes U_n$, each addend $A_1^{\alpha_1}\otimes\cdots\otimes A_n^{\alpha_n}$ is unitarily equivalent to the tensor product of the multiplication operators associated with the functions $\lambda_j\mapsto\lambda_j^{\alpha_j}$.
		Second, by Proposition~\ref{prop:tensorproduct_l2},
		\begin{equation}
			\bigotimes_{j=1}^nL^2(\mathcal{M}_j,\Sigma_j,\mu_j)\cong L^2\left(\mathcal{M},\Sigma,\mu\right),\qquad\text{with }\mathcal{M}=\prod_jX_j,\quad\Sigma=\bigotimes_j\Sigma_j,\quad\mu=\bigotimes_j\mu_j,
		\end{equation}
		and, by Proposition~\ref{prop:tensorproduct_multi}, in this representation each addend $A_1^{\alpha_1}\otimes\cdots\otimes A_n^{\alpha_n}$ is unitarily equivalent to the multiplication operator associated with the function $(\lambda_1,\ldots,\lambda_n)\mapsto \lambda_1^{\alpha_1}\cdots\lambda_n^{\alpha_n}$, on its maximal domain, which comprises all functions $\Psi\in L^2\left(\mathcal{M},\Sigma,\mu\right)$ such that
		\begin{equation}\label{eq:polydomain}
			\int_{\mathcal{M}}\lambda_1^{2\alpha_1}\cdots\lambda_n^{2\alpha_n}\left|\Psi(\lambda_1,\dots,\lambda_n)\right|^2\;\mathrm{d}\mu(\lambda_1,\dots,\lambda_n)<\infty,
		\end{equation}
		and on which it is self-adjoint. 				
		Hence each addend $A_1^{\alpha_1}\otimes\cdots\otimes A_n^{\alpha_n}$ is self-adjoint. 
		
		It remains to consider $p(A_1,\dots,A_n)=\sum_{|\alpha|\leq d}c_\alpha A_1^{\alpha_1}\otimes\cdots\otimes A_n^{\alpha_n}$. This is unitarily equivalent to an operator acting as multiplication by the polynomial function $p(\lambda_1,\dots,\lambda_n)$, on the domain
		\begin{equation}
			\Dom p(A_1,\dots,A_n)=\bigcap_{|\alpha|\leq d}\Dom\left( A_1^{\alpha_1}\otimes\cdots\otimes A_n^{\alpha_n}\right),
		\end{equation}
		each domain being, in the $L^2(\mathcal{M},\Sigma,\mu)$-representation, the space of functions $\Psi\in L^2\left(\mathcal{M},\Sigma,\mu\right)$ satisfying the corresponding condition~\eqref{eq:polydomain}. In particular, the subspace $L^2_{\rm c}(\mathcal{M},\Sigma,\mu)$ of functions vanishing outside $\{|\lambda_1|\leq R,\dots,|\lambda_n|\leq R\}$ for some $R>0$ (``compactly supported'' in the spectral variables) always satisfies this condition, whence $\Dom p(A_1,\dots,A_n)\supseteq L^2_{\rm c}(\mathcal{M},\Sigma,\mu)$. But $L^2_{\rm c}(\mathcal{M},\Sigma,\mu)$ is a core for the multiplication operator by any polynomial (more generally, by any function that is bounded on each such set; see e.g. \cite[Section 2.2, pp.~72--73]{teschl} or~\cite{schmuedgen}), whence, since $p(A_1,\dots,A_n)$ is contained in this (self-adjoint) multiplication operator, it is indeed essentially self-adjoint.	
	\end{proof}
	
	A different proof of the self-adjointness of $A_1\otimes A_2$, $A_1\otimes I$ and $I\otimes A_2$, and of the essential self-adjointness of $A_1\otimes I+I\otimes A_2$, based on Nelson's analytic vector theorem, is given in~\cite[Theorem 7.23]{schmuedgen}; the case of polynomials is treated in~\cite[Chapter 7, Exercise 16]{schmuedgen}.
	
	Recall that the distinction between algebraic and Hilbert space tensor product of two Hilbert spaces disappears when one of the two spaces is finite-dimensional (Proposition~\ref{prop:tensorproduct_finitedim}). An analogous statement holds for the tensor product of operators:
	
	\begin{proposition}\label{prop:tensorproduct_finitedim2}
		Let $\hilb_1,\hilb_2$ be two Hilbert spaces, and $\dim\hilb_1=:d<\infty$; let $A:\hilb_1\rightarrow\hilb_1$ be linear and $B:\Dom B\subseteq\hilb_2\rightarrow\hilb_2$ closable. Then, given any orthonormal basis $(e_j)_{j=1,\dots,d}$ of $\hilb_1$, $A\odot B$ is unitarily equivalent to the operator on $(\hilb_2)^{\oplus d}$ acting as
		\begin{equation}\label{eq:mapped}
			(\hilb_2)^{\oplus d}\supseteq	(\Dom B)^{\oplus d}\ni(\psi_1,\ldots,\psi_d)\mapsto\left(\sum_{j=1}^da_{1j}B\psi_j,\ldots,\sum_{j=1}^da_{dj}B\psi_j\right)\in(\hilb_2)^{\oplus d},
		\end{equation}
		where $a_{jk}=\braket{e_j,Ae_k}_{\hilb_1}$. Besides, $A\odot\overline{B}\subseteq A\otimes B:=\overline{A\odot B}$, with equality if $A$ is invertible. In particular, if $A$ is invertible and $B$ is closed, then $A\otimes B=A\odot B$.
	\end{proposition}
	
	\begin{exercise}\label{exerc:finitedim2}
		Prove this. Hint: as shown in Proposition~\ref{prop:tensorproduct_finitedim}, $\hilb_1\odot\hilb_2$ is unitarily equivalent to $(\hilb_2)^{\oplus d}$ via the transformation $\sum_{j=1}^de_j\otimes\psi_j\mapsto(\psi_1,\dots,\psi_d)$, as every $\Psi\in\hilb_1\odot\hilb_2$ can be written as in the left-hand side. An immediate calculation shows that, under the same transformation, $A\odot B$ is mapped into the operator given by Eq.~\eqref{eq:mapped}. You will then have to show that the closure of the latter operator extends the operator
		\begin{equation}\label{eq:mapped2}
			(\hilb_2)^{\oplus d}\supseteq	(\Dom \overline{B})^{\oplus d}\ni(\psi_1,\ldots,\psi_d)\mapsto\left(\sum_{j=1}^da_{1j}\overline{B}\psi_j,\ldots,\sum_{j=1}^da_{dj}\overline{B}\psi_j\right)\in(\hilb_2)^{\oplus d},
		\end{equation}
		and coincides with it if the matrix $(a_{jk})_{j,k}$ is invertible. Why can the invertibility assumption not be dropped?
		\seesol{finitedim2}
	\end{exercise}
	
	\section[The quantum harmonic oscillator and single-mode boson fields]{The quantum harmonic oscillator and\newline single-mode boson fields}\label{sec:singlemode}
	
	Having introduced direct sums and tensor products, we already have all the ingredients needed to define Fock spaces in full generality. It is nevertheless instructive to begin with the case of single-mode (monochromatic) boson fields, and their link with the quantum harmonic oscillator of Example~\ref{examp:qho}. After recalling how the modes of the electromagnetic field in a cavity reduce to independent harmonic oscillators (Section~\ref{subsec:prelude}), we introduce the ladder representation of the quantum harmonic oscillator (Section~\ref{subsec:ladder}), characterize the maximal domains of the ladder and number operators (Section~\ref{subsec:maximal}), and finally abstract the construction to an arbitrary separable Hilbert space (Section~\ref{subsec:abstract}).
	
	\subsection{Prelude: electromagnetic modes as harmonic oscillators}\label{subsec:prelude}
	
	The following derivation is informal; see e.g.~\cite{cohentannoudji1989,gerryknight} for details. Consider the electromagnetic field in a bounded cavity\index{electromagnetic field} $\Omega\subset\mathbb{R}^3$ with perfectly conducting walls, in the absence of sources, and in units with $c=\varepsilon_0=1$. In the Coulomb gauge, the scalar potential can be set to zero, and the field is described by a vector potential $\bm{A}(\bm{x},t)$ satisfying
	\begin{equation}
		\left(\frac{\partial^2}{\partial t^2}-\Delta\right)\bm{A}=\bm{0},\qquad\bm{\nabla}\cdot\bm{A}=0,\qquad\bm{E}=-\frac{\partial\bm{A}}{\partial t},\qquad\bm{B}=\bm{\nabla}\times\bm{A},
	\end{equation}
	with the tangential component of $\bm{A}$ vanishing on $\partial\Omega$. For a sufficiently regular cavity (e.g. a box), the Laplacian with these boundary conditions is a nonnegative self-adjoint operator with compact resolvent on the real Hilbert space of divergence-free fields in $L^2(\Omega;\mathbb{R}^3)$; hence the spectral theorem provides an orthonormal basis of \emph{normal modes}\index{normal modes} $\bm{u}_k$, with $-\Delta\bm{u}_k=\omega_k^2\bm{u}_k$ and $\omega_k>0$. Expanding
	\begin{equation}
		\bm{A}(\bm{x},t)=\sum_kq_k(t)\,\bm{u}_k(\bm{x}),\qquad q_k(t)\in\mathbb{R},
	\end{equation}
	the wave equation becomes $\ddot q_k=-\omega_k^2q_k$, and, by orthonormality and an integration by parts, the energy of the field reads
	\begin{equation}
		\frac{1}{2}\int_\Omega\left(\bm{E}^2+\bm{B}^2\right)\mathrm{d}^3\bm{x}=\frac{1}{2}\sum_k\left(p_k^2+\omega_k^2q_k^2\right),\qquad p_k:=\dot q_k.
	\end{equation}
	The field is thus equivalent to a family of independent harmonic oscillators, one for each mode, with canonical coordinates $q_k,p_k$.
	
	Quantizing this theory, one promotes each pair $q_k,p_k$ to the position and momentum operators on a copy of $L^2(\mathbb{R})$. Note that $q_k$ is not the position of a particle in $\Omega$, but the amplitude of the $k$th mode, which can take any real value: the variable of $L^2(\mathbb{R})$ is this amplitude, while the cavity only enters through the mode functions $\bm{u}_k$ and the frequencies $\omega_k$. The $k$th term thus becomes the quantum harmonic oscillator $\frac{1}{2}(p^2+\omega_k^2q^2)$ of Example~\ref{examp:qho}, whose ladder representation is the subject of the next subsection. The quantum Hamiltonian of the field should then be the sum of all these oscillators. Making sense of this sum is, however, far from obvious: for infinitely many modes, one cannot simply take the tensor product of infinitely many copies of $L^2(\mathbb{R})$, and in unbounded regions---e.g. in free space---the modes become plane waves labeled by a continuous wavevector, which are not square-integrable, so that the sum turns into an integral. The appropriate framework is the Fock space, discussed in Section~\ref{sec:sq}.
	
	In many situations of interest, a single mode suffices. When a two-level system is coupled to the field in a cavity, and its transition frequency is close to the frequency $\omega_k$ of one of the modes, this resonant mode dominates the dynamics, and the others can often be neglected. This is the physical setting of the single-mode spin--boson models of Section~\ref{sec:singlemodesb}; their multimode counterparts, in which no single mode dominates, behave very differently, as we will see in Sections~\ref{subsec:multidephasing} and~\ref{subsec:multijc}. Let us therefore first consider a single mode, i.e. a single quantum harmonic oscillator.
	
	\subsection{Ladder representation of the quantum harmonic oscillator}\label{subsec:ladder}\index{harmonic oscillator!ladder representation}
	
	Let $\hilb=L^2(\mathbb{R})$, and let $H_0=\frac{1}{2}(p^2+\omega^2q^2)$, with $\Dom H_0=\mathcal{S}(\mathbb{R})$, be the quantum harmonic oscillator of Example~\ref{examp:qho} (see also \cite[Section 8.3]{teschl}). Recall that $H_0$ is essentially self-adjoint, and that its eigenvectors, the Hermite functions $\phi_n$ of Eq.~\eqref{eq:hermite}, form an orthonormal basis of $\hilb$, with $H_0\phi_n=\omega\left(n+\frac{1}{2}\right)\phi_n$; in particular, all eigenvalues of $H_0$ are simple.
	
	One can conveniently recast $H_0$ in terms of two operators $a,a^\dag$---the \emph{ladder operators}---satisfying $[a,a^\dag]=1$. Let us revisit this construction carefully.
	
	\begin{definition}[Ladder operators]\label{def:a}\index{ladder operators}
		Let $\hilb=L^2(\mathbb{R})$. We define the operators $a_0:\Dom a_0\subseteq\hilb\rightarrow\hilb$, $a_0^\dag:\Dom a_0^\dag\subseteq\hilb\rightarrow\hilb$ on the domain $\Dom a_0=\Dom a_0^\dag=\mathcal{S}(\mathbb{R})$ by
		\begin{equation}
			a_0=\sqrt{\frac{\omega}{2}}q+\ii\frac{1}{\sqrt{2\omega}}p,\qquad a_0^\dag=\sqrt{\frac{\omega}{2}}q-\ii\frac{1}{\sqrt{2\omega}}p.
		\end{equation}
	\end{definition}
	
	Since $\mathcal{S}(\mathbb{R})$ is invariant under multiplication by $x$ and under differentiation (Definition~\ref{def:schwartz}), it is invariant under $a_0$ and $a_0^\dag$: $a_0\mathcal{S}(\mathbb{R})\subseteq\mathcal{S}(\mathbb{R})$ and $a_0^\dag\mathcal{S}(\mathbb{R})\subseteq\mathcal{S}(\mathbb{R})$. Consequently, all products and powers of $a_0$ and $a_0^\dag$, such as $a_0^\dag a_0$ or $(a_0^\dag)^n$, are well-defined on $\mathcal{S}(\mathbb{R})$. Let us gather some well-known properties of $a_0,a_0^\dag$:
	
	\begin{proposition}\label{prop:aprop}
		Let $a_0,a_0^\dag$ be as in Definition~\ref{def:a}. Then
		\begin{itemize}
			\item[(i)] $a_0,a_0^\dag$ satisfy
			\begin{equation}\label{eq:aadag1}
				\braket{\psi,a_0\varphi}=\braket{a_0^\dag\psi,\varphi}\qquad\forall\psi,\varphi\in\mathcal{S}(\mathbb{R})
			\end{equation}
			and are therefore closable operators, with $a_0\subseteq (a_0^\dag)^*$ and $a_0^\dag\subseteq a_0^*$;
			\item[(ii)] $[a_0,a_0^\dag]=1$, in the following sense:
			\begin{equation}\label{eq:aadag3}
				[a_0,a_0^\dag]\psi=\psi\qquad\text{for all }\psi\in\mathcal{S}(\mathbb{R});
			\end{equation}
			\item[(iii)] The quantum harmonic oscillator $H_0$ with $\Dom H_0=\mathcal{S}(\mathbb{R})$ satisfies
			\begin{equation}\label{eq:aadag4}
				H_0=\omega\left(a_0^\dag a_0+\frac{1}{2}\right).
			\end{equation}
		\end{itemize}
	\end{proposition}
	
	\begin{proof}
		The equalities~\eqref{eq:aadag1},~\eqref{eq:aadag3} and~\eqref{eq:aadag4} all follow from direct computations. Furthermore, the inclusions $a_0\subseteq(a_0^\dag)^*$ and $a_0^\dag\subseteq a_0^*$ follows directly from Eq.~\eqref{eq:aadag1}.
	\end{proof}
	
	As $\mathcal{S}(\mathbb{R})$ is invariant under $a_0,a_0^\dag$, all the usual commutator identities following from $[a_0,a_0^\dag]=1$ remain valid, as long as all operators are understood to act on vectors $\psi\in\mathcal{S}(\mathbb{R})$. For example:
	
	\begin{lemma}\label{lem:commn}
		For all $k\geq1$ and $\psi\in\mathcal{S}(\mathbb{R})$,
		\begin{equation}
			\left[a_0,(a_0^\dag)^k\right]\psi=k(a_0^\dag)^{k-1}\psi.
		\end{equation}
	\end{lemma}
	
	\begin{exercise}\label{exerc:commn}
		Prove Lemma~\ref{lem:commn}. Hint: use $[a_0,a_0^\dag]\psi=\psi$ and induction.
		\seesol{commn}
	\end{exercise}
	
	This allows us to prove the following:
	
	\begin{proposition}\label{prop:ladderaction}
		Let $\phi_n$ be the $n$th normalized eigenvector of the harmonic oscillator $H_0$. The following properties hold for all integers $n\geq0$:
		\begin{equation}\label{eq:ladderaction}
			\phi_n=\frac{1}{\sqrt{n!}}\left(a_0^\dag\right)^n\phi_0,\qquad a_0\phi_n=\sqrt{n}\phi_{n-1},\qquad a_0^\dag\phi_n=\sqrt{n+1}\phi_{n+1},
		\end{equation}
		with the convention $\phi_{-1}=0$ used above.
	\end{proposition}
	
	\begin{proof}
		We first prove $a_0\phi_0=0$. We then check that, for all $n$, $(a_0^\dag)^n\phi_0$ is a nonzero multiple $c_n\phi_n$ of the $n$th eigenvector $\phi_n$, and we prove $c_n=\sqrt{n!}$. Finally, we use this to show $a_0\phi_n=\sqrt{n}\phi_{n-1}$ (for $n\geq1$) and $a_0^\dag\phi_n=\sqrt{n+1}\phi_{n+1}$.
		
		\medskip
		\emph{Step 1}: $a_0\phi_0=0$. Indeed,
		\begin{equation}
			\|a_0\phi_0\|^2=\braket{a_0\phi_0,a_0\phi_0}=\braket{\phi_0,a_0^\dag a_0\phi_0}=\Braket{\phi_0,\left(\frac{H_0}{\omega}-\frac{1}{2}\right)\phi_0}=0,
		\end{equation}
		since $H_0\phi_0=\frac{\omega}{2}\phi_0$.
		
		\medskip
		\emph{Step 2}: $(a_0^\dag)^n\phi_0=\sqrt{n!}\phi_n$. Define $\tilde{\phi}_n:=(a_0^\dag)^n\phi_0$. 
		Using the property $[a_0,(a_0^\dag)^n]=n(a_0^\dag)^{n-1}$ (Lemma~\ref{lem:commn}),
		\begin{align}
			\left(\frac{H_0}{\omega}-\frac{1}{2}\right)\tilde{\phi}_n&=a_0^\dag a_0\tilde{\phi}_n=a_0^\dag a_0(a_0^\dag)^n\phi_0=a_0^\dag\left((a_0^\dag)^na_0+n(a_0^\dag)^{n-1}\right)\phi_0\nonumber\\
			&=(a_0^\dag)^{n+1}\cancel{a_0\phi_0}+n(a_0^\dag)^n\phi_0=n\tilde{\phi}_n,
		\end{align}
		that is, $\tilde{\phi}_n$ is in the $n$th eigenspace of $H_0$. Since each eigenvalue of $H_0$ is simple, necessarily $\tilde{\phi}_n=c_n\phi_n$ for some constant $c_n$. With the normalization of the Hermite functions chosen above, $c_n$ is positive. Indeed, in the variable $y=\sqrt{\omega}x$, the operator $a_0^\dag$ acts as $\frac{1}{\sqrt{2}}\left(y-\frac{\mathrm{d}}{\mathrm{d}y}\right)$, and, for every polynomial $p$,
		\begin{equation}
			\left(y-\frac{\mathrm{d}}{\mathrm{d}y}\right)\left[\e^{-y^2/2}p(y)\right]=\e^{-y^2/2}\left(2yp(y)-p'(y)\right):
		\end{equation}
		the degree of the polynomial increases by one, and its leading coefficient is doubled. Starting from $\phi_0\propto\e^{-y^2/2}$, by induction $\tilde\phi_n=\e^{-y^2/2}q_n(y)$, with $q_n$ a polynomial of degree $n$ with positive leading coefficient. As $H_n$ has leading coefficient $2^n>0$, comparing leading coefficients in $\tilde\phi_n=c_n\phi_n$ yields $c_n>0$, and therefore $c_n=\|\tilde\phi_n\|$. We now compute $\|\tilde\phi_n\|$ in an abstract way, only relying on the commutation relation $[a_0,a_0^\dag]=1$ and on $a_0\phi_0=0$. We claim that $\|\tilde\phi_n\|=\sqrt{n!}$, so that $c_n=\sqrt{n!}$. Indeed:
		\begin{align}
			\|\tilde\phi_n\|^2&=\left\|(a_0^\dag)^n\phi_0\right\|^2=\Braket{(a_0^\dag)^n\phi_0,(a_0^\dag)^n\phi_0}=\Braket{\phi_0,a_0^n(a_0^\dag)^n\phi_0}.
		\end{align}
		By repeated use of Lemma~\ref{lem:commn}, one can prove that (Exercise~\ref{exerc:commn2} below)
		\begin{equation}\label{eq:anadagn}
			a_0^n(a_0^\dag)^n\phi_0=n!\phi_0.
		\end{equation}
		
		Using Eq.~\eqref{eq:anadagn}, we finally have $\|\tilde\phi_n\|^2=\braket{\phi_0,a_0^n(a_0^\dag)^n\phi_0}=n!$, whence indeed $c_n=\sqrt{n!}$. This completes the proof of the first identity in Eq.~\eqref{eq:ladderaction}.
		
		\medskip
		\emph{Step 3}: proof of the remaining identities in Eq.~\eqref{eq:ladderaction}. We use the first one and, once again, Lemma~\ref{lem:commn}:
		\begin{align}
			a_0\phi_n&=\frac{1}{\sqrt{n!}}a_0(a_0^\dag)^n\phi_0=\frac{1}{\sqrt{n!}}\left[(a_0^\dag)^n\cancel{a_0\phi_0}+n(a_0^\dag)^{n-1}\phi_0\right]=\frac{1}{\sqrt{n!}}n\sqrt{(n-1)!}\phi_{n-1}=\sqrt{n}\phi_{n-1};\\
			a_0^\dag\phi_n&=\frac{1}{\sqrt{n!}}a_0^\dag(a_0^\dag)^n\phi_0=\frac{1}{\sqrt{n!}}(a_0^\dag)^{n+1}\phi_0=\frac{1}{\sqrt{n!}}\sqrt{(n+1)!}\phi_{n+1}=\sqrt{n+1}\phi_{n+1},
		\end{align}
		finishing the proof.
	\end{proof}
	
	\begin{exercise}\label{exerc:commn2}
		Prove Eq.~\eqref{eq:anadagn}, thus completing the proof of Proposition~\ref{prop:ladderaction}. Hint: by Lemma~\ref{lem:commn} and $a_0\phi_0=0$, one has $a_0(a_0^\dag)^n\phi_0=n(a_0^\dag)^{n-1}\phi_0$; iterate.
		\seesol{commn2}
	\end{exercise}
	
	It is also convenient to give a name to the operator $a_0^\dag a_0$.
	
	\begin{definition}[Number operator]\label{def:numberop}\index{number operator}
		Let $\hilb=L^2(\mathbb{R})$. We define the \emph{number operator} $\mathcal{N}_0:\Dom\mathcal{N}_0\subseteq\hilb\rightarrow\hilb$ by $\Dom\mathcal{N}_0=\mathcal{S}(\mathbb{R})$ and $\mathcal{N}_0\psi=a_0^\dag a_0\psi$.
	\end{definition}
	As a direct consequence of Proposition~\ref{prop:ladderaction}, $\mathcal{N}_0\phi_n=n\phi_n$ for all $n\in\mathbb{N}$, which explains its name; by Eq.~\eqref{eq:aadag4}, the harmonic oscillator can then be written as $H_0=\omega(\mathcal{N}_0+\frac{1}{2})$.
	Since $\mathcal{N}_0$ is just a rescaled and shifted version of $H_0$, it is essentially self-adjoint.
	
	The following estimate will be used repeatedly:
	
	\begin{proposition}\label{prop:asharpvsn}
		Let $a_0,a_0^\dag,\mathcal{N}_0$ be the ladder operators and number operator associated with the harmonic oscillator. Then, for all $k\in\mathbb{N}$,
		\begin{equation}
			\big\|A_1 A_2\dots A_k \psi\big\| 
			\leq \big\| \sqrt{(\mathcal{N}_0 +1)(\mathcal{N}_0 +2) \dots (\mathcal{N}_0 +k)}\psi \big\|
		\end{equation}
		for any \(\psi \in \mathcal{S}(\mathbb R)\), where \(A_i \in \{a_0,\, a_0^\dagger\}\) for each \(i = 1,\dots, k\).
		
	\end{proposition}
	
	\begin{exercise}\label{exerc:asharpvsn}
		Prove this. Hint: write $\psi\in\mathcal{S}(\mathbb{R})$ as $\psi=\sum_{n\in\mathbb{N}}c_n\phi_n$ and recall $a_0\phi_n=\sqrt{n}\phi_{n-1}$ and $a_0^\dag\phi_n=\sqrt{n+1}\phi_{n+1}$.
		\seesol{asharpvsn}
	\end{exercise}
	
	\begin{remark}[Choice of the initial domain]\label{rem:initialdomain}
		In the previous discussion, we defined $a_0,a_0^\dag,\mathcal{N}_0$ on the Schwartz space $\mathcal{S}(\mathbb{R})$. This is a convenient choice, since $\mathcal{S}(\mathbb{R})$ is a core for the harmonic oscillator and it is invariant under the action of all operators involved. An equally valid choice would have been the linear span $\hilb_{\rm fin}:=\operatorname{Span}(\phi_n)_{n\in\mathbb{N}}\subset\mathcal{S}(\mathbb{R})$ of the Hermite functions (under the identification $L^2(\mathbb{R})\cong\ell^2=\oplus_{n\in\mathbb{N}}\mathbb{C}$ induced by the Hermite basis, this is precisely the space $\hilb_{\rm fin}$ of Definition~\ref{def:hfin}, whence the notation): being the linear span of the eigenvectors of the harmonic oscillator, $\hilb_{\rm fin}$ is also a core for it (Proposition~\ref{prop:eigenvectors}), and it is clearly invariant under the action of $a_0,a_0^\dag,H_0,\mathcal{N}_0$. All properties proven so far thus remain valid if one replaces $\mathcal{S}(\mathbb{R})$ with $\hilb_{\rm fin}$.
		
		The two choices can be compared in terms of the expansion coefficients $c_n=\braket{\phi_n,\psi}$ of a vector $\psi=\sum_nc_n\phi_n$. Clearly, $\psi\in\hilb_{\rm fin}$ if and only if $c_n=0$ for all sufficiently large $n$. The analogous characterization of $\mathcal{S}(\mathbb{R})$ is provided by the so-called \emph{$\N$-representation theorem}\index{N-representation theorem@$\N$-representation theorem}~\cite[Theorem~V.13]{reedsimon1}:
		\begin{equation}\label{eq:nrep}
			\mathcal{S}(\mathbb{R})=\left\{\psi=\sum_{n\in\mathbb{N}}c_n\phi_n:\;\sup_{n\in\mathbb{N}}n^\alpha|c_n|<\infty\;\text{for all }\alpha\in\mathbb{N}\right\}.
		\end{equation}
		That is, the coefficients of vectors in $\mathcal{S}(\mathbb{R})$ may be nonzero for all $n$, but decay faster than any power as $n\to\infty$.
	\end{remark}
	
	\subsection{Maximal domains of the ladder and number operators}\label{subsec:maximal}
	
	By Proposition~\ref{prop:aprop}, $a_0,a_0^\dag,\mathcal{N}_0$ are closable, on either of the initial domains discussed in Remark~\ref{rem:initialdomain}; we now study their closures.
	
	\begin{definition}\label{defin:aadagnclose}
		We define
		\begin{equation}
			a:=\overline{a_0},\qquad a^\dag:=\overline{a_0^\dag},\qquad\mathcal{N}:=\overline{\mathcal{N}_0}.
		\end{equation}
	\end{definition}
	With a slight abuse of notation, we will still refer to $a,a^\dag,\mathcal{N}$ as the ladder operators (or annihilation/creation operators) and the number operator. 	
	The next step is to characterize the domains of these operators. For the number operator $\mathcal{N}$, this is immediate:
	\begin{proposition}\label{prop:numberselfadjoint}
		$\mathcal{N}$ is a self-adjoint operator with domain
		\begin{equation}\label{eq:domn}
			\Dom\mathcal{N}=\left\{\psi\in \hilb:\sum_{n\in\mathbb{N}}n^2|\braket{\psi,\phi_n}|^2<\infty\right\},
		\end{equation}
		or equivalently
		\begin{equation}\label{eq:domn2}
			\Dom\mathcal{N}=\left\{\psi=\sum_{n\in\mathbb{N}}c_n\phi_n\in\hilb:\sum_{n\in\mathbb{N}}n^2|c_n|^2<\infty\right\}.
		\end{equation}
	\end{proposition}
	
	\begin{proof}
		Being the closure of the essentially self-adjoint operator $\mathcal{N}_0$, $\mathcal{N}$ is a self-adjoint operator with spectrum $\sigma(\mathcal{N})=\mathbb{N}$ and eigenvectors $(\phi_n)_{n\in\mathbb{N}}$. Hence the spectral theorem (Theorem~\ref{thm:spectraltheorem1}) implies that the domain of $\mathcal{N}$ is given by Eq.~\eqref{eq:domn}, which is equivalent to Eq.~\eqref{eq:domn2} since  the coefficients $c_n$ in the expansion of any vector $\psi\in\hilb$ in an orthonormal basis are given by $c_n=\braket{\psi,\phi_n}$.
	\end{proof}
	
	Again by the spectral theorem, we can immediately define powers of $\mathcal{N}$ as follows: given any $\alpha\geq0$, $\mathcal{N}^\alpha$ is the self-adjoint operator given by
	\begin{align*}
		\Dom\mathcal{N}^\alpha&=\left\{\psi=\sum_{n\in\mathbb{N}}c_n\phi_n:\sum_{n\in\mathbb{N}}n^{2\alpha}|c_n|^{2}<\infty\right\},\\
		\N^\alpha\sum_{n\in\mathbb{N}}c_n\phi_n&=\sum_{n\in\mathbb{N}}n^\alpha c_n\phi_n.
	\end{align*}
	All these operators have $(\phi_n)_{n\in\mathbb{N}}$ as eigenvectors, whence $\hilb_{\rm fin}$ and, a fortiori, $\mathcal{S}(\mathbb{R})$ are cores for them.
	
	Let us now turn to $a,a^\dag$.
	
	\begin{proposition}\label{prop:aprop2}
		Let $a,a^\dag$ be as in Definition~\ref{defin:aadagnclose}. Then
		\begin{itemize}
			\item[(i)] $a,a^\dag$ have domains $\Dom a=\Dom a^\dag=\Dom \mathcal{N}^{1/2}$ and are mutually adjoint: $a=(a^\dag)^*$ and $a^\dag=a^*$;
			\item[(ii)] $[a,a^\dag]=1$, in the following sense:
			\begin{equation}\label{eq:aadagclosed}
				[a,a^\dag]\psi=\psi\qquad\text{for all }\psi\in\Dom\N.
			\end{equation}
			\item[(iii)] The operator $H=\overline{H_0}$, i.e. the closure of the harmonic oscillator $H_0$ initially defined on $\Dom H_0=\mathcal{S}(\mathbb{R})$, satisfies
			\begin{equation}\label{eq:hclosed}
				H=\omega\left(a^\dag a+\frac{1}{2}\right)=\omega\left(\N+\frac{1}{2}\right).
			\end{equation}
		\end{itemize}
	\end{proposition}
	This is the counterpart of Proposition~\ref{prop:aprop} for the closures of the operators considered there.
	
	\begin{proof}
		The main difficulty lies in proving (i), which we split into two parts.
		
		\medskip
		\emph{Part 1}: $\Dom a=\Dom a^\dag=\Dom\mathcal{N}^{1/2}$. By definition, $\Dom a$ is the space of all vectors $\psi\in\hilb$ such that there exists $(\psi_n)_{n\in\mathbb{N}}\subset\Dom a_0=\mathcal{S}(\mathbb{R})$ with $\psi_n\to\psi$ and $a_0\psi_n\to\xi$ for some $\xi\in\hilb$; equivalently, $\Dom a=\overline{\Dom a_0}^{\|\cdot\|_{a_0}}$, i.e. it is the closure of $\Dom a_0=\mathcal{S}(\mathbb{R})$ with respect to the graph norm\index{graph norm} $\|\psi\|_{a_0}^2:=\|\psi\|^2+\|a_0\psi\|^2$ induced by $a_0$, since a sequence $(\psi_n)_{n\in\mathbb{N}}\subset\Dom a_0$ converges in this norm if and only if both $(\psi_n)_{n\in\mathbb{N}}$ and $(a_0\psi_n)_{n\in\mathbb{N}}$ converge in the usual norm.
		
		On the other hand, recalling $a_0^\dag\phi_n=\sqrt{n+1}\phi_{n+1}$,
		\begin{align}
			\|\psi\|^2=\sum_{n\in\mathbb{N}}|\braket{\psi,\phi_n}|^2,\qquad \|a_0\psi\|^2&=\sum_{n\in\mathbb{N}}|\braket{a_0\psi,\phi_n}|^2=\sum_{n\in\mathbb{N}}|\braket{\psi,a_0^\dag\phi_n}|^2\nonumber\\
			&=\sum_{n\in\mathbb{N}}(n+1)|\braket{\psi,\phi_{n+1}}|^2=\sum_{n\in\mathbb{N}}n|\braket{\psi,\phi_n}|^2=\|\N^{1/2}\psi\|^2,
		\end{align}
		so that $\|\psi\|^2_{a_0}=\|\psi\|^2+\|\N^{1/2}\psi\|^2$, i.e. the graph norm of $a_0$ coincides with the graph norm $\|\cdot\|_{\N^{1/2}}$ of $\N^{1/2}$. Since $\Dom a_0$ is a core for $\N^{1/2}$, as shown above, we conclude
		\begin{equation}
			\Dom a=\overline{\Dom a_0}^{\|\cdot\|_{a_0}}=\overline{\Dom a_0}^{\|\cdot\|_{\N^{1/2}}}=\Dom\mathcal{N}^{1/2}.
		\end{equation}		
		The proof for $a^\dag$ is analogous: we have
		\begin{align}
			\|a_0^\dag\psi\|^2&=\sum_{n\in\mathbb{N}}|\braket{a^\dag_0\psi,\phi_n}|^2=\sum_{n\in\mathbb{N}}|\braket{\psi,a_0\phi_n}|^2=\sum_{n\in\mathbb{N}}n|\braket{\psi,\phi_{n-1}}|^2=\sum_{n\in\mathbb{N}}(n+1)|\braket{\psi,\phi_n}|^2\nonumber\\
			&=\|(\N+1)^{1/2}\psi\|^2,
		\end{align}
		whence $\|\psi\|^2_{a_0^\dag}=\|\psi\|^2+\|(\mathcal{N}+1)^{1/2}\psi\|^2$, so that the graph norm of $a_0^\dag$ coincides with the graph norm of $(\mathcal{N}+1)^{1/2}$. The same reasoning as before proves $\Dom a^\dag=\Dom(\mathcal{N}+1)^{1/2}=\Dom\mathcal{N}^{1/2}$.
		
		\medskip
		\emph{Part 2}: $a=(a^\dag)^*$ and $a^\dag=a^*$. It will suffice to prove e.g. $a^\dag=a^*$; if so, taking adjoints and recalling that $a$ is closed by definition, the other equality immediately follows.		
		We already know from Proposition~\ref{prop:aprop} that $a_0^\dag\subseteq a_0^*$. Taking closures, $a^\dag\subseteq a^*$, so we only need to show the converse inclusion $a^*\subseteq a^\dag$, i.e. $\Dom a^*\subseteq\Dom a^\dag=\Dom\N^{1/2}$.
		
		Let $\psi\in\Dom a^*$. By definition, there exists $\tilde{\psi}\in\hilb$ such that, for all $\varphi\in\Dom a=\Dom\N^{1/2}$,
		\begin{equation}\label{eq:adja}
			\Braket{\tilde{\psi},\varphi}=\braket{\psi,a\varphi}.
		\end{equation}
		We claim $\psi\in\Dom a^\dag=\Dom\N^{1/2}$. To this end, write $\varphi=\sum_{n\in\mathbb{N}}c_n\phi_n$, $\psi=\sum_{n\in\mathbb{N}}d_n\phi_n$. Our claim becomes $\sum_{n\in\mathbb{N}}n|d_n|^2<\infty$, which is left as an exercise (Exercise~\ref{exerc:adjointa} below).
		
		\medskip
		\emph{Proof of (ii)}. Eq.~\eqref{eq:aadagclosed} holds for all $\psi\in\mathcal{S}(\mathbb{R})$ by Proposition~\ref{prop:aprop}, and we must show that it extends to all $\psi\in\Dom\N$.		
		Given $\psi\in\Dom\N$, since $\mathcal{S}(\mathbb{R})$ is a core for $\N$ there is $(\psi_n)_{n\in\mathbb{N}}\subset\mathcal{S}(\mathbb{R})$ such that $\psi_n\to\psi$ and $\N\psi_n\to\N\psi$. On $\mathcal{S}(\mathbb{R})$ we have $a^\dag a\psi_n=\N\psi_n$ and $aa^\dag\psi_n=(\N+1)\psi_n$. Besides, $a\psi_n\to a\psi$ and $a^\dag\psi_n\to a^\dag\psi$, since the graph norms of $a,a^\dag$ are dominated by the graph norm of $\N^{1/2}$ (Part 1), and hence by that of $\N$. Since $a,a^\dag$ are closed,
		\begin{gather}
			a^\dag(a\psi_n)=\N\psi_n\to\N\psi\implies a^\dag a\psi=\N\psi,\\
			a(a^\dag\psi_n)=(\N+1)\psi_n\to(\N+1)\psi\implies aa^\dag\psi=(\N+1)\psi,
		\end{gather}
		whence $[a,a^\dag]\psi=(\N+1)\psi-\N\psi=\psi$.
		
		\medskip
		\emph{Proof of (iii)}. This is immediate:
		\begin{equation}
			H=\overline{H_0}=\overline{\omega\left(\N_0+\frac{1}{2}\right)}=\omega\left(\overline{\N_0}+\frac{1}{2}\right)=\omega\left(\N+\frac{1}{2}\right).
		\end{equation}
		Finally, $a^\dag a=a^*a$ is self-adjoint (a classical result of von Neumann \cite[Satz 3]{vonneumann1932}, valid for every closed densely defined operator; see also \cite[Theorem X.25]{reedsimon2} or \cite[Proposition 3.18]{schmuedgen}), and we just showed $a^\dag a\supseteq\N$. Since a self-adjoint operator has no proper self-adjoint extensions, $a^\dag a=\N$.
		
	\end{proof}
	
	\begin{exercise}\label{exerc:adjointa}
		Prove that $\sum_{n\in\mathbb{N}}n|d_n|^2<\infty$, thus completing Part 2 of the proof of Proposition~\ref{prop:aprop2}. Hint: choose $\varphi=\phi_m$ in Eq.~\eqref{eq:adja} to compute the coefficients of $\tilde{\psi}$ in the basis $(\phi_m)_{m\in\mathbb{N}}$, and use Parseval's identity.
		\seesol{adjointa}
	\end{exercise}
	
	\subsection{Ladder and number operators on an arbitrary Hilbert space}\label{subsec:abstract}
	
	We introduced $a,a^\dag,\mathcal{N}$ as operators on $\hilb=L^2(\mathbb{R})$ defined in terms of position and momentum (Definitions~\ref{def:a},~\ref{def:numberop} and~\ref{defin:aadagnclose}), and proved (Propositions~\ref{prop:ladderaction} and~\ref{prop:aprop2}) that they act on the Hermite functions---the eigenfunctions of the harmonic oscillator---as in Eq.~\eqref{eq:ladderaction}, with $\mathcal{N}\phi_n=n\phi_n$. Once these relations are established, however, none of the subsequent results refers to the explicit expressions in terms of $q$ and $p$, to the Hermite polynomials, or to the fact that we started from $L^2(\mathbb{R})$.
	
	We can thus abstract from the specific representation we started from, and reformulate everything in an abstract setting:
	
	\begin{definition}[Ladder operators and number operator: abstract definition]\label{def:abstractladder}\index{ladder operators!abstract definition}
		Let $\hilb$ be a complex separable Hilbert space, and $(\phi_n)_{n\in\mathbb{N}}$ a complete orthonormal set in $\hilb$. The \emph{number operator} is defined by
		\begin{align}
			\Dom\N&=\left\{\psi=\sum_{n\in\mathbb{N}}c_n\phi_n:\sum_{n\in\mathbb{N}}n^2|c_n|^2<\infty\right\},\\
			\N\sum_{n\in\mathbb{N}}c_n\phi_n&=\sum_{n\in\mathbb{N}}n\,c_n\phi_n,
		\end{align}
		and the \emph{ladder operators} $a,a^\dag$ (also called annihilation and creation operators, respectively) are given by
		\begin{align}
			\Dom a=\Dom a^\dag=\Dom\N^{1/2}=\left\{\psi=\sum_{n\in\mathbb{N}}c_n\phi_n:\sum_{n\in\mathbb{N}}n|c_n|^2<\infty\right\},
		\end{align}
		\begin{equation}
			a\sum_{n\in\mathbb{N}}c_n\phi_n:=\sum_{n\in\mathbb{N}}\sqrt{n+1}\,c_{n+1}\phi_n,\qquad a^\dag\sum_{n\in\mathbb{N}}c_n\phi_n:=\sum_{n\in\mathbb{N}}\sqrt{n}\,c_{n-1}\phi_n
		\end{equation}
		(with $c_{-1}:=0$), the series converging since $\sum_nn|c_n|^2<\infty$. In particular, $a\phi_n=\sqrt{n}\phi_{n-1}$ and $a^\dag\phi_n=\sqrt{n+1}\phi_{n+1}$.
	\end{definition}
	
	Different choices of $\hilb$ and $(\phi_n)_{n\in\mathbb{N}}$ yield unitarily equivalent operators; in particular, if $\hilb=L^2(\mathbb{R})$ and $(\phi_n)_{n\in\mathbb{N}}$ is the Hermite basis, we recover the operators of Section~\ref{subsec:maximal}.	
	All properties established above thus remain valid in the abstract setting, and, whenever we refer to them in the following, they are understood to hold independently of the particular choice of $\hilb$ and $(\phi_n)_{n\in\mathbb{N}}$. In particular, $\mathcal{N}$ is self-adjoint, $a$ and $a^\dag$ are mutually adjoint, $H=\omega(a^\dag a+\frac{1}{2})$ is self-adjoint (Proposition~\ref{prop:aprop2}), and $[a,a^\dag]\psi=\psi$ for all $\psi\in\Dom\N$. Besides, the span $\hilb_{\rm fin}$ of the vectors $\phi_n$, as well as the space of all vectors whose coefficients $c_n=\braket{\phi_n,\psi}$ decay faster than any power, as in Eq.~\eqref{eq:nrep} (which we still denote by $\mathcal{S}$), are cores for $a,a^\dag,\N$.
	
	\section{Single-mode spin--boson models}\label{sec:singlemodesb}
	
	As discussed in Section~\ref{subsec:prelude}, the quantum harmonic oscillator---which we can now write in its \emph{abstract ladder representation} $H=\omega(a^\dag a+\frac{1}{2})$ on an arbitrary separable Hilbert space $\hilb$---provides an effective description of a single mode of a quantized boson field, such as the electromagnetic one. For later convenience, we redefine this operator by dropping the constant term $\frac{\omega}{2}$, i.e. we set $H=\omega a^\dag a$. This simply corresponds to shifting the ground state energy to $0$. This entails no loss of generality, as the constant $\frac{\omega}{2}$ only contributes a global phase to the dynamics generated by $H$.
	
	Physically, $\phi_n$ is interpreted as the state in which the field contains $n$ particles, so that $a$ and $a^\dag$ respectively annihilate and create one particle (e.g. a photon), while $\N$ ``counts'' the particles. We can thus describe situations in which the number of particles, rather than being a fixed parameter of the system, is a physical observable whose value depends on the state in which the system is prepared.
	
	In this section, we couple a single-mode field to a two-level system. We define single-mode spin--boson models and prove their self-adjointness (Section~\ref{subsec:singlemodesb}); we then solve exactly two important special cases, the pure dephasing model (Section~\ref{subsec:dephasing}) and the Jaynes--Cummings model (Section~\ref{subsec:jc}), and finally discuss the quantum Rabi model and its relation with the Jaynes--Cummings model (Section~\ref{subsec:rabi}).
	
	\subsection{Definition and self-adjointness}\label{subsec:singlemodesb}
	
	We can now introduce models describing the interaction of such a single-mode field with matter. We consider the simplest example: a two-level system (qubit) linearly interacting with a single-mode boson field; the multimode case will be discussed in Section~\ref{subsec:multimodesb}. The Hilbert space of such a system is $\hfrak=\mathbb{C}^2\otimes\hilb\simeq\hilb\oplus\hilb$ (Proposition~\ref{prop:tensorproduct_finitedim}), whose generic element can be written as
	\begin{equation}
		\Psi=e_+\otimes\psi_++e_-\otimes\psi_-\simeq\begin{bmatrix}
			\psi_+\\\psi_-
		\end{bmatrix},\qquad \psi_\pm\in\hilb,\;e_+=\begin{bmatrix}
			1\\0
		\end{bmatrix},e_-=\begin{bmatrix}
			0\\1
		\end{bmatrix}.
	\end{equation} 
	
	\begin{definition}[Single-mode spin--boson model]\label{def:singlemodesb}\index{spin--boson model!single-mode}
		Let $\hilb$ be a complex separable Hilbert space, with $a,a^\dag$ as in Definition~\ref{def:abstractladder}. A \emph{single-mode spin--boson model} is an operator on $\hfrak=\mathbb{C}^2\otimes\hilb$
		defined by
		\begin{align}
			H&=\frac{\omega_0}{2}\sigma_z\otimes I+I\otimes\omega a^\dag a+B\otimes a^\dag+B^*\otimes a,
		\end{align}
		where $\omega_0\geq0$, $\omega>0$, and $B\in\mathbb{C}^{2\times 2}$.
	\end{definition}
	Above, $I$ denotes the identity on either $\mathbb{C}^2$ or $\hilb$, and $\sigma_j$, $j\in\{x,y,z\}$ denotes the $j$th Pauli matrix\index{Pauli matrices}:
	\begin{equation}
		\sigma_x=\begin{bmatrix}
			0&1\\1&0
		\end{bmatrix},\qquad\sigma_y=\begin{bmatrix}
			0&-\ii\\\ii&0
		\end{bmatrix},\qquad\sigma_z=\begin{bmatrix}
			1&0\\0&-1
		\end{bmatrix}.
	\end{equation}	
	In the literature, one often leaves the tensor products and the identities understood, and just writes
	\begin{equation}
		H=\frac{\omega_0}{2}\sigma_z+\omega a^\dag a+Ba^\dag+B^*a.
	\end{equation}
	
	Some physical comments are in order. Here $\omega_0$ represents the energy splitting between the two energy levels of the qubit: we tacitly fixed our energy offset so that the two energy levels are $\pm\frac{\omega_0}{2}$, which we can again do without loss of generality. Requiring $\omega_0\geq0$ is not restrictive either: the sign of $\omega_0$ can be flipped by exchanging the roles of $e_+$ and $e_-$, i.e. by conjugating $H$ with the unitary $\sigma_x\otimes I$, which only replaces $B$ with $\sigma_xB\sigma_x$. The frequency $\omega$, instead, is the energy of a single excitation of the field mode (e.g. a laser frequency), and is therefore positive; this will also be needed later, e.g. to divide by $\omega$. Finally, $B$ is a $2\times2$ matrix that encodes the specific type of qubit--field interaction. When modeling an actual qubit--field interaction from first principles, one typically obtains
	\begin{equation}
		B=\bm{g}\cdot\bm{\sigma}=g_x\sigma_x+g_y\sigma_y+g_z\sigma_z,
	\end{equation}
	with $g_j$, $j\in\{x,y,z\}$, being coupling constants measuring the strength of the interaction in the different ``directions''. With our choice of reference frame, coupling in the \emph{transversal} directions $x,y$, measured by the coupling constants $g_x,g_y$, induces decay, while coupling in the \emph{longitudinal} direction $z$, measured by the coupling constant $g_z$, induces decoherence. We will return to this point later.
	
	Let us first justify that the operator $H$ of Definition~\ref{def:singlemodesb} is well-defined. The first term $\frac{\omega_0}{2}\sigma_z\otimes I$ is clearly a bounded operator. As for the three unbounded terms: since $\mathbb{C}^2$ is finite-dimensional (recall Proposition~\ref{prop:tensorproduct_finitedim2}) and $\Dom a=\Dom a^\dag=\Dom \N^{1/2}$ (Proposition~\ref{prop:aprop2}), $\Dom(I\otimes\omega a^\dag a)=\mathbb{C}^2\odot\Dom\N$ and
	\begin{equation}
		\Dom(B\otimes a^\dag)\supseteq\mathbb{C}^2\odot\Dom\N^{1/2},\qquad\Dom(B^*\otimes a)\supseteq\mathbb{C}^2\odot\Dom\N^{1/2},
	\end{equation}
	with equalities if $B$ is invertible.
	
	As $\Dom\N^{1/2}\supseteq\Dom\N$, $H$ is a well-defined operator on
	\begin{equation}
		\Dom H=\mathbb{C}^2\odot\Dom\N\cong\left\{\begin{bmatrix}
			\psi_+\\\psi_-
		\end{bmatrix}:\psi_\pm\in\Dom\N\right\},
	\end{equation}
	whose action can be represented as
	\begin{equation}
		H\cong\begin{bmatrix}
			\frac{\omega_0}{2}+\omega a^\dag a&0\\0&-\frac{\omega_0}{2}+\omega a^\dag a
		\end{bmatrix}+\begin{bmatrix}
			b_{11}a^\dag+b_{11}^*a&b_{12}a^\dag+b_{21}^*a\\b_{12}^*a+b_{21}a^\dag&b_{22}a^\dag+b_{22}^*a
		\end{bmatrix}.
	\end{equation}	
	The next question is whether $H$ is self-adjoint---and this is where the Kato--Rellich theorem (Theorem~\ref{thm:katorellich}) will be used for the first time. 	
	To simplify the notation, let us write $H=H_{\rm free}+V$, where
	\begin{align}
		H_{\rm free}=\frac{\omega_0}{2}\sigma_z\otimes I+I\otimes\omega a^\dag a,\qquad V=B\otimes a^\dag+B^*\otimes a.
	\end{align}
	
	\begin{lemma}
		$H_{\rm free}$ is a self-adjoint operator on the Hilbert space $\hfrak=\mathbb{C}^2\otimes\hilb$.
	\end{lemma}
	\begin{proof}
		$\omega a^\dag a$ is a self-adjoint operator on $\hilb$ (cf. Proposition~\ref{prop:numberselfadjoint}), and $\frac{\omega_0}{2}\sigma_z$ is clearly self-adjoint on $\mathbb{C}^2$. Recalling that the tensor product of self-adjoint operators is self-adjoint (see Proposition~\ref{prop:polytensor} and discussion thereafter), the two operators $\frac{\omega_0}{2}\sigma_z\otimes I$ and $I\otimes\omega a^\dag a$ are both self-adjoint. 
		Furthermore, since the first one is a bounded self-adjoint operator, their sum is self-adjoint as well.
	\end{proof}
	
	\begin{remark}
		The boundedness of $\frac{\omega_0}{2}\sigma_z$ was used to conclude self-adjointness. If we replaced the qubit with an infinite-dimensional system with unbounded energy, e.g. a second field:
		\begin{equation}
			H_{\rm free}=\omega_{1}a_1^\dag a_1\otimes I+I\otimes \omega_2a_2^\dag a_2,
		\end{equation}
		then Proposition~\ref{prop:polytensor} would only guarantee essential self-adjointness of the sum on the intersection of the domains. Self-adjointness on that domain needs an extra argument: here it holds because both terms are nonnegative (in the joint spectral representation, $(\lambda_1+\lambda_2)^2\geq\lambda_1^2,\lambda_2^2$), but it fails e.g. for $\omega_{1}a_1^\dag a_1\otimes I-I\otimes \omega_2a_2^\dag a_2$.
	\end{remark}
	
	\begin{lemma}\label{lem:reltensor}
		Let $A$ be a closed and $B$ a closable operator on $\hilb$, and let $\mathcal{K}$ be a $d$-dimensional Hilbert space. Assume that $B$ is $A$-bounded. Then $I\otimes B$ is $I\otimes A$-bounded with the same relative bound.
	\end{lemma}
	
	\begin{proof}
		Let $(e_j)_{j=1}^d$ be an orthonormal basis of $\mathcal{K}$. Since $\mathcal{K}$ is finite-dimensional and the identity is invertible, $I\otimes A=I\odot A$ and $I\otimes B=I\odot\overline{B}$ (Proposition~\ref{prop:tensorproduct_finitedim2}). Every $\Psi\in\Dom(I\otimes A)=\mathcal{K}\odot\Dom A$ can then be decomposed as $\Psi=\sum_{j=1}^d e_j\otimes\psi_j$, with $\psi_j\in\Dom A$ and $\|\Psi\|^2=\sum_{j=1}^d\|\psi_j\|^2$ by the orthogonality of the vectors $e_j$. Similarly,
		\begin{equation}
			(I\otimes A)\Psi=\sum_{j=1}^de_j\otimes A\psi_j,\qquad\|(I\otimes A)\Psi\|^2=\sum_{j=1}^d\|A\psi_j\|^2,
		\end{equation}
		\begin{equation}
			(I\otimes B)\Psi=\sum_{j=1}^de_j\otimes B\psi_j,\qquad\|(I\otimes B)\Psi\|^2=\sum_{j=1}^d\|B\psi_j\|^2.
		\end{equation}
		As $B$ is $A$-bounded, there exist $a,b\geq0$ such that $\|B\psi_j\|^2\leq a^2\|A\psi_j\|^2+b^2\|\psi_j\|^2$ for every $j=1,\dots,d$ (recall Proposition~\ref{prop:squared_relbound}), and therefore
		\begin{equation}
			\|(I\otimes B)\Psi\|^2=\sum_{j=1}^d\|B\psi_j\|^2\leq\sum_{j=1}^d\left(a^2\|A\psi_j\|^2+b^2\|\psi_j\|^2\right)=a^2\|(I\otimes A)\Psi\|^2+b^2\|\Psi\|^2.
		\end{equation}
		Thus the $(I\otimes A)$-bound of $I\otimes B$ is not larger than the $A$-bound of $B$; conversely, choosing $\Psi=e_1\otimes\psi$ shows that it is not smaller either.
	\end{proof}
	
	\begin{proposition}\label{prop:visrelbound}
		The operator $V=B\otimes a^\dag+B^*\otimes a$ is infinitesimally $H_{\rm free}$-bounded, and therefore $H$ is self-adjoint.
	\end{proposition}
	
	\begin{proof}
		Recall that, by Proposition~\ref{prop:asharpvsn},
		\begin{equation}
			\|a\psi\|\leq\|(\N+1)^{1/2}\psi\|,\qquad\|a^\dag\psi\|\leq\|(\N+1)^{1/2}\psi\|.
		\end{equation}
		Therefore, recalling Lemma~\ref{lem:reltensor}, for all $\Psi\in\mathbb{C}^2\otimes\Dom\N$ we have
		\begin{equation}
			\|(I\otimes a)\Psi\|\leq\|I\otimes(\N+1)^{1/2}\Psi\|,\qquad\|(I\otimes a^\dag)\Psi\|\leq\|I\otimes(\N+1)^{1/2}\Psi\|,
		\end{equation}
		whence
		\begin{equation}
			\|(B\otimes a^\dag)\Psi\|=\|(B\otimes I)(I\otimes a^\dag)\Psi\|\leq\|B\otimes I\|\|(I\otimes a^\dag)\Psi\|\leq \|B\|\|I\otimes(\N+1)^{1/2}\Psi\|.
		\end{equation}		
		An analogous argument yields $\|(B^*\otimes a)\Psi\|\leq\|B^*\|\|I\otimes(\N+1)^{1/2}\Psi\|$, whence
		\begin{align}
			\|V\Psi\|\leq\left(\|B\|+\|B^*\|\right)\|I\otimes(\N+1)^{1/2}\Psi\|=2\|B\|\|I\otimes(\N+1)^{1/2}\Psi\|.
		\end{align}
		
		We proved that $V$ is relatively bounded with respect to $I\otimes(\N+1)^{1/2}=(I\otimes(\N+1))^{1/2}$, and thus also with respect to $(I\otimes\N)^{1/2}$ as this operator only differs from the former by a bounded term. 		
		But recall that every symmetric operator is infinitesimally bounded with respect to its square (Example~\ref{examp:wrtsquare}). Therefore, $(I\otimes\N)^{1/2}$ is infinitesimally bounded with respect to $I\otimes\N$, or equivalently with respect to $I\otimes\omega\N$, and so is $V$: for all $\epsilon>0$, there exists $b_\epsilon\geq0$ such that
		\begin{equation}
			\|V\Psi\|\leq\epsilon\|(I\otimes\omega\N)\Psi\|+b_\epsilon\|\Psi\|.
		\end{equation}
		But $I\otimes\omega\N=H_{\rm free}-\frac{\omega_0}{2}\sigma_z\otimes I$, whence, using the triangle inequality and $\|\sigma_z\|=1$,
		\begin{equation}
			\|V\Psi\|\leq\epsilon\left\|\left(H_{\rm free}-\frac{\omega_0}{2}\sigma_z\otimes I\right)\Psi\right\|+b_\epsilon\|\Psi\|\leq\epsilon\|H_{\rm free}\Psi\|+\left(b_\epsilon+\frac{\epsilon|\omega_0|}{2}\right)\|\Psi\|,
		\end{equation}
		which shows that $V$ is indeed infinitesimally $H_{\rm free}$-bounded. Besides, $V$ is symmetric on $\mathbb{C}^2\odot\Dom\N^{1/2}\supseteq\Dom H_{\rm free}$, since $a$ and $a^\dag$ are mutually adjoint on $\Dom\N^{1/2}$ (Proposition~\ref{prop:aprop2}): for all $u,v\in\mathbb{C}^2$ and $\varphi,\psi\in\Dom\N^{1/2}$,
		\begin{equation}
			\Braket{u\otimes\varphi,(B\otimes a^\dag)(v\otimes\psi)}=\braket{B^*u,v}\braket{a\varphi,\psi}=\Braket{(B^*\otimes a)(u\otimes\varphi),v\otimes\psi},
		\end{equation}
		and the same holds, by linearity, on the whole $\mathbb{C}^2\odot\Dom\N^{1/2}$. By the Kato--Rellich theorem (Theorem~\ref{thm:katorellich}), $H=H_{\rm free}+V$ is therefore self-adjoint.
	\end{proof}
	
	We can now put the tools developed so far to use, and study specific spin--boson models, corresponding to specific choices of the matrix $B$. We consider three cases: $B\propto\sigma_z$, $B\propto\sigma_-$ (the \emph{Jaynes--Cummings model}~\cite{jaynescummings1963,shoreknight1993}), and $B\propto\sigma_x$ (the \emph{quantum Rabi model}, named after Rabi's semiclassical model of a spin in a rotating magnetic field~\cite{rabi1937}). As we will see, the first two models are exactly solvable: one can determine their spectrum exactly and thus, in principle, their dynamics. A comprehensive, physics-oriented account of these models and of their many generalizations can be found in~\cite{larsonmavrogordatos}. The key observation is the following: the model with $B\propto\sigma_z$ decomposes into the direct sum of two operators on $\hilb$ that can be solved exactly, while the model with $B\propto\sigma_-$ decomposes into the direct sum of countably many $2\times 2$ matrices. In keeping with the spirit of these notes, all computations will be carried out explicitly.
	
	\subsection{The pure dephasing spin--boson model}\label{subsec:dephasing}\index{spin--boson model!pure dephasing}
	
	We start with $B=g\sigma_z$ (longitudinal coupling) with $g\in\mathbb{R}$. Our model becomes
	\begin{align}
		H&=\frac{\omega_0}{2}\sigma_z\otimes I+I\otimes\omega a^\dag a+g\sigma_z\otimes(a+a^\dag)\nonumber\\
		&\cong\begin{bmatrix}
			\frac{\omega_0}{2}+\omega a^\dag a+g(a+a^\dag)&0\\0&-\frac{\omega_0}{2}+\omega a^\dag a-g(a+a^\dag)
		\end{bmatrix}.
	\end{align}
	This model is manifestly diagonal with respect to the computational basis of $\mathbb{C}^2$:
	\begin{equation}\label{eq:dephasing-decomp}
		H\cong H_+\oplus H_-,\qquad H_\pm:=\omega a^\dag a\pm\left(g(a+a^\dag)+\frac{\omega_0}{2}\right),
	\end{equation}
	with each $H_\pm:\Dom\N\subseteq\hilb\rightarrow\hilb$ being self-adjoint. Hence there is no coupling between the two energy levels of the qubit, which evolve independently: $\e^{-\ii tH}\cong\e^{-\ii tH_+}\oplus\e^{-\ii tH_-}$.
	
	Let us justify the name ``pure dephasing'' (see also~\cite{palma1996,breuerpetruccione}).	
	Consider a spin--boson system initially prepared in a (possibly mixed) state of the form (Section~\ref{subsec:mixed} and Example~\ref{examp:qubitdensity})
	\begin{equation}
		\varrho=\rho\otimes\ket{\Omega}\!\bra{\Omega},\qquad \rho=\sum_{s,s'=\pm}\rho_{ss'}\ket{s}\!\bra{s'}=\begin{bmatrix}
			\rho_{++}&\rho_{+-}\\\rho_{-+}&\rho_{--}
		\end{bmatrix},\qquad\operatorname{tr}\rho=\rho_{++}+\rho_{--}=1,
	\end{equation}
	with $\Omega\in\hilb$ a unit vector. Thus
	\begin{equation}
		\varrho\cong\begin{bmatrix}
			\rho_{++}\ket{\Omega}\!\bra{\Omega}&\rho_{+-}\ket{\Omega}\!\bra{\Omega}\\\rho_{-+}\ket{\Omega}\!\bra{\Omega}&\rho_{--}\ket{\Omega}\!\bra{\Omega}
		\end{bmatrix}.
	\end{equation}
	The evolution of this state under $H$ is given by
	\begin{align}
		\varrho(t)&:=\e^{-\ii tH}\varrho\e^{+\ii tH}\nonumber\\
		&\cong\begin{bmatrix}
			\e^{-\ii tH_+}&\\&\e^{-\ii tH_-}
		\end{bmatrix}\begin{bmatrix}
			\rho_{++}\ket{\Omega}\!\bra{\Omega}&\rho_{+-}\ket{\Omega}\!\bra{\Omega}\\\rho_{-+}\ket{\Omega}\!\bra{\Omega}&\rho_{--}\ket{\Omega}\!\bra{\Omega}
		\end{bmatrix}\begin{bmatrix}
			\e^{+\ii tH_+}&\\&\e^{+\ii tH_-}
		\end{bmatrix}\nonumber\\
		&=\begin{bmatrix}
			\rho_{++}\e^{-\ii tH_+}\ket{\Omega}\!\bra{\Omega}\e^{+\ii t H_+}&	\rho_{+-}	\e^{-\ii tH_+}\ket{\Omega}\!\bra{\Omega}\e^{+\ii tH_-}\\	\rho_{-+}\e^{-\ii tH_-}\ket{\Omega}\!\bra{\Omega}	\e^{+\ii tH_+}&\rho_{--}	\e^{-\ii tH_-}\ket{\Omega}\!\bra{\Omega}	\e^{+\ii tH_-}
		\end{bmatrix}.
	\end{align}
	The reduced evolution of the spin density matrix $\rho$ is defined by $\rho(t):=\operatorname{tr}_{\hilb}\varrho(t)$, with $\operatorname{tr}_\hilb(\cdot)$ the partial trace over $\hilb$ (Proposition~\ref{prop:partialtrace}), which here simply amounts to taking the trace of each entry of the $2\times2$ $\mathcal{B}(\hilb)$-valued matrix above:
	\begin{equation}\label{eq:dephasingchannel}
		\rho(t)=\begin{bmatrix}
			\rho_{++}&\rho_{+-}\varphi(t)\\\rho_{-+}\overline{\varphi(t)}&\rho_{--}
		\end{bmatrix},\qquad \varphi(t)=\Braket{\Omega,\e^{+\ii tH_-}\e^{-\ii tH_+}\Omega}.
	\end{equation}
	This is a \emph{dephasing qubit channel}\index{dephasing channel} (see e.g. \cite[Section 8.3]{nielsenchuang}). Here, a (quantum) \emph{channel}\index{quantum channel} is a map sending density operators into density operators, and the reduced evolution of a system which interacts with an environment, starting from a product state as above, is a typical example of it; see~\cite[Chapter 8]{nielsenchuang}. In the dephasing channel, the diagonal elements of the qubit density matrix---the populations of each energy level---are unchanged, while the off-diagonal elements---the coherences---are multiplied by a time-dependent factor $\varphi(t)$, the dephasing function. Clearly, $|\varphi(t)|\leq1$.
	
	The above calculation holds for any Hamiltonian on $\mathbb{C}^2\otimes\hilb$ of the form $H\cong H_+\oplus H_-$. Next we show that, in our specific case, the model is exactly solvable. This amounts to showing that the two Hamiltonians $H_\pm$ of Eq.~\eqref{eq:dephasing-decomp} are separately solvable.
	
	\begin{lemma}\label{lem:predispl}
		The operators $a+a^\dag$ and $\ii(a-a^\dag)$, with domain $\Dom\N^{1/2}$, are essentially self-adjoint.
	\end{lemma}
	
	\begin{proof}
		The simplest way to prove this is to go back to the Hermite representation: the restrictions of $a+a^\dag$ and $\ii(a-a^\dag)$ to $\mathcal{S}(\mathbb{R})$ are just (multiples of) the position and momentum operators on $\mathcal{S}(\mathbb{R})$, namely $\sqrt{2\omega}\,q$ and $-\sqrt{2/\omega}\,p$, which are essentially self-adjoint there. As $\Dom\N^{1/2}\supseteq\mathcal{S}(\mathbb{R})$, and both operators are symmetric on $\Dom\N^{1/2}$ ($a$ and $a^\dag$ being mutually adjoint there), a fortiori they are essentially self-adjoint on $\Dom\N^{1/2}$.
	\end{proof}
	
	\begin{definition}[Weyl displacement operator]\label{def:weyl}\index{Weyl displacement operator}
		Let $\lambda\in\mathbb{R}$. We define the \emph{Weyl displacement operator} $W(\lambda)$ by
		\begin{equation}
			W(\lambda)=\exp\left(-\ii\lambda\;\overline{\ii(a^\dag-a)}\right),
		\end{equation}
		where the exponential must be interpreted in the sense of the spectral theorem (Theorem~\ref{thm:spectraltheorem1}) for $\overline{\ii(a^\dag-a)}$.
	\end{definition}
	
	This definition makes sense since, by Lemma~\ref{lem:predispl}, the operator $\overline{\ii(a^\dag-a)}=-\overline{\ii(a-a^\dag)}$ is self-adjoint. Moreover:
	\begin{lemma}\label{lem:displ}
		For all $\lambda\in\mathbb{R}$, $W(\lambda)$ is unitary, with $W(\lambda)^{-1}=W(\lambda)^*=W(-\lambda)$.
	\end{lemma}
	
	\begin{proof}
		Direct consequence of the spectral theorem.
	\end{proof}
	
	Since all realizations of $a,a^\dag$ are unitarily equivalent (Section~\ref{subsec:abstract}), and $W(\lambda)$ is defined in terms of $a,a^\dag$ alone, we may compute with $W(\lambda)$ in any of them. In the Hermite representation, $W(\lambda)$ has a transparent meaning: it is a translation.
	
	\begin{lemma}\label{lem:weyltranslation}
		Let $\hilb=L^2(\mathbb{R})$, with $(\phi_n)_{n\in\mathbb{N}}$ the Hermite functions, and $\lambda\in\mathbb{R}$. Then $W(\lambda)=\e^{-\ii c_\lambda p}$, with $p$ the momentum operator (Example~\ref{examp:momentum}) and $c_\lambda:=\sqrt{2/\omega}\,\lambda$; that is,
		\begin{equation}\label{eq:weyltranslation}
			(W(\lambda)\psi)(x)=\psi(x-c_\lambda)\qquad\text{for all }\psi\in L^2(\mathbb{R}).
		\end{equation}
		In particular, $W(\lambda)\mathcal{S}(\mathbb{R})=\mathcal{S}(\mathbb{R})$.
	\end{lemma}
	
	\begin{proof}
		By (the proof of) Lemma~\ref{lem:predispl}, the restriction of $\ii(a^\dag-a)$ to $\mathcal{S}(\mathbb{R})$ is $\sqrt{2/\omega}\,p$, whose closure is the self-adjoint operator $\sqrt{2/\omega}\,p$ on $\mathrm{H}^1(\mathbb{R})$. The latter is thus contained in the symmetric operator $\overline{\ii(a^\dag-a)}$; since self-adjoint operators admit no proper symmetric extensions, $\overline{\ii(a^\dag-a)}=\sqrt{2/\omega}\,p$, whence $W(\lambda)=\e^{-\ii c_\lambda p}$. Now let $\mathcal{F}$ be the Fourier transform, so that $\mathcal{F}p\mathcal{F}^*$ is the multiplication by the variable $k$ (Example~\ref{examp:momentum}). By the spectral theorem, $\mathcal{F}\e^{-\ii c_\lambda p}\mathcal{F}^*$ is the multiplication by $\e^{-\ii c_\lambda k}$, which is precisely the Fourier transform of the translation by $c_\lambda$: the Fourier transform of $\psi(\cdot-c)$ is $\e^{-\ii ck}\hat\psi(k)$. This proves Eq.~\eqref{eq:weyltranslation}. The last claim follows since translations map Schwartz functions into Schwartz functions, and $W(\lambda)^{-1}=W(-\lambda)$ is a translation as well.
	\end{proof}
	
	\begin{proposition}\label{prop:displacement}
		For all $\lambda\in\mathbb{R}$,
		\begin{equation}
			W(\lambda)aW(\lambda)^*=a-\lambda,\qquad W(\lambda)a^\dag W(\lambda)^*=a^\dag-\lambda.
		\end{equation}
	\end{proposition}
	
	\begin{proof}
		The two equalities are adjoint to each other, so that it suffices to prove the first one. By unitary equivalence, we can work in the Hermite representation, where $a$ is the closure of $a_0=\sqrt{\omega/2}\,q+\ii p/\sqrt{2\omega}$, defined on $\mathcal{S}(\mathbb{R})$ (Definitions~\ref{def:a} and~\ref{defin:aadagnclose}). By Lemma~\ref{lem:weyltranslation}, $W(\lambda)$ is the translation by $c_\lambda$, which commutes with $p$ and shifts the position: for all $\psi\in\mathcal{S}(\mathbb{R})$,
		\begin{equation}
			\left(W(\lambda)\,q\,W(\lambda)^*\psi\right)(x)=(x-c_\lambda)\psi(x),\qquad W(\lambda)\,p\,W(\lambda)^*\psi=p\psi.
		\end{equation}
		Since $\sqrt{\omega/2}\,c_\lambda=\lambda$, we get $W(\lambda)a_0W(\lambda)^*\psi=(a_0-\lambda)\psi$ for all $\psi\in\mathcal{S}(\mathbb{R})$. Taking closures on both sides yields the claim. On the one hand, $\overline{a_0-\lambda}=a-\lambda$. On the other hand, since $W(\lambda)^*\mathcal{S}(\mathbb{R})=\mathcal{S}(\mathbb{R})$ by Lemma~\ref{lem:weyltranslation}, the operator $W(\lambda)a_0W(\lambda)^*$ is defined on $\mathcal{S}(\mathbb{R})$, and its closure is $W(\lambda)\overline{a_0}\,W(\lambda)^*=W(\lambda)aW(\lambda)^*$, since conjugation by a unitary operator preserves graph limits.
	\end{proof}
	
	\begin{remark}[Coherent states]\label{rem:coherent}
		In particular, $W(\lambda)$ acts as follows on the \emph{vacuum} state $\phi_0$ (the zeroth eigenvector of $\N$):
		\begin{equation}
			W(\lambda)\phi_0=\e^{-\lambda^2/2}\sum_{m=0}^\infty\frac{\lambda^m}{\sqrt{m!}}\phi_m,
		\end{equation}
		i.e. it generates the \emph{coherent states}\index{coherent states} familiar from quantum optics (see e.g.~\cite{gerryknight}; the matrix elements of $W(\lambda)$ between arbitrary number states are computed in~\cite{cahillglauber1969}). Indeed, replacing $\lambda$ with $-\lambda$ in Proposition~\ref{prop:displacement} gives $W(\lambda)^*aW(\lambda)=a+\lambda$. Since $a\phi_0=0$, the vector $\psi_\lambda:=W(\lambda)\phi_0$ is in $\Dom a$, and $a\psi_\lambda=W(\lambda)(a+\lambda)\phi_0=\lambda\psi_\lambda$. Writing $\psi_\lambda=\sum_mc_m\phi_m$, we get $\sqrt{m+1}\,c_{m+1}=\braket{a^\dag\phi_m,\psi_\lambda}=\braket{\phi_m,a\psi_\lambda}=\lambda c_m$, whence $c_m=c_0\lambda^m/\sqrt{m!}$. Normalization gives $|c_0|=\e^{-\lambda^2/2}$; besides, by Lemma~\ref{lem:weyltranslation}, $c_0=\braket{\phi_0,W(\lambda)\phi_0}$ is the overlap of two Gaussians centered at $0$ and $c_\lambda$, and is therefore positive (explicitly, $c_0=\e^{-\omega c_\lambda^2/4}=\e^{-\lambda^2/2}$). Weyl operators and coherent states for fields with arbitrarily many modes are discussed in Section~\ref{subsec:multidephasing} (Definition~\ref{def:coherentmulti} and Proposition~\ref{prop:weylmulti}).
	\end{remark}
	
	The reason for this detour is the following. Recall that $H\cong H_+\oplus H_-$, with $H_\pm$ as in Eq.~\eqref{eq:dephasing-decomp}.
	
	\begin{proposition}\label{prop:weyldisplaced}
		Let $H_\pm$ be as in Eq.~\eqref{eq:dephasing-decomp}. Then
		\begin{equation}
			W_\pm H_\pm W_\pm^* =\omega a^\dag a-\frac{g^2}{\omega}\pm\frac{\omega_0}{2},
		\end{equation}
		where $W_\pm:=W\left(\pm\frac{g}{\omega}\right)$.
	\end{proposition}
	
	\begin{exercise}\label{exerc:weyl}
		Prove this. Hint: this is a direct application of the equalities $W(\lambda)aW(\lambda)^*=a-\lambda$ and $W(\lambda)a^\dag W(\lambda)^*=a^\dag-\lambda$ (Proposition~\ref{prop:displacement}).
		\seesol{weyl}
	\end{exercise}
	
	Hence, introducing $W:=W_+\oplus W_-$, Proposition~\ref{prop:weyldisplaced} yields
	\begin{equation}
		WHW^* = \left(\omega a^\dag a-\frac{g^2}{\omega}+\frac{\omega_0}{2}\right)\oplus\left(\omega a^\dag a-\frac{g^2}{\omega}-\frac{\omega_0}{2}\right).
	\end{equation}
	That is, the pure dephasing spin--boson model is unitarily equivalent to the direct sum of two displaced harmonic oscillators.
	
	Each of the two operators is exactly solvable: 
	\begin{corollary}\label{coroll:dephasingspectrum}
		Both operators $H_\pm$ have a complete set of eigenvectors $(\phi_n^\pm)_{n\in\mathbb{N}}$, with associated eigenvalues given by
		\begin{equation}
			E_n^\pm=\omega n-\frac{g^2}{\omega}\pm\frac{\omega_0}{2}.
		\end{equation}
	\end{corollary}
	
	\begin{exercise}\label{exerc:dephasingdiag}
		Fill in the details. Hint: define $\phi_n^\pm:=W_\pm^*\phi_n$. Why do these vectors form a complete orthonormal set?
		\seesol{dephasingdiag}
	\end{exercise}
	
	\begin{remark}[The dephasing function for the vacuum]\label{rem:dephasingvacuum}
		Proposition~\ref{prop:weyldisplaced} also allows us to compute the dephasing function $\varphi(t)$ explicitly. Take for instance $\Omega=\phi_0$, the vacuum of the field. Since $\e^{-\ii tH_\pm}=W_\pm^*\,\e^{-\ii t(\omega a^\dag a-g^2/\omega\pm\omega_0/2)}\,W_\pm$, each vector $\e^{-\ii tH_\pm}\phi_0$ is, up to a phase, a coherent state (Remark~\ref{rem:coherent}), and a direct computation using the standard properties of coherent states (see e.g.~\cite{gerryknight}) gives (see also Corollary~\ref{coroll:multidephasing} below, of which this is the special case $J=g^2\delta_\omega$)
		\begin{equation}\label{eq:dephasingvacuum}
			\varphi(t)=\e^{-\ii\omega_0t}\exp\left(-\frac{8g^2}{\omega^2}\sin^2\frac{\omega t}{2}\right).
		\end{equation}
		The modulus $|\varphi(t)|$ decreases from $1$ to its minimum $\e^{-8g^2/\omega^2}$ at $t=\pi/\omega$, and returns to $1$ at $t=2\pi/\omega$: the coherence of the qubit is periodically lost and fully recovered. This is a characteristic feature of a single field mode: since the spectra of $H_\pm$ are contained in $\omega\mathbb{N}+\mathrm{const}$ (Corollary~\ref{coroll:dephasingspectrum}), $\e^{\ii tH_-}\e^{-\ii tH_+}$ is periodic in $t$, with period $2\pi/\omega$, up to the phase $\e^{-\ii\omega_0t}$, for every choice of $\Omega$. An irreversible loss of coherence requires a field with infinitely many modes, as we will see in Section~\ref{subsec:multidephasing} (Remark~\ref{rem:irreversible}).
	\end{remark}
	
	\subsection{The Jaynes--Cummings model}\label{subsec:jc}\index{Jaynes--Cummings model}
	
	Let us now consider the case $B=g\sigma_-$, where
	\begin{equation}
		\sigma_+=\begin{bmatrix}
			0&1\\0&0
		\end{bmatrix},\qquad\sigma_-=\begin{bmatrix}
			0&0\\1&0
		\end{bmatrix},\qquad \sigma_\pm=\sigma_\mp^*.
	\end{equation} 
	These two matrices play the role of ladder operators for the spin, as $\sigma_+e_-=e_+$ and $\sigma_-e_+=e_-$, thus implementing transitions between the two energy levels.
	
	The corresponding spin--boson model reads
	\begin{align}
		H&=\frac{\omega_0}{2}\sigma_z\otimes I+I\otimes\omega a^\dag a+g\left(\sigma_+\otimes a+\sigma_-\otimes a^\dag\right)\nonumber\\
		&\cong\begin{bmatrix}
			\frac{\omega_0}{2}+\omega a^\dag a&ga\\ga^\dag&-\frac{\omega_0}{2}+\omega a^\dag a
		\end{bmatrix}.
	\end{align}
	Unlike in the pure dephasing case (Section~\ref{subsec:dephasing}), the matrix now has off-diagonal terms, signaling that the model induces transitions between the two spin levels. 	
	To solve the pure dephasing model, we crucially used the fact that it is block diagonal or, equivalently, that it has a conserved quantity (a symmetry), namely $S=\sigma_z\otimes I\cong I\oplus(-I)$.
	To solve the Jaynes--Cummings model, we likewise look for a symmetry.
	
	To this end, consider the two interaction terms $\sigma_+\otimes a$ and $\sigma_-\otimes a^\dag$ in the Jaynes--Cummings model. The first term corresponds to an excitation of the spin (transition $e_-\rightarrow e_+$) together with the annihilation of a particle; the second term corresponds to a decay of the spin (transition $e_+\rightarrow e_-$) together with the creation of a particle. That is, $H$ preserves the number of excitations of the spin--boson system, defined as the number of boson particles plus either $1$ or $0$, depending on whether the spin is in $e_+$ or $e_-$. Let us formalize this.
	
	\begin{definition}[Number of excitations]\label{def:ntot}\index{number of excitations}
		Let $\hfrak=\mathbb{C}^2\otimes\hilb$. The \emph{number of excitations} of the spin--boson system is the operator defined on $\Dom\N_{\rm tot}=\mathbb{C}^2\odot\Dom\N$ by
		\begin{equation}
			\N_{\rm tot}:=\sigma_+\sigma_-\otimes I+I\otimes\N\cong\begin{bmatrix}
				\N+1&0\\0&\N
			\end{bmatrix}.
		\end{equation}
	\end{definition}
	
	The second expression follows from the fact that $\sigma_+\sigma_-$ is the projector associated with the excited state $e_+$ of the spin:
	\begin{equation}
		\sigma_+\sigma_-=\begin{bmatrix}
			0&1\\0&0
		\end{bmatrix}\begin{bmatrix}
			0&0\\1&0
		\end{bmatrix}=\begin{bmatrix}
			1&0\\0&0
		\end{bmatrix}.
	\end{equation}
	
	As we will see, this operator has a pure point spectrum consisting of all natural numbers, just like $\N$, with an important difference: each $n\in\mathbb{N}$ is a simple eigenvalue of $\N$, with eigenspace spanned by $\phi_n$, while, as an eigenvalue of $\N_{\rm tot}$, each $n\geq1$ has a two-dimensional eigenspace (only $n=0$ is simple):
	
	\begin{proposition}\label{prop:hfrakn}
		Let $n\in\mathbb{N}$. Then:
		\begin{itemize}
			\item For $n=0$, the space 
			\begin{equation}
				\mathcal{E}_0:=\operatorname{Span}\{e_-\otimes\phi_0\}=\operatorname{Span}\left\{\begin{bmatrix}
					0\\\phi_0
				\end{bmatrix}\right\}
			\end{equation}
			is an eigenspace for $\N_{\rm tot}$ with eigenvalue $0$;
			\item For $n\geq1$, the space 
			\begin{equation}
				\mathcal{E}_n:=\operatorname{Span}\{e_+\otimes\phi_{n-1},e_-\otimes\phi_n\}=\operatorname{Span}\left\{\begin{bmatrix}
					\phi_{n-1}\\0
				\end{bmatrix},\begin{bmatrix}
					0\\\phi_n
				\end{bmatrix}\right\}
			\end{equation}
			is an eigenspace for $\N_{\rm tot}$ with eigenvalue $n$.
		\end{itemize}
	\end{proposition}
	
	\begin{proof}
		This follows immediately from the matrix representation of $\N_{\rm tot}$ in Definition~\ref{def:ntot}.
	\end{proof}
	
	For $n\geq1$, $\mathcal{E}_n$ is thus a two-dimensional space with orthonormal basis $\{e_+\otimes\phi_{n-1},e_-\otimes\phi_n\}$, and every element of $\mathcal{E}_n$ is uniquely determined by two complex coefficients $c_n^\pm$:
	\begin{equation}
		\mathcal{E}_n\ni\Psi=c_n^+\begin{bmatrix}
			\phi_{n-1}\\0
		\end{bmatrix}+c_n^-\begin{bmatrix}
			0\\\phi_n
		\end{bmatrix},
	\end{equation}
	whence $\mathcal{E}_n\cong\mathbb{C}^2$ via $\Psi\mapsto\begin{bmatrix}
		c_n^+&c_n^-
	\end{bmatrix}^\intercal$. The most general $\Psi\in\hfrak$ can be uniquely represented as
	\begin{equation}\label{eq:jcpsi}
		\Psi\cong\left[\begin{array}{c}
			c_0\\\hline c_1^+\\c_1^-\\\hline c_2^+\\c_2^-\\\hline\vdots
		\end{array}\right]
	\end{equation}
	and, in this representation,
	\begin{equation}\label{eq:jcntot}
		\N_{\rm tot}\cong\left[\begin{array}{c|c c|c c|c}
			0& & & & & \\\hline
			&1& & & & \\
			& &1& & & \\\hline
			& & &2& & \\
			& & & &2& \\ \hline
			& & & & &\ddots
		\end{array}\right].
	\end{equation}
	
	The fact that the Jaynes--Cummings Hamiltonian $H$ preserves the number of excitations can now be phrased as follows:
	
	\begin{proposition}\label{prop:solving_jc1}
		Let $H$ be the Jaynes--Cummings Hamiltonian on $\hfrak=\mathbb{C}^2\otimes\hilb$, and $\mathcal{E}_n$ the $n$th eigenspace of $\N_{\rm tot}$. Then each $\mathcal{E}_n$ is an invariant subspace for $H$, i.e. $H\mathcal{E}_n\subseteq\mathcal{E}_n$. In particular, under the isomorphism $\mathcal{E}_n\cong\mathbb{C}^2$ described after Proposition~\ref{prop:hfrakn}, 	
		for $n\geq1$ the action of $H$ on $\mathcal{E}_n$ is represented by the $2\times2$ matrix 
		\begin{equation}
			H_n=\begin{bmatrix}
				\frac{\omega_0}{2}+\omega(n-1)&g\sqrt{n}\\g\sqrt{n}&-\frac{\omega_0}{2}+\omega n
			\end{bmatrix}.
		\end{equation}
	\end{proposition}
	
	\begin{exercise}\label{exerc:jcinvariant}
		Prove this. Hint: this follows from a direct computation of $H\,e_+\otimes\phi_{n-1}$ and $H\,e_-\otimes\phi_n$.
		\seesol{jcinvariant}
	\end{exercise}
	
	That is: in the basis of Eqs.~\eqref{eq:jcpsi}--\eqref{eq:jcntot}, the Jaynes--Cummings Hamiltonian is represented by the block-diagonal infinite matrix
	\begin{equation}
		H\cong\left[\begin{array}{c|c|c|c}
			-\frac{\omega_0}{2}& & &   \\\hline
			&H_1& &  \\\hline
			& &H_2&   \\\hline
			& & &\ddots  
		\end{array}\right],
	\end{equation}
	with $H_n$ as in Proposition~\ref{prop:solving_jc1}. Solving the Jaynes--Cummings model thus reduces to finding the eigenvalues and eigenvectors of $2\times 2$ matrices: an infinite-dimensional problem has been reduced to a sequence of much simpler two-dimensional ones.
	
	\begin{proposition}\label{prop:solving_jc2}
		Let $n\geq1$ and $g>0$. The eigenvalues $E_n^\pm$ of the $2\times2$ matrix $H_n$ are given by
		\begin{equation}
			E_n^\pm=\omega\left(n-\frac{1}{2}\right)\pm\sqrt{\frac{(\omega_0-\omega)^2}{4}+g^2n},
		\end{equation}
		with the corresponding eigenvectors being given by
		\begin{equation}
			\begin{bmatrix}
				\cos\theta_n\\\sin\theta_n
			\end{bmatrix}\;(\text{for }E_n^+),\qquad \begin{bmatrix}
				-\sin\theta_n\\\cos\theta_n
			\end{bmatrix}\;(\text{for }E_n^-),
		\end{equation}
		where $\theta_n\in(0,\pi/2)$ is uniquely defined by
		\begin{equation}
			\cos2\theta_n=\frac{\omega_0-\omega}{R_n},\qquad\sin2\theta_n=\frac{2g\sqrt{n}}{R_n},\qquad R_n=\sqrt{(\omega_0-\omega)^2+4g^2n}.
		\end{equation}
		In particular, $\theta_n=\pi/4$ for $\omega_0=\omega$.
	\end{proposition}
	\begin{proof}
		This is an elementary exercise in linear algebra.
	\end{proof}
	
	Correspondingly, the Jaynes--Cummings Hamiltonian $H$ has eigenvalues $E_0=-\frac{\omega_0}{2}$ and $E_n^\pm$, $n\geq1$, as in Proposition~\ref{prop:solving_jc2}, and the corresponding eigenstates (\emph{dressed states}\index{dressed states}) are
	\begin{equation}
		\Phi_0\cong\begin{bmatrix}
			0\\\phi_0
		\end{bmatrix},\qquad\Phi_n^+\cong\begin{bmatrix}
			\cos\theta_n\,\phi_{n-1}\\\sin\theta_n\,\phi_n
		\end{bmatrix},\qquad \Phi_n^-\cong\begin{bmatrix}
			-\sin\theta_n\,\phi_{n-1}\\\cos\theta_n\,\phi_n
		\end{bmatrix}\qquad(n\geq1),
	\end{equation}
	with $\theta_n$ as in Proposition~\ref{prop:solving_jc2}. In particular, in the \emph{resonant} case $\omega_0=\omega$, $\theta_n=\pi/4$ and the eigenstates $\Phi_n^\pm$ are maximally entangled states. The dynamics induced by $H$ on $\mathcal{E}_n$ consists of oscillations between the bare states $e_+\otimes\phi_{n-1}$ and $e_-\otimes\phi_n$ (the eigenstates $\Phi_n^\pm$ only acquire phases), with frequency $E_n^+-E_n^-=R_n$, known as \emph{Rabi oscillations}\index{Rabi oscillations}. This completes the solution of the model.
	
	We previously showed (Section~\ref{subsec:dephasing}) that the spin--boson model with $B\propto\sigma_z$ is associated with a dephasing qubit channel: given an initial state $\rho\otimes\ket{\Omega}\!\bra{\Omega}$, with $\rho$ a $2\times2$ density matrix and $\Omega\in\hilb$ any unit vector, the reduced state $\rho(t)$ of the qubit is given by Eq.~\eqref{eq:dephasingchannel}.	
	In a similar way, the Jaynes--Cummings model defines a \emph{qubit amplitude-damping channel}\index{amplitude-damping channel} (see again \cite[Section 8.3]{nielsenchuang}):
	
	\begin{exercise}\label{exerc:ampdamp}
		Let $H$ be the Jaynes--Cummings model on $\mathbb{C}^2\otimes\hilb$, and $\phi_0\in\hilb$ the vacuum state (zeroth eigenvector of $\N$). Show that, given any $2\times 2$ density matrix $\rho$,
		\begin{equation}
			\rho(t):=\operatorname{tr}_\hilb\left(\e^{-\ii tH}\rho\otimes\ket{\phi_0}\!\bra{\phi_0}\e^{\ii tH}\right)=\begin{bmatrix}
				|\alpha(t)|^2	\rho_{++}&\alpha(t)\rho_{+-}\\\overline{\alpha(t)}\rho_{-+}&\rho_{--}+(1-|\alpha(t)|^2)\rho_{++}
			\end{bmatrix}
		\end{equation}
		for some function $\alpha(t)$. Hint: use the fact (Proposition~\ref{prop:solving_jc1}) that $H$ preserves the eigenspaces of the total number of excitations; in particular, $e_-\otimes\phi_0$ is an eigenvector of $H$, and $e_+\otimes\phi_0$ evolves inside the two-dimensional space spanned by $e_+\otimes\phi_0$ and $e_-\otimes\phi_1$. This allows you to recover the claimed expression for $\rho(t)$ without solving the Schr\"odinger equation.
		\seesol{ampdamp}
	\end{exercise}
	
	\begin{remark}[Vacuum Rabi oscillations]\label{rem:vacuumrabi}\index{Rabi oscillations!vacuum}
		We can now compute explicitly the function $\alpha(t)$ of the amplitude-damping channel of Exercise~\ref{exerc:ampdamp}. The vector $e_+\otimes\phi_0$ belongs to the invariant subspace $\mathcal{E}_1$, on which, in the basis $\{e_+\otimes\phi_0,e_-\otimes\phi_1\}$, $H$ acts as the matrix (cf. Proposition~\ref{prop:solving_jc1} with $n=1$)
		\begin{equation}
			H_1=\frac{\omega}{2}I+\frac{1}{2}M,\qquad M:=\begin{bmatrix}\Delta&2g\\2g&-\Delta\end{bmatrix},\qquad\Delta:=\omega_0-\omega.
		\end{equation}
		Since $M^2=R_1^2I$, with $R_1=\sqrt{\Delta^2+4g^2}$ as in Proposition~\ref{prop:solving_jc2}, one has
		\begin{equation}
			\e^{-\ii tH_1}=\e^{-\ii\omega t/2}\left(\cos\frac{R_1t}{2}\,I-\ii\sin\frac{R_1t}{2}\,\frac{M}{R_1}\right),
		\end{equation}
		whence (cf. the solution of Exercise~\ref{exerc:ampdamp})
		\begin{equation}\label{eq:vacuumrabi}
			\alpha(t)=\e^{-\ii(\omega+\omega_0)t/2}\left(\cos\frac{R_1t}{2}-\ii\frac{\Delta}{R_1}\sin\frac{R_1t}{2}\right),\qquad|\alpha(t)|^2=1-\frac{4g^2}{R_1^2}\sin^2\frac{R_1t}{2}.
		\end{equation}
		That is, the population of the excited state oscillates between $1$ and $1-4g^2/R_1^2$ (between $1$ and $0$ at resonance, where $|\alpha(t)|^2=\cos^2(gt)$): the qubit periodically emits a single boson and reabsorbs it. These are the \emph{vacuum Rabi oscillations}. As in the pure dephasing case (Remark~\ref{rem:dephasingvacuum}), the loss is not irreversible; a genuine, exponential decay of the excited state only arises for a continuum of modes, as in the Wigner--Weisskopf theory of spontaneous emission (see Section~\ref{subsec:multijc}).
	\end{remark}
	
	\subsection{The quantum Rabi model}\label{subsec:rabi}\index{quantum Rabi model}
	
	Finally, we consider the case $B=g\sigma_x$, whence
	\begin{align}
		H_{\rm QR}&=\frac{\omega_0}{2}\sigma_z\otimes I+I\otimes\omega a^\dag a+g\sigma_x\otimes(a+a^\dag)\nonumber\\&\cong\begin{bmatrix}
			\frac{\omega_0}{2}+\omega a^\dag a&g(a+a^\dag)\\g(a+a^\dag)&-\frac{\omega_0}{2}+\omega a^\dag a
		\end{bmatrix}.
	\end{align}
	This is the quantum Rabi model, which describes, under some simplifications, the interaction of a two-level system with a single mode of the electric field.
	
	Since $\sigma_x=\sigma_++\sigma_-$, the model can be written as
	\begin{equation}
		H_{\rm QR}=\overbrace{\frac{\omega_0}{2}\sigma_z\otimes I+I\otimes\omega a^\dag a+g\left(\sigma_+\otimes a+\sigma_-\otimes a^\dag\right)}^{=H_{\rm JC}}+g\left(\sigma_+\otimes a^\dag+\sigma_-\otimes a\right),
	\end{equation}
	that is, $H_{\rm QR}$ differs from the Jaynes--Cummings model by two additional interaction terms, $\sigma_+\otimes a^\dag$ and $\sigma_-\otimes a$, that implement two new transitions: the first term corresponds to an excitation of the spin (transition $e_-\rightarrow e_+$) together with the creation of a particle, and the second term corresponds to a decay of the spin (transition $e_+\rightarrow e_-$) together with the annihilation of a particle.
	These terms change the total number of excitations of the system, thus breaking the symmetry observed in the Jaynes--Cummings case.	
	Consequently, Propositions~\ref{prop:solving_jc1}--\ref{prop:solving_jc2} cease to be valid: $H_{\rm QR}$ does not commute with $\N_{\rm tot}$ (i.e., it does not share eigenspaces with it), and cannot be solved as easily; an analytical characterization of its spectrum is known~\cite{braak2011,larsonmavrogordatos}, but it is considerably more involved.
	
	However, the situation simplifies in the regime of large $\omega$, at fixed coupling $g$ and fixed detuning $\Delta=\omega_0-\omega$. The heuristic argument goes as follows. In the interaction picture\index{interaction picture} with respect to the Hamiltonian $\frac{\omega}{2}\sigma_z+\omega a^\dag a$, the Hamiltonians $H_{\rm QR}$ and $H_{\rm JC}$ respectively become
	\begin{align}
		H_{\rm QR}^{\rm I}(t)&=\frac{\Delta}{2}\sigma_z+g\left(\sigma_+a+\sigma_-a^\dag\right)+g\left(\e^{2\ii\omega t}\sigma_+a^\dag+\e^{-2\ii\omega t}\sigma_-a\right);\\
		H_{\rm JC}^{\rm I}(t)&=\frac{\Delta}{2}\sigma_z+g\left(\sigma_+a+\sigma_-a^\dag\right),
	\end{align}
	That is, in the interaction picture the two models differ by two terms with time-dependent oscillating coefficients $\e^{\pm2\ii\omega t}$. As $\omega\to\infty$, these terms oscillate faster and faster, and are thus expected to average out: going back to the original picture, the dynamics generated by the two Hamiltonians should become close, in a suitable sense, as $\omega$ grows. This would enable one to discard the additional terms and approximate $H_{\rm QR}$ with the (considerably easier) Jaynes--Cummings model---the so-called \emph{rotating-wave approximation}\index{rotating-wave approximation} (RWA).
	
	While this relatively easy argument has been known for decades (see e.g.~\cite{larsonmavrogordatos}), a rigorous and quantitative assessment of the RWA is relatively recent~\cite{burgarth}; a proof of this convergence, together with an explicit error bound, is presented in Appendix~\ref{app:rwa}.
	
	\begin{remark}[Beyond quadratic Hamiltonians]\label{rem:quadratic}
		All models considered so far---the harmonic oscillator and the single-mode spin--boson models---are polynomials of degree at most two in the ladder operators $a,a^\dag$, with matrix coefficients acting on the spin: the free part $\omega a^\dag a$ is quadratic, and the interaction is linear in $a,a^\dag$. Purely bosonic Hamiltonians of this kind generate \emph{Gaussian} unitaries, which play a central role in quantum optics and continuous-variable quantum information~\cite{weedbrook2012}. It is no coincidence that all these operators turned out to be essentially self-adjoint on natural domains such as the span $\hilb_{\rm fin}$ of the number states: every symmetric polynomial of degree at most two in $a,a^\dag$ is essentially self-adjoint on $\hilb_{\rm fin}$. This can be shown, for instance, by means of Nelson's analytic vector theorem\index{Nelson's analytic vector theorem} \cite[Theorem X.39]{reedsimon2}: by Proposition~\ref{prop:asharpvsn}, the norm of the $m$th power of such a polynomial applied to a number state grows at most like $C^m m!$ for some constant $C$ (depending on the number state), so that number states are analytic vectors.
		
		The situation changes drastically for polynomials of higher degree, such as the higher-order squeezing operators $\ii(a^{\dag k}-a^k)$, $k\geq3$, or the $k$-photon generalizations of the quantum Rabi model, in which the spin is coupled to $a^k$ and $a^{\dag k}$. Operators of this kind are of interest in continuous-variable quantum computing, since Gaussian operations alone do not suffice for universal quantum computation~\cite{lloydbraunstein1999}. However, they generally fail to be essentially self-adjoint: they admit infinitely many self-adjoint extensions, each defining a different dynamics, so that the formal expression of the Hamiltonian no longer determines the physics uniquely~\cite{fischer2026,fischerknapp2026}. Whether a given symmetric polynomial in $a,a^\dag$ (equivalently, in position and momentum) defines a unique dynamics, infinitely many, or none at all can in fact be decided from its coefficients alone~\cite{fischer2026hamiltonian}. This has concrete consequences: for instance, finite-dimensional truncations of the Fock space, routinely used in numerical simulations, need not converge to a unique dynamics~\cite{ashhab2026}.
	\end{remark}
	
	\section{Fock spaces and second quantization}\label{sec:sq}
	
	We have seen how to handle single-mode light--matter interaction models rigorously. We now turn to the more general and realistic case of boson fields with multiple, possibly continuously many, modes.
	
	It is worth putting this step in perspective. In ordinary quantum mechanics, a system of $N$ particles in a region $\Omega\subseteq\mathbb{R}^d$ is described on the Hilbert space $L^2(\Omega^N)$---or on its subspace of totally symmetric or antisymmetric wavefunctions, for identical bosons or fermions---while fields only enter the Hamiltonian as classical external backgrounds: the number $N$ of particles is a fixed parameter. This rules out all systems with a variable number of particles: quantum fields, such as the electromagnetic one, collective excitations described by quasi-particles (phonons, magnons, spin waves\dots), grand canonical ensembles in quantum statistical mechanics. The single-mode field of Section~\ref{sec:singlemode} already escapes this limitation, its number of particles being an observable rather than a parameter (Section~\ref{subsec:singlemodesb}). Fock spaces do the same for fields with arbitrarily many modes, and the corresponding procedure is referred to, for historical reasons, as \emph{second quantization} (see e.g.~\cite{fetterwalecka,negeleorland} for the standard physics treatment).
	
	In the single-mode case (Section~\ref{sec:singlemode}), we considered a Hilbert space spanned by a family of vectors $(\phi_n)_{n\in\mathbb{N}}$, with each $\phi_n$ representing the state in which the field has $n$ particles. On this space, we constructed the annihilation and creation operators $a,a^\dag$, acting as in Eq.~\eqref{eq:ladderaction}, and all vectors $\phi_n$ (and thus, by linearity and density, all states) can be constructed by applying creation operators to the vacuum as in Eq.~\eqref{eq:ladderaction}; in the physicists' bra-ket notation, $\ket{n}=(n!)^{-1/2}(a^\dag)^n\ket{0}$.
	
	We now need to extend this formalism to multi-mode fields. Heuristically, we want our Hilbert space to be spanned by vectors of the form
	\begin{equation}
		\ket{n_{k_1},n_{k_2},\ldots,n_{k_N}}\propto\left(a^\dag(k_1)\right)^{n_{k_1}}\left(a^\dag(k_2)\right)^{n_{k_2}}\cdots\left(a^\dag(k_N)\right)^{n_{k_N}}\ket{0},
	\end{equation}
	with $k_1,\dots,k_N$ ranging in some, possibly uncountable, momentum space. This would represent a state in which the field has $n_{k_1}$ excitations with momentum $k_1$, $n_{k_2}$ excitations with momentum $k_2$, and so on, up to $n_{k_N}$ excitations with momentum $k_N$.
	This should be achieved through a family of annihilation and creation ``operators'' $a(k),a^\dag(k)$ satisfying suitable commutation or anticommutation relations.
	
	To make mathematical sense of such expressions in full generality, we first need to construct Hilbert spaces for quantum systems with a variable number of particles: the Fock spaces. We define them in Section~\ref{subsec:fock}; we then introduce the second quantization of single-particle operators (Section~\ref{subsec:dgamma}) and of unitary operators (Section~\ref{subsec:gammau}), and finally annihilation and creation operators (Section~\ref{subsec:aadag}). The symmetry constraints for identical bosons and fermions will be implemented in Section~\ref{sec:bf}.
	
	\subsection{The Fock space}\label{subsec:fock}
	
	The construction of this section goes back to Fock~\cite{fock1932}; more extensive treatments can be found e.g. in~\cite{reedsimon2,brattelirobinson2,derezinskigerard,arai}. 	
	Hereafter, $\sph$ denotes the \emph{single-particle} Hilbert space, describing a single excitation of the field (e.g. a single photon); the symbol $\hilb$ is reserved for generic Hilbert spaces.
	
	\begin{definition}[Fock space]\label{def:fock}\index{Fock space}\index{vacuum}
		Let $\sph$ be a Hilbert space. We define the Hilbert space $\Gamma(\sph)$ via
		\begin{equation}
			\Gamma(\sph):=\bigoplus_{n\in\mathbb{N}}\sph^{\otimes n},\qquad\text{where }\sph^{\otimes n}=\bigotimes_{j=1}^n\sph,
		\end{equation}
		with the convention $\sph^{\otimes0}=\mathbb{C}$. 
		This is the \emph{Fock space} with \emph{single-particle Hilbert space} $\sph$. The vector $\varepsilon_0=(1,0,0,\ldots)$ is the \emph{vacuum state} of $\Gamma(\sph)$.
	\end{definition}
	
	The tensor product above is a Hilbert space tensor product (Definition~\ref{def:hs_tensorproduct}). Its most general element is a sequence\footnote{Hereafter, with an abuse of notation, we use the same symbol $\|\cdot\|$ for the distinct norms on all spaces $\sph^{\otimes n}$.}
	\begin{equation}
		\psi=(\psi_n)_{n\in\mathbb{N}},\qquad \psi_n\in\sph^{\otimes n},\qquad \sum_{n\in\mathbb{N}}\|\psi_n\|^2<\infty,
	\end{equation}
	with $\psi_n$ representing an $n$-particle state of the field.
	
	Being a direct sum of Hilbert spaces, $\Gamma(\sph)$ is itself a Hilbert space with scalar product and norm
	\begin{equation}
		\braket{\psi,\phi}=\sum_{n\in\mathbb{N}}\braket{\psi_n,\phi_n},\qquad\|\psi\|^2=\sum_{n\in\mathbb{N}}\|\psi_n\|^2.
	\end{equation}
	
	\begin{example}\label{examp:fockC}
		Let $\sph=\mathbb{C}$. Then $\sph^{\otimes n}=\mathbb{C}$ as well, whence
		\begin{equation}
			\Gamma(\mathbb{C})=\bigoplus_{n\in\mathbb{N}}\mathbb{C}=\ell^2,
		\end{equation}
		i.e. $\ell^2$ itself is a Fock space with single-particle space $\mathbb{C}$.
		As every infinite-dimensional separable Hilbert space is isomorphic to $\ell^2$, every such space can be seen as a Fock space.	This also shows that all Fock spaces (except in the trivial case $\sph=\{0\}$) are infinite-dimensional, even if $\sph$ is not.
	\end{example}

	The results of Section~\ref{sec:directsum} allow us to prove the following properties directly:
	
	\begin{proposition}\label{prop:fockspace_separable}
		Let $\sph$ be a separable Hilbert space, with orthonormal basis $(e^{(n)})_{n\in\mathbb{N}}$. Then $\Gamma(\sph)$ is separable, with orthonormal basis 
		\begin{equation}
			\{\varepsilon_0\}\cup(\varepsilon_1^{(n)})_{n\in\mathbb{N}}\cup(\varepsilon_2^{(n_1,n_2)})_{n_1,n_2\in\mathbb{N}}\cup(\varepsilon_3^{(n_1,n_2,n_3)})_{n_1,n_2,n_3\in\mathbb{N}}\cup\ldots,
		\end{equation}
		where $\varepsilon_0$ is the vacuum, and
		\begin{align}
			(\varepsilon_1^{(n)})_{n\in\mathbb{N}}&=\left(0,e^{(n)},0,0,\ldots\right);\\
			(\varepsilon_2^{(n_1,n_2)})_{n_1,n_2\in\mathbb{N}}&=\left(0,0,e^{(n_1)}\otimes e^{(n_2)},0,\ldots\right);\\
			(\varepsilon_3^{(n_1,n_2,n_3)})_{n_1,n_2,n_3\in\mathbb{N}}&=\left(0,0,0,e^{(n_1)}\otimes e^{(n_2)}\otimes e^{(n_3)},\ldots\right).
		\end{align}
	\end{proposition}
	\begin{proof}
		Direct consequence of Propositions~\ref{prop:tensorproductbases} (on the tensor product of separable spaces) and~\ref{prop:directsumbases} (on the direct sum of separable spaces).
	\end{proof}
	
	\begin{proposition}\label{prop:finitefock_dense}\index{finite-particle vectors}
		Let $\sph$ be a Hilbert space, and $\Gamma_{\rm fin}(\sph)$ be the subspace of $\Gamma(\sph)$ given by
		\begin{equation}
			\Gamma_{\rm fin}(\sph):=\bigcup_{N\in\mathbb{N}}\bigoplus_{n=0}^N\sph^{\otimes n}=\left\{\psi\in\Gamma(\sph):\;\exists N\in\mathbb{N}\text{ s.t. }\,\psi_n=0\,\forall n>N\right\}.
		\end{equation}
		Then $\Gamma_{\rm fin}(\sph)$ is dense in $\Gamma(\sph)$.
	\end{proposition}
	
	\begin{proof}
		Direct consequence of Proposition~\ref{prop:hfin}, applied to $\hilb_n=\sph^{\otimes n}$.
	\end{proof}
	
	Recall that the algebraic tensor product $\sph^{\odot n}=\operatorname{Span}\{\phi_1\otimes\cdots\otimes\phi_n:\;\phi_j\in\sph\}$ is dense in $\sph^{\otimes n}$ by definition.
	
	\begin{proposition}\label{prop:finitefock_dense2}
		Let $\sph$ be a Hilbert space, and $\Gamma_{0}(\sph)$ be the subspace of $\Gamma(\sph)$ given by
		\begin{equation}
			\Gamma_{0}(\sph):=\bigcup_{N\in\mathbb{N}}\bigoplus_{n=0}^N\sph^{\odot n}=\left\{\psi\in\Gamma(\sph):\;\exists N\in\mathbb{N}\text{ s.t. }\,\psi_n\in\sph^{\odot n}\:\forall n\leq N,\text{ and }\psi_n=0\,\forall n>N\right\}.
		\end{equation}
		Then $\Gamma_{0}(\sph)$ is dense in $\Gamma(\sph)$.
	\end{proposition}
	
	\begin{proof}
		By Proposition~\ref{prop:finitefock_dense}, $\Gamma_{\rm fin}(\sph)$ is dense in $\Gamma(\sph)$. At the same time, $\sph^{\odot n}$ is dense in $\sph^{\otimes n}$, which implies that $\Gamma_0(\sph)$ is dense in $\Gamma_{\rm fin}(\sph)$. Therefore, $\Gamma_0(\sph)$ is dense in $\Gamma(\sph)$ as well.
	\end{proof}
	
	When dealing with Fock spaces---or more generally, as we have seen in Section~\ref{sec:directsum}, with direct sums of Hilbert spaces---one usually identifies
	\begin{equation}
		\Gamma(\sph)\ni(0,0,\ldots,0,\psi_n,0,\ldots)\leftrightarrow \psi_n\in\sph^{\otimes n}.
	\end{equation}
	With this abuse of notation, the two dense subspaces $\Gamma_{\rm fin}(\sph)$ and $\Gamma_{0}(\sph)$ can be more compactly written as
	\begin{align}
		\Gamma_{\rm fin}(\sph)&=\operatorname{Span}\left\{\psi_n:\;n\in\mathbb{N},\,\psi_n\in\sph^{\otimes n}\right\};\\
		\Gamma_0(\sph)&=\operatorname{Span}\left\{\phi_1\otimes\phi_2\otimes\cdots\otimes\phi_n:\;n\in\mathbb{N},\,\phi_1,\dots,\phi_n\in\sph\right\}.
	\end{align}
	In particular, vectors of the form $\phi_1\otimes\cdots\otimes\phi_n$ for some finite but arbitrarily large $n$ span a dense subspace of $\Gamma(\sph)$, as expected.
	
	The space $\Gamma_0(\sph)$ is the counterpart of the linear span $\hilb_{\rm fin}$ of the number states $\phi_n$ in the single-mode case (Section~\ref{subsec:maximal}).
	
	\begin{example}\label{examp:fockl2}
		Let $(\mathcal{M},\Sigma,\mu)$ be a $\sigma$-finite measure space, and $\sph=L^2(\mathcal{M},\Sigma,\mu)$ the corresponding $L^2$ space. Then, by Proposition~\ref{prop:tensorproduct_l2} (on the tensor product of $L^2$ spaces),
		\begin{equation}
			\sph^{\otimes n}=L^2(\mathcal{M},\Sigma,\mu)^{\otimes n}\cong L^2\left(\mathcal{M}^n,\Sigma^{\otimes n},\mu^{\otimes n}\right),
		\end{equation} 
		whence, with the convention $L^2(\mathcal{M}^0,\Sigma^{\otimes 0},\mu^{\otimes 0})=\mathbb{C}$,
		\begin{equation}
			\Gamma\left(\sph\right)\cong\bigoplus_{n\in\mathbb{N}}L^2\left(\mathcal{M}^n,\Sigma^{\otimes n},\mu^{\otimes n}\right).
		\end{equation}		
		That is, each element of the Fock space is a sequence $\psi=(\psi_n)_{n\in\mathbb{N}}$, where $\psi_n$ is an $n$-particle wavefunction (either in the position or momentum representation) and
		\begin{equation}
			\psi_n:\mathcal{M}^n\rightarrow\mathbb{C},\qquad\sum_{n\in\mathbb{N}}\int_{\mathcal{M}^n}|\psi_n(x_1,\dots,x_n)|^2\;\mathrm{d}^n\mu<\infty.
		\end{equation}
		Note that, at this stage, we are not yet requiring $\psi_n$ to satisfy any symmetry constraint with respect to particle exchange, i.e. any permutation of the variables $x_1,\dots,x_n$.
	\end{example}
	
	\subsection{Second quantization of operators}\label{subsec:dgamma}
	
	We now construct operators on the Fock space, such as Hamiltonians, out of single-particle ones:
	
	\begin{definition}\label{def:secondquantization}\index{second quantization}
		Let $\sph$ be a Hilbert space, and $A:\Dom A\subseteq\sph\rightarrow\sph$ an operator which is either self-adjoint or bounded.\footnote{These are the only cases needed in these notes, and in both of them the sums below are well-defined and closable: if $A$ is self-adjoint, each of them is essentially self-adjoint (Proposition~\ref{prop:polytensor}); if $A$ is bounded, each of them is a bounded, everywhere defined operator, and thus already closed. For a general closed operator $A$, closability of the sums is not automatic.} The \emph{second quantization} of $A$ is the operator $\mathrm{d}\Gamma(A):\Dom\mathrm{d}\Gamma(A)\subseteq\Gamma(\sph)\rightarrow\Gamma(\sph)$ defined by $\mathrm{d}\Gamma(A):=\bigoplus_{n\in\mathbb{N}}A_n$, where $A_0=0$ and, for $n\geq1$,
		\begin{equation}
			A_n=\overline{\sum_{j=1}^n I\otimes\cdots\otimes I\otimes \overbrace{A}^{j\text{th}}\otimes I\otimes\cdots\otimes I}.
		\end{equation}
	\end{definition}
	The reason for the notation $\mathrm{d}\Gamma(A)$ will become clear in Proposition~\ref{prop:gammaexp}. Recall that $\otimes$ here denotes the Hilbert space tensor product of operators (Definition~\ref{def:hs_tensorproduct_op}), i.e. the closure of the algebraic one $I\odot\cdots\odot A\odot\cdots\odot I$.
	
	\begin{proposition}\label{prop:props_2q}
		Let $A:\Dom A\subseteq\sph\rightarrow\sph$ be either self-adjoint or bounded, and $\mathrm{d}\Gamma(A)$ its second quantization. Then
		\begin{enumerate}
			\item[(1)] $\mathrm{d}\Gamma(A)$ is closed;
			\item[(2)] If $A$ is self-adjoint, $\mathrm{d}\Gamma(A)$ is self-adjoint. Besides, the space
			\begin{equation}\label{eq:coreD}
				\mathcal{D}=\operatorname{Span}\left\{\phi_1\otimes\phi_2\otimes\cdots\otimes\phi_n:\;n\in\mathbb{N},\;\phi_1,\dots,\phi_n\in\Dom A\right\}
			\end{equation}
			is a core for $\mathrm{d}\Gamma(A)$.
		\end{enumerate}
	\end{proposition}
	
	\begin{proof}
		(1) follows directly from the fact that the direct sum of closed operators is closed (Proposition~\ref{prop:closure_directsum}).
		
		(2): self-adjointness of $\mathrm{d}\Gamma(A)$ follows from the fact that each $A_n$ is self-adjoint (Proposition~\ref{prop:polytensor}) and the fact that the direct sum of self-adjoint operators is self-adjoint (Proposition~\ref{prop:adjoint_directsum}).		
		We are left with proving that $\mathcal{D}$ is a core. By Proposition~\ref{prop:directsum_esssa} (direct sum of essentially self-adjoint operators is essentially self-adjoint), it suffices to prove that, for each $n\geq1$, the space
		\begin{equation}
			\mathcal{D}_n=\operatorname{Span}\left\{\phi_1\otimes\phi_2\otimes\cdots\otimes\phi_n:\;\phi_1,\ldots,\phi_n\in\Dom A\right\}=\bigodot_{j=1}^n\Dom A
		\end{equation}
		is a core for $A_n$. We use the representation constructed in the proof of Proposition~\ref{prop:polytensor}: up to a unitary transformation, $\sph^{\otimes n}=L^2(\mathcal{M},\Sigma,\mu)$, $A_n$ is the multiplication by $\lambda_1+\ldots+\lambda_n$, and the subspace $L^2_{\rm c}(\mathcal{M},\Sigma,\mu)$ of functions vanishing outside $[-R,R]^n$ for some $R>0$ is a core for $A_n$. It thus suffices to approximate every $\Psi\in L^2_{\rm c}(\mathcal{M},\Sigma,\mu)$ in the graph norm of $A_n$ by elements of $\mathcal{D}_n$.
		
		To this end, let $P_R:=P_A([-R,R])$ be the spectral projection of $A$ on $[-R,R]$ (Theorem~\ref{thm:spectraltheorem1}), so that $\Ran P_R\subseteq\Dom A$. In the above representation, $P_R^{\otimes n}$ is the multiplication by the characteristic function of $[-R,R]^n$ (Proposition~\ref{prop:tensorproduct_multi}). Hence, given $\Psi\in L^2_{\rm c}(\mathcal{M},\Sigma,\mu)$, we have $\Psi=P_R^{\otimes n}\Psi$ for $R$ large enough, and $\|A_n\Phi\|\leq nR\|\Phi\|$ for all $\Phi\in\Ran P_R^{\otimes n}$. Now take any sequence $(\chi_k)_{k\in\mathbb{N}}\subset\sph^{\odot n}$ converging to $\Psi$, and set $\Phi_k:=P_R^{\otimes n}\chi_k$. Then $\Phi_k\in\bigodot_{j=1}^n\Ran P_R\subseteq\mathcal{D}_n$, $\Phi_k\to P_R^{\otimes n}\Psi=\Psi$ and, since $\Phi_k-\Psi\in\Ran P_R^{\otimes n}$,
		\begin{equation}
			\|A_n\Phi_k-A_n\Psi\|\leq nR\|\Phi_k-\Psi\|\to0,
		\end{equation}
		which is what we wanted to show.
	\end{proof}
	
	For bounded operators, point (2) simplifies as follows: 
	
	\begin{corollary}
		Let $A\in\mathcal{B}(\sph)$ be a self-adjoint bounded operator. Then $\Gamma_0(\sph)$ (Proposition~\ref{prop:finitefock_dense2}) is a core for $\mathrm{d}\Gamma(A)$.
	\end{corollary}
	
	\begin{proof}
		Immediate consequence of Proposition~\ref{prop:props_2q} and $\Dom A=\sph$.
	\end{proof}
	
	\begin{example}[Second quantization of multiplication operators]\label{examp:2q_multiplication}
		Let $\sph=L^2(\mathcal{M},\Sigma,\mu)$, $\omega:\mathcal{M}\rightarrow\mathbb{R}$ be a measurable function, and $M_\omega:\Dom M_\omega\subseteq\sph\rightarrow\sph$ be the corresponding multiplication operator; to simplify the notation, we denote it by $\omega$ as well. By Definition~\ref{def:secondquantization}, $\mathrm{d}\Gamma(\omega)=\oplus_n\omega_n$, with $\omega_n$ the closure of $\sum_{j=1}^nI\otimes\cdots\otimes\omega\otimes\cdots\otimes I$, the operator $\omega$ acting on the $j$th factor. Recall that $L^2(\mathcal{M},\Sigma,\mu)^{\otimes n}\cong L^2\left(\mathcal{M}^n,\Sigma^{\otimes n},\mu^{\otimes n}\right)$ and that, in this representation, the tensor product of multiplication operators is the multiplication operator by the product function (Proposition~\ref{prop:tensorproduct_multi}). Hence, the operator acting as $\omega$ on the $j$th factor and trivially elsewhere is the multiplication by the function $(k_1,\dots,k_n)\mapsto1\cdots1\cdot\omega(k_j)\cdot1\cdots1=\omega(k_j)$. Consequently, the sum of these operators acts as the multiplication by $\sum_j\omega(k_j)$ on its domain. Its closure $\omega_n$ is self-adjoint (Proposition~\ref{prop:polytensor}) and contained in the self-adjoint multiplication operator by $\sum_j\omega(k_j)$ on its maximal domain; since self-adjoint operators admit no proper symmetric extensions, $\omega_n$ is precisely the multiplication operator associated with the function $(k_1,\dots,k_n)\mapsto\sum_{j=1}^n\omega(k_j)$, that is,
		\begin{equation}\label{eq:omegan}
			\left(\omega_n\psi_n\right)(k_1,\dots,k_n)=\sum_{j=1}^n\omega(k_j)\psi_n(k_1,\dots,k_n).
		\end{equation}
	\end{example}
	
	\begin{example}[Number operator]\label{examp:number_fock}\index{number operator!on Fock space}
		The second quantization of the identity on $\sph$ is simply
		\begin{equation}
			\mathrm{d}\Gamma(I)=\bigoplus_{n\in\mathbb{N}}\sum_{j=1}^nI\otimes \cdots\otimes I=\bigoplus_{n\in\mathbb{N}}nI,
		\end{equation}
		that is,
		\begin{align}
			\Dom\mathrm{d}\Gamma(I)&=\left\{\psi\in\Gamma(\sph):\sum_{n\in\mathbb{N}}n^2\|\psi_n\|^2<\infty\right\};\\
			\left(\mathrm{d}\Gamma(I)\psi\right)_n&=n\psi_n,
		\end{align}
		i.e. the operator that multiplies by $n$ the $n$-particle component of $\psi$.
		
		This clearly generalizes the number operator of Section~\ref{sec:singlemode}, and we will call it the number operator as well, with the notation $\N:=\mathrm{d}\Gamma(I)$. Note that, by Proposition~\ref{prop:props_2q}, the space $\Gamma_0(\sph)$ (Proposition~\ref{prop:finitefock_dense2}) is a core for $\N$. By construction, each space $\sph^{\otimes n}$ is an eigenspace of $\N$: $\N\psi_n=n\psi_n$ for all $\psi_n\in\sph^{\otimes n}$. Powers of $\N$ can be immediately defined: given $\alpha\geq0$, we have
		\begin{align}
			\Dom\N^\alpha&=\left\{\psi\in\Gamma(\sph):\sum_{n\in\mathbb{N}}n^{2\alpha}\|\psi_n\|^2<\infty\right\};\\
			\left(\N^\alpha\psi\right)_n&=n^{\alpha}\psi_n.
		\end{align}
		In particular, $\Dom\N^{1/2}$ consists of all $\psi\in\Gamma(\sph)$ with $\sum_nn\|\psi_n\|^2<\infty$. Again, $\Gamma_0(\sph)$ is a core for all these operators.
	\end{example}
	
	The number operator also shows that the second quantization of a bounded operator can be unbounded; in fact, one easily shows that $\mathrm{d}\Gamma(A)$ is unbounded for every $A\neq0$.
	
	\begin{remark}
		We have shown how to second quantize single-particle operators. The construction can be generalized, e.g. to two-particle operators. Let $A:\Dom A\subseteq\sph^{\otimes 2}\rightarrow\sph^{\otimes 2}$ and, for all $n\geq2$ and $i,j=1,\dots,n$, let $A_{ij}^{(n)}:\Dom A_{ij}^{(n)}\subseteq\sph^{\otimes n}\rightarrow\sph^{\otimes n}$ be the closure of the operator acting as $A$ on the $i$th and $j$th copies of $\sph$, and as the identity elsewhere. 		
		Then, with the convention $A_{ij}^{(n)}=0$ for $n=0,1$, one can define
		\begin{equation}
			\mathrm{d}\Gamma^{(2)}(A)=\bigoplus_{n\in\mathbb{N}}\sum_{1\leq i<j\leq n}A_{ij}^{(n)}.
		\end{equation}
		The same can be done with $k$-particle Hamiltonians.
	\end{remark}
	
	\subsection{Second quantization of unitary operators}\label{subsec:gammau}
	
	We now explain the origin of the notation $\mathrm{d}\Gamma(A)$, and introduce a tool that will be used repeatedly in the following: the second quantization of unitary operators.
	
	\begin{definition}\label{defin:gammau}\index{second quantization!of unitary operators}
		Let $U\in\mathcal{B}(\sph)$ be a unitary operator on $\sph$. We define the operator $\Gamma(U)\in\mathcal{B}(\Gamma(\sph))$ via
		\begin{equation}
			\Gamma(U):=\bigoplus_{n\in\mathbb{N}}U^{\otimes n},
		\end{equation}
		that is, $\left(\Gamma(U)\psi\right)_n=U\otimes U\otimes\cdots \otimes U\,\psi_n$ for every $\psi\in\Gamma(\sph)$.
	\end{definition}
	Being the direct sum of unitary operators, $\Gamma(U)$ is itself unitary.	
	More generally, given a unitary operator $U:\sph\rightarrow\mathfrak{k}$ between two Hilbert spaces, the same construction yields a unitary operator $\Gamma(U):=\oplus_nU^{\otimes n}:\Gamma(\sph)\rightarrow\Gamma(\mathfrak{k})$, where $U^{\otimes n}:\sph^{\otimes n}\rightarrow\mathfrak{k}^{\otimes n}$ is the unitary operator acting as $U^{\otimes n}\phi_1\otimes\cdots\otimes\phi_n=U\phi_1\otimes\cdots\otimes U\phi_n$ (cf. Lemma~\ref{lem:unitary}). Second quantization is covariant under such unitaries:
	
	\begin{lemma}\label{lem:gammacov}
		Let $U:\sph\rightarrow\mathfrak{k}$ be unitary, and $A:\Dom A\subseteq\sph\rightarrow\sph$ either self-adjoint or bounded. Then
		\begin{equation}
			\Gamma(U)\,\mathrm{d}\Gamma(A)\,\Gamma(U)^*=\mathrm{d}\Gamma(UAU^*).
		\end{equation}
		Besides, $\Gamma(U)$ maps the $n$-particle sector $\sph^{\otimes n}$ onto $\mathfrak{k}^{\otimes n}$; in particular, $\Gamma(U)\N^s\Gamma(U)^*=\N^s$ for all $s\geq0$.
	\end{lemma}
	
	\begin{proof}
		The last claim is clear, since $\Gamma(U)=\oplus_nU^{\otimes n}$ acts sector by sector. As both $\Gamma(U)$ and $\mathrm{d}\Gamma(\cdot)$ act sector by sector, it suffices to show $U^{\otimes n}A_n(U^{\otimes n})^*=(UAU^*)_n$ for $n\geq1$, with $A_n$ as in Definition~\ref{def:secondquantization}. For $\phi_1,\dots,\phi_n\in\Dom A$,
		\begin{equation}
			U^{\otimes n}\left(I\otimes\cdots\otimes\overbrace{A}^{j\text{th}}\otimes\cdots\otimes I\right)\phi_1\otimes\cdots\otimes\phi_n=\left(I\otimes\cdots\otimes\overbrace{UAU^*}^{j\text{th}}\otimes\cdots\otimes I\right)U^{\otimes n}\phi_1\otimes\cdots\otimes\phi_n.
		\end{equation}
		Hence $U^{\otimes n}$ maps the restriction of $A_n$ to $\bigodot_{j=1}^n\Dom A$ into the restriction of $(UAU^*)_n$ to $\bigodot_{j=1}^n\Dom(UAU^*)=U^{\otimes n}\bigodot_{j=1}^n\Dom A$. These subspaces are cores for $A_n$ and $(UAU^*)_n$: by Proposition~\ref{prop:props_2q} if $A$ (and thus $UAU^*$) is self-adjoint, and trivially if $A$ is bounded, since then $A_n$ is bounded and $\sph^{\odot n}$ is dense. As conjugation by a unitary operator preserves closures, the claim follows.
	\end{proof}
	
	\begin{remark}[Unitary covariance and the spectral theorem]\label{rem:covariance}\index{second quantization!unitary covariance}
		Lemma~\ref{lem:gammacov} will be used repeatedly in the following, often in conjunction with the spectral theorem in its multiplicative form (Theorem~\ref{thm:spectraltheorem2}). Given a self-adjoint operator $A$ on $\sph$, let $U:\sph\rightarrow L^2(\mathcal{M},\Sigma,\mu)$ be a unitary operator such that $UAU^*$ is a multiplication operator. By Lemma~\ref{lem:gammacov}, $\Gamma(U)$ maps $\Gamma(\sph)$ unitarily onto $\Gamma(L^2(\mathcal{M},\Sigma,\mu))$, and $\mathrm{d}\Gamma(A)$ into $\mathrm{d}\Gamma(UAU^*)$, which acts on each $n$-particle sector $L^2(\mathcal{M}^n,\Sigma^{\otimes n},\mu^{\otimes n})$ as a multiplication operator (Example~\ref{examp:2q_multiplication}). Hence, as far as properties invariant under unitary equivalence are concerned---self-adjointness, domains, norm estimates, spectra---we can assume without loss of generality that $\sph=L^2(\mathcal{M},\Sigma,\mu)$ and that $A$ is a multiplication operator, thus replacing abstract operator-theoretic arguments by explicit computations with functions. Annihilation and creation operators transform covariantly as well (Lemma~\ref{lem:gammacov2}), so that the same strategy can be applied to them. We will use it, for instance, in the proofs of Propositions~\ref{prop:domega_bb} and~\ref{prop:a(f)domega} and throughout Section~\ref{sec:singular}. Even when no operator has to be diagonalized, the fact that every separable Hilbert space is unitarily equivalent to some $L^2$ space allows us to work with wavefunctions (see e.g. the proof of Proposition~\ref{prop:ccr/car}).
	\end{remark}
	
	\begin{proposition}\label{prop:gammaexp}
		Let $A:\Dom A\subseteq\sph\rightarrow\sph$ be self-adjoint. Then, for all $t\in\mathbb{R}$,
		\begin{equation}
			\Gamma(\e^{-\ii tA})=\e^{-\ii t\mathrm{d}\Gamma(A)}.
		\end{equation}
	\end{proposition}
	
	\begin{proof}
		As $\mathrm{d}\Gamma(A)$ is self-adjoint, $\e^{-\ii t\mathrm{d}\Gamma(A)}$ is well-defined and unitary. Since both operators are bounded, it suffices to prove that they agree on a dense subspace of $\Gamma(\sph)$, a convenient choice being the space $\mathcal{D}$ of Eq.~\eqref{eq:coreD}. So let $\phi_1,\dots,\phi_n\in\Dom A$. By definition, $\Gamma(\e^{-\ii tA})\phi_1\otimes\cdots\otimes\phi_n=\e^{-\ii tA}\phi_1\otimes\cdots\otimes\e^{-\ii tA}\phi_n$. Differentiating in $t$ by the product rule, one readily sees that the vector $\Phi(t):=\Gamma(\e^{-\ii tA})\phi_1\otimes\cdots\otimes\phi_n$ satisfies
		\begin{equation}
			\ii\frac{\mathrm{d}}{\mathrm{d}t}\Phi(t)=\left(\sum_{j=1}^n I\otimes\dots\otimes \overbrace{A}^{j\text{th}}\otimes\cdots\otimes I\right)\Phi(t)=A_n\Phi(t)=\mathrm{d}\Gamma(A)\Phi(t),
		\end{equation}
		with $A_n$ as in Definition~\ref{def:secondquantization}. That is, $\Phi(t)$ solves the Schr\"odinger equation generated by $\mathrm{d}\Gamma(A)$ with initial condition $\Phi(0)=\phi_1\otimes\cdots\otimes\phi_n$, whose unique solution is $\e^{-\ii t\mathrm{d}\Gamma(A)}\phi_1\otimes\cdots\otimes\phi_n$ (Theorem~\ref{thm:stone}). This proves the claim.
	\end{proof}
	
	As a first application of unitary covariance, we prove the following estimate, which will be needed for spin--boson models (Section~\ref{sec:multimodesb}).
	
	\begin{proposition}\label{prop:domega_bb}
		Let $A:\Dom A\subseteq\sph\rightarrow\sph$ be self-adjoint and \emph{strictly positive}, i.e. there exists $m>0$ such that $A\geq m$, that is, $\braket{\psi,(A-m)\psi}\geq0$ for all $\psi\in\Dom A$. Then, for all $s\geq0$, $\Dom\mathrm{d}\Gamma(A)^s\subseteq\Dom\N^s$ and, for all $\psi\in\Dom\mathrm{d}\Gamma(A)^s$,
		\begin{equation}
			\left\|\mathrm{d}\Gamma(A)^s\psi\right\|\geq m^{s}\|\N^s\psi\|.
		\end{equation}
	\end{proposition}
	
	\begin{proof}
		By the spectral theorem (Theorem~\ref{thm:spectraltheorem2}) and the unitary covariance of second quantization (Lemma~\ref{lem:gammacov} and Remark~\ref{rem:covariance}), we can assume without loss of generality that $A$ is a multiplication operator on $\sph=L^2(\mathcal{M},\Sigma,\mu)$.
		
		So let $k\mapsto\omega(k)$ be a real-valued measurable function; as usual, we denote the corresponding multiplication operator by $\omega$ as well. By assumption, $\omega(k)\geq m>0$
		almost everywhere. As discussed in Example~\ref{examp:2q_multiplication}, $\mathrm{d}\Gamma(\omega)=\oplus_n\omega_n$, with $\omega_n$ as in Eq.~\eqref{eq:omegan}. Given $s\geq0$, let $\psi=(\psi_n)_{n\in\mathbb{N}}\in\Dom\mathrm{d}\Gamma(\omega)^s$. Then
		\begin{equation}
			\infty>\sum_{n\in\mathbb{N}}\int\left(\sum_{j=1}^n\omega(k_j)\right)^{2s}\left|\psi_n(k_1,\dots,k_n)\right|^2\;\mathrm{d}^n\mu\geq m^{2s}\sum_{n\in\mathbb{N}}n^{2s}\int\left|\psi_n\right|^2\;\mathrm{d}^n\mu=m^{2s}\left\|\N^s\psi\right\|^2,
		\end{equation}
		which proves $\psi\in\Dom\N^s$ and the claimed inequality.
	\end{proof}
	
	\subsection{Annihilation and creation operators}\label{subsec:aadag}
	
	We now introduce the main ingredient for constructing multi-mode field theories: operators that ``create'' or ``destroy'' a field excitation with a given wavefunction $f\in\sph$. In the single-mode case, we first defined the ladder operators on a dense invariant subspace (Definition~\ref{def:a} and Remark~\ref{rem:initialdomain}), and then took their closures (Definition~\ref{defin:aadagnclose} and Proposition~\ref{prop:aprop2}). We now follow the same strategy, with an important novelty: the Hilbert space of a single excitation is no longer one-dimensional, but a whole, generally infinite-dimensional, space of possible excitations. Consequently, instead of a single pair $a,a^\dag$ of annihilation and creation operators, we will have a whole family $(a(f))_{f\in\sph}$, $(a^\dag(f))_{f\in\sph}$.
	
	\begin{definition}[Annihilation and creation operators]\label{def:a0(f)}\index{annihilation and creation operators}
		Let $f\in\sph$. We define the operators $a_0(f),a_0^\dag(f)$ via $\Dom a_0(f)=\Dom a_0^\dag(f)=\Gamma_0(\sph)$
		and
		\begin{align}
			a_0(f)\phi_1\otimes\phi_2\otimes\cdots\otimes\phi_n&:=\sqrt{n}\braket{f,\phi_n}\phi_1\otimes\cdots\otimes\phi_{n-1};\\
			a^\dag_0(f)\phi_1\otimes\phi_2\otimes\cdots\otimes\phi_n&:=\sqrt{n+1}\phi_1\otimes\phi_2\otimes\cdots\otimes\phi_n\otimes f\,,
		\end{align} 
		with the convention $a_0(f)\sph^{\otimes 0}=0$.
	\end{definition}
	These definitions are consistent: although a vector of $\sph^{\odot n}$ admits many representations as a linear combination of pure tensors, the right-hand sides only depend on the vector itself. Indeed, recalling that pure tensors are conjugate-multilinear forms (Definition~\ref{def:algtensorhs}), $(\phi_1\otimes\cdots\otimes\phi_n)(u_1,\dots,u_n)=\prod_j\braket{u_j,\phi_j}$, one checks on pure tensors, and thus by linearity for all $\Psi\in\sph^{\odot n}$, that
	\begin{align}
		(a_0(f)\Psi)(u_1,\dots,u_{n-1})&=\sqrt{n}\,\Psi(u_1,\dots,u_{n-1},f),\\
		(a_0^\dag(f)\Psi)(u_1,\dots,u_{n+1})&=\sqrt{n+1}\,\Psi(u_1,\dots,u_n)\braket{u_{n+1},f},
	\end{align}
	which manifestly depend on $\Psi$ only.
	
	By definition, $a_0(f)$ and $a_0^\dag(f)$ map $\sph^{\odot n}$ into $\sph^{\odot(n-1)}$ and $\sph^{\odot(n+1)}$ respectively; in particular, both operators leave $\Gamma_0(\sph)$ invariant.
	
	\begin{proposition}
		The mappings $\sph\ni f\mapsto a_0(f)$ and $\sph\ni f\mapsto a_0^\dag(f)$ are respectively antilinear and linear, in the following sense: for all $f,g\in\sph$ and $\alpha,\beta\in\mathbb{C}$,
		\begin{equation}
			a_0(\alpha f+\beta g)=\overline{\alpha}a_0(f)+\overline{\beta}a_0(g),\qquad a_0^\dag(\alpha f+\beta g)=\alpha a_0^\dag(f)+\beta a_0^\dag(g).
		\end{equation}
	\end{proposition}
	
	\begin{proof}
		Direct consequence of Definition~\ref{def:a0(f)}.
	\end{proof}
	
	\begin{remark}
		One should not mistake the antilinearity of the mapping $f\mapsto a_0(f)$ for antilinearity of the operator $a_0(f)$ itself: like all operators in these notes, $a_0(f)$ is linear.
	\end{remark}
	
	Some basic properties generalize immediately from the single-mode case, starting with property (i) of Proposition~\ref{prop:aprop}:
	
	\begin{proposition}\label{prop:a(f)prop}
		
		For every $f\in\sph$, $a_0(f),a_0^\dag(f)$ satisfy
		\begin{equation}
			\Braket{\psi,a_0(f)\varphi}=\Braket{a_0^\dag(f)\psi,\varphi}\qquad\forall\psi,\varphi\in\Gamma_0(\sph),
		\end{equation}
		and therefore are closable operators, with $a_0(f)\subseteq(a_0^\dag(f))^*$ and $a_0^\dag(f)\subseteq(a_0(f))^*$.
	\end{proposition}
	
	\begin{exercise}\label{exerc:a0fprop}
		Prove Proposition~\ref{prop:a(f)prop}. Hint: by linearity, it suffices to take $\psi=\psi_1\otimes\cdots\otimes\psi_{m}$ and $\varphi=\varphi_1\otimes\cdots\otimes\varphi_n$; compute both sides with Definition~\ref{def:a0(f)}, and note that both vanish unless $m=n-1$. Closability then follows from Proposition~\ref{prop:closableiffadjointdense}.
		\seesol{a0fprop}
	\end{exercise}
	
	As in the single-mode case, the operators $\a{f},\adag{f}$ will be defined as the closures of these operators. We first need the following estimate:
	
	\begin{proposition}\label{prop:a0(f)vsn}
		Let $f\in\sph$. Then, for all $\psi\in\Gamma_0(\sph)$,
		\begin{equation}
			\left\|a_0(f)\psi\right\|\leq\|f\|\left\|\N^{1/2}\psi\right\|\qquad\text{and}\qquad	\left\|a_0^\dag(f)\psi\right\|\leq\|f\|\left\|(\N+1)^{1/2}\psi\right\|.
		\end{equation}
	\end{proposition}
	
	\begin{proof}
		Write $\psi=(\psi_n)_{n\in\mathbb{N}}$ with $\psi_n\in\sph^{\odot n}$. Since $a_0(f)$ and $a_0^\dag(f)$ map $\sph^{\odot n}$ respectively into $\sph^{\odot(n-1)}$ and $\sph^{\odot(n+1)}$, which are mutually orthogonal for different $n$,
		\begin{equation}
			\|a_0(f)\psi\|^2=\sum_{n\in\mathbb{N}}\|a_0(f)\psi_n\|^2,\qquad\|a_0^\dag(f)\psi\|^2=\sum_{n\in\mathbb{N}}\|a_0^\dag(f)\psi_n\|^2,
		\end{equation}
		while $\|\N^{1/2}\psi\|^2=\sum_nn\|\psi_n\|^2$ and $\|(\N+1)^{1/2}\psi\|^2=\sum_n(n+1)\|\psi_n\|^2$. Hence it suffices to prove
		\begin{equation}
			\|a_0(f)\psi_n\|\leq\sqrt{n}\|f\|\|\psi_n\|,\qquad\|a_0^\dag(f)\psi_n\|\leq\sqrt{n+1}\|f\|\|\psi_n\|\qquad\text{for all }\psi_n\in\sph^{\odot n}.
		\end{equation}
		Note that it does not suffice to check these for pure tensors $\phi_1\otimes\cdots\otimes\phi_n$ and then use linearity and the triangle inequality: the latter gives $\|a_0(f)\sum_r\psi^{(r)}\|\leq\sqrt{n}\|f\|\sum_r\|\psi^{(r)}\|$, and $\sum_r\|\psi^{(r)}\|$ can be much larger than $\|\sum_r\psi^{(r)}\|$.
		
		For $a_0(f)$, assume $f\neq0$ and complete $e_1:=f/\|f\|$ to an orthonormal basis $(e_k)_{k\geq1}$ of $\sph$. Given $\psi_n=\sum_{r=1}^Nu_r\otimes v_r$ with $u_r\in\sph^{\odot(n-1)}$ and $v_r\in\sph$, expanding each $v_r$ in this basis gives
		\begin{equation}
			\psi_n=\sum_{k\geq1}\chi_k\otimes e_k,\qquad\chi_k:=\sum_{r=1}^N\braket{e_k,v_r}u_r\in\sph^{\otimes(n-1)},\qquad\|\psi_n\|^2=\sum_{k\geq1}\|\chi_k\|^2,
		\end{equation}
		the last equality following from the orthogonality of the vectors $\chi_k\otimes e_k$ for different $k$. By definition,
		\begin{equation}
			a_0(f)\psi_n=\sqrt{n}\sum_{r=1}^N\braket{f,v_r}u_r=\sqrt{n}\|f\|\chi_1,
		\end{equation}
		whence $\|a_0(f)\psi_n\|=\sqrt{n}\|f\|\|\chi_1\|\leq\sqrt{n}\|f\|\|\psi_n\|$.
		
		For $a_0^\dag(f)$ the computation is easier. By definition, $a_0^\dag(f)\psi_n=\sqrt{n+1}\,\psi_n\otimes f$, and the definition of the scalar product on $\sph^{\odot(n+1)}$ immediately gives $\|\psi_n\otimes f\|=\|\psi_n\|\|f\|$ for every $\psi_n\in\sph^{\odot n}$, whence
		\begin{equation}
			\|a_0^\dag(f)\psi_n\|=\sqrt{n+1}\|f\|\|\psi_n\|.\qedhere
		\end{equation}
	\end{proof}
	
	\begin{definition}[Annihilation and creation operators]\label{def:a(f)}
		Let $f\in\sph$. The \emph{annihilation and creation operators} associated with $f\in\sph$ are the operators $\a{f},\adag{f}$ defined by
		\begin{equation}
			\a{f}:=\overline{a_0(f)},\qquad \adag{f}:=\overline{a_0^\dag(f)}.
		\end{equation}
	\end{definition}
	
	An analogous argument as in the single-mode case (cf. Proposition~\ref{prop:aprop2}) yields the following:
	
	\begin{proposition}\label{prop:a(f)prop2}
		For every $f\in\sph$, $\a{f},\adag{f}$ have domains $\Dom\a{f}\supseteq\Dom\N^{1/2}$ and $\Dom\adag{f}\supseteq\Dom\N^{1/2}$. Besides, for all $\psi\in\Dom\N^{1/2}$,
		\begin{equation}
			\left\|a(f)\psi\right\|\leq\|f\|\left\|\N^{1/2}\psi\right\|\qquad\text{and}\qquad	\left\|a^\dag(f)\psi\right\|\leq\|f\|\left\|(\N+1)^{1/2}\psi\right\|
		\end{equation}
		and, for all $\psi,\phi\in\Dom\N^{1/2}$,
		\begin{equation}
			\Braket{\psi,\a{f}\phi}=\Braket{\adag{f}\psi,\phi}.
		\end{equation}
	\end{proposition}
	
	\begin{proof}
		All statements follow from Proposition~\ref{prop:a0(f)vsn} and the fact that $\Gamma_0(\sph)$ is a core for $\N^{1/2}$ (Example~\ref{examp:number_fock}); we proceed in three steps.
		
		\medskip
		\emph{Step 1}: prove $\Dom\a{f}\supseteq\Dom\N^{1/2}$, $\Dom\adag{f}\supseteq\Dom\N^{1/2}$.		
		We focus on $\a{f}$. By definition, $\Dom\a{f}$ is the closure of $\Dom a_0(f)$ with respect to the graph norm $\|\psi\|^2_{a_0(f)}:=\|\psi\|^2+\|a_0(f)\psi\|^2$ of $a_0(f)$. By Proposition~\ref{prop:a0(f)vsn}, the latter is dominated by the graph norm $\|\psi\|^2_{\N^{1/2}}:=\|\psi\|^2+\|\N^{1/2}\psi\|^2$ of $\N^{1/2}$:
		\begin{equation}
			\|\psi\|^2_{a_0(f)}\leq\|\psi\|^2+\|f\|^2\|\N^{1/2}\psi\|^2\leq C_f^2\|\psi\|^2_{\N^{1/2}},\qquad C_f^2:=\max\{1,\|f\|^2\}.
		\end{equation}
		That is, the graph norm of $\N^{1/2}$ dominates that of $a_0(f)$. This means that every sequence that converges in $\|\cdot\|_{\N^{1/2}}$ also converges, to the same limit, in $\|\cdot\|_{a_0(f)}$:
		\begin{equation}
			\overline{\Dom a_0(f)}^{\|\cdot\|_{\N^{1/2}}}\subseteq\overline{\Dom a_0(f)}^{\|\cdot\|_{a_0(f)}}=\Dom a(f).
		\end{equation}
		But $\Dom a_0(f)=\Gamma_0(\sph)$ and $\Gamma_0(\sph)$ is a core for $\N^{1/2}$, so $\overline{\Dom a_0(f)}^{\|\cdot\|_{\N^{1/2}}}=\Dom\N^{1/2}$ and $\Dom\N^{1/2}\subseteq\Dom a(f)$.
		Same argument for $\adag{f}$.
		
		Note that, in the single-mode case (proof of Proposition~\ref{prop:aprop2}), we had equality between the two norms, $\|\cdot\|_{a_0}=\|\cdot\|_{\N^{1/2}}$, which enabled us to conclude $\Dom a=\Dom\N^{1/2}$ without further arguments. Here we only have an inequality.
		
		\medskip
		\emph{Step 2}: prove 	$\|a(f)\psi\|\leq\|f\|\|\N^{1/2}\psi\|$, $\|a^\dag(f)\psi\|\leq\|f\|\|(\N+1)^{1/2}\psi\|$.		
		We use a density argument. By Proposition~\ref{prop:a0(f)vsn}, these inequalities hold for $\psi\in\Gamma_0(\sph)$. Let $\psi\in\Dom\N^{1/2}$: since $\Gamma_0(\sph)$ is a core for $\N^{1/2}$, there exists $(\psi_n)_{n\in\mathbb{N}}\subset\Gamma_0(\sph)$ such that $\psi_n\to\psi$ and $\N^{1/2}\psi_n\to\N^{1/2}\psi$, and, for every $n$, $\|\a{f}\psi_n\|\leq\|f\|\|\N^{1/2}\psi_n\|$.
		By Step 1, $\psi\in\Dom\a{f}$ as well; moreover, since $\psi_n\to\psi$ in the graph norm of $\N^{1/2}$, which dominates the graph norm of $a_0(f)$, the sequence $(\psi_n)_{n\in\mathbb{N}}$ is Cauchy in the latter, and therefore, $\a{f}$ being closed, $\a{f}\psi_n\to\a{f}\psi$.
		Taking limits in the above inequality, we obtain $	\|\a{f}\psi\|\leq\|f\|\|\N^{1/2}\psi\|$.
		Same procedure for $\adag{f}$.
		
		\medskip
		\emph{Step 3}: prove 	$\Braket{\psi,\a{f}\phi}=\braket{\adag{f}\psi,\phi}$ for all $\psi,\phi\in\Dom\N^{1/2}$.		
		The argument is analogous to that of Step 2: given $\psi,\phi\in\Dom\N^{1/2}$, there are sequences $(\psi_n)_{n\in\mathbb{N}},(\phi_n)_{n\in\mathbb{N}}\subset\Gamma_0(\sph)$ such that $\psi_n\to\psi,\phi_n\to\phi,\N^{1/2}\phi_n\to\N^{1/2}\phi,\N^{1/2}\psi_n\to\N^{1/2}\psi$. Using this and $\Dom\N^{1/2}\subseteq\Dom\a{f},\Dom\adag{f}$, one proves the claimed equalities (Exercise~\ref{exerc:afstep3} below).
	\end{proof}
	
	\begin{exercise}\label{exerc:afstep3}
		Fill in the missing details of Step 3. Hint: for the approximating sequences, $\braket{\psi_n,\a{f}\phi_n}=\braket{\adag{f}\psi_n,\phi_n}$ by Proposition~\ref{prop:a(f)prop}; pass to the limit $n\to\infty$, using the estimates of Step 2 to show that $\a{f}\phi_n\to\a{f}\phi$ and $\adag{f}\psi_n\to\adag{f}\psi$.
		\seesol{afstep3}
	\end{exercise}
	
	\begin{example}[Annihilation and creation operators on $\Gamma(L^2(\mathcal{M},\Sigma,\mu))$]\label{examp:afl2}
		Let $\sph=L^2(\mathcal{M},\Sigma,\mu)$, so that $\Gamma(\sph)\cong\oplus_nL^2(\mathcal{M}^n,\Sigma^{\otimes n},\mu^{\otimes n})$ (Example~\ref{examp:fockl2}). Given $f\in \sph$, by definition the annihilation and creation operators $\a{f},\adag{f}$ act as follows:
		\begin{align}\label{eq:afl2}
			(\a{f}\psi_n)(x_1,x_2,\dots,x_{n-1})&=\sqrt{n}\int_\mathcal{M}\overline{f(x_n)}\psi_n(x_1,x_2,\ldots,x_{n-1},x_n)\;\mathrm{d}\mu(x_n);\\
			(\adag{f}\psi_n)(x_1,x_2,\ldots,x_n,x_{n+1})&=\sqrt{n+1}\,\psi_n(x_1,x_2,\ldots,x_n)f(x_{n+1})
		\end{align}
		at least for $\psi=(\psi_n)_{n\in\mathbb{N}}\in\Gamma_0(\sph)$. By a density argument combined with closedness, one can show that these expressions remain valid for all $\psi\in\Dom\N^{1/2}$.		
		Note that $\a{f}$ (resp. $\adag{f}$) converts an $n$-particle wavefunction into an $(n-1)$-particle (resp. $(n+1)$-particle) one.
	\end{example}
	
	\section{Bosonic and fermionic Fock spaces}\label{sec:bf}
	
	So far, we have not taken into account any symmetry constraint. Given $\phi_1,\phi_2\in\sph$, the vectors $\phi_1\otimes\phi_2$ and $\phi_2\otimes\phi_1$ are generally distinct, and thus represent distinct two-particle states; equivalently, if $\sph=L^2(\mathcal{M},\Sigma,\mu)$, $n$-particle wavefunctions generally change under permutations of their variables. However, quantum mechanics imposes a further constraint: the wavefunction describing a system of $n$ \emph{indistinguishable particles}\index{indistinguishable particles} must be either \emph{completely symmetric (bosonic) or completely antisymmetric (fermionic)} under any permutation of the particles. That is, for every permutation $\sigma$ of $\{1,\dots,n\}$,
	\begin{align}\label{eq:bosonsfermions}
		\psi_n\left(x_{\sigma(1)},x_{\sigma(2)},\ldots,x_{\sigma(n)}\right)&=\phantom{\sgn(\sigma)}\psi_n(x_1,x_2,\ldots,x_n)\qquad\text{(for bosons)};\nonumber\\
		\psi_n\left(x_{\sigma(1)},x_{\sigma(2)},\ldots,x_{\sigma(n)}\right)&=\sgn(\sigma)\psi_n(x_1,x_2,\ldots,x_n)\qquad\text{(for fermions)},
	\end{align}
	where $\sgn(\sigma)$ is the sign of the permutation (Definition~\ref{def:sign} below). In this section, we implement these constraints on the Fock space. Although only bosonic fields enter the spin--boson models of Sections~\ref{sec:multimodesb} and~\ref{sec:singular}, we treat bosons and fermions in parallel: the comparison is instructive---for instance, fermionic annihilation and creation operators are bounded, while bosonic ones are not (Proposition~\ref{prop:fermionsbounded})---and fermionic Fock spaces are the natural setting for many-body systems of electrons and other fermions. We first collect the necessary facts about permutations and their action on tensor products (Section~\ref{subsec:permutations}); we then construct the subspaces of symmetric and antisymmetric tensors (Section~\ref{subsec:symtensors}), the bosonic and fermionic Fock spaces (Section~\ref{subsec:fockpm}), and the corresponding annihilation and creation operators (Sections~\ref{subsec:apm}--\ref{subsec:dgammaladder}).
	
	\subsection{Permutations and their action on tensor products}\label{subsec:permutations}
	
	\begin{definition}[Permutations]\label{def:permutations}\index{permutation}
		Let $n\geq1$. A \emph{permutation} of $\{1,2,\ldots,n\}$ is a bijection $\sigma:\{1,2,\dots,n\}\rightarrow\{1,2,\ldots,n\}$. The set $\mathfrak{S}_n$ of all permutations of $\{1,2,\ldots,n\}$, which has $n!$ elements, is a group with respect to the composition $(\sigma\tau)(i):=\sigma(\tau(i))$, the \emph{symmetric group}\index{symmetric group}; its identity element is the trivial permutation $\mathrm{id}$.
	\end{definition}
	
	A permutation $\sigma\in\mathfrak{S}_n$ is often represented in the two-line notation
	\begin{equation}
		\sigma=\left(
		\begin{array}{ccccc}
			1&2&3&\cdots&n\\
			\sigma(1)&\sigma(2)&\sigma(3)&\cdots&\sigma(n)
		\end{array}
		\right).
	\end{equation}
	
	\begin{example}\label{examp:permutation}
		The permutation
		\begin{equation}\label{eq:permexample}
			\sigma=\left(
			\begin{array}{cccc}
				1&2&3&4\\
				3&2&4&1
			\end{array}
			\right)\in\mathfrak{S}_4
		\end{equation}
		is defined by $\sigma(1)=3$, $\sigma(2)=2$, $\sigma(3)=4$, $\sigma(4)=1$. Its inverse is obtained by exchanging the two lines and reordering the columns: $\sigma^{-1}(1)=4$, $\sigma^{-1}(2)=2$, $\sigma^{-1}(3)=1$, $\sigma^{-1}(4)=3$.
	\end{example}
	
	\begin{definition}[Sign of a permutation]\label{def:sign}\index{permutation!sign}
		Let $\sigma\in\mathfrak{S}_n$. A pair $(i,j)$ with $1\leq i<j\leq n$ is an \emph{inversion} of $\sigma$ if $\sigma(i)>\sigma(j)$. The \emph{sign} of $\sigma$ is
		\begin{equation}\label{eq:sign}
			\sgn(\sigma):=(-1)^{\#\left\{(i,j):\;1\leq i<j\leq n\text{ and }\sigma(i)>\sigma(j)\right\}},
		\end{equation}
		and $\sigma$ is said to be \emph{even} if $\sgn(\sigma)=1$, and \emph{odd} if $\sgn(\sigma)=-1$.
	\end{definition}
	
	\begin{example}\label{examp:sign}
		For the permutation~\eqref{eq:permexample}, the pairs $(\sigma(i),\sigma(j))$ with $i<j$ are $(3,2)$, $(3,4)$, $(3,1)$, $(2,4)$, $(2,1)$, $(4,1)$. Four of them, namely $(3,2)$, $(3,1)$, $(2,1)$, $(4,1)$, are inverted, so $\sigma$ is even. As another example, let $1\leq k<l\leq n$, and let $t\in\mathfrak{S}_n$ be the \emph{transposition}\index{permutation!transposition} exchanging $k$ and $l$, i.e. $t(k)=l$, $t(l)=k$, and $t(i)=i$ otherwise. Its inversions are $(k,l)$, the $l-k-1$ pairs $(k,i)$, and the $l-k-1$ pairs $(i,l)$, with $k<i<l$; their number $2(l-k)-1$ is odd, so every transposition is odd.
	\end{example}
	
	\begin{proposition}\label{prop:sign}
		Let $n\geq1$. Then $\sgn(\mathrm{id})=1$ and, for all $\sigma,\tau\in\mathfrak{S}_n$,
		\begin{equation}\label{eq:sgnmult}
			\sgn(\sigma\tau)=\sgn(\sigma)\sgn(\tau),\qquad\sgn(\sigma^{-1})=\sgn(\sigma),\qquad\sum_{\sigma\in\mathfrak{S}_n}\sgn(\sigma)=0\quad(n\geq2),
		\end{equation}
		the last identity meaning that, for $n\geq2$, exactly half of the permutations in $\mathfrak{S}_n$ are even.
	\end{proposition}
	
	\begin{proof}
		Clearly $\sgn(\mathrm{id})=1$. Since every inversion contributes a factor $-1$, one has $\sgn(\sigma)=\prod_{i<j}\frac{\sigma(j)-\sigma(i)}{j-i}$. Hence
		\begin{equation}
			\sgn(\sigma\tau)=\prod_{i<j}\frac{\sigma(\tau(j))-\sigma(\tau(i))}{\tau(j)-\tau(i)}\prod_{i<j}\frac{\tau(j)-\tau(i)}{j-i}=\sgn(\sigma)\sgn(\tau),
		\end{equation}
		since each factor of the first product is unchanged if $i$ and $j$ are exchanged, and the pairs $\{\tau(i),\tau(j)\}$ range over all unordered pairs of distinct elements of $\{1,\dots,n\}$, so that the first product equals $\sgn(\sigma)$. Applying this to $\sigma\sigma^{-1}=\mathrm{id}$ gives $\sgn(\sigma)\sgn(\sigma^{-1})=1$, i.e. $\sgn(\sigma^{-1})=\sgn(\sigma)$. Finally, if $n\geq2$, let $t$ be any transposition: by Example~\ref{examp:sign} and Eq.~\eqref{eq:sgnmult}, $\sigma\mapsto t\sigma$ is a bijection of $\mathfrak{S}_n$ which maps even permutations into odd ones and vice versa, whence the last identity in Eq.~\eqref{eq:sgnmult}.
	\end{proof}
	
	We will also need the following decomposition of the permutations of $n+1$ elements, in which we regard $\mathfrak{S}_n$ as the subgroup of $\mathfrak{S}_{n+1}$ of all permutations fixing $n+1$.
	
	\begin{lemma}\label{lem:taudecomp}
		Let $n\geq1$ and, for $j\in\{1,\dots,n+1\}$, let $\tau_j\in\mathfrak{S}_{n+1}$ be the permutation
		\begin{equation}\label{eq:tauj}
			\tau_j=\left(
			\begin{array}{ccccccccc}
				1&\cdots&j-1&j&j+1&\cdots&n&n+1\\
				1&\cdots&j-1&j+1&j+2&\cdots&n+1&j
			\end{array}
			\right).
		\end{equation}
		Then $\sgn(\tau_j)=(-1)^{n+1-j}$. Besides, every $\sigma'\in\mathfrak{S}_{n+1}$ can be written in a unique way as $\sigma'=\tau_j\sigma$, with $j\in\{1,\dots,n+1\}$ and $\sigma\in\mathfrak{S}_n$; in particular, $\sgn(\sigma')=(-1)^{n+1-j}\sgn(\sigma)$.
	\end{lemma}
	
	\begin{proof}
		The inversions of $\tau_j$ are precisely the $n+1-j$ pairs $(i,n+1)$ with $j\leq i\leq n$, since $\tau_j(i)=i+1>j=\tau_j(n+1)$, while $\tau_j$ preserves the relative order of $1,\dots,n$; hence $\sgn(\tau_j)=(-1)^{n+1-j}$. Given $\sigma'\in\mathfrak{S}_{n+1}$, if $\sigma'=\tau_j\sigma$ with $\sigma(n+1)=n+1$, then necessarily $j=\tau_j(n+1)=\sigma'(n+1)$ and $\sigma=\tau_j^{-1}\sigma'$; conversely, with this choice of $j$, the permutation $\tau_j^{-1}\sigma'$ fixes $n+1$. The last claim follows from Eq.~\eqref{eq:sgnmult}.
	\end{proof}
	
	In terms of variables, $(x_{\tau_j(1)},\dots,x_{\tau_j(n+1)})=(x_1,\dots,x_{j-1},x_{j+1},\dots,x_{n+1},x_j)$: the permutation $\tau_j$ moves the variable $x_j$ to the last slot, and shifts the variables $x_{j+1},\dots,x_{n+1}$ one slot to the left, without changing their relative order.
	
	Permutations act on tensor products by permuting the factors:
	
	\begin{definition}[Permutation operators]\label{def:usigma}\index{permutation!operator}
		Let $n\geq1$ and $\sigma\in\mathfrak{S}_n$. The \emph{permutation operator} $U_\sigma\in\mathcal{B}(\sph^{\otimes n})$ is the unique bounded operator such that
		\begin{equation}\label{eq:usigma}
			U_\sigma(\phi_1\otimes\phi_2\otimes\cdots\otimes\phi_n)=\phi_{\sigma^{-1}(1)}\otimes\phi_{\sigma^{-1}(2)}\otimes\cdots\otimes\phi_{\sigma^{-1}(n)}\qquad\text{for all }\phi_1,\dots,\phi_n\in\sph.
		\end{equation}
	\end{definition}
	
	This definition is well-posed: since the map~\eqref{eq:usigma} preserves the scalar products of pure tensors, its linear extension to $\sph^{\odot n}$ is well-defined and isometric (cf. the proof of Proposition~\ref{prop:welldef_tensorproduct}), and thus extends uniquely to an isometry on $\sph^{\otimes n}$. In words, $U_\sigma$ moves the $i$th factor to the slot $\sigma(i)$. The inverse in Eq.~\eqref{eq:usigma} ensures $U_\sigma U_\tau=U_{\sigma\tau}$ rather than $U_\sigma U_\tau=U_{\tau\sigma}$:
	
	\begin{lemma}\label{lem:usigma}
		$U_\mathrm{id}=I$ and, for all $\sigma,\tau\in\mathfrak{S}_n$,
		\begin{equation}\label{eq:usigmaprops}
			U_\sigma U_\tau=U_{\sigma\tau},\qquad U_\sigma^*=U_{\sigma^{-1}};
		\end{equation}
		that is, the family of operators $(U_\sigma)_{\sigma\in\mathfrak{S}_n}$ provides a unitary representation of $\mathfrak{S}_n$ on $\sph^{\otimes n}$.
	\end{lemma}
	
	\begin{exercise}\label{exerc:usigma}
		Prove Lemma~\ref{lem:usigma}. Hint: since all operators involved are bounded, it suffices to check the identities on pure tensors $\phi_1\otimes\cdots\otimes\phi_n$, using Eq.~\eqref{eq:usigma}; for the adjoint, compute $\braket{U_\sigma\phi_1\otimes\cdots\otimes\phi_n,\chi_1\otimes\cdots\otimes\chi_n}$ and relabel the factors.
		\seesol{usigma}
	\end{exercise}
	
	\subsection{Symmetric and antisymmetric tensors}\label{subsec:symtensors}
	
	The constraints~\eqref{eq:bosonsfermions} can be implemented by averaging over all permutations:
	
	\begin{definition}[Symmetrizers]\label{def:symmetrizers}\index{symmetrizer}
		Let $n\geq2$. The \emph{complete symmetrizer} and \emph{complete antisymmetrizer} are respectively the two operators $S_n^+,S_n^-\in\mathcal{B}(\sph^{\otimes n})$ defined by
		\begin{equation}\label{eq:symmetrizers}
			S_n^+:=\frac{1}{n!}\sum_{\sigma\in\mathfrak{S}_n}U_\sigma,\qquad S_n^-:=\frac{1}{n!}\sum_{\sigma\in\mathfrak{S}_n}\sgn(\sigma)U_\sigma,
		\end{equation}
		that is, for all $\phi_1,\dots,\phi_n\in\sph$,
		\begin{align}
			S_n^+\phi_1\otimes\phi_2\otimes\cdots\otimes\phi_n&=\frac{1}{n!}\sum_{\sigma\in\mathfrak{S}_n}\phi_{\sigma^{-1}(1)}\otimes\phi_{\sigma^{-1}(2)}\otimes\cdots\otimes\phi_{\sigma^{-1}(n)};\\
			S_n^-\phi_1\otimes\phi_2\otimes\cdots\otimes\phi_n&=\frac{1}{n!}\sum_{\sigma\in\mathfrak{S}_n}\sgn(\sigma)\phi_{\sigma^{-1}(1)}\otimes\phi_{\sigma^{-1}(2)}\otimes\cdots\otimes\phi_{\sigma^{-1}(n)}.
		\end{align}
	\end{definition}
	
	\begin{proposition}\label{prop:symmetrizers}
		Let $n\geq2$. The symmetrizers $S_n^\pm$ are mutually orthogonal self-adjoint projections on $\sph^{\otimes n}$:
		\begin{equation}
			\left(S_n^\pm\right)^2=S_n^\pm=(S_n^\pm)^*,\qquad S_n^+S_n^-=S_n^-S_n^+=0.
		\end{equation}
	\end{proposition}
	
	\begin{proof}
		We use Lemma~\ref{lem:usigma}, Eq.~\eqref{eq:sgnmult}, and the fact that, $\mathfrak{S}_n$ being a group, $\tau\mapsto\sigma\tau$ and $\sigma\mapsto\sigma^{-1}$ are bijections of $\mathfrak{S}_n$, which allows us to change summation variables.
		
		\medskip
		\emph{Fact 1}: $(S_n^\pm)^2=S_n^\pm$. By Eqs.~\eqref{eq:usigmaprops} and~\eqref{eq:sgnmult}, and changing variables $\tau\mapsto\pi:=\sigma\tau$ in the inner sum,
		\begin{equation}
			\left(S_n^-\right)^2=\frac{1}{(n!)^2}\sum_{\sigma,\tau\in\mathfrak{S}_n}\sgn(\sigma\tau)U_{\sigma\tau}=\frac{1}{(n!)^2}\sum_{\sigma\in\mathfrak{S}_n}\sum_{\pi\in\mathfrak{S}_n}\sgn(\pi)U_\pi=\frac{1}{n!}\sum_{\pi\in\mathfrak{S}_n}\sgn(\pi)U_\pi=S_n^-,
		\end{equation}
		and analogously, without the signs, $(S_n^+)^2=S_n^+$.
		
		\medskip
		\emph{Fact 2}: $S_n^+S_n^-=S_n^-S_n^+=0$. This is proven analogously; see Exercise~\ref{exerc:spsm}.
		
		\medskip
		\emph{Fact 3}: $(S_n^\pm)^*=S_n^\pm$. By Eqs.~\eqref{eq:usigmaprops} and~\eqref{eq:sgnmult}, and changing variables $\tau:=\sigma^{-1}$,
		\begin{equation}
			(S_n^-)^*=\frac{1}{n!}\sum_{\sigma\in\mathfrak{S}_n}\sgn(\sigma)U_{\sigma^{-1}}=\frac{1}{n!}\sum_{\tau\in\mathfrak{S}_n}\sgn(\tau^{-1})U_\tau=\frac{1}{n!}\sum_{\tau\in\mathfrak{S}_n}\sgn(\tau)U_\tau=S_n^-,
		\end{equation}
		and analogously, without the signs, $(S_n^+)^*=S_n^+$.
	\end{proof}
	
	\begin{exercise}\label{exerc:spsm}
		Prove that $S_n^+S_n^-=S_n^-S_n^+=0$, thus completing the proof of Proposition~\ref{prop:symmetrizers}. Hint: write $S_n^+S_n^-$ as a double sum over $\sigma,\tau\in\mathfrak{S}_n$, change variables as in the proof of Fact 1, and use Eq.~\eqref{eq:sgnmult}.
		\seesol{spsm}
	\end{exercise}
	
	The symmetrizers interact with permutation operators in a simple way, which makes precise in which sense they produce completely symmetric or antisymmetric vectors:
	
	\begin{lemma}\label{lem:sunu}
		Let $n\geq2$. For all $\sigma\in\mathfrak{S}_n$,
		\begin{equation}\label{eq:sunu}
			S_n^+U_\sigma=U_\sigma S_n^+=S_n^+,\qquad S_n^-U_\sigma=U_\sigma S_n^-=\sgn(\sigma)S_n^-.
		\end{equation}
		Consequently, $\Psi\in S_n^+\sph^{\otimes n}$ if and only if $U_\sigma\Psi=\Psi$ for all $\sigma\in\mathfrak{S}_n$, and $\Psi\in S_n^-\sph^{\otimes n}$ if and only if $U_\sigma\Psi=\sgn(\sigma)\Psi$ for all $\sigma\in\mathfrak{S}_n$.
	\end{lemma}
	
	\begin{proof}
		By Eqs.~\eqref{eq:usigmaprops} and~\eqref{eq:sgnmult}, changing variables $\pi:=\tau\sigma$,
		\begin{equation}
			S_n^-U_\sigma=\frac{1}{n!}\sum_{\tau\in\mathfrak{S}_n}\sgn(\tau)U_{\tau\sigma}=\frac{1}{n!}\sum_{\pi\in\mathfrak{S}_n}\sgn(\pi\sigma^{-1})U_\pi=\sgn(\sigma)S_n^-,
		\end{equation}
		and analogously $U_\sigma S_n^-=\sgn(\sigma)S_n^-$, as well as the bosonic identities. If $\Psi=S_n^\pm\Phi$, then $U_\sigma\Psi=U_\sigma S_n^\pm\Phi$ is given by Eq.~\eqref{eq:sunu}; conversely, if $U_\sigma\Psi=\Psi$ (resp. $U_\sigma\Psi=\sgn(\sigma)\Psi$) for all $\sigma$, then Eq.~\eqref{eq:symmetrizers} and $\sgn(\sigma)^2=1$ give $S_n^\pm\Psi=\Psi$.
	\end{proof}
	
	Since $S_n^\pm$ are mutually orthogonal self-adjoint projections, the spaces $S_n^\pm\sph^{\otimes n}$ are mutually orthogonal closed subspaces of $\sph^{\otimes n}$, and thus Hilbert spaces with the norm inherited from $\sph^{\otimes n}$.
	
	\begin{proposition}\label{prop:dense_symasym}
		The spaces $S_n^\pm\sph^{\odot n}=\operatorname{Span}\left\{S_n^\pm(\phi_1\otimes\cdots\otimes\phi_n):\phi_1,\dots,\phi_n\in\sph\right\}$ are dense in $S_n^\pm\sph^{\otimes n}$.
	\end{proposition}
	
	\begin{proof}
		This follows from the density of $\sph^{\odot n}$ in $\sph^{\otimes n}$ and the boundedness of $S_n^\pm$.
	\end{proof}
	
	While the operators $S_n^\pm$ are formally similar, the spaces $S_n^\pm\sph^{\otimes n}$ have strikingly different properties. To explain this, let us fix an orthonormal basis $(e_j)_{j\geq1}$ of $\sph$, so that the vectors $e_{j_1}\otimes\cdots\otimes e_{j_n}$, with $j_1,\dots,j_n\geq1$, form an orthonormal basis of $\sph^{\otimes n}$ (Proposition~\ref{prop:tensorproductbases}), and their (anti)symmetrizations $S_n^\pm e_{j_1}\otimes\cdots\otimes e_{j_n}$ span a dense subspace of $S_n^\pm\sph^{\otimes n}$ (Proposition~\ref{prop:dense_symasym}). The latter vectors, however, are neither orthonormal nor, in general, distinct or nonzero: their behavior depends on the repetitions in the string of indices $(j_1,\dots,j_n)$, i.e. on whether two or more indices have the same value, in a way which is very different in the two cases.
	
	\begin{proposition}\label{prop:repetitions}
		Let $(e_j)_{j\geq1}$ be an orthonormal basis of $\sph$, $n\geq2$, and $j_1,\dots,j_n\geq1$. Then:
		\begin{itemize}
			\item[(i)] for all $\sigma\in\mathfrak{S}_n$,
			\begin{equation}
				S_n^+e_{j_{\sigma(1)}}\otimes\cdots\otimes e_{j_{\sigma(n)}}=S_n^+e_{j_1}\otimes\cdots\otimes e_{j_n},\qquad S_n^-e_{j_{\sigma(1)}}\otimes\cdots\otimes e_{j_{\sigma(n)}}=\sgn(\sigma)\,S_n^-e_{j_1}\otimes\cdots\otimes e_{j_n};
			\end{equation}
			\item[(ii)] if $j_k=j_l$ for some $k\neq l$, then $S_n^-e_{j_1}\otimes\cdots\otimes e_{j_n}=0$.
		\end{itemize}
	\end{proposition}
	
	\begin{proof}
		(i) By Eq.~\eqref{eq:usigma}, $e_{j_{\sigma(1)}}\otimes\cdots\otimes e_{j_{\sigma(n)}}=U_{\sigma^{-1}}\,e_{j_1}\otimes\cdots\otimes e_{j_n}$, so the claim follows from Lemma~\ref{lem:sunu} and $\sgn(\sigma^{-1})=\sgn(\sigma)$. (ii) Let $t$ be the transposition exchanging $k$ and $l$. Since $j_k=j_l$, the vector $e_{j_{t(1)}}\otimes\cdots\otimes e_{j_{t(n)}}$ coincides with $e_{j_1}\otimes\cdots\otimes e_{j_n}$, whence, by (i) and Example~\ref{examp:sign}, $S_n^-e_{j_1}\otimes\cdots\otimes e_{j_n}=-S_n^-e_{j_1}\otimes\cdots\otimes e_{j_n}$.
	\end{proof}
	
	In the symmetric case, repetitions are thus allowed, and, by (i), the vector $S_n^+e_{j_1}\otimes\cdots\otimes e_{j_n}$ only depends on the unordered multiset of indices $\{j_1,\dots,j_n\}$, i.e. on the family of integers $m_1,m_2,m_3,\ldots$, with $\sum_km_k=n$, where $m_k$ is the multiplicity---or, in physicists' jargon, \emph{occupation number}\index{occupation number}---of the index $k$ (the number of times $e_k$ appears in the tensor product). In the antisymmetric case, by (ii), repetitions are instead forbidden: no two particles can occupy the same state. In physicists' jargon, this is \emph{Pauli's exclusion principle}\index{Pauli exclusion principle}.
	
	\begin{example}\label{examp:repetitions}
		Let $n=3$, and consider the vectors $S_3^\pm e_1\otimes e_2\otimes e_1$ and $S_3^\pm e_1\otimes e_1\otimes e_2$, corresponding to the strings $121$ and $112$, in which the index $1$ appears twice. The six permutations of either string yield each of the strings
		\begin{equation}\label{eq:strings}
			112,\qquad121,\qquad211
		\end{equation}
		twice. In the bosonic case, this gives
		\begin{equation}
			S_3^+e_1\otimes e_2\otimes e_1=\frac{1}{3}\left(e_1\otimes e_1\otimes e_2+e_1\otimes e_2\otimes e_1+e_2\otimes e_1\otimes e_1\right)=S_3^+e_1\otimes e_1\otimes e_2,
		\end{equation}
		both vectors corresponding to the occupation numbers $m_1=2$, $m_2=1$. In the fermionic case, the two copies of each string in Eq.~\eqref{eq:strings} come with opposite signs, since the corresponding permutations differ by the transposition exchanging the two copies of $e_1$, so that all terms in $S_3^-e_1\otimes e_2\otimes e_1$ cancel in pairs.
	\end{example}
	
	Proposition~\ref{prop:repetitions} allows us to construct orthonormal bases of $S_n^\pm\sph^{\otimes n}$ and, in particular, to compute their dimensions when $\dim\sph=d<\infty$. In the antisymmetric (fermionic) case, repetitions must be avoided, and it suffices to consider strictly increasing strings of indices:
	
	\begin{proposition}\label{prop:basis_antisym}
		The family of vectors
		\begin{equation}
			\sqrt{n!}\,S^-_n e_{j_1}\otimes e_{j_2}\otimes\cdots\otimes e_{j_n},\qquad 1\leq j_1<j_2<\ldots<j_n,
		\end{equation}
		is an orthonormal basis of $S_n^-\sph^{\otimes n}$. In particular, if $\dim\sph=d<\infty$, then $\dim S_n^-\sph^{\otimes n}$ is the number of combinations without repetitions of $d$ objects in groups of $n$:
		\begin{equation}
			\dim S_n^-\sph^{\otimes n}=\binom{d}{n}=\frac{d!}{n!(d-n)!},
		\end{equation}
		with the convention $\binom{d}{n}=0$ for $d<n$.
	\end{proposition}
	
	\begin{exercise}\label{exerc:basisantisym}
		Prove Proposition~\ref{prop:basis_antisym}. Hint: by Proposition~\ref{prop:repetitions}, every vector $S_n^-e_{k_1}\otimes\cdots\otimes e_{k_n}$ is either zero or $\pm$ one of the vectors of Proposition~\ref{prop:basis_antisym}, so these span a dense subspace of $S_n^-\sph^{\otimes n}$. For orthonormality, expand the scalar product of two of them using Definition~\ref{def:symmetrizers}, and note that $e_{j_{\sigma(1)}}\otimes\cdots\otimes e_{j_{\sigma(n)}}$ and $e_{k_{\tau(1)}}\otimes\cdots\otimes e_{k_{\tau(n)}}$ are orthogonal unless they coincide.
		\seesol{basisantisym}
	\end{exercise}
	
	In the symmetric (bosonic) case, it is more convenient to use the occupation number representation. By Proposition~\ref{prop:repetitions}(i), every vector $S_n^+e_{j_1}\otimes\cdots\otimes e_{j_n}$ is uniquely determined by the occupation numbers $m_1,m_2,\ldots\in\mathbb{N}$, with $\sum_km_k=n$; using the bra-ket notation, we denote it by $\ket{m_1,m_2,\ldots}$, the indices $(j_1,\dots,j_n)$ being any family containing $m_1$ copies of $1$, $m_2$ copies of $2$, and so on.
	
	\begin{proposition}\label{prop:basis_sym}
		The family of vectors
		\begin{equation}
			\sqrt{\frac{n!}{m_1!m_2!\cdots}}\ket{m_1,m_2,\ldots},\qquad m_1,m_2,\ldots\in\mathbb{N},\qquad\sum_km_k=n,
		\end{equation}
		is an orthonormal basis of $S_n^+\sph^{\otimes n}$.  In particular, if $\dim\sph=d<\infty$, then $\dim S_n^+\sph^{\otimes n}$ is the number of combinations with repetitions of $d$ objects in groups of $n$:
		\begin{equation}
			\dim S_n^+\sph^{\otimes n}=\binom{d+n-1}{n}=\frac{(d+n-1)!}{n!(d-1)!}.
		\end{equation}
	\end{proposition}
	
	\begin{exercise}\label{exerc:basissym}
		Prove Proposition~\ref{prop:basis_sym}. Hint: proceed as in Exercise~\ref{exerc:basisantisym}. To compute the norm of $\ket{m_1,m_2,\ldots}$, note that, among the $n!$ permutations of a string of indices with occupation numbers $m_1,m_2,\ldots$, exactly $m_1!m_2!\cdots$ leave the string unchanged.
		\seesol{basissym}
	\end{exercise}
	
	The bosonic subspaces are thus much larger than the fermionic ones: for any fixed $d$, the dimension of $S_n^+\sph^{\otimes n}$ grows without bound (polynomially), while that of $S_n^-\sph^{\otimes n}$ vanishes for $n>d$:
	\begin{equation}
		\dim S_n^+\sph^{\otimes n}\sim\frac{n^{d-1}}{(d-1)!}\quad\text{as }n\to\infty,\qquad \dim S_n^-\sph^{\otimes n}=0\quad\text{for }n>d.
	\end{equation}
	
	\begin{example}\label{examp:symL2}
		Let $(\mathcal{M},\Sigma,\mu)$ be a $\sigma$-finite measure space, and $\sph=L^2(\mathcal{M},\Sigma,\mu)$, so that $\sph^{\otimes n}\cong L^2(\mathcal{M}^n,\Sigma^{\otimes n},\mu^{\otimes n})$ (Proposition~\ref{prop:tensorproduct_l2}). In this representation, $U_\sigma$ acts as $(U_\sigma\psi)(x_1,\dots,x_n)=\psi(x_{\sigma(1)},\dots,x_{\sigma(n)})$, as one checks on products of single-particle functions by means of Eq.~\eqref{eq:usigma}. Consequently, the symmetrizers act as
		\begin{equation}
			(S_n^\pm\psi)(x_1,x_2,\dots,x_n)=\frac{1}{n!}\sum_{\sigma\in\mathfrak{S}_n}(\pm1)^{\sigma}\,\psi\left(x_{\sigma(1)},x_{\sigma(2)},\ldots,x_{\sigma(n)}\right),
		\end{equation}
		where $(+1)^\sigma:=1$ and $(-1)^\sigma:=\sgn(\sigma)$, and, by Lemma~\ref{lem:sunu}, the space $S_n^\pm L^2(\mathcal{M},\Sigma,\mu)^{\otimes n}$ is identified with the space $L^2_\pm(\mathcal{M}^n,\Sigma^{\otimes n},\mu^{\otimes n})$ of $n$-particle wavefunctions satisfying, respectively, the bosonic and fermionic constraints~\eqref{eq:bosonsfermions}.
	\end{example}
	
	\subsection{Definition of the bosonic and fermionic Fock spaces}\label{subsec:fockpm}
	
	Coming back to Fock spaces, we can now define the following:
	
	\begin{definition}[Bosonic and fermionic Fock spaces]\label{def:fockpm}\index{Fock space!bosonic}\index{Fock space!fermionic}
		Let $\sph$ be a Hilbert space. The \emph{bosonic Fock space} $\Gamma^+(\sph)$ and the \emph{fermionic Fock space} $\Gamma^-(\sph)$ with single-particle space $\sph$ are defined via
		\begin{equation}
			\Gamma^\pm(\sph):=\bigoplus_{n\in\mathbb{N}}S_n^\pm\sph^{\otimes n},
		\end{equation}
		with the convention $S^\pm_0\sph^{\otimes0}=\mathbb{C}$ and $S^\pm_1\sph^{\otimes 1}=\sph$. Equivalently, $\Gamma^\pm(\sph)=S^\pm\Gamma(\sph)$, where $S^\pm=\oplus_{n\in\mathbb{N}}S_n^\pm$.
	\end{definition}
	
	Both $\Gamma^\pm(\sph)$ are Hilbert spaces (as each $S_n^\pm\sph^{\otimes n}$ is) with the scalar product and norm inherited from $\Gamma(\sph)$, and share its vacuum $\varepsilon_0=(1,0,0,\ldots)$.
	
	A first clear difference between the two spaces $\Gamma^\pm(\sph)$ emerges when $\dim\sph=d<\infty$: excluding the trivial case $d=0$, the bosonic Fock space $\Gamma^+(\sph)$ is infinite-dimensional (just as $\Gamma(\sph)$), while the fermionic Fock space $\Gamma^-(\sph)$ is finite-dimensional, since $S_n^-\sph^{\otimes n}=0$ for all $n>d$. Precisely, by the binomial theorem,
	\begin{equation}
		\dim \Gamma^-(\sph)=\sum_{n=0}^d\dim S_n^-\sph^{\otimes n}=\sum_{n=0}^d\binom{d}{n}=2^d.
	\end{equation}
	The fermionic Fock space is infinite-dimensional only if $\dim\sph=\infty$; even then, however, the ``finiteness'' of fermions is reflected in other properties, as we will see.
	
	Most properties proven for $\Gamma(\sph)$ go through for $\Gamma^\pm(\sph)$:
	
	\begin{proposition}
		Let $\sph$ be a separable Hilbert space. Then $\Gamma^\pm(\sph)$ is separable.
	\end{proposition}
	
	\begin{proof}
		Analogous to the case of $\Gamma(\sph)$ (Proposition~\ref{prop:fockspace_separable}).
	\end{proof}
	
	An explicit basis of $\Gamma^\pm(\sph)$ can then be obtained from a single-particle basis $(e_j)_{j\geq1}$ by constructing bases of $S_n^\pm\sph^{\otimes n}$ as in Propositions~\ref{prop:basis_sym}--\ref{prop:basis_antisym} and ``piecing together'' the bases for all $n\in\mathbb{N}$. Specifically, up to normalization factors, $\Gamma^+(\sph)$ has a basis made of the vectors $\ket{m_1,m_2,m_3,\ldots}$, with $m_1,m_2,\ldots\in\mathbb{N}$ and only finitely many of them nonzero, each belonging to the $(\sum_jm_j)$-particle subspace; and $\Gamma^-(\sph)$ has a basis made of the vectors $S_n^-e_{j_1}\otimes \cdots\otimes e_{j_n}$, with $n\in\mathbb{N}$ and $1\leq j_1<\ldots<j_n$.
	
	Recall that $\Gamma_{\rm fin}(\sph)$ and $\Gamma_0(\sph)$ (cf. Propositions~\ref{prop:finitefock_dense}--\ref{prop:finitefock_dense2}) are dense in $\Gamma(\sph)$. The same holds for their symmetric and antisymmetric counterparts:
	
	\begin{proposition}
		Let $\sph$ be a Hilbert space, and $\Gamma^\pm(\sph)$ the corresponding bosonic and fermionic Fock spaces. The spaces
		\begin{align}
			\Gamma^\pm_{\rm fin}(\sph)&=\operatorname{Span}\left\{\psi_n:n\in\mathbb{N},\;\psi_n\in S_n^\pm\sph^{\otimes n}\right\};\\
			\Gamma^\pm_0(\sph)&=\operatorname{Span}\left\{S_n^\pm(\phi_1\otimes\phi_2\otimes\cdots\otimes\phi_n):n\in\mathbb{N},\;\phi_1,\dots,\phi_n\in\sph\right\}
		\end{align}
		are dense in $\Gamma^\pm(\sph)$.
	\end{proposition}
	
	\begin{proof}
		The denseness of $\Gamma^\pm_{\rm fin}(\sph)$ is proven as in the non-(anti)symmetrized case (Proposition~\ref{prop:finitefock_dense}), and that of $\Gamma^\pm_0(\sph)$ as in Proposition~\ref{prop:finitefock_dense2}, taking into account that $S_n^\pm\sph^{\odot n}$ is dense in $S_n^\pm\sph^{\otimes n}$ (Proposition~\ref{prop:dense_symasym}).
	\end{proof}
	
	The next logical step is to define operators on $\Gamma^\pm(\sph)$. On $\Gamma(\sph)$, we defined the second quantization $\mathrm{d}\Gamma(A)$ of single-particle operators $A:\Dom A\subseteq\sph\rightarrow\sph$, as well as the annihilation and creation operators $\a{f},\adag{f}$.
	These operators can be restricted to $\Gamma^\pm(\sph)$, but the restrictions must also map $\Gamma^\pm(\sph)$ into itself, i.e. preserve the (anti)symmetry of vectors.	
	For second quantized operators, this is indeed true:
	
	\begin{proposition}\label{prop:invariant}
		Let $A:\Dom A\subseteq\sph\rightarrow\sph$ be a self-adjoint operator on $\sph$, and $\mathrm{d}\Gamma(A)$ its second quantization. Then the projections $S^\pm=\oplus_nS_n^\pm$ commute with $\mathrm{d}\Gamma(A)$: $S^\pm\Dom\mathrm{d}\Gamma(A)\subseteq\Dom\mathrm{d}\Gamma(A)$ and $\mathrm{d}\Gamma(A)S^\pm\psi=S^\pm\mathrm{d}\Gamma(A)\psi$ for all $\psi\in\Dom\mathrm{d}\Gamma(A)$. In particular, the space $\Dom\mathrm{d}\Gamma(A)\cap\Gamma^\pm(\sph)$
		is an invariant subspace for $\mathrm{d}\Gamma(A)$, that is, $\mathrm{d}\Gamma(A)\psi\in\Gamma^\pm(\sph)$ for every $\psi\in\Dom\mathrm{d}\Gamma(A)\cap\Gamma^\pm(\sph)$.
	\end{proposition}
	
	\begin{proof}
		Recall that $\mathrm{d}\Gamma(A)=\oplus_n A_n$, with $A_n$ as in Definition~\ref{def:secondquantization}, and that $S_n^\pm$ are linear combinations of the permutation operators $U_\sigma$ (Eq.~\eqref{eq:symmetrizers}). We claim that, for all $n\in\mathbb{N}$ and $\sigma\in\mathfrak{S}_n$, $A_nU_\sigma=U_\sigma A_n$. 		
		It suffices to prove this equality for vectors of the form $\phi_1\otimes\cdots\otimes\phi_n$ with $\phi_1,\dots,\phi_n\in\Dom A$: they span a core $\mathcal{D}_n$ for $A_n$ (Proposition~\ref{prop:props_2q}), and $U_\sigma$ maps $\mathcal{D}_n$ onto itself, so that the equality extends to $\Dom A_n$ by the closedness of $A_n$. Since $U_\sigma$ permutes the factors (Eq.~\eqref{eq:usigma}), while $A_n$ applies $A$ to each factor in turn, both $A_nU_\sigma\phi_1\otimes\cdots\otimes\phi_n$ and $U_\sigma A_n\phi_1\otimes\cdots\otimes\phi_n$ are equal to
		\begin{equation}
			\sum_{j=1}^n\phi_{\sigma^{-1}(1)}\otimes\cdots\otimes A\phi_{\sigma^{-1}(j)}\otimes\cdots\otimes\phi_{\sigma^{-1}(n)},
		\end{equation}
		which proves the claim. But then, by Eq.~\eqref{eq:symmetrizers},
		\begin{equation}
			A_nS_n^+=\frac{1}{n!}\sum_{\sigma\in\mathfrak{S}_n}A_nU_\sigma=\frac{1}{n!}\sum_{\sigma\in\mathfrak{S}_n}U_\sigma A_n=S_n^+A_n,
		\end{equation}
		and analogously $A_nS_n^-=S_n^-A_n$. Taking direct sums (and using $\|A_nS_n^\pm\psi_n\|=\|S_n^\pm A_n\psi_n\|\leq\|A_n\psi_n\|$) concludes the proof.
	\end{proof}
	
	This allows us to define the restrictions $\mathrm{d}\Gamma(A)\restriction(\Dom\mathrm{d}\Gamma(A)\cap\Gamma^\pm(\sph))$, which are densely defined operators on the Hilbert spaces $\Gamma^\pm(\sph)$. 
	
	Let us recall a general property of self-adjoint operators (cf. \cite[Section 2.5]{teschl}):
	
	\begin{lemma}\label{lem:invariant}
		Let $T:\Dom T\subseteq\hfrak\rightarrow\hfrak$ be a self-adjoint operator on a Hilbert space $\hfrak$, and $\mathcal{K}\subset\hfrak$ a closed subspace whose orthogonal projection $P$ commutes with $T$, i.e. $P\Dom T\subseteq\Dom T$ and $TP\psi=PT\psi$ for all $\psi\in\Dom T$. Then the restrictions of $T$ to $\Dom T\cap\mathcal{K}$ and $\Dom T\cap\mathcal{K}^\perp$ are self-adjoint operators on $\mathcal{K}$ and $\mathcal{K}^\perp$, respectively.
	\end{lemma}
	
	\begin{exercise}\label{exerc:invariantlemma}
		Prove this. Hint: under the assumption, both $\mathcal{K}$ and $\mathcal{K}^\perp$ are invariant under $T$, and $T$ decomposes as the direct sum of its restrictions to $\mathcal{K}$ and $\mathcal{K}^\perp$, which are then self-adjoint by Proposition~\ref{prop:adjoint_directsum}. (Mere invariance of $\Dom T\cap\mathcal{K}$ is not enough in general: this intersection could even be $\{0\}$).
		\seesol{invariantlemma}
	\end{exercise}
	
	\begin{proposition}\label{prop:domega_symm_sa}
		Let $A:\Dom A\subseteq\sph\rightarrow\sph$ be a self-adjoint operator on $\sph$, and $\mathrm{d}\Gamma(A)$ its second quantization. The restrictions of $\mathrm{d}\Gamma(A)$ to $\Gamma^\pm(\sph)$ are self-adjoint operators on $\Gamma^\pm(\sph)$.
	\end{proposition}
	
	\begin{proof}
		Direct consequence of Proposition~\ref{prop:invariant} and Lemma~\ref{lem:invariant}. 
	\end{proof}
	
	In the same way, one proves that both spaces are invariant under $\Gamma(U)$ (Definition~\ref{defin:gammau}):
	
	\begin{proposition}\label{prop:gammaUinvariant}
		Let $U\in\mathcal{B}(\sph)$ be a unitary operator on $\sph$, and $\Gamma(U)\in\mathcal{B}(\Gamma(\sph))$ as per Definition~\ref{defin:gammau}. The spaces $\Gamma^\pm(\sph)$ are invariant subspaces for $\Gamma(U)$.
	\end{proposition}
	
	\begin{exercise}\label{exerc:gammaU}
		Prove this. Hint: use the same strategy as in the proof of Proposition~\ref{prop:invariant}, showing that $U^{\otimes n}$ commutes with every permutation operator $U_\sigma$, and hence with $S_n^\pm$.
		\seesol{gammaU}
	\end{exercise}
	
	\subsection{Bosonic and fermionic annihilation and creation operators}\label{subsec:apm}
	
	We now turn to annihilation and creation operators. Unlike for $\mathrm{d}\Gamma(A)$, here (anti)symmetry has to be enforced by hand, as the following example shows.
	
	\begin{example}\label{examp:adagnotsym}
		Take the symmetric vector $\phi\otimes\phi\in S_2^+\sph^{\otimes 2}$: then $\adag{f}\phi\otimes\phi=\sqrt{3}\phi\otimes\phi\otimes f$, which is not symmetric unless $f$ and $\phi$ are linearly dependent.
	\end{example}
	
	\begin{definition}[Bosonic and fermionic annihilation and creation operators]\label{def:apm}\index{annihilation and creation operators!bosonic}\index{annihilation and creation operators!fermionic}
		Let $f\in\sph$. The \emph{bosonic annihilation and creation operators} $a_+(f)$ and the \emph{fermionic annihilation and creation operators} $a_-(f)$ are defined by
		\begin{align}
			a_\pm(f)&:=S^\pm\a{f}\restriction\left(\Dom\a{f}\cap\Gamma^\pm(\sph)\right),\label{eq:apm-a}\\
			a^\dag_\pm(f)&:=\overline{S^\pm\adag{f}\restriction\left(\Dom\adag{f}\cap\Gamma^\pm(\sph)\right)}.\label{eq:apm-adag}
		\end{align}
	\end{definition}
	Since $\a{f}$ already maps $\Dom\a{f}\cap\Gamma^\pm(\sph)$ into $\Gamma^\pm(\sph)$ (see the proof of Proposition~\ref{prop:apm(f)prop} below), the projection $S^\pm$ in Eq.~\eqref{eq:apm-a} acts trivially; we keep it for symmetry with Eq.~\eqref{eq:apm-adag}, where it is essential (Example~\ref{examp:adagnotsym}).
	
	\begin{proposition}\label{prop:apm(f)prop}
		Let $f\in\sph$. The operators $a_\pm(f),a^\dag_\pm(f)$ satisfy the following properties:
		\begin{itemize}
			\item[(i)] $a_\pm(f),a^\dag_\pm(f)$ are closed operators on $\Gamma^\pm(\sph)$;
			\item[(ii)] For all $\psi\in\Dom\N^{1/2}\cap\Gamma^\pm(\sph)$,
			\begin{equation}
				\|a_\pm(f)\psi\|\leq\|f\|\|\N^{1/2}\psi\|,\qquad \|a^\dag_\pm(f)\psi\|\leq\|f\|\|(\N+1)^{1/2}\psi\|;
			\end{equation}
			\item[(iii)] For all $\psi,\varphi\in\Dom\N^{1/2}\cap\Gamma^\pm(\sph)$,
			\begin{equation}
				\Braket{\psi,a_\pm(f)\varphi}=\Braket{a_\pm^\dag(f)\psi,\varphi}.
			\end{equation}
		\end{itemize}
	\end{proposition}
	
	\begin{proof}
		(i) $\a{f}$ maps $\Dom\a{f}\cap\Gamma^\pm(\sph)$ into $\Gamma^\pm(\sph)$: annihilation preserves the (anti)symmetry. Indeed, $\a{f}$ acts sector by sector, mapping $\sph^{\otimes n}$ into $\sph^{\otimes(n-1)}$ as a bounded operator (Proposition~\ref{prop:a0(f)vsn}). Given $\sigma\in\mathfrak{S}_{n-1}$, let $\tilde\sigma\in\mathfrak{S}_n$ act as $\sigma$ on $\{1,\dots,n-1\}$ and fix $n$, so that $\sgn(\tilde\sigma)=\sgn(\sigma)$. Since $U_{\tilde\sigma}$ leaves the last factor untouched, Definition~\ref{def:a0(f)} and Eq.~\eqref{eq:usigma} give $\a{f}U_{\tilde\sigma}=U_\sigma\a{f}$ on $\sph^{\odot n}$, and thus on $\sph^{\otimes n}$ by continuity. Hence, if $\psi_n\in S_n^\pm\sph^{\otimes n}$, Lemma~\ref{lem:sunu} yields $U_\sigma\a{f}\psi_n=\a{f}U_{\tilde\sigma}\psi_n$, which equals $\a{f}\psi_n$ in the bosonic case and $\sgn(\sigma)\a{f}\psi_n$ in the fermionic one, for all $\sigma\in\mathfrak{S}_{n-1}$; by Lemma~\ref{lem:sunu} again, $\a{f}\psi_n\in S^\pm_{n-1}\sph^{\otimes(n-1)}$. Therefore $a_\pm(f)$ is the restriction of the closed operator $\a{f}$ to the closed subspace $\Gamma^\pm(\sph)$, hence closed. Instead, $a^\dag_\pm(f)$ is closed by definition; the closure exists since $S^\pm\adag{f}\restriction(\Dom\adag{f}\cap\Gamma^\pm(\sph))$ is contained in the adjoint of the densely defined operator $a_\pm(f)\restriction(\Dom\N^{1/2}\cap\Gamma^\pm(\sph))$. Indeed, taking closures in $a_0(f)\subseteq a_0^\dag(f)^*$ (Proposition~\ref{prop:a(f)prop}) gives $\a{f}\subseteq\adag{f}^*$, whence, for all $\psi\in\Dom\adag{f}\cap\Gamma^\pm(\sph)$ and $\varphi\in\Dom\N^{1/2}\cap\Gamma^\pm(\sph)$, using $S^\pm\varphi=\varphi$,
		\begin{equation}
			\Braket{S^\pm\adag{f}\psi,\varphi}=\Braket{\adag{f}\psi,\varphi}=\Braket{\psi,\a{f}\varphi}=\Braket{\psi,a_\pm(f)\varphi}.
		\end{equation}
		
		(ii) Since $\|S^\pm\|\leq1$, for all $\psi\in\Dom\N^{1/2}\cap\Gamma^\pm(\sph)$ we have $\|a_\pm(f)\psi\|=\|S^\pm a(f)\psi\|\leq\|\a{f}\psi\|$ and $\|a^\dag_\pm(f)\psi\|\leq\|\adag{f}\psi\|$, and the claim follows from the corresponding estimates for $\a{f},\adag{f}$ (Proposition~\ref{prop:a(f)prop2}).
		
		(iii) For all $\psi,\varphi\in\Dom\N^{1/2}\cap\Gamma^\pm(\sph)$, using $S^\pm\psi=\psi$, $S^\pm\varphi=\varphi$, the self-adjointness of $S^\pm$, and Proposition~\ref{prop:a(f)prop2},
		\begin{equation}
			\Braket{\psi,a_\pm(f)\varphi}=\Braket{S^\pm\psi,a(f)\varphi}=\Braket{\psi,a(f)\varphi}=\Braket{\adag{f}\psi,\varphi}=\Braket{\adag{f}\psi,S^\pm\varphi}=\Braket{a_\pm^\dag(f)\psi,\varphi}.
		\end{equation}
	\end{proof}
	
	\begin{example}[Single-mode fields are bosonic]\label{examp:singlemodebosonic}
		Take $\sph=\mathbb{C}$, i.e. a single mode. Then $S_n^+\mathbb{C}^{\otimes n}=\mathbb{C}^{\otimes n}=\mathbb{C}$ for all $n$, so that $\Gamma^+(\mathbb{C})=\Gamma(\mathbb{C})=\ell^2$ (Example~\ref{examp:fockC}), with orthonormal basis $(1^{\otimes n})_{n\in\mathbb{N}}$, where $1^{\otimes n}$ is the $n$-particle vector $1\otimes\cdots\otimes1$ (and $1^{\otimes0}=\varepsilon_0$). By Definitions~\ref{def:a0(f)} and~\ref{def:apm}, the bosonic operators associated with the unit vector $f=1$ act as
		\begin{equation}
			a_+(1)1^{\otimes n}=\sqrt{n}\,1^{\otimes(n-1)},\qquad a_+^\dag(1)1^{\otimes n}=\sqrt{n+1}\,1^{\otimes(n+1)},\qquad\N1^{\otimes n}=n\,1^{\otimes n}.
		\end{equation}
		Identifying $1^{\otimes n}$ with $\phi_n$, these are precisely the ladder and number operators of Section~\ref{sec:singlemode} (cf. Eq.~\eqref{eq:ladderaction} and Definition~\ref{def:abstractladder}): the single-mode field is the bosonic Fock space over a one-dimensional single-particle space, which is why we referred to single-mode fields as boson fields. By contrast, $S_n^-\mathbb{C}^{\otimes n}=0$ for all $n\geq2$, so that the fermionic Fock space reduces to $\Gamma^-(\mathbb{C})=\mathbb{C}\oplus\mathbb{C}\cong\mathbb{C}^2$, spanned by the vacuum $\varepsilon_0$ and by the one-particle state $1$: a single fermionic mode can only be either empty or occupied. In the basis $\{1,\varepsilon_0\}$, one finds that $a_-(1)$ and $a_-^\dag(1)$ act as the matrices $\sigma_-$ and $\sigma_+$ of Section~\ref{subsec:jc}: a single fermionic mode is a two-level system.
	\end{example}
	
	\begin{remark}
		The closure in the definition of $a^\dag_\pm(f)$ cannot be omitted. Let $f\neq0$. The proof of Proposition~\ref{prop:a0(f)vsn} in fact gives $\|a_0^\dag(f)\psi\|=\|f\|\|(\N+1)^{1/2}\psi\|$, so that the graph norms of $a_0^\dag(f)$ and $\N^{1/2}$ are equivalent, and $\Dom\adag{f}=\Dom\N^{1/2}$. In the bosonic case, however, take a unit vector $g\perp f$, and $\psi:=\sum_{n\geq1}\frac{1}{n}g^{\otimes n}\in\Gamma^+(\sph)$. On the one hand, $\psi\notin\Dom\N^{1/2}$, since $\sum_nn\cdot\frac{1}{n^2}=\infty$. On the other hand, the vectors $S^+\adag{f}g^{\otimes n}=\sqrt{n+1}\,S_{n+1}^+(g^{\otimes n}\otimes f)$ are mutually orthogonal, with norm $\|f\|$; hence the partial sums of $\psi$ converge in the graph norm of $S^+\adag{f}$, and $\psi$ belongs to the domain of the closure $a_+^\dag(f)$, but not to $\Dom\adag{f}\cap\Gamma^+(\sph)$.
	\end{remark}
	
	As for second quantized operators (Lemma~\ref{lem:gammacov}), annihilation and creation operators are covariant under unitary transformations of the single-particle space:
	
	\begin{lemma}\label{lem:gammacov2}
		Let $U:\sph\rightarrow\mathfrak{k}$ be unitary, and $f\in\sph$. Then $\Gamma(U)$ maps $\Gamma^\pm(\sph)$ onto $\Gamma^\pm(\mathfrak{k})$, and
		\begin{equation}
			\Gamma(U)a^\sharp(f)\Gamma(U)^*=a^\sharp(Uf),\qquad\Gamma(U)a^\sharp_\pm(f)\Gamma(U)^*=a^\sharp_\pm(Uf),
		\end{equation}
		where $a^\sharp$ stands for either $a$ or $a^\dag$.
	\end{lemma}
	
	\begin{proof}
		For all $\phi_1,\dots,\phi_n\in\sph$, since $\braket{f,\phi_n}=\braket{Uf,U\phi_n}$,
		\begin{equation}
			U^{\otimes(n-1)}a_0(f)\phi_1\otimes\cdots\otimes\phi_n=\sqrt{n}\braket{Uf,U\phi_n}U\phi_1\otimes\cdots\otimes U\phi_{n-1}=a_0(Uf)U^{\otimes n}\phi_1\otimes\cdots\otimes\phi_n,
		\end{equation}
		and analogously $U^{\otimes(n+1)}a_0^\dag(f)\phi_1\otimes\cdots\otimes\phi_n=a_0^\dag(Uf)U^{\otimes n}\phi_1\otimes\cdots\otimes\phi_n$. By linearity, $\Gamma(U)a_0^\sharp(f)=a_0^\sharp(Uf)\Gamma(U)$ on $\Gamma_0(\sph)$; since $\Gamma(U)\Gamma_0(\sph)=\Gamma_0(\mathfrak{k})$, and conjugation by a unitary operator preserves closures, $\Gamma(U)a^\sharp(f)\Gamma(U)^*=a^\sharp(Uf)$. Furthermore, $U^{\otimes n}U_\sigma=U_\sigma U^{\otimes n}$ for all $\sigma\in\mathfrak{S}_n$ (as in the solution of Exercise~\ref{exerc:gammaU}), whence $\Gamma(U)S^\pm=S^\pm\Gamma(U)$, and thus $\Gamma(U)\Gamma^\pm(\sph)=\Gamma^\pm(\mathfrak{k})$. The claim for $a^\sharp_\pm(f)$ then follows from Definition~\ref{def:apm}, since restrictions, compositions with $S^\pm$ and closures are all preserved under unitary conjugation.
	\end{proof}
	
	Let us finally see how these operators act when $\sph=L^2(\mathcal{M},\Sigma,\mu)$. In this case, the unsymmetrized annihilation and creation operators act as in Eq.~\eqref{eq:afl2}, and $S^\pm_n\sph^{\otimes n}$ is the space of $n$-particle wavefunctions satisfying, respectively, the bosonic and fermionic constraints~\eqref{eq:bosonsfermions} (Example~\ref{examp:symL2}).
	
	\begin{proposition}\label{prop:a(f)explicit}
		Let $\sph=L^2(\mathcal{M},\Sigma,\mu)$ and $f\in\sph$. Then, for all $n\in\mathbb{N}$ and $\psi_n\in S_n^\pm\sph^{\otimes n}$, the operators $a_\pm(f),a_\pm^\dag(f)$ act as follows (for the creation operator, the term with $j=n+1$ is understood as $\psi_n(x_1,\dots,x_{n})f(x_{n+1})$):
		\begin{align}
			\left(a_\pm(f)\psi_n\right)(x_1,\dots,x_{n-1})&=\sqrt{n}\int\overline{f(x_n)}\psi_n(x_1,\dots,x_{n-1},x_n)\;\mathrm{d}\mu(x_n);\\
			\left(a^\dag_\pm(f)\psi_n\right)(x_1,\dots,x_{n-1},x_n,x_{n+1})&=\frac{1}{\sqrt{n+1}}\sum_{j=1}^{n+1}(\pm1)^{n+1-j}\psi_n\left(x_1,\dots,\cancel{x_j},\dots,x_n,x_{n+1}\right)f(x_j).
		\end{align}
	\end{proposition}
	
	The annihilation operator acts as usual: integrating over one variable does not change the (anti)symmetry in the remaining ones, so no (anti)symmetrization is required. When applying the creation operator, instead, we need to restore (anti)symmetry. As we will see, this amounts to summing $n+1$ copies of the function $\psi_n(x_1,\dots,x_n)f(x_{n+1})$, the $j$th of which is obtained by moving the variable $x_j$ from $\psi_n$ to $f$, in the place of $x_{n+1}$, and putting $x_{n+1}$ as the last variable of $\psi_n$ (so that the relative order of the variables inside $\psi_n$ is not changed); in the antisymmetric case, each term is additionally weighted by $(-1)^{n+1-j}$ to preserve antisymmetry.
	
	\begin{proof}
		Given $\psi_n\in S_n^\pm\sph^{\otimes n}$, by definition we have $a_\pm(f)\psi_n=S^\pm a(f)\psi_n$, thus, by Eq.~\eqref{eq:afl2}, $a_\pm(f)\psi_n$ is $\sqrt{n}$ times the (anti)symmetrization of the $(n-1)$-particle function $(x_1,\dots,x_{n-1})\mapsto \int\overline{f(x_n)}\psi_n(x_1,\ldots,x_{n-1},x_{n})\,\mathrm{d}\mu(x_n)$. But the latter function is already completely symmetric/antisymmetric, as expected from the general case. Indeed, any permutation of $x_1,\dots,x_{n-1}$ can be regarded as a permutation of $x_1,\dots,x_{n-1},x_n$ such that $\sigma(n)=n$, and so the new wavefunction ``inherits'' the symmetry of $\psi_n$.
		
		Let us focus on the creation operator. By definition,
		\begin{equation}\label{eq:adagpm-def}
			\left(a_\pm^\dag(f)\psi_n\right)(x_1,\dots,x_{n+1})=\sqrt{n+1}\,S^\pm_{n+1}\psi_n(x_1,\dots,x_n)f(x_{n+1}),
		\end{equation}
		which involves a sum over all $(n+1)!$ permutations of the variables $x_1,\dots,x_{n+1}$ (Example~\ref{examp:symL2}). By Lemma~\ref{lem:taudecomp}, every $\sigma'\in\mathfrak{S}_{n+1}$ can be uniquely written as $\sigma'=\tau_j\sigma$, with $j\in\{1,\dots,n+1\}$ and $\sigma\in\mathfrak{S}_n$, and $\sgn(\sigma')=(-1)^{n+1-j}\sgn(\sigma)$: that is, by first moving $x_j$ to the last slot (the argument of $f$), and then permuting the remaining variables. For each $j$, denote by $y^{(j)}=(y^{(j)}_1,\dots,y^{(j)}_n):=(x_1,\dots,x_{j-1},x_{j+1},\dots,x_{n+1})$ the variables remaining once $x_j$ has been removed; writing $(\pm1)^\sigma$ as in Example~\ref{examp:symL2}, and using $(n+1)!=(n+1)\,n!$, we get
		\begin{align}
			S^\pm_{n+1}\psi_n(x_1,\dots,x_n)f(x_{n+1})&=\frac{1}{(n+1)!}\sum_{j=1}^{n+1}\sum_{\sigma\in\mathfrak{S}_n}(\pm1)^{n+1-j}(\pm1)^\sigma\,\psi_n\left(y^{(j)}_{\sigma(1)},\dots,y^{(j)}_{\sigma(n)}\right)f(x_j)\nonumber\\
			&=\frac{1}{n+1}\sum_{j=1}^{n+1}(\pm1)^{n+1-j}\left(S^\pm_n\psi_n\right)\left(y^{(j)}_1,\dots,y^{(j)}_n\right)f(x_j)\nonumber\\
			&=\frac{1}{n+1}\sum_{j=1}^{n+1}(\pm1)^{n+1-j}\psi_n\left(x_1,\dots,x_{j-1},x_{j+1},\dots,x_{n+1}\right)f(x_j),
		\end{align}
		where, in the last step, we used $S^\pm_n\psi_n=\psi_n$, since $\psi_n\in S_n^\pm\sph^{\otimes n}$. That is, because of the (anti)symmetry of $\psi_n$, only the variable moved to the last slot matters when summing over permutations. Plugging this into Eq.~\eqref{eq:adagpm-def} yields the claim.
	\end{proof}
	
	\begin{example}\label{examp:adagpm3}
		For $n=3$, Proposition~\ref{prop:a(f)explicit} reads
		\begin{align}
			\left(a_\pm(f)\psi_3\right)(x_1,x_2)&=\sqrt{3}\int\overline{f(x_3)}\psi_3(x_1,x_2,x_3)\;\mathrm{d}\mu(x_3),\nonumber\\
			\left(a^\dag_\pm(f)\psi_3\right)(x_1,x_2,x_3,x_4)&=\frac{1}{2}\Bigl(\pm\psi_3(x_2,x_3,x_4)f(x_1)+\psi_3(x_1,x_3,x_4)f(x_2)\nonumber\\
			&\qquad\pm\psi_3(x_1,x_2,x_4)f(x_3)+\psi_3(x_1,x_2,x_3)f(x_4)\Bigr).
		\end{align}
	\end{example}
	
	\subsection{Canonical commutation and anticommutation relations}\label{subsec:ccr}
	
	So far, $a_\pm(f)$ and $a^\dag_\pm(f)$ behave much like their non-symmetrized counterparts. The first important differences emerge in their commutation and anticommutation relations.
	
	Recall that, in the single-mode case, we had $[a,a^\dag]=1$ (to be understood as an identity on $\Dom\N$; see Proposition~\ref{prop:aprop2}). We now consider the multimode case.
	
	\begin{lemma}\label{lem:domn}
		The following properties hold:
		\begin{itemize}
			\item[(i)] For all $f\in\sph$, $\a{f}\Dom\N\subseteq\Dom\N^{1/2}$ and $\adag{f}\Dom\N\subseteq\Dom\N^{1/2}$;
			\item[(ii)] For all $f,g\in\sph$, the following operators are well-defined on $\Dom\N$:
			\begin{equation}
				\a{f}\a{g},\qquad \a{f}\adag{g},\qquad \adag{f}\a{g},\qquad \adag{f}\adag{g}.
			\end{equation}
		\end{itemize}
		The same properties hold if one replaces $\a{f},\adag{f}$, etc., with $a_\pm(f),a^\dag_\pm(f)$, and $\Dom\N$, $\Dom\N^{1/2}$ with their intersections with $\Gamma^\pm(\sph)$.
	\end{lemma}
	
	\begin{proof}
		\emph{(i)} Let $\psi=(\psi_n)_{n\in\mathbb{N}}\in\Dom\N\subseteq\Dom\N^{1/2}$. First note that $\a{f}$ acts sector by sector on $\psi$, i.e. $(\a{f}\psi)_n=\a{f}\psi_{n+1}$: the truncations $(\psi_0,\dots,\psi_N,0,0,\dots)$ converge to $\psi$ in the graph norm of $\N^{1/2}$, and hence, by Proposition~\ref{prop:a(f)prop2}, $\a{f}$ applied to them converges to $\a{f}\psi$. Therefore, $\a{f}\psi$ is in $\Dom\N^{1/2}$ if and only if
		\begin{equation}
			\sum_{n\in\mathbb{N}}n\left\|\left(\a{f}\psi\right)_n\right\|^2=\sum_{n\in\mathbb{N}}n\|\a{f}\psi_{n+1}\|^2<\infty.
		\end{equation}
		Since $\|\a{f}\psi\|\leq\|f\|\|\N^{1/2}\psi\|$ for all $\psi\in\Dom\N^{1/2}$ (Proposition~\ref{prop:a(f)prop2}), for every $n\in\mathbb{N}$ we have
		\begin{equation}
			\|\a{f}\psi_{n+1}\|\leq\|f\|\|\N^{1/2}\psi_{n+1}\|=(n+1)^{1/2}\|f\|\|\psi_{n+1}\|,
		\end{equation}
		whence
		\begin{equation}
			\sum_{n\in\mathbb{N}}n\|\a{f}\psi_{n+1}\|^2\leq\|f\|^2\sum_{n\in\mathbb{N}}n(n+1)\|\psi_{n+1}\|^2\leq\|f\|^2\sum_{n\geq1}n^2\|\psi_n\|^2=\|f\|^2\|\N\psi\|^2<\infty.
		\end{equation}
		The argument for $\adag{f}$ is similar; altogether, for all $\psi\in\Dom\N$,
		\begin{equation}\label{eq:domn-bounds}
			\left\|\N^{1/2}\a{f}\psi\right\|\leq\|f\|\|\N\psi\|,\qquad\left\|\N^{1/2}\adag{f}\psi\right\|\leq\|f\|\|(\N+1)\psi\|.
		\end{equation}
		The proof for $a_\pm(f)$ and $a^\dag_\pm(f)$ proceeds in the same way, since they satisfy the same estimates as $\a{f}$ and $\adag{f}$ (Proposition~\ref{prop:apm(f)prop}(ii)), and Eq.~\eqref{eq:domn-bounds} holds for them as well.
		
		(ii) is a direct consequence of (i).		
	\end{proof}
	
	\begin{proposition}[Canonical commutation/anticommutation relations]\label{prop:ccr/car}\index{canonical commutation relations}\index{canonical anticommutation relations}
		Let $f,g\in\sph$. For all $\psi\in\Dom\N\cap\Gamma^+(\sph)$, the \emph{canonical commutation relations (CCR)} hold:
		\begin{equation}
			\left[a_+(f),a_+^\dag(g)\right]\psi=\braket{f,g}\psi,\qquad\left[a_+(f),a_+(g)\right]\psi=0.
		\end{equation}
		For all $\psi\in\Dom\N\cap\Gamma^-(\sph)$, the \emph{canonical anticommutation relations (CAR)} hold:
		\begin{equation}
			\left\{a_-(f),a_-^\dag(g)\right\}\psi=\braket{f,g}\psi,\qquad\left\{a_-(f),a_-(g)\right\}\psi=0.
		\end{equation}
		Above, $[A,B]:=AB-BA$ and $\{A,B\}:=AB+BA$ are respectively the \emph{commutator} and \emph{anticommutator}.
	\end{proposition}
	
	\begin{proof}
		We prove this in the case $\sph=L^2(\mathcal{M},\Sigma,\mu)$. This is no loss of generality: every separable Hilbert space is unitarily equivalent to an $L^2$ space (e.g. to $\ell^2$, i.e. $L^2$ on $\mathbb{N}$ with the counting measure), and, if $U:\sph\rightarrow L^2(\mathcal{M},\Sigma,\mu)$ is unitary, then $\Gamma(U)$ maps $\Gamma^\pm(\sph)$ onto $\Gamma^\pm(L^2(\mathcal{M},\Sigma,\mu))$ and $\Gamma(U)a^\sharp_\pm(f)\Gamma(U)^*=a^\sharp_\pm(Uf)$, with $a^\sharp_\pm\in\{a_\pm,a^\dag_\pm\}$ (Lemma~\ref{lem:gammacov2}); besides, $\Gamma(U)$ commutes with $\N$ (Lemma~\ref{lem:gammacov}), and $\braket{Uf,Ug}=\braket{f,g}$. In this case, $a_\pm(f)$ and $a^\dag_\pm(f)$ act as in Proposition~\ref{prop:a(f)explicit}.
		
		Let $n\in\mathbb{N}$ and $\psi_n\in S_n^\pm\sph^{\otimes n}$. Then
		\begin{align}
			\left(a_\pm(f)\psi_n\right)(x_1,\dots,x_{n-1})&=\sqrt{n}\int\overline{f(x'_n)}\psi_n(x_1,\dots,x_{n-1},x'_n)\;\mathrm{d}\mu(x'_n).
		\end{align}
		Here we denoted the $n$th variable---the one being integrated out---as $x'_n$ instead of $x_n$; the reason will become clear in a moment. Applying $a^\dag_\pm(g)$ to the $(n-1)$-particle wavefunction above,
		\begin{align}
			\left(a^\dag_\pm(g)a_\pm(f)\psi_n\right)(x_1,\dots,x_n)&=\frac{1}{\sqrt{n}}\sum_{j=1}^n(\pm1)^{n-j}\left(a_\pm(f)\psi_n\right)(x_1,\dots,\cancel{x_j},\dots,x_n)g(x_j)\nonumber\\
			&=\sum_{j=1}^n(\pm1)^{n-j}\int\overline{f(x'_{n})}g(x_j)\psi_n(x_1,\dots,\cancel{x_j},\dots,x_{n},x'_{n})\;\mathrm{d}\mu(x'_n).
		\end{align}
		Indeed, recall that the creation operator is the sum of all terms in which, for each $j$, the variable $x_j$ is ``deleted'' from $\psi_n$ and ``transferred'' to $g$, while all subsequent variables change place accordingly: $x_{j+1}$ becomes the $j$th variable, $x_{j+2}$ becomes the $(j+1)$th one, and so on, and the new variable---here, $x_n$---goes in the last place. The integration variable $x'_n$ is not involved in this process and, being a dummy variable, can be renamed freely; choosing $x'_n=:x_{n+1}$,
		
		\begin{multline}\label{eq:ccr-adaga}
			\left(a^\dag_\pm(g)a_\pm(f)\psi_n\right)(x_1,\dots,x_n)\\=\sum_{j=1}^n(\pm1)^{n-j}\int\overline{f(x_{n+1})}g(x_j)\psi_n(x_1,\dots,\cancel{x_j},\dots,x_{n},x_{n+1})\;\mathrm{d}\mu(x_{n+1}).
		\end{multline}		
		Vice versa, separating the term with $j=n+1$ in Proposition~\ref{prop:a(f)explicit},
		\begin{multline}
			\left(a_\pm^\dag(g)\psi_n\right)(x_1,\dots,x_{n+1})\\=\frac{1}{\sqrt{n+1}}\sum_{j=1}^{n}(\pm1)^{n+1-j}\psi_n\left(x_1,\dots,\cancel{x_j},\dots,x_{n+1}\right)g(x_j)+\frac{1}{\sqrt{n+1}}\psi_n(x_1,\dots,x_n)g(x_{n+1}),
		\end{multline}
		and thus, applying $a_\pm(f)$ to the $(n+1)$-particle function above,
		\begin{align}
			\left(a_\pm(f)a_\pm^\dag(g)\psi_n\right)(x_1,\dots,x_n)&=\sqrt{n+1}\int\overline{f(x_{n+1})}\left(a^\dag_\pm(g)\psi_n\right)(x_1,\dots,x_{n+1})\,\mathrm{d}\mu(x_{n+1})	\nonumber\\
			&=\sum_{j=1}^{n+1}(\pm1)^{n+1-j}\int\overline{f(x_{n+1})}g(x_j)\psi_n(x_1,\dots,\cancel{x_j},\dots,x_{n+1})\,\mathrm{d}\mu(x_{n+1})\nonumber\\
			&=\pm\sum_{j=1}^{n}(\pm1)^{n-j}\int\overline{f(x_{n+1})}g(x_j)\psi_n(x_1,\dots,\cancel{x_j},\dots,x_{n+1})\,\mathrm{d}\mu(x_{n+1})\nonumber\\
			&\;\;+\underbrace{\left(\int\overline{f(x_{n+1})}g(x_{n+1})\;\mathrm{d}\mu(x_{n+1})\right)}_{=\braket{f,g}}\psi_n(x_1,\dots,x_n),\label{eq:ccr-aadag}
		\end{align}
		where, in the last step, we separated the terms with $j=1,\dots,n$, in which one of the variables of $\psi_n$ is integrated, from the term with $j=n+1$, in which the variable of $g$ is integrated against $\overline{f}$. Comparing Eqs.~\eqref{eq:ccr-adaga} and~\eqref{eq:ccr-aadag}, we conclude 
		\begin{equation}
			a_\pm(f)a_\pm^\dag(g)\psi_n=\pm a^\dag_\pm(g)a_\pm(f)\psi_n+\braket{f,g}\psi_n,
		\end{equation}
		which proves the first relations of Proposition~\ref{prop:ccr/car} for $\psi=\psi_n\in S_n^\pm\sph^{\otimes n}$. By linearity, they hold for all $\psi\in\Dom\N\cap\Gamma^\pm(\sph)$ with finitely many nonzero components. Finally, let $\psi\in\Dom\N\cap\Gamma^\pm(\sph)$ be arbitrary, and $\psi^{(m)}:=(\psi_0,\dots,\psi_m,0,0,\dots)$. By Eq.~\eqref{eq:domn-bounds} (and its analogue for $a_\pm^\sharp$) and Proposition~\ref{prop:apm(f)prop}(ii), there is $C\geq0$ such that $\|a^\sharp_\pm(f)a^\sharp_\pm(g)\varphi\|\leq C\|(\N+1)\varphi\|$ for all $\varphi\in\Dom\N\cap\Gamma^\pm(\sph)$ and all choices $a^\sharp_\pm\in\{a_\pm,a_\pm^\dag\}$. Since $\psi^{(m)}\to\psi$ in the graph norm of $\N$, both sides of the relations evaluated at $\psi^{(m)}$ converge to the corresponding ones evaluated at $\psi$, which proves the claim for $\psi$.
		
		The remaining relations are proven analogously (Exercise~\ref{exerc:ccr/car}).
	\end{proof}
	
	\begin{exercise}\label{exerc:ccr/car}
		Complete the proof by showing the remaining relations in Proposition~\ref{prop:ccr/car}. Hint: compute $a_\pm(f)a_\pm(g)\psi_n$ by applying Proposition~\ref{prop:a(f)explicit} twice, and show that exchanging $f$ and $g$ amounts to exchanging the last two variables of $\psi_n$, which, by its (anti)symmetry, only produces a factor $\pm1$.
		\seesol{ccr/car}
	\end{exercise}
	
	Another fundamental difference between bosonic and fermionic operators is the following:
	
	\begin{proposition}\label{prop:fermionsbounded}
		Let $f\in\sph$, and let $a_-(f),a_-^\dag(f)$ be the corresponding fermionic annihilation and creation operators. Then $a_-(f),a_-^\dag(f)$ are bounded, and extend uniquely to bounded operators on the whole fermionic Fock space $\Gamma^-(\sph)$, with norms $\|a_-(f)\|=\|a^\dag_-(f)\|=\|f\|$.
	\end{proposition}
	
	This, once again, is a mathematical reflection of Pauli's exclusion principle: the number of fermionic excitations that can be created in a given state is finite.
	
	\begin{proof}
		Let $f\in\sph$ and $\psi\in\Dom\N\cap\Gamma^-(\sph)$. Recall that $a_-(f),a_-^\dag(f)$ are mutually adjoint on $\Dom\N^{1/2}$ (Proposition~\ref{prop:apm(f)prop}), that $a_-(f)\Dom\N\subseteq\Dom\N^{1/2}$ and $a_-^\dag(f)\Dom\N\subseteq\Dom\N^{1/2}$ (Lemma~\ref{lem:domn}), and that they satisfy the CAR, $\{a_-(f),a_-^\dag(f)\}\psi=\|f\|^2\psi$ (Proposition~\ref{prop:ccr/car}). Using these three properties, for all $\psi\in\Dom\N\cap\Gamma^-(\sph)$ we have
		\begin{equation}
			\left\|a_-(f)\psi\right\|^2=\Braket{\psi,a^\dag_-(f)a_-(f)\psi}=\|f\|^2\|\psi\|^2-\Braket{\psi,a_-(f)a^\dag_-(f)\psi}=\|f\|^2\|\psi\|^2-\left\|a^\dag_-(f)\psi\right\|^2,
		\end{equation}
		whence both $\|a_-(f)\psi\|^2$ and $\|a^\dag_-(f)\psi\|^2$ are bounded by $\|f\|^2\|\psi\|^2$. This shows $\|a_-(f)\psi\|\leq\|f\|\|\psi\|$ and $\|a^\dag_-(f)\psi\|\leq\|f\|\|\psi\|$ on the dense subspace $\Dom\N\cap\Gamma^-(\sph)$ of $\Gamma^-(\sph)$. Therefore, by the continuous linear extension theorem, $a_-(f),a_-^\dag(f)$ uniquely extend to two bounded operators on the whole $\Gamma^-(\sph)$, 		with operator norms $\|a_-(f)\|\leq\|f\|$ and $\|a_-^\dag(f)\|\leq\|f\|$. Equality holds since $a_-(f)f=\braket{f,f}\varepsilon_0$, whence $\|a_-(f)f\|=\|f\|\|f\|$, and $a_-^\dag(f)\varepsilon_0=f$, whence $\|a_-^\dag(f)\varepsilon_0\|=\|f\|\|\varepsilon_0\|$, $\varepsilon_0$ being the vacuum.
	\end{proof}
	
	The argument above fails for bosonic operators: the CCR yield $\|a^\dag_+(f)\psi\|^2=\|f\|^2\|\psi\|^2+\|a_+(f)\psi\|^2$ instead. In fact, the estimates of Proposition~\ref{prop:apm(f)prop}(ii) are sharp in the bosonic case, in the sense that some vectors saturate them:
	
	\begin{example}\label{examp:saturate}
		Let $f\in\sph$. Then, for every $n\in\mathbb{N}$, the $n$-particle state $f^{\otimes n}=f\otimes f\otimes\cdots\otimes f$ is in $\Gamma^+(\sph)$, since all permutation operators leave it invariant (Lemma~\ref{lem:sunu}), and satisfies $\|f^{\otimes n}\|=\|f\|^n$ and $\N^{1/2}f^{\otimes n}=\sqrt{n}f^{\otimes n}$. Then $a_+(f)f^{\otimes n}=\sqrt{n}\braket{f,f}f^{\otimes(n-1)}$, whence
		\begin{equation}
			\|a_+(f)f^{\otimes n}\|=\sqrt{n}\|f\|^2\|f\|^{n-1}=\sqrt{n}\|f\|\|f^{\otimes n}\|=\|f\|\left\|\N^{1/2}f^{\otimes n}\right\|.
		\end{equation}		
		Analogously, $a^\dag_+(f)f^{\otimes n}=\sqrt{n+1}S^+_{n+1}f^{\otimes(n+1)}=\sqrt{n+1}f^{\otimes(n+1)}$, whence
		\begin{equation}
			\left\|a^\dag_+(f)f^{\otimes n}\right\|=\sqrt{n+1}\|f\|^{n+1}=\sqrt{n+1}\|f\|\|f^{\otimes n}\|=\|f\|\|(\N+1)^{1/2}f^{\otimes n}\|.
		\end{equation}
	\end{example}
	
	\subsection{Second-quantized operators in terms of annihilation and creation operators}\label{subsec:dgammaladder}
	
	We conclude by reconciling our definition of second quantized operators with the one usually encountered in quantum field theory:
	\begin{equation}
		H=\sum_{k\in\mathbb{N}}\omega_ka^\dag_{k}a_k\qquad\text{or}\qquad H=\int\omega(k)a^\dag(k)a(k)\;\mathrm{d}k,
	\end{equation}
	for discrete and continuous fields respectively, where one imposes
	\begin{equation}
		[a_k,a_{k'}^\dag]=\delta_{kk'},\;[a_k,a_{k'}]=0\qquad\text{resp.}\qquad\left[a(k),a^\dag(k')\right]=\delta(k-k'),\;\left[a(k),a(k')\right]=0
	\end{equation}
	in the bosonic case, or
	\begin{equation}
		\{a_k,a_{k'}^\dag\}=\delta_{kk'},\;\left\{a_k,a_{k'}\right\}=0\qquad\text{resp.}\qquad\left\{a(k),a^\dag(k')\right\}=\delta(k-k'),\;\left\{a(k),a(k')\right\}=0
	\end{equation}
	in the fermionic one.	
	How should these equalities be properly interpreted? To begin with, fix an orthonormal basis $(e_k)_{k\in\mathbb{N}}\subset\sph$. Then, as a direct consequence of the CCR and CAR (Proposition~\ref{prop:ccr/car}) and of $\braket{e_k,e_{k'}}=\delta_{kk'}$, the operators $a_k:=a_\pm(e_k)$ and $a^\dag_k:=a^\dag_\pm(e_k)$ satisfy the discrete relations above, as identities on $\Dom\N\cap\Gamma^\pm(\sph)$.
	
	\begin{lemma}\label{lem:adaga_sq}
		Let $(e_k)_{k\in\mathbb{N}}\subset\sph$ be an orthonormal basis. Then, for all $\psi\in\Gamma^\pm_0(\sph)$,
		\begin{equation}
			a^\dag_\pm(e_k)a_\pm(e_{k'})\psi=\mathrm{d}\Gamma\Bigl(\ket{e_k}\bra{e_{k'}}\Bigr)\psi,
		\end{equation}
		i.e. $a^\dag_\pm(e_k)a_\pm(e_{k'})$ coincides with the second quantization of the rank-one operator $\ket{e_k}\bra{e_{k'}}$.
	\end{lemma}
	
	\begin{proof}
		Again, we prove the claim in the case $\sph=L^2(\mathcal{M},\Sigma,\mu)$; as in the proof of Proposition~\ref{prop:ccr/car}, this is no loss of generality, since in addition $\Gamma(U)\,\mathrm{d}\Gamma(B)\,\Gamma(U)^*=\mathrm{d}\Gamma(UBU^*)$ for every bounded $B$ (Lemma~\ref{lem:gammacov}). Let $\psi_n\in S_n^\pm\sph^{\otimes n}$. By the same computation leading to Eq.~\eqref{eq:ccr-adaga},
		\begin{multline}\label{eq:one}
			\left(a^\dag_\pm(e_k)a_\pm(e_{k'})\psi_n\right)(x_1,\dots,x_n)\\=\sum_{j=1}^n(\pm1)^{n-j}\int\overline{e_{k'}(x_{n+1})}e_k(x_j)\psi_n(x_1,\dots,\cancel{x_j},\dots,x_{n},x_{n+1})\;\mathrm{d}\mu(x_{n+1}).
		\end{multline}
		Besides, applying $I\otimes\cdots\otimes\ket{e_{k}}\!\bra{e_{k'}}\otimes\cdots\otimes I$, with the rank-one operator in the $j$th slot, means multiplying $\psi_n(x_1,\dots,x_n)$ by $\overline{e_{k'}(x_j)}$, integrating over $x_j$ (which we can harmlessly rename $x_{n+1}$), and then multiplying the resulting $(n-1)$-particle function by $e_k$, evaluated at the ``new'' variable $x_j$. Hence, by Definition~\ref{def:secondquantization},
		\begin{multline}\label{eq:two}
			\left(\mathrm{d}\Gamma\left(\ket{e_k}\!\bra{e_{k'}}\right)\psi_n\right)(x_1,\dots,x_n)\\=\sum_{j=1}^n\int\overline{e_{k'}(x_{n+1})}e_k(x_j)\psi_n(x_1,\dots,x_{j-1},x_{n+1},x_{j+1},\dots,x_n)\;\mathrm{d}\mu(x_{n+1}).
		\end{multline}
		The two expressions differ in two respects. First, in Eq.~\eqref{eq:one} the arguments of $\psi_n$ are $x_1,\dots,x_{j-1},x_{j+1},\dots,x_n,x_{n+1}$, while in Eq.~\eqref{eq:two} they are $x_1,\dots,x_{j-1},x_{n+1},x_{j+1},\dots,x_n$: in both cases, $x_j$ is removed and a new variable $x_{n+1}$ is inserted without changing the relative order of the old ones, but at different positions. Second, the sign $(\pm1)^{n-j}$ is absent in Eq.~\eqref{eq:two}. The two orderings are related by a permutation which moves $x_{n+1}$ from the $j$th to the last slot, in the same way as the permutations $\tau_j$ of Lemma~\ref{lem:taudecomp}; as there, its sign is $(-1)^{n-j}$, since the $n-j$ variables $x_{j+1},\dots,x_n$ are moved to the left of $x_{n+1}$. Since $\psi_n\in S_n^\pm\sph^{\otimes n}$, Eq.~\eqref{eq:bosonsfermions} then gives
		\begin{equation}
			\psi_n(x_1,\dots,x_{j-1},x_{n+1},x_{j+1},\dots,x_n)=(\pm1)^{n-j}\psi_n(x_1,\dots,\cancel{x_j},\dots,x_{n},x_{n+1}).
		\end{equation}
		Plugging this equality into Eq.~\eqref{eq:two} yields Eq.~\eqref{eq:one}, i.e. the claimed equality for $\psi_n\in S_n^\pm\sph^{\otimes n}$. By linearity, the same holds for all $\psi\in\Gamma^\pm_0(\sph)$.
	\end{proof}
	
	\begin{remark}
		The same equality does not hold for the non-symmetrized annihilation and creation operators. Indeed, given $\psi_n=\phi_1\otimes\cdots\otimes\phi_{n-1}\otimes\phi_n$, one has $a(e_{k'})\psi_n=\sqrt{n}\braket{e_{k'},\phi_n}\phi_1\otimes\cdots\otimes\phi_{n-1}$, and thus
		\begin{equation}
			a^\dag(e_k)a(e_{k'})\psi_n=n\braket{e_{k'},\phi_n}\phi_1\otimes\cdots\otimes\phi_{n-1}\otimes e_k=n\left(I\otimes \cdots\otimes I\otimes\ket{e_k}\!\bra{e_{k'}}\right)\psi_n,
		\end{equation}
		i.e. the rank-one operator $\ket{e_k}\!\bra{e_{k'}}$ is only applied to the last factor, while $\mathrm{d}\Gamma(\ket{e_k}\!\bra{e_{k'}})$ applies it to each factor in turn (Definition~\ref{def:secondquantization}).
	\end{remark}

	\begin{proposition}\label{prop:dgammaladder}
		Let $A:\Dom A\subseteq\sph\rightarrow\sph$ be a self-adjoint operator on $\sph$, and $(e_k)_{k\in\mathbb{N}}$ an orthonormal basis contained in the domain of $A$. Then the equality
		\begin{equation}
			\mathrm{d}\Gamma(A)\psi=\sum_{k,k'\in\mathbb{N}}\Braket{e_k,Ae_{k'}}a_\pm^{\dag}(e_k) a_\pm(e_{k'})\psi
		\end{equation}
		holds for all $\psi\in S^\pm\mathcal{D}$, with $\mathcal{D}$ as in Eq.~\eqref{eq:coreD} (by Propositions~\ref{prop:props_2q} and~\ref{prop:invariant}, $S^\pm\mathcal{D}$ is a core for the restriction of $\mathrm{d}\Gamma(A)$ to $\Gamma^\pm(\sph)$), the double series being understood as an iterated one (first over $k'$, then over $k$). 
		
		Besides, if $(e_k)_{k\in\mathbb{N}}$ is a basis of eigenvectors of $A$, with $\alpha_k$ being the $k$th eigenvalue of $A$, then, for all $\psi\in S^\pm\mathcal{D}$,
		\begin{equation}
			\mathrm{d}\Gamma(A)\psi=\sum_{k\in\mathbb{N}}\alpha_k a^\dag_\pm(e_k) a_\pm(e_k)\psi,\qquad\N\psi=\sum_{k\in\mathbb{N}}a_\pm^\dag(e_k) a_\pm(e_k)\psi,
		\end{equation}
		the latter being the special case $A=I$, i.e. $\N=\mathrm{d}\Gamma(I)$, for which $S^\pm\mathcal{D}=\Gamma_0^\pm(\sph)$.
	\end{proposition}

	\begin{proof}
		Let $(e_k)_{k\in\mathbb{N}}\subset\Dom A$, $n\in\mathbb{N}$, and $\psi_n\in S_n^\pm\mathcal{D}_n$, with $\mathcal{D}_n=\bigodot_{j=1}^n\Dom A$ as in the proof of Proposition~\ref{prop:props_2q}; note that $S_n^\pm\mathcal{D}_n\subseteq\mathcal{D}_n$, since each $U_\sigma$ maps $\mathcal{D}_n$ into itself (Eq.~\eqref{eq:usigma}). Therefore $\mathrm{d}\Gamma(A)\psi_n$ is the sum over $j$ of the operators acting as $A$ on the $j$th factor of $\psi_n$ (Definition~\ref{def:secondquantization}). But, using the completeness relation twice, the equality
		\begin{equation}\label{eq:compl}
			A=\sum_{k,k'\in\mathbb{N}}\ket{e_k}\Braket{e_k,Ae_{k'}}\bra{e_{k'}}
		\end{equation}
		holds strongly on $\Dom A$; applying it in the $j$th slot, which is legitimate since $\psi_n\in\mathcal{D}_n$, we get
		\begin{align}
			\mathrm{d}\Gamma(A)\psi_n&=\sum_{k,k'\in\mathbb{N}}\Braket{e_k,Ae_{k'}}\sum_{j=1}^n\Bigl(I\otimes\cdots\otimes\overbrace{\ket{e_k}\!\bra{e_{k'}}}^{j\text{th}}\otimes\cdots\otimes I\Bigr)\psi_n\nonumber\\
			&=\sum_{k,k'\in\mathbb{N}}\Braket{e_k,Ae_{k'}}\mathrm{d}\Gamma\Bigl(\ket{e_k}\!\bra{e_{k'}}\Bigr)\psi_n=\sum_{k,k'\in\mathbb{N}}\Braket{e_k,Ae_{k'}}a_\pm^\dag(e_k)a_\pm(e_{k'})\psi_n,
		\end{align}
		where we recognized the second quantization of $\ket{e_k}\!\bra{e_{k'}}$, and used the equality $\mathrm{d}\Gamma(\ket{e_k}\!\bra{e_{k'}})\psi_n=a_\pm^\dag(e_k)a_\pm(e_{k'})\psi_n$ proven in Lemma~\ref{lem:adaga_sq}.
		
		In particular, if $(e_k)_{k\in\mathbb{N}}$ is an eigenbasis of $A$, Eq.~\eqref{eq:compl} simplifies as 
		\begin{equation}\label{eq:compl2}
			A=\sum_{k\in\mathbb{N}}\alpha_k\ket{e_k}\!\bra{e_k},
		\end{equation}
		and one proves the desired equality.
	\end{proof}
	
	This allows us to expand the second quantization of every self-adjoint operator $A$ admitting an orthonormal basis of eigenvectors (pure point spectrum) as expected:
	\begin{equation}
		\mathrm{d}\Gamma(A)=\sum_{k\in\mathbb{N}}\alpha_k a^\dag_\pm(e_k)a_\pm(e_k).
	\end{equation}
	Physically, this describes a fermionic or bosonic quantum field with a discrete (finite or countable) set of modes---e.g. an electromagnetic field confined in a cavity.
	
	\begin{remark}[Continuous modes and Gelfand triples]\label{rem:gelfand}\index{Gelfand triple}
		If $A$ has a purely continuous spectrum (e.g. the Laplacian on $\mathbb{R}^d$), one would heuristically like to replace the eigenbasis by a ``continuous basis'' $(e(k))_{k\in\mathbb{R}}$ with $Ae(k)=\alpha(k)e(k)$ and $\braket{e(k),e(k')}=\delta(k-k')$, and Eq.~\eqref{eq:compl2} by $A=\int\alpha(k)\ket{e(k)}\!\bra{e(k)}\,\mathrm{d}k$, thus obtaining the familiar expressions
		\begin{equation}\label{eq:formalcontinuous}
			\mathrm{d}\Gamma(A)=\int\alpha(k)a^\dag(k)a(k)\;\mathrm{d}k,\qquad [a(k),a^\dag(k')]=\delta(k-k'),\quad[a(k),a(k')]=0
		\end{equation}
		(with anticommutators in the fermionic case). As they stand, these expressions are ill-defined, since no operator on a separable Hilbert space admits an uncountable family of orthonormal eigenvectors. They can be given a rigorous meaning, at least in the simplest cases, by means of \emph{Gelfand triples} (or \emph{rigged Hilbert spaces})~\cite{gelfandvilenkin4,delamadrid2005}. One considers a dense subspace $\mathcal{D}\subset\sph$ of ``well-behaved'' vectors carrying a finer topology than that of $\sph$, e.g. the Schwartz space $\mathcal{S}(\mathbb{R})\subset L^2(\mathbb{R})$, and its dual $\mathcal{D}'$, so that $\mathcal{D}\subset\sph\cong\sph'\subset\mathcal{D}'$ by the Riesz representation theorem. The generalized eigenvectors $e(k)$, as well as the Dirac delta $\delta[f]:=f(0)$, are elements of $\mathcal{D}'$, and the expressions~\eqref{eq:formalcontinuous} can be interpreted in the sense of distributions.
	\end{remark}
	
	\section{Multimode spin--boson models}\label{sec:multimodesb}
	
	We can finally return to spin--boson models, now coupling the qubit to a boson field with arbitrarily many, possibly continuously many, modes, by means of the Fock space formalism of Sections~\ref{sec:sq} and~\ref{sec:bf}. We first define these models and establish their self-adjointness (Section~\ref{subsec:multimodesb}); we then show that the two exactly solvable single-mode models of Section~\ref{sec:singlemodesb} have multimode counterparts, the pure dephasing model (Section~\ref{subsec:multidephasing}) and the Jaynes--Cummings model (Section~\ref{subsec:multijc}), which can still be solved, and which display a genuinely new phenomenon: irreversibility.
	
	\subsection{Definition and self-adjointness}\label{subsec:multimodesb}
	
	In the single-mode case (Section~\ref{subsec:singlemodesb}), we defined spin--boson models as the self-adjoint operators $H=\omega_0\frac{\sigma_z}{2}+\omega a^\dag a+Ba^\dag+B^* a$ on $\hfrak=\mathbb{C}^2\otimes L^2(\mathbb{R})$, with $\omega_0\geq0$, $\omega>0$ and $B\in\mathbb{C}^{2\times 2}$. Their multimode counterpart is, formally,
	\begin{equation}\label{eq:sbformal}
		H=\omega_0\frac{\sigma_z}{2}+\int\omega(k)a^\dag(k)a(k)\,\mathrm{d}\mu(k)+\int\left[ B\,f(k)a^\dag(k)+B^*\,\overline{f(k)}a(k)\right]\mathrm{d}\mu(k),
	\end{equation}
	where the variable $k$ plays the role of a (quasi)momentum indexing the boson modes, possibly together with discrete labels such as polarization, $\omega$ is the \emph{dispersion relation}\index{dispersion relation} of the field, $f$ is a function---the \emph{form factor}\index{form factor}---that modulates the coupling between the spin and the boson mode $k$, and $[a(k),a^\dag(k')]=\delta(k-k')$, $[a(k),a(k')]=0$ (formally, or in the sense of distributions; cf. Remark~\ref{rem:gelfand}). Models of this kind are ubiquitous in the physics of open quantum systems~\cite{leggett1987,breuerpetruccione}, and have been extensively studied in mathematical physics, see e.g.~\cite{spohn1989,hubnerspohn1995,araihirokawa1997,haslerherbst2011}.
	
	\begin{definition}[Spin--boson model]\label{def:spinboson}\index{spin--boson model}
		Let $\sph$ be a complex separable Hilbert space, and $\Gamma^+(\sph)$ the corresponding bosonic Fock space. A \emph{spin--boson model} is an operator on $\hfrak=\mathbb{C}^2\otimes\Gamma^+(\sph)$ defined by 
		\begin{align}
			\Dom H&:=\mathbb{C}^2\odot\Dom\mathrm{d}\Gamma(\omega);\\
			H&:=\omega_0\frac{\sigma_z}{2}\otimes I+I\otimes\mathrm{d}\Gamma(\omega)+B\otimes a_+^\dag(f)+B^*\otimes a_+(f),
		\end{align}
		where $\omega_0\geq0$, $B\in\mathbb{C}^{2\times 2}$, $f\in\sph$, and $\omega:\Dom\omega\subseteq\sph\rightarrow\sph$ is a self-adjoint operator on the single-particle space $\sph$.
	\end{definition}
	
	Here $a_+(f),a_+^\dag(f)$ are the bosonic annihilation and creation operators associated with $f$ (Definition~\ref{def:apm}); we keep the subscript $+$ throughout, to distinguish them from the non-symmetrized operators $\a{f},\adag{f}$ of Section~\ref{subsec:aadag}. The formal expression~\eqref{eq:sbformal} corresponds to the case where $\sph=L^2(\mathcal{M},\Sigma,\mu)$, and $\omega$ is (the multiplication operator associated with) a real-valued, measurable function on $\mathcal{M}$.
	
	To prove that $H$ is self-adjoint, we must control the interaction term. In the single-mode case, this was done by comparing $a,a^\dag$ with $\N^{1/2}$ (Proposition~\ref{prop:visrelbound}). Here we can do the same thanks to Proposition~\ref{prop:apm(f)prop}, but then we need to compare $\N$ with $\mathrm{d}\Gamma(\omega)$, and this is possible only if $\omega$ is strictly positive (Proposition~\ref{prop:domega_bb}): physically, the field must be \emph{massive}\index{massive field}. This would rule out \emph{massless}\index{massless field} fields, for which $\omega$ is not bounded away from zero---among them, some of the most relevant physical situations, e.g. the electromagnetic field. (For the latter, the boson modes are labeled by a wavevector $\bm{k}\in\mathbb{R}^3$ and one of the two transverse polarizations, i.e. $\mathcal{M}=\mathbb{R}^3\times\{1,2\}$, and $\omega(\bm{k},\lambda)=|\bm{k}|$). A sharper estimate, Proposition~\ref{prop:a(f)domega} below, allows us to compare $a_+(f),a_+^\dag(f)$ with $\mathrm{d}\Gamma(\omega)^{1/2}$ directly, without passing through $\N$. With a slight abuse of notation, here and in the following we still denote by $\mathrm{d}\Gamma(\omega)$ its restriction to $\Gamma^+(\sph)$, which is self-adjoint by Proposition~\ref{prop:domega_symm_sa}.
	
	The following estimates are standard (see e.g.~\cite{derezinskigerard,arai,lonigro2022}). They involve the operator $\omega^{-1/2}$, defined via the spectral theorem: although $\lambda\mapsto\lambda^{-1/2}$ is singular at $\lambda=0$, which may well belong to $\sigma(\omega)$ (e.g. for $\omega(k)=|k|$), the assumption $\Ker\omega=\{0\}$ means precisely that $P_\omega(\{0\})=0$, so that $\omega^{-1/2}$ is a well-defined, generally unbounded, self-adjoint operator. The meaning of the condition $f\in\Dom\omega^{-1/2}$ is discussed in Remark~\ref{rem:uvir} below.
	
	\begin{proposition}\label{prop:a(f)domega}
		Let $\omega:\Dom\omega\subseteq\sph\rightarrow\sph$ be a nonnegative self-adjoint operator with $\Ker\omega=\{0\}$, and let $f\in\Dom\omega^{-1/2}$. Then:
		\begin{itemize}
			\item[(i)] $\Dom\mathrm{d}\Gamma(\omega)^{1/2}\subseteq\Dom a_+(f)\cap\Dom a_+^\dag(f)$ and, for all $\psi\in\Dom\mathrm{d}\Gamma(\omega)^{1/2}$,
			\begin{equation}\label{eq:a(f)domega}
				\left\|a_+(f)\psi\right\|\leq\left\|\omega^{-1/2}f\right\|\left\|\mathrm{d}\Gamma(\omega)^{1/2}\psi\right\|,\qquad\left\|a_+^\dag(f)\psi\right\|\leq\left\|\omega^{-1/2}f\right\|\left\|\mathrm{d}\Gamma(\omega)^{1/2}\psi\right\|+\|f\|\|\psi\|;
			\end{equation}
			\item[(ii)] For all $\psi,\varphi\in\Dom\mathrm{d}\Gamma(\omega)^{1/2}$,
			\begin{equation}
				\Braket{\psi,a_+(f)\varphi}=\Braket{a_+^\dag(f)\psi,\varphi}.
			\end{equation}
		\end{itemize}
	\end{proposition}
	
	\begin{proof}
		As in the proof of Proposition~\ref{prop:domega_bb}, by the spectral theorem (Theorem~\ref{thm:spectraltheorem2}) and Remark~\ref{rem:covariance} we can assume without loss of generality that $\sph=L^2(\mathcal{M},\Sigma,\mu)$ and that $\omega$ is the multiplication operator by a measurable function with $\omega(k)>0$ for almost all $k\in \mathcal{M}$. Indeed, if $U:\sph\rightarrow L^2(\mathcal{M},\Sigma,\mu)$ is the unitary operator given by the spectral theorem, then, by Lemmas~\ref{lem:gammacov} and~\ref{lem:gammacov2}, $\Gamma(U)$ maps $\Gamma^+(\sph)$ onto $\Gamma^+(L^2(\mathcal{M},\Sigma,\mu))$, $\Gamma(U)\mathrm{d}\Gamma(\omega)\Gamma(U)^*=\mathrm{d}\Gamma(U\omega U^*)$ and $\Gamma(U)a^\sharp_+(f)\Gamma(U)^*=a^\sharp_+(Uf)$, with $\|(U\omega U^*)^{-1/2}Uf\|=\|\omega^{-1/2}f\|$ and $\|Uf\|=\|f\|$. By Example~\ref{examp:2q_multiplication}, $\mathrm{d}\Gamma(\omega)$ acts on the $n$-particle sector as the multiplication by $\omega_n(k_1,\dots,k_n)=\sum_{j=1}^n\omega(k_j)$, whence
		\begin{align}
			\Dom\mathrm{d}\Gamma(\omega)^{1/2}&=\left\{\psi\in\Gamma^+(\sph):\sum_{n\in\mathbb{N}}\int\omega_n|\psi_n|^2\;\mathrm{d}^n\mu<\infty\right\},\\
			\left\|\mathrm{d}\Gamma(\omega)^{1/2}\psi\right\|^2&=\sum_{n\in\mathbb{N}}\int\omega_n|\psi_n|^2\;\mathrm{d}^n\mu.
		\end{align}
		
		\medskip
		\emph{Step 1}: finite-particle vectors. Let $\mathcal{D}_{\rm fin}$ be the subspace of all $\psi\in\Dom\mathrm{d}\Gamma(\omega)^{1/2}$ such that $\psi_n=0$ for all $n$ large enough. Clearly $\mathcal{D}_{\rm fin}\subset\Dom\N\cap\Gamma^+(\sph)$, so $a_+(f)$ and $a_+^\dag(f)$ are defined on it. Let $\psi\in\mathcal{D}_{\rm fin}$ and $n\geq1$. By Proposition~\ref{prop:a(f)explicit}, for almost all $(k_1,\dots,k_{n-1})\in \mathcal{M}^{n-1}$,
		\begin{align}
			\left(a_+(f)\psi_n\right)(k_1,\dots,k_{n-1})&=\sqrt{n}\int\overline{f(k_n)}\,\psi_n(k_1,\dots,k_{n-1},k_n)\;\mathrm{d}\mu(k_n)\nonumber\\
			&=\sqrt{n}\int\overline{\omega(k_n)^{-1/2}f(k_n)}\,\omega(k_n)^{1/2}\psi_n(k_1,\dots,k_{n-1},k_n)\;\mathrm{d}\mu(k_n),
		\end{align}
		whence, by the Cauchy--Schwarz inequality,
		\begin{equation}
			\left|\left(a_+(f)\psi_n\right)(k_1,\dots,k_{n-1})\right|^2\leq n\left\|\omega^{-1/2}f\right\|^2\int\omega(k_n)\left|\psi_n(k_1,\dots,k_{n-1},k_n)\right|^2\mathrm{d}\mu(k_n).
		\end{equation}
		Integrating over $(k_1,\dots,k_{n-1})$, we get
		\begin{equation}\label{eq:a(f)domega_step1}
			\left\|a_+(f)\psi_n\right\|^2\leq n\left\|\omega^{-1/2}f\right\|^2\int\omega(k_n)\left|\psi_n(k_1,\dots,k_n)\right|^2\mathrm{d}^n\mu.
		\end{equation}
		Now we use the symmetry of $\psi_n$: as $|\psi_n|^2$ is invariant under permutations of its variables, the integral on the right-hand side does not change if $\omega(k_n)$ is replaced by $\omega(k_j)$, for any $j=1,\dots,n$. Therefore
		\begin{equation}
			n\int\omega(k_n)|\psi_n|^2\;\mathrm{d}^n\mu=\sum_{j=1}^n\int\omega(k_j)|\psi_n|^2\;\mathrm{d}^n\mu=\int\omega_n|\psi_n|^2\;\mathrm{d}^n\mu.
		\end{equation}
		Since $a_+(f)$ maps the $n$-particle sector into the $(n-1)$-particle one, summing Eq.~\eqref{eq:a(f)domega_step1} over $n$ yields
		\begin{equation}
			\left\|a_+(f)\psi\right\|^2=\sum_{n\geq1}\left\|a_+(f)\psi_n\right\|^2\leq\left\|\omega^{-1/2}f\right\|^2\sum_{n\in\mathbb{N}}\int\omega_n|\psi_n|^2\;\mathrm{d}^n\mu=\left\|\omega^{-1/2}f\right\|^2\left\|\mathrm{d}\Gamma(\omega)^{1/2}\psi\right\|^2.
		\end{equation}
		As for the creation operator: since $\psi\in\Dom\N\cap\Gamma^+(\sph)$, by Lemma~\ref{lem:domn} both $a_+(f)\psi$ and $a^\dag_+(f)\psi$ belong to $\Dom\N^{1/2}$, so that, by Proposition~\ref{prop:apm(f)prop}(iii) and the CCR (Proposition~\ref{prop:ccr/car}),
		\begin{equation}
			\left\|a_+^\dag(f)\psi\right\|^2=\Braket{\psi,a_+(f)a_+^\dag(f)\psi}=\Braket{\psi,\left(a_+^\dag(f)a_+(f)+\lVert f\rVert^2\right)\psi}=\left\|a_+(f)\psi\right\|^2+\|f\|^2\|\psi\|^2,
		\end{equation}
		whence $\|a_+^\dag(f)\psi\|\leq\|a_+(f)\psi\|+\|f\|\|\psi\|$, and the second inequality in Eq.~\eqref{eq:a(f)domega} follows from the first one. Finally, (ii) holds on $\mathcal{D}_{\rm fin}$ by Proposition~\ref{prop:apm(f)prop}(iii).
		
		\medskip
		\emph{Step 2}: approximation. Let $\psi\in\Dom\mathrm{d}\Gamma(\omega)^{1/2}$, and let $\psi^{(N)}:=(\psi_0,\psi_1,\dots,\psi_N,0,0,\dots)\in\mathcal{D}_{\rm fin}$. Then $\psi^{(N)}\to\psi$ and $\mathrm{d}\Gamma(\omega)^{1/2}\psi^{(N)}\to\mathrm{d}\Gamma(\omega)^{1/2}\psi$ as $N\to\infty$, since
		\begin{equation}
			\left\|\psi-\psi^{(N)}\right\|^2=\sum_{n>N}\|\psi_n\|^2,\qquad\left\|\mathrm{d}\Gamma(\omega)^{1/2}\left(\psi-\psi^{(N)}\right)\right\|^2=\sum_{n>N}\int\omega_n|\psi_n|^2\;\mathrm{d}^n\mu
		\end{equation}
		are tails of convergent series. Applying Step 1 to $\psi^{(N)}-\psi^{(M)}\in\mathcal{D}_{\rm fin}$,
		\begin{equation}
			\left\|a_+(f)\left(\psi^{(N)}-\psi^{(M)}\right)\right\|\leq\left\|\omega^{-1/2}f\right\|\left\|\mathrm{d}\Gamma(\omega)^{1/2}\left(\psi^{(N)}-\psi^{(M)}\right)\right\|,
		\end{equation}
		so $(a_+(f)\psi^{(N)})_{N\in\mathbb{N}}$ is a Cauchy sequence, and thus converges. As $a_+(f)$ is closed (Proposition~\ref{prop:apm(f)prop}(i)), we conclude that $\psi\in\Dom a_+(f)$ and $a_+(f)\psi=\lim_Na_+(f)\psi^{(N)}$; taking the limit $N\to\infty$ in the inequality of Step 1 for $\psi^{(N)}$ yields the first inequality in~\eqref{eq:a(f)domega}. The very same argument applies to the closed operator $a_+^\dag(f)$. Finally, given $\psi,\varphi\in\Dom\mathrm{d}\Gamma(\omega)^{1/2}$, (ii) holds for $\psi^{(N)},\varphi^{(N)}$, and taking the limit $N\to\infty$ on both sides proves it for $\psi,\varphi$.
	\end{proof}
	
	\begin{remark}[The role of the bosonic symmetry]
		The symmetry of $\psi_n$ was crucial in Step 1, and indeed the estimate fails for the unsymmetrized operator $\a{f}$ on $\Gamma(\sph)$, which only annihilates in the last tensor slot (here $\mathrm{d}\Gamma(\omega)$ is understood on the whole $\Gamma(\sph)$). To see this, let $f\in\Dom\omega^{-1}$, take $v=\omega^{-1}f$, so that $\braket{f,v}=\braket{v,\omega v}=\|\omega^{-1/2}f\|^2$, and take $u\in\Dom\omega^{1/2}$ such that $\braket{u,\omega u}\|v\|^2<\|u\|^2\braket{v,\omega v}$ (which is possible, e.g., whenever $\braket{v,\omega v}/\|v\|^2>\inf\sigma(\omega)$). Then $\psi=u\otimes v\in\sph^{\otimes2}$ satisfies $\a{f}\psi=\sqrt{2}\braket{f,v}u$, whence
		\begin{align}
			\left\|\a{f}\psi\right\|^2&=2\left\|\omega^{-1/2}f\right\|^2\|u\|^2\braket{v,\omega v}\nonumber\\
			&>\left\|\omega^{-1/2}f\right\|^2\left(\braket{u,\omega u}\|v\|^2+\|u\|^2\braket{v,\omega v}\right)=\left\|\omega^{-1/2}f\right\|^2\left\|\mathrm{d}\Gamma(\omega)^{1/2}\psi\right\|^2.
		\end{align}
		The same proof works for the fermionic operator $a_-(f)$, since $|\psi_n|^2$ is symmetric for antisymmetric $\psi_n$ as well; but this is of little interest, since $a_-(f)$ is bounded anyway.
	\end{remark}
	
	\begin{remark}[The massive case]\label{rem:a(f)domega_massive}
		If $\omega\geq m>0$, then $\Ker\omega=\{0\}$ and $\omega^{-1/2}$ is bounded with $\|\omega^{-1/2}\|\leq m^{-1/2}$, so every $f\in\sph$ satisfies the assumptions of Proposition~\ref{prop:a(f)domega}, with $\|\omega^{-1/2}f\|\leq m^{-1/2}\|f\|$. In this case, a weaker version of Proposition~\ref{prop:a(f)domega} follows directly from the estimates in terms of $\N$, with no need for the explicit computation of Step 1 and the approximation argument of Step 2. Indeed, by Proposition~\ref{prop:domega_bb}, $\Dom\mathrm{d}\Gamma(\omega)^{1/2}\subseteq\Dom\N^{1/2}$ with $\|\N^{1/2}\psi\|\leq m^{-1/2}\|\mathrm{d}\Gamma(\omega)^{1/2}\psi\|$; therefore, by Proposition~\ref{prop:apm(f)prop}, $\Dom\mathrm{d}\Gamma(\omega)^{1/2}\subseteq\Dom a_+(f)\cap\Dom a_+^\dag(f)$, property (ii) holds, and, using $\sqrt{n+1}\leq\sqrt{n}+1$,
		\begin{align}
			\left\|a_+(f)\psi\right\|&\leq\|f\|\left\|\N^{1/2}\psi\right\|\leq m^{-1/2}\|f\|\left\|\mathrm{d}\Gamma(\omega)^{1/2}\psi\right\|,\\
			\left\|a_+^\dag(f)\psi\right\|&\leq\|f\|\left\|(\N+1)^{1/2}\psi\right\|\leq m^{-1/2}\|f\|\left\|\mathrm{d}\Gamma(\omega)^{1/2}\psi\right\|+\|f\|\|\psi\|.
		\end{align}
		These are the estimates~\eqref{eq:a(f)domega} with $\|\omega^{-1/2}f\|$ replaced by the larger constant $m^{-1/2}\|f\|$. The two constants can be very different: $\|\omega^{-1/2}f\|$ is much smaller than $m^{-1/2}\|f\|$ whenever $f$ is concentrated on modes with energy much larger than $m$.
	\end{remark}
	
	We can now prove the self-adjointness of spin--boson models.
	
	\begin{proposition}\label{prop:sb_sa}
		Assume $\omega\geq0$, $\Ker\omega=\{0\}$, and $f\in\Dom\omega^{-1/2}$. The operator $H$ is well-defined and self-adjoint.
	\end{proposition}
	
	\begin{proof}
		The operator $H_{\rm free}:=\frac{\omega_0}{2}\sigma_z\otimes I+I\otimes\mathrm{d}\Gamma(\omega)$ is clearly well-defined and self-adjoint, since $\omega$ is self-adjoint and thus (cf. Proposition~\ref{prop:domega_symm_sa}) $\mathrm{d}\Gamma(\omega)$ is self-adjoint on $\Gamma^+(\sph)$ as well. We must show that $V:=B\otimes a_+^\dag(f)+B^*\otimes a_+(f)$ is a symmetric operator with $\Dom V\supseteq\Dom H_{\rm free}$, infinitesimally bounded with respect to $H_{\rm free}$.
		
		By Proposition~\ref{prop:a(f)domega}(i), $a_+(f)$ and $a_+^\dag(f)$ are relatively bounded with respect to $\mathrm{d}\Gamma(\omega)^{1/2}$, and thus, by Lemma~\ref{lem:reltensor}, $I\otimes a_+(f)$ and $I\otimes a_+^\dag(f)$ are relatively bounded with respect to $I\otimes\mathrm{d}\Gamma(\omega)^{1/2}$. Following the same computations as in the single-mode case (cf. Proposition~\ref{prop:visrelbound}), for all $\Psi\in\mathbb{C}^2\otimes\Dom\mathrm{d}\Gamma(\omega)^{1/2}$ we have
		\begin{equation}
			\|V\Psi\|\leq\|B\|\left(\left\|(I\otimes a_+^\dag(f))\Psi\right\|+\left\|(I\otimes a_+(f))\Psi\right\|\right),
		\end{equation}
		so that $V$ is relatively bounded with respect to $I\otimes\mathrm{d}\Gamma(\omega)^{1/2}=(I\otimes\mathrm{d}\Gamma(\omega))^{1/2}$, which in turn is infinitesimally bounded with respect to $I\otimes\mathrm{d}\Gamma(\omega)$ (recall that every symmetric operator is infinitesimally bounded with respect to its square, cf. Example~\ref{examp:wrtsquare}). Hence $V$ is infinitesimally bounded with respect to $I\otimes\mathrm{d}\Gamma(\omega)$, and thus with respect to $H_{\rm free}$, which only differs from $I\otimes\mathrm{d}\Gamma(\omega)$ by a bounded term. Besides, by Proposition~\ref{prop:a(f)domega}(ii), $V$ is symmetric on $\mathbb{C}^2\otimes\Dom\mathrm{d}\Gamma(\omega)^{1/2}\supseteq\Dom H_{\rm free}$.
		
		Applying the Kato--Rellich theorem (Theorem~\ref{thm:katorellich}) completes the proof.
	\end{proof}
	
	\begin{remark}[The massive case]
		If $\omega\geq m>0$, the assumptions of Proposition~\ref{prop:sb_sa} are satisfied by every $f\in\sph$, and the proof can be carried out by only using the estimates in terms of $\N$ (cf. Remark~\ref{rem:a(f)domega_massive}):
		\begin{enumerate}
			\item[(i)] By Proposition~\ref{prop:domega_bb}, $\mathrm{d}\Gamma(\omega)$ is a nonnegative operator with $\Dom\mathrm{d}\Gamma(\omega)\subseteq\Dom\mathrm{d}\Gamma(\omega)^{1/2}\subseteq\Dom\N^{1/2}$ and, for all $\psi\in\Dom\mathrm{d}\Gamma(\omega)^{1/2}$,
			\begin{equation}
				\left\|\mathrm{d}\Gamma(\omega)^{1/2}\psi\right\|\geq\sqrt{m}\left\|\N^{1/2}\psi\right\|;
			\end{equation}
			\item[(ii)] By Proposition~\ref{prop:apm(f)prop}, $\Dom\N^{1/2}\subseteq\Dom a_+(f),\Dom a_+^\dag(f)$ and, for all $\psi\in\Dom\N^{1/2}$,
			\begin{equation}
				\left\|a_+(f)\psi\right\|\leq\|f\|\left\|\N^{1/2}\psi\right\|\leq\|f\|\left\|(\N+1)^{1/2}\psi\right\|,\qquad \left\|a_+^\dag(f)\psi\right\|\leq\|f\|\left\|(\N+1)^{1/2}\psi\right\|.
			\end{equation}
		\end{enumerate}
		By (ii), exactly as in the single-mode case, $V$ is relatively bounded with respect to $I\otimes\N^{1/2}$; by (i), $I\otimes\N^{1/2}$ is relatively bounded with respect to $I\otimes\mathrm{d}\Gamma(\omega)^{1/2}$, and the proof is concluded as above.
	\end{remark}
	
	\begin{remark}\label{rem:uvir}
		Note that we had to impose two conditions on the form factor: $f\in\sph$, which is needed to define $a_+(f),a_+^\dag(f)$ in the first place, and $\omega^{-1/2}f\in\sph$. If $\sph=L^2(\mathcal{M},\Sigma,\mu)$, they read
		\begin{equation}
			\int|f(k)|^2\;\mathrm{d}\mu(k)<\infty,\qquad\int\frac{|f(k)|^2}{\omega(k)}\;\mathrm{d}\mu(k)<\infty.
		\end{equation}
		Given the first condition, the second one only constrains the behavior of $f$ on the modes with small energy $\omega(k)$: it is an \emph{infrared condition}\index{infrared condition}, automatically satisfied by massive fields, and only relevant for massless ones. For instance, for a scalar massless field, with $\omega(k)=|k|$ on $\mathbb{R}^3$, it requires $f$ not to be too singular as $k\to0$. The first condition, instead, also constrains the high-energy modes, and is not implied by the second one when $\omega$ is unbounded. When it fails, the model of Definition~\ref{def:spinboson}---as well as more complex field--matter models---exhibits \emph{UV divergences}\index{UV divergence}, and the construction of Proposition~\ref{prop:sb_sa} breaks down. We will see in Section~\ref{sec:singular} how to go beyond it.
	\end{remark}
	
	\subsection{The multimode pure dephasing model}\label{subsec:multidephasing}\index{spin--boson model!pure dephasing}
	
	The two exactly solvable single-mode models of Section~\ref{sec:singlemodesb} have natural multimode counterparts, obtained by choosing $B=\sigma_z$ and $B=\sigma_-$ in Definition~\ref{def:spinboson} (the coupling constant being absorbed in $f$). Their solution follows the same strategy as in the single-mode case, and reveals a genuinely new phenomenon: for a continuum of modes, the loss of coherence (this section) and the decay of the excited state (Section~\ref{subsec:multijc}) become irreversible. In both cases, as in Section~\ref{subsec:multimodesb}, we take $\sph=L^2(\mathcal{M},\Sigma,\mu)$, $\omega$ the multiplication by a measurable function with $\omega(k)>0$ for almost all $k$, and $f\in\Dom\omega^{-1/2}$, so that the models are self-adjoint by Proposition~\ref{prop:sb_sa}. It is convenient to introduce the \emph{spectral density}\index{spectral density} of the coupling, i.e. the measure $J$ on $(0,\infty)$ given by
	\begin{equation}
		J(E):=\int_{\{k:\,\omega(k)\in E\}}|f(k)|^2\;\mathrm{d}\mu(k)\qquad(E\subseteq(0,\infty)\text{ Borel}),
	\end{equation}
	so that $\int\eta(\omega(k))|f(k)|^2\,\mathrm{d}\mu(k)=\int\eta(\lambda)\,\mathrm{d}J(\lambda)$ for every nonnegative or bounded measurable $\eta$. For a single mode of frequency $\omega$ and coupling $g$, $J=g^2\delta_\omega$ is a point mass.
	
	To solve the pure dephasing model, we need the multimode counterpart of the Weyl operators of Section~\ref{subsec:dephasing}. We proceed in parallel with the single-mode case, starting from the counterpart of Lemma~\ref{lem:predispl}.
	
	\begin{lemma}\label{lem:predisplmulti}
		Let $h\in\sph$. The operator $\ii(a_+^\dag(h)-a_+(h))$, with domain $\Gamma^+_{\rm fin}(\sph)$, is essentially self-adjoint. Besides, $\Dom\N^{1/2}\cap\Gamma^+(\sph)$ is contained in the domain of its closure, which acts on it as $\ii(a_+^\dag(h)-a_+(h))$.
	\end{lemma}
	
	\begin{proof}
		The operator is symmetric by Proposition~\ref{prop:apm(f)prop}(iii), and maps $\Gamma^+_{\rm fin}(\sph)$ into itself. By Proposition~\ref{prop:apm(f)prop}(ii), $\|a_+^\sharp(h)\psi_n\|\leq\|h\|\sqrt{n+1}\,\|\psi_n\|$ for every $n$-particle vector $\psi_n$, with $a_+^\sharp\in\{a_+,a_+^\dag\}$. Hence, applying the operator $m$ times to $\psi_n$ yields a sum of $2^m$ vectors, each with norm at most $\|h\|^m\sqrt{(n+1)\cdots(n+m)}\,\|\psi_n\|$, and thus, for all $t>0$,
		\begin{equation}
			\sum_{m\in\mathbb{N}}\frac{t^m}{m!}\left\|\left(\ii(a_+^\dag(h)-a_+(h))\right)^m\psi_n\right\|\leq\|\psi_n\|\sum_{m\in\mathbb{N}}\frac{(2t\|h\|)^m}{m!}\sqrt{\frac{(n+m)!}{n!}}<\infty
		\end{equation}
		by the ratio test. That is, finite-particle vectors are analytic vectors for $\ii(a_+^\dag(h)-a_+(h))$, which is therefore essentially self-adjoint on them by Nelson's analytic vector theorem (cf. Remark~\ref{rem:quadratic}). Finally, given $\psi\in\Dom\N^{1/2}\cap\Gamma^+(\sph)$, its truncations $\psi^{(N)}:=(\psi_0,\dots,\psi_N,0,0,\dots)\in\Gamma^+_{\rm fin}(\sph)$ converge to $\psi$ in the graph norm of $\N^{1/2}$; by Proposition~\ref{prop:apm(f)prop}(ii), $\ii(a_+^\dag(h)-a_+(h))\psi^{(N)}\to\ii(a_+^\dag(h)-a_+(h))\psi$, which proves the last claim.
	\end{proof}
	
	\begin{definition}[Weyl operator]\label{def:weylmulti}\index{Weyl displacement operator!on the Fock space}
		Let $h\in\sph$. We define the \emph{Weyl operator} $W(h)$ by
		\begin{equation}
			W(h):=\exp\left(-\ii\,\overline{\ii\left(a_+^\dag(h)-a_+(h)\right)}\right).
		\end{equation}
	\end{definition}
	
	For $\sph=\mathbb{C}$ and $h=\lambda\in\mathbb{R}$, this is the operator $W(\lambda)$ of Definition~\ref{def:weyl} (cf. Example~\ref{examp:singlemodebosonic}). As in Lemma~\ref{lem:displ}, by the spectral theorem $W(h)$ is unitary, with $W(h)^{-1}=W(h)^*=W(-h)$. More generally, since $\ii(a_+^\dag(th)-a_+(th))=t\,\ii(a_+^\dag(h)-a_+(h))$ for $t\in\mathbb{R}$, the operators $W(th)$, $t\in\mathbb{R}$, form the strongly continuous one-parameter unitary group generated by $\overline{\ii(a_+^\dag(h)-a_+(h))}$ (Proposition~\ref{prop:unitarygroup}).
	
	In the single-mode case, the action of $W(\lambda)$ was made explicit in the Hermite representation, where $W(\lambda)$ is a translation (Lemma~\ref{lem:weyltranslation}). This representation has no direct multimode counterpart; instead, we will make the action of $W(h)$ explicit on a convenient family of vectors.
	
	\begin{definition}[Exponential vectors and coherent states]\label{def:coherentmulti}\index{exponential vectors}\index{coherent states}
		Let $f\in\sph$. The \emph{exponential vector} associated with $f$ is
		\begin{equation}
			\varepsilon(f):=\left(1,f,\frac{f^{\otimes 2}}{\sqrt{2!}},\frac{f^{\otimes 3}}{\sqrt{3!}},\ldots\right)\in\Gamma^+(\sph),
		\end{equation}
		and the \emph{coherent state} associated with $f$ is $\psi_f:=\e^{-\|f\|^2/2}\varepsilon(f)$. Given a subspace $\mathcal{V}\subseteq\sph$, we denote by $\operatorname{Exp}(\mathcal{V}):=\operatorname{Span}\{\varepsilon(f):f\in\mathcal{V}\}$ the span of the corresponding exponential vectors.
	\end{definition}
	
	Note that $\varepsilon(0)=\varepsilon_0$ is the vacuum.
	
	\begin{lemma}\label{lem:expvectors}
		Let $f,g\in\sph$. Then:
		\begin{itemize}
			\item[(i)] $\varepsilon(f)\in\Dom\N^s$ for all $s\geq0$, and $\braket{\varepsilon(f),\varepsilon(g)}=\e^{\braket{f,g}}$;
			\item[(ii)] $a_+(g)\varepsilon(f)=\braket{g,f}\varepsilon(f)$; besides, the function $s\in\mathbb{R}\mapsto\varepsilon(f+sg)\in\Gamma^+(\sph)$ is differentiable, with
			\begin{equation}
				\frac{\mathrm{d}}{\mathrm{d}s}\varepsilon(f+sg)=a_+^\dag(g)\varepsilon(f+sg);
			\end{equation}
			\item[(iii)] $\Gamma(U)\varepsilon(f)=\varepsilon(Uf)$ for every unitary operator $U$ on $\sph$; besides, if $A$ is a self-adjoint operator on $\sph$ and $f\in\Dom A$, then $\varepsilon(f)\in\Dom\mathrm{d}\Gamma(A)$ and $\mathrm{d}\Gamma(A)\varepsilon(f)=a_+^\dag(Af)\varepsilon(f)$;
			\item[(iv)] if $\mathcal{V}\subseteq\sph$ is a dense subspace, then $\operatorname{Exp}(\mathcal{V})$ is dense in $\Gamma^+(\sph)$, and is a core for $a_+(g)$ and $a_+^\dag(g)$; besides, if $A$ is a self-adjoint operator on $\sph$, then $\operatorname{Exp}(\Dom A)$ is a core for $\mathrm{d}\Gamma(A)$.
		\end{itemize}
	\end{lemma}
	
	\begin{proof}
		(i) Since $\|f^{\otimes n}\|=\|f\|^n$, we have $\sum_nn^{2s}\|f\|^{2n}/n!<\infty$, so that $\varepsilon(f)\in\Dom\N^s$ (Example~\ref{examp:number_fock}); and, since $\braket{f^{\otimes n},g^{\otimes n}}=\braket{f,g}^n$, we have $\braket{\varepsilon(f),\varepsilon(g)}=\sum_n\braket{f,g}^n/n!=\e^{\braket{f,g}}$.
		
		(ii) and (iii) are left as an exercise (Exercise~\ref{exerc:expvectors} below).
		
		(iv) We first show that, given $f\in\sph$ and $n\in\mathbb{N}$, the vector $f^{\otimes n}$ can be approximated by linear combinations of the exponential vectors $\varepsilon(zf)$, $z\in\mathbb{C}$, in a strong sense. Let $M:=n+1$, $\zeta:=\e^{2\pi\ii/M}$ and, for $0<r\leq1$,
		\begin{equation}
			\Psi_r:=\frac{\sqrt{n!}}{Mr^n}\sum_{j=0}^{M-1}\zeta^{-jn}\varepsilon(r\zeta^jf).
		\end{equation}
		Since $\frac{1}{M}\sum_{j=0}^{M-1}\zeta^{j(m-n)}$ equals $1$ if $m-n$ is a multiple of $M$, and $0$ otherwise, only the $m$-particle components with $m=n+lM$, $l\in\mathbb{N}$, survive in $\Psi_r$:
		\begin{equation}
			\Psi_r-f^{\otimes n}=\sum_{l\geq1}r^{lM}\sqrt{\frac{n!}{(n+lM)!}}\,f^{\otimes(n+lM)}.
		\end{equation}
		Consequently, $\|(\N+1)^{1/2}(\Psi_r-f^{\otimes n})\|^2\leq r^{2M}\sum_{m}(m+1)\,n!\,\|f\|^{2m}/m!\to0$ as $r\to0$. Similarly, if $f\in\Dom A$, then $\mathrm{d}\Gamma(A)f^{\otimes m}=\sum_{j=1}^mf^{\otimes(j-1)}\otimes Af\otimes f^{\otimes(m-j)}$ has norm at most $m\|Af\|\|f\|^{m-1}$, whence $\|\mathrm{d}\Gamma(A)(\Psi_r-f^{\otimes n})\|\to0$ as well.
		
		Now let $\mathcal{V}\subseteq\sph$ be a dense subspace. By the polarization identity\index{polarization identity} (Exercise~\ref{exerc:polarization} below)
		\begin{equation}\label{eq:polarization}
			S_n^+(\phi_1\otimes\cdots\otimes\phi_n)=\frac{1}{2^nn!}\sum_{s_1,\dots,s_n=\pm1}s_1\cdots s_n\left(s_1\phi_1+\ldots+s_n\phi_n\right)^{\otimes n},
		\end{equation}
		the span of the vectors $f^{\otimes n}$ with $f\in\mathcal{V}$ is $S_n^+\mathcal{V}^{\odot n}$, which is dense in $S_n^+\sph^{\otimes n}$, since $\mathcal{V}^{\odot n}$ is dense in $\sph^{\otimes n}$ (Exercise~\ref{exerc:denseindense}) and $S_n^+$ is bounded. As $\N$ is bounded on each $n$-particle sector, by the previous step every vector of $\Gamma^+_{\rm fin}(\sph)$ is the limit of a sequence in $\operatorname{Exp}(\mathcal{V})$ in the graph norm of $(\N+1)^{1/2}$. In particular, $\operatorname{Exp}(\mathcal{V})$ is dense. Moreover, as $a_+^\sharp(g)$ is relatively bounded with respect to $(\N+1)^{1/2}$ (Proposition~\ref{prop:apm(f)prop}), the closure of the restriction of $a_+^\sharp(g)$ to $\operatorname{Exp}(\mathcal{V})$ extends its restriction to $\Gamma^+_{\rm fin}(\sph)$, which is a core for $a_+^\sharp(g)$. Indeed, let $P_N$ be the orthogonal projection onto the sectors with at most $N$ particles; since $a_+(g)$ is the closure of an operator mapping $n$-particle vectors into $(n-1)$-particle ones, one readily checks that $a_+(g)P_N\psi=P_{N-1}a_+(g)\psi$ for all $\psi\in\Dom a_+(g)$, whence $P_N\psi\to\psi$ in the graph norm of $a_+(g)$; the same holds for $a_+^\dag(g)$, with $P_{N+1}$ in place of $P_{N-1}$. Hence $\operatorname{Exp}(\mathcal{V})$ is a core for $a_+^\sharp(g)$.
		
		Finally, let $A$ be self-adjoint. By Proposition~\ref{prop:props_2q}, the space $\mathcal{D}$ of Eq.~\eqref{eq:coreD} is a core for $\mathrm{d}\Gamma(A)$; since $S^+$ commutes with $\mathrm{d}\Gamma(A)$ (Proposition~\ref{prop:invariant}), $S^+\mathcal{D}$ is a core for its restriction to $\Gamma^+(\sph)$. By Eq.~\eqref{eq:polarization}, $S^+\mathcal{D}$ is spanned by the vectors $f^{\otimes n}$ with $f\in\Dom A$ and $n\in\mathbb{N}$, each of which, by the first step of the proof, is the limit of a sequence in $\operatorname{Exp}(\Dom A)$ in the graph norm of $\mathrm{d}\Gamma(A)$. This concludes the proof.
	\end{proof}
	
	\begin{exercise}\label{exerc:expvectors}
		Prove points (ii) and (iii) of Lemma~\ref{lem:expvectors}. Hint: by Definitions~\ref{def:a0(f)} and~\ref{def:apm}, $a_+(g)f^{\otimes n}=\sqrt{n}\braket{g,f}f^{\otimes(n-1)}$ and $a_+^\dag(g)f^{\otimes(n-1)}=\frac{1}{\sqrt{n}}\sum_{j=1}^nf^{\otimes(j-1)}\otimes g\otimes f^{\otimes(n-j)}$, the latter sum being the derivative of $s\mapsto(f+sg)^{\otimes n}$ at $s=0$. To prove the differentiability of $s\mapsto\varepsilon(f+sg)$, estimate the remainder of the first-order Taylor expansion of $(f+sg)^{\otimes n}$ sector by sector.
		\seesol{expvectors}
	\end{exercise}
	
	\begin{exercise}\label{exerc:polarization}
		Prove the polarization identity~\eqref{eq:polarization}. Hint: expand $(s_1\phi_1+\ldots+s_n\phi_n)^{\otimes n}$ by multilinearity, and show that, after summing over $s_1,\dots,s_n=\pm1$, only the terms in which each $\phi_j$ appears exactly once survive.
		\seesol{polarization}
	\end{exercise}
	
	By Lemma~\ref{lem:expvectors}, the coherent state $\psi_h$ is normalized and, for all $g\in\sph$, $a_+(g)\psi_h=\braket{g,h}\psi_h$: as in the single-mode case, coherent states are common eigenvectors of all annihilation operators. For $\sph=\mathbb{C}$ and $h=\lambda\in\mathbb{R}$, $\psi_\lambda$ is the coherent state $W(\lambda)\phi_0$ of Remark~\ref{rem:coherent}. We can now compute the action of Weyl operators on exponential vectors: just like $W(\lambda)$ translates wavefunctions in the Hermite representation (Lemma~\ref{lem:weyltranslation}), $W(h)$ translates exponential vectors, up to a scalar factor.
	
	\begin{lemma}\label{lem:weylexp}
		For all $f,h\in\sph$,
		\begin{equation}\label{eq:weylexp}
			W(h)\varepsilon(f)=\e^{-\|h\|^2/2-\braket{h,f}}\,\varepsilon(f+h).
		\end{equation}
		In particular, if $\mathcal{V}\subseteq\sph$ is a subspace and $h\in\mathcal{V}$, then $W(h)\operatorname{Exp}(\mathcal{V})=\operatorname{Exp}(\mathcal{V})$.
	\end{lemma}
	
	\begin{proof}
		Let $T:=\overline{\ii(a_+^\dag(h)-a_+(h))}$, so that $W(th)=\e^{-\ii tT}$ for $t\in\mathbb{R}$, and let
		\begin{equation}
			\Psi(t):=\e^{-t^2\|h\|^2/2-t\braket{h,f}}\,\varepsilon(f+th).
		\end{equation}
		By Lemma~\ref{lem:expvectors}(i), $\Psi(t)\in\Dom\N^{1/2}$; thus, by Lemma~\ref{lem:predisplmulti}, $\Psi(t)\in\Dom T$, and $T$ acts on it as $\ii(a_+^\dag(h)-a_+(h))$. By Lemma~\ref{lem:expvectors}(ii) and the product rule, $\Psi(t)$ is differentiable, with
		\begin{equation}
			\frac{\mathrm{d}}{\mathrm{d}t}\Psi(t)=\left(a_+^\dag(h)-\braket{h,f+th}\right)\Psi(t)=\left(a_+^\dag(h)-a_+(h)\right)\Psi(t)=-\ii T\Psi(t).
		\end{equation}
		That is, $\Psi(t)$ solves the Schr\"odinger equation generated by $T$ with initial condition $\Psi(0)=\varepsilon(f)$, whose unique solution is $\e^{-\ii tT}\varepsilon(f)$ (cf. the proof of Proposition~\ref{prop:gammaexp}; indeed, the derivative of $\e^{\ii tT}\Psi(t)$ vanishes). Setting $t=1$ yields Eq.~\eqref{eq:weylexp}. The last claim follows since $\varepsilon(f+h)\in\operatorname{Exp}(\mathcal{V})$ for all $f\in\mathcal{V}$, and $W(h)^{-1}=W(-h)$.
	\end{proof}
	
	The following properties generalize Lemma~\ref{lem:displ}, Proposition~\ref{prop:displacement} and Remark~\ref{rem:coherent} (see also~\cite[Section X.7]{reedsimon2} and~\cite{derezinskigerard,arai}).
	
	\begin{proposition}\label{prop:weylmulti}
		For all $g,h\in\sph$:
		\begin{itemize}
			\item[(i)] $W(h)$ is unitary, with $W(h)^*=W(-h)$, and $W(g)W(h)=\e^{-\ii\Im\braket{g,h}}W(g+h)$\index{Weyl displacement operator!Weyl relations};
			\item[(ii)] $W(h)a_+(g)W(h)^*=a_+(g)-\braket{g,h}$ and $W(h)a_+^\dag(g)W(h)^*=a_+^\dag(g)-\braket{h,g}$;
			\item[(iii)] $W(h)\varepsilon_0$ is the coherent state $\psi_h$ of Definition~\ref{def:coherentmulti}; in particular, $\braket{\varepsilon_0,W(h)\varepsilon_0}=\e^{-\|h\|^2/2}$;
			\item[(iv)] $\Gamma(U)W(h)\Gamma(U)^*=W(Uh)$ for every unitary $U$ on $\sph$; in particular, by Proposition~\ref{prop:gammaexp}, $\e^{-\ii t\mathrm{d}\Gamma(\omega)}W(h)\e^{\ii t\mathrm{d}\Gamma(\omega)}=W(\e^{-\ii t\omega}h)$.
		\end{itemize}
	\end{proposition}
	
	\begin{proof}
		(i) Unitarity and $W(h)^*=W(-h)$ were discussed above. Given $f\in\sph$, by Eq.~\eqref{eq:weylexp} applied twice,
		\begin{equation}
			W(g)W(h)\varepsilon(f)=\e^{-\|h\|^2/2-\braket{h,f}}\e^{-\|g\|^2/2-\braket{g,f+h}}\varepsilon(f+g+h)=\e^{\frac{1}{2}(\braket{h,g}-\braket{g,h})}W(g+h)\varepsilon(f),
		\end{equation}
		since $\|g+h\|^2=\|g\|^2+\|h\|^2+\braket{g,h}+\braket{h,g}$; besides, $\frac{1}{2}(\braket{h,g}-\braket{g,h})=-\ii\Im\braket{g,h}$. Both sides being bounded operators, the equality extends from $\operatorname{Exp}(\sph)$, which is dense by Lemma~\ref{lem:expvectors}(iv), to the whole $\Gamma^+(\sph)$.
		
		(ii) Let $f\in\sph$ and $c:=\e^{-\|h\|^2/2-\braket{h,f}}$. By Eq.~\eqref{eq:weylexp} and Lemma~\ref{lem:expvectors}(ii),
		\begin{equation}
			a_+(g)W(h)\varepsilon(f)=c\braket{g,f+h}\varepsilon(f+h)=W(h)\left(a_+(g)+\braket{g,h}\right)\varepsilon(f).
		\end{equation}
		Besides, differentiating the identity $W(h)\varepsilon(f+sg)=c\,\e^{-s\braket{h,g}}\varepsilon(f+h+sg)$ at $s=0$, and using that the bounded operator $W(h)$ commutes with derivatives, by Lemma~\ref{lem:expvectors}(ii) we get
		\begin{equation}
			W(h)a_+^\dag(g)\varepsilon(f)=c\left(a_+^\dag(g)-\braket{h,g}\right)\varepsilon(f+h)=\left(a_+^\dag(g)-\braket{h,g}\right)W(h)\varepsilon(f).
		\end{equation}
		By linearity, and since $W(h)^*\operatorname{Exp}(\sph)=\operatorname{Exp}(\sph)$ by Lemma~\ref{lem:weylexp}, the two equalities in (ii) hold on $\operatorname{Exp}(\sph)$. As in the proof of Proposition~\ref{prop:displacement}, we now take closures on both sides. As $\operatorname{Exp}(\sph)$ is a core for $a_+^\sharp(g)$ (Lemma~\ref{lem:expvectors}(iv)), the closures of the right-hand sides are $a_+(g)-\braket{g,h}$ and $a_+^\dag(g)-\braket{h,g}$; for the same reason, and since $W(h)^*\operatorname{Exp}(\sph)=\operatorname{Exp}(\sph)$, the closures of the left-hand sides are $W(h)a_+(g)W(h)^*$ and $W(h)a_+^\dag(g)W(h)^*$, since conjugation by a unitary operator preserves graph limits.
		
		(iii) By Eq.~\eqref{eq:weylexp} with $f=0$, $W(h)\varepsilon_0=\e^{-\|h\|^2/2}\varepsilon(h)=\psi_h$; thus, by Lemma~\ref{lem:expvectors}(i), $\braket{\varepsilon_0,W(h)\varepsilon_0}=\e^{-\|h\|^2/2}\braket{\varepsilon(0),\varepsilon(h)}=\e^{-\|h\|^2/2}$.
		
		(iv) is left as an exercise (Exercise~\ref{exerc:weylcov} below).
	\end{proof}
	
	\begin{exercise}\label{exerc:weylcov}
		Prove Proposition~\ref{prop:weylmulti}(iv). Hint: use Lemma~\ref{lem:expvectors}(iii) and Eq.~\eqref{eq:weylexp} to show that both sides act in the same way on exponential vectors.
		\seesol{weylcov}
	\end{exercise}
	
	Let now $B=\sigma_z$. Exactly as in Section~\ref{subsec:dephasing}, the model decomposes as $H\cong H_+\oplus H_-$, with
	\begin{equation}
		H_\pm=\mathrm{d}\Gamma(\omega)\pm\left(a_+^\dag(f)+a_+(f)+\frac{\omega_0}{2}\right)
	\end{equation}
	on $\Gamma^+(\sph)$, with domain $\Dom\mathrm{d}\Gamma(\omega)$; since $H$ is self-adjoint, so are $H_\pm$ (Proposition~\ref{prop:adjoint_directsum}). The reduced dynamics of the qubit is again a dephasing channel, with dephasing function $\varphi(t)=\braket{\Omega,\e^{\ii tH_-}\e^{-\ii tH_+}\Omega}$: the derivation in Section~\ref{subsec:dephasing} only used the decomposition $H\cong H_+\oplus H_-$. Each $H_\pm$ is a \emph{van Hove Hamiltonian}\index{van Hove model}~\cite{derezinski2003}, which can be diagonalized by completing the square as in Proposition~\ref{prop:weyldisplaced}:
	
	\begin{proposition}\label{prop:multidephasing}
		Assume, in addition, $f\in\Dom\omega^{-1}$, and let $W_\pm:=W(\pm\omega^{-1}f)$. Then
		\begin{equation}
			W_\pm H_\pm W_\pm^*=\mathrm{d}\Gamma(\omega)-\left\|\omega^{-1/2}f\right\|^2\pm\frac{\omega_0}{2}.
		\end{equation}
	\end{proposition}
	
	\begin{proof}
		As in the single-mode case (Exercise~\ref{exerc:weyl}), we complete the square by means of Proposition~\ref{prop:weylmulti}(ii); we only need some care with domains. Let $h:=\pm\omega^{-1}f$, so that $W_\pm=W(h)$, and $E_f:=\|\omega^{-1/2}f\|^2$. Note that $h\in\Ran\omega^{-1}=\Dom\omega$, with $\omega h=\pm f$ and $\braket{f,h}=\pm E_f$. The space $\operatorname{Exp}(\Dom\omega)$ is a core for $\mathrm{d}\Gamma(\omega)$ by Lemma~\ref{lem:expvectors}(iv), and is mapped onto itself by $W_\pm^*=W(-h)$ by Lemma~\ref{lem:weylexp}; in particular, $W_\pm^*\operatorname{Exp}(\Dom\omega)\subseteq\Dom\mathrm{d}\Gamma(\omega)=\Dom H_\pm$.
		
		Let $k\in\Dom\omega$ and $k':=k-h\in\Dom\omega$, so that $W_\pm^*\varepsilon(k)$ is a multiple of $\varepsilon(k')$ by Eq.~\eqref{eq:weylexp}. By Lemma~\ref{lem:expvectors}(iii), $\mathrm{d}\Gamma(\omega)\varepsilon(k')=a_+^\dag(\omega k')\varepsilon(k')$; as $\omega k'\pm f=\omega k$,
		\begin{equation}
			H_\pm\varepsilon(k')=\left(a_+^\dag(\omega k)\pm a_+(f)\pm\frac{\omega_0}{2}\right)\varepsilon(k').
		\end{equation}
		Therefore, by Proposition~\ref{prop:weylmulti}(ii) and Lemma~\ref{lem:expvectors}(ii)--(iii),
		\begin{align}
			W_\pm H_\pm W_\pm^*\varepsilon(k)&=\left(a_+^\dag(\omega k)-\braket{h,\omega k}\pm a_+(f)\mp\braket{f,h}\pm\frac{\omega_0}{2}\right)\varepsilon(k)\nonumber\\
			&=\left(a_+^\dag(\omega k)-E_f\pm\frac{\omega_0}{2}\right)\varepsilon(k)=\left(\mathrm{d}\Gamma(\omega)-E_f\pm\frac{\omega_0}{2}\right)\varepsilon(k),
		\end{align}
		where we used $\braket{h,\omega k}=\braket{\omega h,k}=\pm\braket{f,k}$, $a_+(f)\varepsilon(k)=\braket{f,k}\varepsilon(k)$, and $\braket{f,h}=\pm E_f$. Thus, the self-adjoint operator $W_\pm H_\pm W_\pm^*$ extends the restriction of the self-adjoint operator $\mathrm{d}\Gamma(\omega)-E_f\pm\omega_0/2$ to the core $\operatorname{Exp}(\Dom\omega)$. Being closed, it extends $\mathrm{d}\Gamma(\omega)-E_f\pm\omega_0/2$ itself; since self-adjoint operators admit no proper symmetric extensions, the two operators coincide.
	\end{proof}
	
	\begin{corollary}\label{coroll:multidephasing}
		Under the assumptions of Proposition~\ref{prop:multidephasing}, the dephasing function for the vacuum $\Omega=\varepsilon_0$ is
		\begin{equation}\label{eq:multidephasing}
			\varphi(t)=\e^{-\ii\omega_0t}\exp\left(-8\int\frac{\sin^2(\lambda t/2)}{\lambda^2}\;\mathrm{d}J(\lambda)\right).
		\end{equation}
	\end{corollary}
	
	\begin{proof}
		Let $h:=\omega^{-1}f$, $h_t:=\e^{-\ii t\omega}h$ and $E_f:=\|\omega^{-1/2}f\|^2$. By Proposition~\ref{prop:multidephasing} and Proposition~\ref{prop:weylmulti}(iv), since $\e^{-\ii t\mathrm{d}\Gamma(\omega)}\varepsilon_0=\varepsilon_0$,
		\begin{equation}
			\e^{-\ii tH_\pm}\varepsilon_0=\e^{-\ii t(\pm\omega_0/2-E_f)}\,W(\mp h)\,\e^{-\ii t\mathrm{d}\Gamma(\omega)}W(\pm h)\varepsilon_0=\e^{-\ii t(\pm\omega_0/2-E_f)}\,W(\mp h)W(\pm h_t)\varepsilon_0.
		\end{equation}
		By Proposition~\ref{prop:weylmulti}(i), $W(\mp h)W(\pm h_t)=\e^{\ii\Im\braket{h,h_t}}W(\pm u_t)$, with $u_t:=h_t-h$; the phase is the same for both signs. Therefore, using (i) and (iii) once more,
		\begin{align}
			\varphi(t)&=\Braket{\e^{-\ii tH_-}\varepsilon_0,\e^{-\ii tH_+}\varepsilon_0}=\e^{-\ii\omega_0t}\Braket{W(-u_t)\varepsilon_0,W(u_t)\varepsilon_0}\nonumber\\
			&=\e^{-\ii\omega_0t}\Braket{\varepsilon_0,W(2u_t)\varepsilon_0}=\e^{-\ii\omega_0t}\e^{-2\|u_t\|^2},
		\end{align}
		and $\|u_t\|^2=\int|\e^{-\ii t\omega(k)}-1|^2\omega(k)^{-2}|f(k)|^2\,\mathrm{d}\mu(k)=4\int\lambda^{-2}\sin^2(\lambda t/2)\,\mathrm{d}J(\lambda)$.
	\end{proof}
	
	For a single mode, $J=g^2\delta_\omega$, and Eq.~\eqref{eq:multidephasing} reduces to Eq.~\eqref{eq:dephasingvacuum}.
	
	\begin{remark}[Irreversible decoherence]\label{rem:irreversible}\index{decoherence}
		The right-hand side of Eq.~\eqref{eq:multidephasing} is finite for every $f\in\sph$, since $\sin^2(\lambda t/2)/\lambda^2\leq t^2/4$. Approximating $f$ by $f_\epsilon:=f\chi_{\{\omega>\epsilon\}}\in\Dom\omega^{-1}$ and arguing as in the proof of Theorem~\ref{thm:singularsb}(iii), one shows that the formula remains valid for all $f\in\Dom\omega^{-1/2}$: that proof only uses $\|\omega^{-1/2}(f-f_\epsilon)\|\to0$ and $\|\omega^{-1/2}f_\epsilon\|\leq\|\omega^{-1/2}f\|$, which hold for this infrared cutoff as well (by dominated convergence), and the resulting norm resolvent convergence implies the strong convergence of the unitary groups (Section~\ref{subsec:spectrum}). Since $8\sin^2(\lambda t/2)=4(1-\cos\lambda t)$,
		\begin{equation}
			|\varphi(t)|=\exp\left(-4\int\frac{1-\cos\lambda t}{\lambda^2}\;\mathrm{d}J(\lambda)\right).\label{eq:absphi}
		\end{equation}
		For a single mode, $|\varphi(t)|$ is periodic, as found in Remark~\ref{rem:dephasingvacuum}. If instead $J$ is absolutely continuous, as for a continuum of modes, and $\|\omega^{-1}f\|^2=\int\lambda^{-2}\,\mathrm{d}J(\lambda)<\infty$, the Riemann--Lebesgue lemma (Appendix~\ref{app:measure}) yields
		\begin{equation}
			\lim_{t\to\infty}|\varphi(t)|=\e^{-4\|\omega^{-1}f\|^2}<1:
		\end{equation}
		coherence is partially and irreversibly lost, without revivals. If $J$ is absolutely continuous and $\|\omega^{-1}f\|=\infty$---an infrared divergence, occurring e.g. for an \emph{ohmic}\index{spectral density!ohmic} spectral density $\mathrm{d}J(\lambda)\propto\lambda\,\e^{-\lambda/\Lambda}\,\mathrm{d}\lambda$---then $|\varphi(t)|\to0$: the qubit decoheres completely~\cite{palma1996,breuerpetruccione}.
	\end{remark}
	
	\begin{remark}[Exponential decoherence and flat spectral densities]\label{rem:expdephasing}\index{spectral density!flat}\index{quantum regression}
		In Markovian descriptions of dephasing, coherences decay exponentially in time. This behavior can be obtained exactly, at all times, from Eq.~\eqref{eq:absphi} for a \emph{flat} spectral density: if $\mathrm{d}J(\lambda)=J_0\,\mathrm{d}\lambda$ on $(0,\infty)$, then, since $\int_0^\infty(1-\cos\lambda t)\lambda^{-2}\,\mathrm{d}\lambda=\pi t/2$ for $t\geq0$, one gets $|\varphi(t)|=\e^{-2\pi J_0t}$. The same holds, with a doubled decay rate, for a flat spectral density on the whole real line, which requires a dispersion relation unbounded from below. In both cases, however, $\|f\|^2=\int\mathrm{d}J(\lambda)=\infty$: the form factor is not square-integrable, and the model falls outside the scope of Definition~\ref{def:spinboson}. Exponential decoherence for such models, and the validity of the quantum regression theorem in this setting, are discussed in~\cite{lonigrochruscinski2022dephasing}.
	\end{remark}
	
	\subsection{The multimode Jaynes--Cummings model}\label{subsec:multijc}\index{Jaynes--Cummings model!multimode}
	
	Let now $B=\sigma_-$, so that
	\begin{equation}
		H=\frac{\omega_0}{2}\sigma_z\otimes I+I\otimes\mathrm{d}\Gamma(\omega)+\sigma_+\otimes a_+(f)+\sigma_-\otimes a_+^\dag(f).
	\end{equation}
	This is the multimode counterpart of the Jaynes--Cummings model of Section~\ref{subsec:jc}. In the literature, it is also referred to as the \emph{rotating-wave spin--boson model}\index{rotating-wave spin--boson model}: it is obtained from the spin--boson model with $B=\sigma_x$ by discarding the counter-rotating terms $\sigma_+\otimes a_+^\dag(f)$ and $\sigma_-\otimes a_+(f)$, just as the Jaynes--Cummings model is obtained from the quantum Rabi model via the rotating-wave approximation (Section~\ref{subsec:rabi} and Appendix~\ref{app:rwa}).
	As in the single-mode case (Proposition~\ref{prop:solving_jc1}), $H$ preserves the total number of excitations $\N_{\rm tot}=\sigma_+\sigma_-\otimes I+I\otimes\N$: one checks that $H$ commutes with the orthogonal projections onto the eigenspaces of $\N_{\rm tot}$, so that its restrictions to them are self-adjoint (Lemma~\ref{lem:invariant}). The vector $e_-\otimes\varepsilon_0$ is an eigenvector of $H$ with eigenvalue $-\omega_0/2$, while the one-excitation eigenspace is
	\begin{equation}
		\mathcal{E}_1=\left\{c\,e_+\otimes\varepsilon_0+e_-\otimes g:\;c\in\mathbb{C},\;g\in\sph\right\}\cong\mathbb{C}\oplus\sph,
	\end{equation}
	where $g$ is regarded as an element of the one-particle sector of $\Gamma^+(\sph)$. Since $a_+(f)g=\braket{f,g}\varepsilon_0$ and $a_+^\dag(f)\varepsilon_0=f$, on $\mathcal{E}_1$ the operator $H$ acts as
	\begin{equation}
		H\begin{bmatrix}c\\g\end{bmatrix}=\begin{bmatrix}\frac{\omega_0}{2}c+\braket{f,g}\\cf+\left(\omega-\frac{\omega_0}{2}\right)g\end{bmatrix},\qquad c\in\mathbb{C},\;g\in\Dom\omega,
	\end{equation}
	which is the multimode counterpart of the $2\times2$ matrix $H_1$ of Proposition~\ref{prop:solving_jc1}: the single state $e_-\otimes\phi_1$ is replaced by the whole one-particle space. The restriction of $H$ to $\mathcal{E}_1$, a single discrete level coupled to a continuum, is known in the literature as the \emph{Friedrichs} (or \emph{Friedrichs--Lee}) \emph{model}\index{Friedrichs--Lee model}~\cite{friedrichs1948,lee1954}. Consequently, as in Exercise~\ref{exerc:ampdamp}, the reduced dynamics of the qubit, with the field initially in the vacuum, is an amplitude-damping channel, now with
	\begin{equation}
		\alpha(t)=\e^{-\ii\frac{\omega_0}{2}t}\Braket{e_+\otimes\varepsilon_0,\e^{-\ii tH}e_+\otimes\varepsilon_0}.
	\end{equation}
	To compute it, note that $\mathcal{E}_1$ is invariant under $\e^{-\ii tH}$, so that we can write $\e^{-\ii tH}e_+\otimes\varepsilon_0=c(t)\,e_+\otimes\varepsilon_0+e_-\otimes g_t$, with $c(t)\in\mathbb{C}$, $g_t\in\sph$, and $\alpha(t)=\e^{-\ii\frac{\omega_0}{2}t}c(t)$. The Schr\"odinger equation then yields two coupled equations for $c(t)$ and $g_t$; eliminating $g_t$, one finds that $\beta(t):=\e^{\ii\omega_0t}\alpha(t)$ solves the integro-differential equation (Exercise~\ref{exerc:multijc} below)
	\begin{equation}\label{eq:multijc}
		\dot\beta(t)=-\int_0^tK(t-s)\beta(s)\;\mathrm{d}s,\qquad\beta(0)=1,\qquad K(\tau):=\int\e^{-\ii(\lambda-\omega_0)\tau}\;\mathrm{d}J(\lambda).
	\end{equation}
	For a single mode, $K(\tau)=g^2\e^{\ii\Delta\tau}$ with $\Delta=\omega_0-\omega$, and one recovers the vacuum Rabi oscillations of Eq.~\eqref{eq:vacuumrabi}.
	
	\begin{exercise}\label{exerc:multijc}
		Derive Eq.~\eqref{eq:multijc}. Hint: since $e_+\otimes\varepsilon_0\in\Dom H$, the vector $\e^{-\ii tH}e_+\otimes\varepsilon_0$ belongs to $\mathcal{E}_1\cap\Dom H$ and solves the Schr\"odinger equation. Deduce two coupled equations for $c(t)$ and $g_t$, with $c(0)=1$ and $g_0=0$; solve the one for $g_t$ by the variation-of-constants formula, and insert the result into the one for $c(t)$.
		\seesol{multijc}
	\end{exercise}
	
	For a continuum of modes, assume that $J$ has a smooth density, still denoted by $J(\lambda)$, so that $K(\tau)$ decays as $\tau\to\infty$. If this decay is fast compared to the evolution of $\beta$ (weak coupling), one may formally replace $\beta(s)$ by $\beta(t)$ in Eq.~\eqref{eq:multijc} and extend the integral to infinity, obtaining $\beta(t)\simeq\e^{-(\gamma/2+\ii\delta)t}$, with
	\begin{equation}
		\frac{\gamma}{2}+\ii\delta=\int_0^\infty K(\tau)\;\mathrm{d}\tau=\pi J(\omega_0)+\ii\,\mathrm{P}\!\!\int\frac{J(\lambda)}{\omega_0-\lambda}\;\mathrm{d}\lambda,
	\end{equation}
	$\mathrm{P}$ denoting the principal value. Hence $|\alpha(t)|^2\simeq\e^{-\gamma t}$, with $\gamma=2\pi J(\omega_0)$ (Fermi's golden rule\index{Fermi's golden rule}): the excited state decays exponentially and irreversibly, in contrast with the vacuum Rabi oscillations of Remark~\ref{rem:vacuumrabi}. This is the Wigner--Weisskopf theory\index{Wigner--Weisskopf theory} of spontaneous emission~\cite{breuerpetruccione}.
	
	\begin{remark}[Exponential decay and flat spectral densities]\label{rem:expdecay}\index{spectral density!flat}\index{exponential decay}
		The exponential decay predicted by the Wigner--Weisskopf theory is the result of an approximation. It becomes exact, at all times, for a flat spectral density on the whole real line: if $\mathrm{d}J(\lambda)=J_0\,\mathrm{d}\lambda$ on $\mathbb{R}$, then $K(\tau)=2\pi J_0\,\delta(\tau)$, and Eq.~\eqref{eq:multijc} reduces to $\dot\beta(t)=-\pi J_0\beta(t)$ (the delta being centered at the endpoint of the integration interval), whence $|\alpha(t)|^2=\e^{-\gamma t}$ with $\gamma=2\pi J_0$ exactly, and no frequency shift. Such a spectral density requires a dispersion relation unbounded from below---consistently with the fact that a Hamiltonian bounded from below can never produce an exactly exponential decay at all times~\cite{khalfin1958}---and a form factor which is not square-integrable, since $\|f\|^2=\int\mathrm{d}J(\lambda)=\infty$. Spin--boson models of this kind, and the exact validity of the quantum regression theorem for them, are discussed in~\cite{lonigrochruscinski2022regression}.
	\end{remark}
	
	\section{Spin--boson models with singular form factors}\label{sec:singular}
	
	Remarks~\ref{rem:expdephasing} and~\ref{rem:expdecay} show that exactly exponential decoherence and decay, which are usually obtained as the result of a Markovian approximation, can be derived from first principles by considering form factors which are not square-integrable. This raises a natural question: is it possible to define spin--boson models with non-normalizable form factors? In Remark~\ref{rem:uvir} we observed that the self-adjointness of spin--boson models, as established in Proposition~\ref{prop:sb_sa}, requires two conditions on the form factor: $f\in\sph$ and $\omega^{-1/2}f\in\sph$. In this final section we show that, thanks to the KLMN theorem (Theorem~\ref{thm:klmn}), the first condition can be dropped: as long as $\omega^{-1/2}f\in\sph$, a spin--boson model can be defined as a form sum, even if $f$ is not normalizable. To this end, we first extend the annihilation operators, and define the creation operators as forms, for such form factors (Section~\ref{subsec:creationforms}); we then define the corresponding spin--boson models, and show that they are limits of models with an ultraviolet cutoff (Section~\ref{subsec:formsum}). Finally, we briefly discuss what happens for more singular form factors (Section~\ref{subsec:renormalization}), such as those associated with the flat spectral densities of Remarks~\ref{rem:expdephasing} and~\ref{rem:expdecay}. The approach presented here is the starting point of~\cite{lonigro2022,lonigro2023mpag}, where it is applied to a larger class of generalized spin--boson models.
	
	\subsection{Annihilation operators and creation forms}\label{subsec:creationforms}
	
	Here and in the following, until Section~\ref{subsec:renormalization}, we consider form factors satisfying $\omega^{-1/2}f\in\sph$, but not necessarily $f\in\sph$; in this case, as we will see, most constructions can be carried out by means of quadratic forms. The more singular case will be briefly discussed in Section~\ref{subsec:renormalization}. As in the proof of Proposition~\ref{prop:a(f)domega}, we assume without loss of generality that $\sph=L^2(\mathcal{M},\Sigma,\mu)$ and that $\omega$ is the multiplication operator by a measurable function with $\omega(k)>0$ for almost all $k\in \mathcal{M}$. For $s\geq0$, let
	\begin{equation}
		\sph_{-s}:=\left\{f:\mathcal{M}\rightarrow\mathbb{C}\text{ measurable}:\;\left\|\omega^{-s/2}f\right\|^2=\int\frac{|f(k)|^2}{\omega(k)^s}\;\mathrm{d}\mu(k)<\infty\right\},
	\end{equation}
	so that $\sph_0=\sph$, and the natural space of form factors is $\sph_{-1}$; we follow the notation of~\cite{lonigro2022}. If $\omega$ is bounded from below by some $m>0$, then $\sph\subseteq\sph_{-1}$; if $\omega$ is unbounded, then $\sph_{-1}$ contains functions which are not square-integrable.
	
	\begin{example}\label{examp:singularff}
		Let $\mathcal{M}=\mathbb{R}^3$ with the Lebesgue measure, $\omega(k)=|k|$ (a scalar massless field), and consider the form factor $f(k)=1$ for $|k|<1$ and $f(k)=|k|^{-\alpha}$ for $|k|\geq1$, with $\alpha>0$. Using spherical coordinates,
		\begin{equation}
			\int|f(k)|^2\;\mathrm{d}k=4\pi\left(\frac{1}{3}+\int_1^\infty r^{2-2\alpha}\;\mathrm{d}r\right),\qquad\int\frac{|f(k)|^2}{\omega(k)}\;\mathrm{d}k=4\pi\left(\frac{1}{2}+\int_1^\infty r^{1-2\alpha}\;\mathrm{d}r\right).
		\end{equation}
		The first integral is finite if and only if $\alpha>3/2$, the second one if and only if $\alpha>1$. Hence, for $1<\alpha\leq3/2$, we have $f\in\sph_{-1}$ but $f\notin\sph$.
	\end{example}
	
	The annihilation operator remains well-defined for non-normalizable form factors:
	
	\begin{proposition}\label{prop:a(f)singular}
		Let $f\in\sph_{-1}$. For all $\psi\in\Dom\mathrm{d}\Gamma(\omega)^{1/2}$ and $n\geq1$, the integral
		\begin{equation}\label{eq:a(f)singular}
			\left(a_+(f)\psi\right)_{n-1}(k_1,\dots,k_{n-1}):=\sqrt{n}\int\overline{f(k_n)}\,\psi_n(k_1,\dots,k_{n-1},k_n)\;\mathrm{d}\mu(k_n)
		\end{equation}
		converges absolutely for almost all $(k_1,\dots,k_{n-1})\in \mathcal{M}^{n-1}$, and Eq.~\eqref{eq:a(f)singular} defines a linear operator $a_+(f):\Dom\mathrm{d}\Gamma(\omega)^{1/2}\rightarrow\Gamma^+(\sph)$ satisfying
		\begin{equation}\label{eq:a(f)singular_bound}
			\left\|a_+(f)\psi\right\|\leq\left\|\omega^{-1/2}f\right\|\left\|\mathrm{d}\Gamma(\omega)^{1/2}\psi\right\|.
		\end{equation}
		The map $f\mapsto a_+(f)$ is antilinear and, if $f\in\sph\cap\sph_{-1}$, this definition is consistent with Definition~\ref{def:apm}: the operator defined by Eq.~\eqref{eq:a(f)singular} coincides with the restriction of the bosonic annihilation operator $a_+(f)$ to $\Dom\mathrm{d}\Gamma(\omega)^{1/2}$.
	\end{proposition}
	
	\begin{proof}
		By the Cauchy--Schwarz inequality,
		\begin{equation}
			\int\left|f(k_n)\right|\left|\psi_n(k_1,\dots,k_n)\right|\mathrm{d}\mu(k_n)\leq\left\|\omega^{-1/2}f\right\|\left(\int\omega(k_n)\left|\psi_n(k_1,\dots,k_n)\right|^2\mathrm{d}\mu(k_n)\right)^{1/2},
		\end{equation}
		and the right-hand side is finite for almost all $(k_1,\dots,k_{n-1})$, since $\int\omega(k_n)|\psi_n|^2\,\mathrm{d}^n\mu$ is bounded by $\int\omega_n|\psi_n|^2\,\mathrm{d}^n\mu<\infty$. The resulting function is symmetric in its $n-1$ variables, and the computation in Step 1 of the proof of Proposition~\ref{prop:a(f)domega}---which only used $\|\omega^{-1/2}f\|<\infty$ and the symmetry of $\psi_n$, but never $f\in\sph$---yields $\|(a_+(f)\psi)_{n-1}\|^2\leq\|\omega^{-1/2}f\|^2\int\omega_n|\psi_n|^2\,\mathrm{d}^n\mu$. Summing over $n$ proves that $a_+(f)\psi\in\Gamma^+(\sph)$ and Eq.~\eqref{eq:a(f)singular_bound}. Antilinearity is clear. Finally, if $f\in\sph\cap\sph_{-1}$, by Proposition~\ref{prop:a(f)explicit} the operator defined by Eq.~\eqref{eq:a(f)singular} and the operator $a_+(f)$ of Definition~\ref{def:apm} coincide on finite-particle vectors; both satisfy the same bound (Proposition~\ref{prop:a(f)domega}), and finite-particle vectors are dense in $\Dom\mathrm{d}\Gamma(\omega)^{1/2}$ with respect to the graph norm of $\mathrm{d}\Gamma(\omega)^{1/2}$ (Step 2 of the same proof), so they coincide everywhere.
	\end{proof}
	
	Singular annihilation operators can be approximated by regular ones. For $\Lambda>0$, let $f_\Lambda:=f\chi_{\{\omega\leq\Lambda\}}$ be the form factor with an \emph{ultraviolet cutoff}\index{ultraviolet cutoff}, i.e. with all modes of energy larger than $\Lambda$ removed. Then $f_\Lambda\in\sph$, since $\int|f_\Lambda|^2\,\mathrm{d}\mu\leq\Lambda\int|f|^2\omega^{-1}\,\mathrm{d}\mu$, and $\|\omega^{-1/2}(f-f_\Lambda)\|^2=\int_{\{\omega>\Lambda\}}|f|^2\omega^{-1}\,\mathrm{d}\mu\to0$ as $\Lambda\to\infty$ by dominated convergence. By antilinearity and Eq.~\eqref{eq:a(f)singular_bound},
	\begin{equation}\label{eq:a(f)cutoff}
		\left\|\left(a_+(f)-a_+(f_\Lambda)\right)\psi\right\|\leq\left\|\omega^{-1/2}(f-f_\Lambda)\right\|\left\|\mathrm{d}\Gamma(\omega)^{1/2}\psi\right\|\xrightarrow{\Lambda\to\infty}0.
	\end{equation}
	
	The situation is very different for the creation operator. Given $\psi\in\Dom\mathrm{d}\Gamma(\omega)^{1/2}$ and a form factor $f\in\sph\cap\sph_{-1}$, the identity $\|a^\dag_+(f)\psi\|^2=\|a_+(f)\psi\|^2+\|f\|^2\|\psi\|^2$ holds (it was proven on finite-particle vectors in Step 1 of the proof of Proposition~\ref{prop:a(f)domega}, and extends by continuity). If $f\notin\sph$, then $\|f_\Lambda\|\to\infty$ by monotone convergence, while $\|a_+(f_\Lambda)\psi\|$ stays bounded by Eq.~\eqref{eq:a(f)cutoff}; therefore $\|a_+^\dag(f_\Lambda)\psi\|\to\infty$ for every $\psi\neq0$. This is to be expected: formally, $a_+^\dag(f)$ creates a particle with the non-normalizable wavefunction $f$---for instance, $a_+^\dag(f)\varepsilon_0$ ``$=$'' $f\notin\sph$. The creation operator, however, still makes sense as a form: for $f\in\sph\cap\sph_{-1}$ we have $\braket{\varphi,a_+^\dag(f)\psi}=\braket{a_+(f)\varphi,\psi}$ (Proposition~\ref{prop:a(f)domega}), and the right-hand side is well-defined for all $f\in\sph_{-1}$ and $\varphi,\psi\in\Dom\mathrm{d}\Gamma(\omega)^{1/2}$. This is exactly the situation of the delta potential (Example~\ref{examp:delta}), with $v\mapsto v(0)\delta$ replaced by $\psi\mapsto a_+^\dag(f)\psi$.
	
	\subsection{Spin--boson models as form sums}\label{subsec:formsum}
	
	Let $\omega_0\geq0$ and $B\in\mathbb{C}^{2\times2}$, and let $H_{\rm free}=\frac{\omega_0}{2}\sigma_z\otimes I+I\otimes\mathrm{d}\Gamma(\omega)$ on $\hfrak=\mathbb{C}^2\otimes\Gamma^+(\sph)$, as in Section~\ref{sec:multimodesb}. Since $\mathrm{d}\Gamma(\omega)\geq0$, we have $H_{\rm free}\geq-c_0$ with $c_0:=\omega_0/2$. Besides, as $H_{\rm free}$ differs from $I\otimes\mathrm{d}\Gamma(\omega)$ by a bounded operator, the two have the same form domain (cf. Theorem~\ref{thm:klmn} and Remark~\ref{rem:krklmn}):
	\begin{equation}
		\mathcal{Q}(H_{\rm free})=\mathbb{C}^2\otimes\Dom\mathrm{d}\Gamma(\omega)^{1/2}=:\mathcal{Q},
	\end{equation}
	and, writing $K:=I\otimes\mathrm{d}\Gamma(\omega)^{1/2}$ on $\mathcal{Q}$, for all $\Psi\in\mathcal{Q}$,
	\begin{equation}\label{eq:freeformbound}
		\|K\Psi\|^2\leq\left\|(H_{\rm free}+c_0)^{1/2}\Psi\right\|^2\leq\|K\Psi\|^2+2c_0\|\Psi\|^2.
	\end{equation}
	Formally, the interaction is $V=B\otimes a_+^\dag(f)+B^*\otimes a_+(f)$. Following the discussion at the end of Section~\ref{subsec:creationforms}, we replace it by the Hermitian form
	\begin{equation}
		v_f(\Phi,\Psi):=\Braket{(B^*\otimes a_+(f))\Phi,\Psi}+\Braket{\Phi,(B^*\otimes a_+(f))\Psi},\qquad\Phi,\Psi\in\mathcal{Q},
	\end{equation}
	where $B^*\otimes a_+(f)=(B^*\otimes I)(I\otimes a_+(f))$ is well-defined on $\mathcal{Q}$ by Proposition~\ref{prop:a(f)singular}. Hereafter we set $F:=\|B\|\|\omega^{-1/2}f\|$.
	
	\begin{theorem}\label{thm:singularsb}
		Let $f\in\sph_{-1}$. Then:
		\begin{itemize}
			\item[(i)] The form $v_f$ is infinitesimally $H_{\rm free}$-form-bounded. Consequently, there exists a unique self-adjoint operator $H_f$ with $\Dom H_f\subseteq\mathcal{Q}$ and
			\begin{equation}\label{eq:singularsb}
				\braket{\Phi,H_f\Psi}=q_{H_{\rm free}}(\Phi,\Psi)+v_f(\Phi,\Psi)\qquad\text{for all }\Phi\in\mathcal{Q},\;\Psi\in\Dom H_f;
			\end{equation}
			moreover, $\mathcal{Q}(H_f)=\mathcal{Q}$ and $H_f\geq-(c_0+2F^2)$.
			\item[(ii)] If $f\in\sph$, then $H_f$ coincides with the spin--boson model of Definition~\ref{def:spinboson}.
			\item[(iii)] Let $\lambda:=c_0+2F^2+1$. Then, for all $\Lambda>0$,
			\begin{equation}\label{eq:normresolvent}
				\left\|(H_f+\lambda)^{-1}-(H_{f_\Lambda}+\lambda)^{-1}\right\|\leq2\sqrt{2}\,\|B\|\left\|\omega^{-1/2}(f-f_\Lambda)\right\|\xrightarrow{\Lambda\to\infty}0.
			\end{equation}
		\end{itemize}
	\end{theorem}
	
	\begin{proof}
		(i) For $\Psi\in\mathcal{Q}$ we have $v_f(\Psi)=2\operatorname{Re}\braket{(B^*\otimes a_+(f))\Psi,\Psi}$, whence, by Proposition~\ref{prop:a(f)singular} (applied to each component of $\Psi$, as in Lemma~\ref{lem:reltensor}) and $\|B^*\otimes I\|=\|B\|$,
		\begin{equation}
			|v_f(\Psi)|\leq2\|B\|\left\|(I\otimes a_+(f))\Psi\right\|\|\Psi\|\leq2F\|K\Psi\|\|\Psi\|\leq\epsilon\|K\Psi\|^2+\frac{F^2}{\epsilon}\|\Psi\|^2
		\end{equation}
		for every $\epsilon>0$. By Eq.~\eqref{eq:freeformbound}, $\|K\Psi\|^2\leq\|(H_{\rm free}+c_0)^{1/2}\Psi\|^2$, so $v_f$ is infinitesimally $H_{\rm free}$-form-bounded. The remaining claims follow from the KLMN theorem (Theorem~\ref{thm:klmn}) applied with $\epsilon=1/2$.
		
		(ii) Let $f\in\sph$, and let $H=H_{\rm free}+V$ be the self-adjoint operator of Proposition~\ref{prop:sb_sa}, with $\Dom H=\Dom H_{\rm free}\subseteq\mathcal{Q}$. For $\Phi\in\mathcal{Q}$ and $\Psi\in\Dom H_{\rm free}$, Proposition~\ref{prop:a(f)domega}(ii) and Proposition~\ref{prop:a(f)singular} yield $\braket{\Phi,V\Psi}=v_f(\Phi,\Psi)$, and thus $H$ satisfies Eq.~\eqref{eq:singularsb}. By the uniqueness in (i), $H=H_f$.
		
		(iii) Write $R:=(H_f+\lambda)^{-1}$ and $R_\Lambda:=(H_{f_\Lambda}+\lambda)^{-1}$. Since $\|\omega^{-1/2}f_\Lambda\|\leq\|\omega^{-1/2}f\|$, both $H_f$ and $H_{f_\Lambda}$ satisfy $H\geq-(\lambda-1)$, so $\|R\|,\|R_\Lambda\|\leq1$. Let $\varphi,\psi\in\hfrak$, and set $u:=R_\Lambda\varphi\in\Dom H_{f_\Lambda}$ and $w:=R\psi\in\Dom H_f$. Applying Eq.~\eqref{eq:singularsb} twice, once for $H_f$ and once (after complex conjugation) for $H_{f_\Lambda}$,
		\begin{equation}
			\braket{\varphi,w}-\braket{u,\psi}=\Braket{(H_{f_\Lambda}+\lambda)u,w}-\Braket{u,(H_f+\lambda)w}=v_{f_\Lambda}(u,w)-v_f(u,w),
		\end{equation}
		that is, since $R_\Lambda$ is self-adjoint and $f\mapsto v_f$ is real-linear,
		\begin{equation}\label{eq:resolventdiff}
			\Braket{\varphi,(R-R_\Lambda)\psi}=v_{f_\Lambda-f}(u,w)=\Braket{(B^*\otimes a_+(f_\Lambda-f))u,w}+\Braket{u,(B^*\otimes a_+(f_\Lambda-f))w}.
		\end{equation}
		To bound the right-hand side we need a bound on $\|Ku\|$ and $\|Kw\|$ which is uniform in $\Lambda$. As in the proof of the KLMN theorem, $q_{H_{\rm free}}(u)+v_{f_\Lambda}(u)\geq\frac{1}{2}\|Ku\|^2-(c_0+2F^2)\|u\|^2$, whence
		\begin{equation}
			\frac{1}{2}\|Ku\|^2+\|u\|^2\leq q_{H_{\rm free}}(u)+v_{f_\Lambda}(u)+\lambda\|u\|^2=\braket{u,\varphi}\leq\|\varphi\|^2,
		\end{equation}
		so $\|Ku\|\leq\sqrt{2}\|\varphi\|$, and analogously $\|Kw\|\leq\sqrt{2}\|\psi\|$. Using Eq.~\eqref{eq:a(f)singular_bound} in Eq.~\eqref{eq:resolventdiff},
		\begin{align}
			\left|\Braket{\varphi,(R-R_\Lambda)\psi}\right|&\leq\|B\|\left\|\omega^{-1/2}(f-f_\Lambda)\right\|\left(\|Ku\|\|w\|+\|u\|\|Kw\|\right)\nonumber\\
			&\leq2\sqrt{2}\,\|B\|\left\|\omega^{-1/2}(f-f_\Lambda)\right\|\|\varphi\|\|\psi\|,
		\end{align}
		which proves Eq.~\eqref{eq:normresolvent}.
	\end{proof}
	
	Theorem~\ref{thm:singularsb} shows that, if $f\in\sph_{-1}$, the spin--boson model is the norm resolvent limit of the regularized models $H_{f_\Lambda}$ obtained by removing all modes of energy larger than $\Lambda$; in particular, the corresponding unitary groups converge strongly (Section~\ref{subsec:spectrum}). Crucially, removing the cutoff requires no modification of the Hamiltonian. The price to pay is the loss of explicit control on the operator domain. For instance, if $f\notin\sph$ and $Be\neq0$ for some $e\in\mathbb{C}^2$, then $\Psi_0:=e\otimes\varepsilon_0\in\Dom H_{\rm free}$ but $\Psi_0\notin\Dom H_f$: formally, $H_f\Psi_0$ contains the non-normalizable term $Be\otimes f$. Indeed, if $\Psi_0$ were in $\Dom H_f$, then by Eq.~\eqref{eq:singularsb} the map $\Phi\mapsto v_f(\Phi,\Psi_0)=\braket{\Phi,H_f\Psi_0}-\braket{\Phi,H_{\rm free}\Psi_0}$ would be bounded on $\mathcal{Q}$. But, since $a_+(f)\varepsilon_0=0$, for $\Phi=Be\otimes h$ with $h\in\Dom\omega^{1/2}$ in the one-particle sector one finds $v_f(\Phi,\Psi_0)=\|Be\|^2\,\overline{\braket{f,h}}$, with $\braket{f,h}=\int\overline{f}h\,\mathrm{d}\mu$; the boundedness of $h\mapsto\braket{f,h}$ on the dense subspace $\Dom\omega^{1/2}$ would then imply, by the Riesz representation theorem, $f\in\sph$---a contradiction. The characterization of $\Dom H_f$ is a nontrivial problem; see~\cite{lonigro2022,lilllonigro2025}.
	
	In particular, for the massless field of Example~\ref{examp:singularff}, the theorem applies for $1<\alpha\leq3/2$, a range in which Proposition~\ref{prop:sb_sa} does not.
	
	\subsection{Stronger divergences and renormalization}\label{subsec:renormalization}
	
	If $f\notin\sph_{-1}$---e.g. for $\alpha\leq1$ in Example~\ref{examp:singularff}, or for the flat spectral densities of Remarks~\ref{rem:expdephasing} and~\ref{rem:expdecay}---the bound~\eqref{eq:a(f)singular_bound} is no longer available, the form $v_f$ is not even defined on $\mathcal{Q}$, and the strategy of Section~\ref{subsec:formsum} breaks down. The cutoff models $H_{f_\Lambda}$ are still well-defined, but they are not expected to converge as $\Lambda\to\infty$. A simple way to see what goes wrong is the van Hove model\index{van Hove model}, i.e. a boson field linearly coupled to a classical source, formally $\mathrm{d}\Gamma(\omega)+a_+^\dag(f)+a_+(f)$: completing the square as in Proposition~\ref{prop:multidephasing}, one finds that it is unitarily equivalent to $\mathrm{d}\Gamma(\omega)-\|\omega^{-1/2}f\|^2$, at least if $\omega^{-1}f\in\sph$. Thus, as $\Lambda\to\infty$, the bottom of the spectrum of the regularized model diverges to $-\infty$ if $\omega^{-1/2}f\notin\sph$; see~\cite{derezinski2003} for a rigorous and complete discussion of this exactly solvable model.
	
	To obtain a meaningful limit, one must then \emph{renormalize}\index{renormalization} the Hamiltonian: for instance, one subtracts a cutoff-dependent constant $E_\Lambda$, diverging as $\Lambda\to\infty$, and studies the limit of $H_{f_\Lambda}-E_\Lambda$. When such a limit exists, the resulting renormalized Hamiltonian is, in general, no longer a form perturbation of $H_{\rm free}$, and its construction requires methods beyond the ones discussed in these notes. The prototypical example is the Nelson model\index{Nelson model}, describing nonrelativistic particles interacting with a scalar boson field~\cite{nelson1964}. Alternative approaches construct the renormalized Hamiltonian directly, e.g. via \emph{interior boundary conditions}\index{interior boundary conditions}~\cite{schmidt2018,lampartschmidt2019}. For spin--boson models and their generalizations, renormalized Hamiltonians were constructed in~\cite{dammoller2020,lonigro2022,lonigro2023,hinrichs2025,alvarez2025}. This is a very active field of research; an introduction to the different approaches to renormalization in nonrelativistic quantum field theory can be found in \cite[Chapter 1]{lillthesis}.
	
	With renormalization, we have reached questions of current research, which branch into several directions. A first one concerns the physical role of ultraviolet divergences: which features of the regularized models---such as the exactly exponential decoherence and decay of Remarks~\ref{rem:expdephasing} and~\ref{rem:expdecay}---survive the removal of the cutoff, and in which sense the resulting dynamics describes realistic physical systems~\cite{lonigrochruscinski2022dephasing,lonigrochruscinski2022regression}. A second one concerns the control of approximations, such as the rotating-wave approximation of Appendix~\ref{app:rwa}, for fields with infinitely many modes or for several spins~\cite{burgarth,richterdicke}. A third one is the construction and renormalization of more complex models, with richer matter systems and couplings. All these directions rely on the tools developed in these notes, and the reader should now have all the prerequisites to start exploring the corresponding literature.
	
	\phantomsection
	\addcontentsline{toc}{section}{Acknowledgments}
	\section*{Acknowledgments}\label{sec:ack}
	I am grateful to Leonhard Richter for his many contributions to the course from which these notes originated: his feedback on the lecture slides, the design of the original version of Appendix~\ref{app:rwa}, based on a guest lecture he gave in the course, and his help in drafting the exercises and their solutions and in running the exercise classes.
	
	\appendix
	
	\section{A primer on measure theory}\label{app:measure}
	
	This appendix collects, without proofs, the notions and results from measure and integration theory that are used throughout the notes; see e.g. \cite[Appendix A]{teschl} for a concise but complete treatment, and the classic textbooks~\cite{rudin,folland,bogachev} for comprehensive ones.
	
	\paragraph{Measurable spaces and measures.} A \emph{$\sigma$-algebra}\index{sigma-algebra@$\sigma$-algebra} on a set $\mathcal{M}$ is a collection $\Sigma$ of subsets of $\mathcal{M}$ which contains $\mathcal{M}$ and is closed under complements and countable unions; the pair $(\mathcal{M},\Sigma)$ is a \emph{measurable space}, and the elements of $\Sigma$ are its \emph{measurable sets}. The $\sigma$-algebra generated by a family of subsets is the smallest $\sigma$-algebra containing it. The \emph{Borel} $\sigma$-algebra\index{Borel sigma-algebra@Borel $\sigma$-algebra} $\mathfrak{B}(\mathbb{R}^d)$ is the one generated by the open subsets of $\mathbb{R}^d$ or, equivalently, by the rectangles $\prod_{j=1}^d[a_j,b_j]$; for $d=1$, by the intervals. A \emph{measure} on $(\mathcal{M},\Sigma)$ is a map $\mu:\Sigma\rightarrow[0,\infty]$ with $\mu(\emptyset)=0$ which is \emph{countably additive}: $\mu(\cup_nE_n)=\sum_n\mu(E_n)$ for every countable family $(E_n)_n\subset\Sigma$ of disjoint sets. The triple $(\mathcal{M},\Sigma,\mu)$ is a \emph{measure space}; it is \emph{finite} if $\mu(\mathcal{M})<\infty$, and \emph{$\sigma$-finite} if $\mathcal{M}=\cup_{j\in\mathbb{N}}\mathcal{M}_j$ with $\mu(\mathcal{M}_j)<\infty$ for all $j$. Typical examples are the \emph{Lebesgue measure}\index{Lebesgue measure} on $\mathfrak{B}(\mathbb{R}^d)$, the unique measure assigning to each rectangle its volume $\prod_j(b_j-a_j)$; the \emph{counting measure} on $\mathbb{N}$, $\mu(E)=\#E$, for which $L^2(\mathbb{N})=\ell^2$; and the \emph{Dirac measure} (point mass) $\delta_x$ at a point $x$, with $\delta_x(E)=1$ if $x\in E$ and $\delta_x(E)=0$ otherwise. A set $E\in\Sigma$ with $\mu(E)=0$ is a \emph{null set}, and a property holds \emph{$\mu$-almost everywhere} (a.e.) if it holds outside a null set.
	
	\paragraph{Measurable functions and integrals.} A function $f:\mathcal{M}\rightarrow\mathbb{C}$ is \emph{measurable}\index{measurable function} if $f^{-1}(B)\in\Sigma$ for every Borel set $B\subseteq\mathbb{C}$. Continuous functions on $\mathbb{R}^d$ are measurable, and sums, products and pointwise limits of measurable functions are measurable. The \emph{characteristic function} $\chi_E$ of a set $E$, equal to $1$ on $E$ and to $0$ elsewhere, is measurable if and only if $E\in\Sigma$. The \emph{Lebesgue integral}\index{Lebesgue integral} of a measurable $f\geq0$ is defined as the supremum of the integrals $\sum_jc_j\mu(E_j)$ of all \emph{simple functions} $\sum_jc_j\chi_{E_j}\leq f$ (finite sums, $c_j\geq0$, $E_j\in\Sigma$), and takes values in $[0,\infty]$; a measurable $f:\mathcal{M}\rightarrow\mathbb{C}$ is \emph{integrable} if $\int|f|\,\mathrm{d}\mu<\infty$, in which case $\int f\,\mathrm{d}\mu$ is defined by linearity, splitting $f$ into the positive and negative parts of its real and imaginary parts. Functions which coincide a.e. have the same integral. The two most important convergence results are:
	\begin{itemize}
		\item \emph{Monotone convergence}\index{monotone convergence theorem}: if $0\leq f_1\leq f_2\leq\ldots$ and $f_n\to f$ a.e., then $\int f_n\,\mathrm{d}\mu\to\int f\,\mathrm{d}\mu$;
		\item \emph{Dominated convergence}\index{dominated convergence theorem}: if $f_n\to f$ a.e. and $|f_n|\leq g$ a.e. for all $n$, with $g$ integrable, then $f$ is integrable and $\int|f_n-f|\,\mathrm{d}\mu\to0$; in particular, $\int f_n\,\mathrm{d}\mu\to\int f\,\mathrm{d}\mu$.
	\end{itemize}
	
	\paragraph{$L^p$ spaces.} For $1\leq p<\infty$, $L^p(\mathcal{M},\Sigma,\mu)$ is the space of (equivalence classes, modulo a.e. equality, of) measurable functions with $\int|f|^p\,\mathrm{d}\mu<\infty$; it is a Banach space with the norm $\|f\|_p:=(\int|f|^p\,\mathrm{d}\mu)^{1/p}$. $L^\infty(\mathcal{M},\Sigma,\mu)$ is the space of \emph{essentially bounded} functions, with $\|f\|_\infty$ the smallest $C\geq0$ such that $|f|\leq C$ a.e. The space $L^2$ is a Hilbert space (Example~\ref{examp:l2}), and the Cauchy--Schwarz inequality reads $\int|fg|\,\mathrm{d}\mu\leq\|f\|_2\|g\|_2$. For the Lebesgue measure on an open set $\Omega\subseteq\mathbb{R}^d$, the space $\mathrm{C}^\infty_{\rm c}(\Omega)$ of smooth compactly supported functions is dense in $L^2(\Omega)$.
	
	\paragraph{Product measures and Fubini's theorem.} Given two $\sigma$-finite measure spaces $(\mathcal{M}_1,\Sigma_1,\mu_1)$ and $(\mathcal{M}_2,\Sigma_2,\mu_2)$, the \emph{product $\sigma$-algebra} $\Sigma_1\otimes\Sigma_2$ is the $\sigma$-algebra on $\mathcal{M}_1\times\mathcal{M}_2$ generated by the sets $E_1\times E_2$, with $E_1\in\Sigma_1$ and $E_2\in\Sigma_2$, and the \emph{product measure}\index{product measure} $\mu_1\otimes\mu_2$ is the unique measure on it such that $(\mu_1\otimes\mu_2)(E_1\times E_2)=\mu_1(E_1)\mu_2(E_2)$ for all such sets. For instance, the product of the Lebesgue measures on $\mathbb{R}^{d_1}$ and $\mathbb{R}^{d_2}$ is the Lebesgue measure on $\mathbb{R}^{d_1+d_2}$. \emph{Fubini's theorem}\index{Fubini's theorem} states that, if $f:\mathcal{M}_1\times\mathcal{M}_2\rightarrow\mathbb{C}$ is integrable with respect to $\mu_1\otimes\mu_2$, then $f(x,\cdot)$ is $\mu_2$-integrable for $\mu_1$-almost all $x$, the function $x\mapsto\int f(x,y)\,\mathrm{d}\mu_2(y)$ is $\mu_1$-integrable, and
	\begin{equation}
		\int_{\mathcal{M}_1\times\mathcal{M}_2}f\,\mathrm{d}(\mu_1\otimes\mu_2)=\int_{\mathcal{M}_1}\left(\int_{\mathcal{M}_2}f(x,y)\,\mathrm{d}\mu_2(y)\right)\mathrm{d}\mu_1(x)=\int_{\mathcal{M}_2}\left(\int_{\mathcal{M}_1}f(x,y)\,\mathrm{d}\mu_1(x)\right)\mathrm{d}\mu_2(y);
	\end{equation}
	for measurable $f\geq0$, the same identity holds without any integrability assumption, both sides being possibly infinite (\emph{Tonelli's theorem}).
	
	\paragraph{Image measures, densities, and the Riemann--Lebesgue lemma.} Given a measure space $(\mathcal{M},\Sigma,\mu)$ and a measurable function $\omega:\mathcal{M}\rightarrow\mathbb{R}$, the \emph{image measure} $\omega_*\mu$ on $\mathfrak{B}(\mathbb{R})$ is defined by $(\omega_*\mu)(E):=\mu(\omega^{-1}(E))$, and satisfies $\int\eta\,\mathrm{d}(\omega_*\mu)=\int\eta\circ\omega\,\mathrm{d}\mu$ for every measurable $\eta\geq0$ (and for every bounded measurable $\eta$, if $\mu$ is finite). Given a measurable $\rho\geq0$, the formula $\nu(E):=\int_E\rho\,\mathrm{d}\mu$ defines a measure, said to have \emph{density} $\rho$ with respect to $\mu$; for instance, the spectral density $J$ of Section~\ref{subsec:multidephasing} is the image, under the dispersion relation $\omega$, of the measure with density $|f|^2$ with respect to $\mu$. A Borel measure on $\mathbb{R}$ is \emph{absolutely continuous} if it has a density with respect to the Lebesgue measure. The \emph{Riemann--Lebesgue lemma}\index{Riemann--Lebesgue lemma} states that, if $\nu$ is a finite, absolutely continuous Borel measure on $\mathbb{R}$, then $\int\e^{-\ii\lambda t}\,\mathrm{d}\nu(\lambda)\to0$ as $|t|\to\infty$.
	
	\section{The rotating-wave approximation}\label{app:rwa}
	
	In this appendix we give a rigorous and quantitative justification of the rotating-wave approximation (RWA) discussed in Section~\ref{subsec:rabi}, by means of an ``integration-by-parts'' method; see~\cite{burgarth} for this approach, and~\cite{richterdicke} for its extension to the Dicke model\index{Dicke model}, in which an arbitrary number of spins interacts with the same field mode.
	
	\subsection{Setting and heuristics}
	
	Recall the Jaynes--Cummings model (Section~\ref{subsec:jc})
	\begin{equation}\label{eq:rwa-hjc}
		H_{\rm JC}=\frac{\omega_0}{2}\sigma_z\otimes I+I\otimes\omega a^\dag a+g\left(\sigma_+\otimes a+\sigma_-\otimes a^\dag\right)
	\end{equation}
	on $\Dom H_{\rm JC}=\mathbb{C}^2\odot\Dom\N$, i.e. the spin--boson model with $B=g\sigma_-$, and the quantum Rabi model
	\begin{equation}\label{eq:rwa-hqr}
		H_{\rm QR}=\frac{\omega_0}{2}\sigma_z\otimes I+I\otimes\omega a^\dag a+g\left(\sigma_++\sigma_-\right)\otimes\left(a+a^\dag\right)
	\end{equation}
	on the same domain $\Dom H_{\rm QR}=\mathbb{C}^2\odot\Dom\N$, i.e. the spin--boson model with $B=g(\sigma_++\sigma_-)=g\sigma_x$ (Section~\ref{subsec:rabi}). Both operators are self-adjoint by Proposition~\ref{prop:visrelbound}. Here, as in Definition~\ref{def:singlemodesb}, $a$ and $a^\dag$ are the abstract ladder operators of Definition~\ref{def:abstractladder}, associated with a fixed orthonormal basis $(\phi_n)_{n\in\mathbb{N}}$ of $\hilb$: they do not depend on $\omega$, $\omega_0$ and $g$, which only enter the Hamiltonians as coefficients. This matters for the limit $\omega\to\infty$ considered below, since in the Hermite representation of Section~\ref{sec:singlemode} the operator $a$ and the Hermite functions themselves depend on $\omega$. The free Hamiltonian $H_{\rm free}=\frac{\omega_0}{2}\sigma_z\otimes I+I\otimes\omega a^\dag a$ is the same for both models, while the interactions
	\begin{equation}
		V_{\rm JC}=g\left(\sigma_+\otimes a+\sigma_-\otimes a^\dag\right),\qquad V_{\rm QR}=g\left(\sigma_++\sigma_-\right)\otimes\left(a+a^\dag\right)=V_{\rm JC}+g\left(\sigma_+\otimes a^\dag+\sigma_-\otimes a\right)
	\end{equation}
	differ by the so-called \emph{counter-rotating terms}\index{counter-rotating terms} $\sigma_+\otimes a^\dag$ and $\sigma_-\otimes a$. Because of these terms, the quantum Rabi model does not preserve the number of excitations, and cannot be written in block-diagonal form as easily as the Jaynes--Cummings model. In short, $H_{\rm QR}$ is more natural for describing matter--field interactions, but much harder to diagonalize than $H_{\rm JC}$. Are there situations in which it is justified to replace $H_{\rm QR}$ with $H_{\rm JC}$? This replacement is precisely the rotating-wave approximation.
	
	Let $\Delta:=\omega_0-\omega$ be the \emph{detuning}\index{detuning}, and let
	\begin{equation}
		H_0:=\frac{\omega}{2}\sigma_z\otimes I+I\otimes\omega a^\dag a
	\end{equation}
	be the free Hamiltonian on resonance, i.e. with both components having the same frequency $\omega$. In the interaction picture with respect to $H_0$, the two models become
	\begin{align}
		\tilde{H}_{\rm QR}(t)&:=\e^{\ii H_0t}(H_{\rm QR}-H_0)\e^{-\ii H_0t}\nonumber\\
		&\phantom{:}=\frac{\Delta}{2}\sigma_z\otimes I+g\left(\sigma_-\otimes a^\dag+\sigma_+\otimes a+\e^{-2\ii\omega t}\sigma_-\otimes a+\e^{2\ii\omega t}\sigma_+\otimes a^\dag\right),\label{eq:rwa-hqr-int}\\
		\tilde{H}_{\rm JC}(t)&:=\e^{\ii H_0t}(H_{\rm JC}-H_0)\e^{-\ii H_0t}\nonumber\\
		&\phantom{:}=\frac{\Delta}{2}\sigma_z\otimes I+g\left(\sigma_-\otimes a^\dag+\sigma_+\otimes a\right).\label{eq:rwa-hjc-int}
	\end{align}
	
	\begin{exercise}\label{exerc:rwa-interaction}
		Prove Eqs.~\eqref{eq:rwa-hqr-int}--\eqref{eq:rwa-hjc-int} on $\mathbb{C}^2\odot\Dom\N$. Hint: show that, for all $\phi\in\Dom\N^{1/2}$ and $v\in\mathbb{C}^2$,
		\begin{equation}
			\e^{\ii\omega t\N}a^\dag\e^{-\ii\omega t\N}\phi=\e^{\ii\omega t}a^\dag\phi,\qquad\e^{\ii\omega t\N}a\,\e^{-\ii\omega t\N}\phi=\e^{-\ii\omega t}a\phi,\qquad \e^{\ii\frac{\omega t}{2}\sigma_z}\sigma_\pm\e^{-\ii\frac{\omega t}{2}\sigma_z}v=\e^{\pm\ii\omega t}\sigma_\pm v.
		\end{equation}
		\seesol{rwa-interaction}
	\end{exercise}
	
	The two interaction-picture Hamiltonians differ by the terms with coefficients $\e^{\pm2\ii\omega t}$. In particular, $\tilde{H}_{\rm JC}$ is, at least formally, the time average of $\tilde{H}_{\rm QR}(t)$ over one period $T=\pi/\omega$, since $\int_0^T\e^{\pm2\ii\omega t}\,\mathrm{d}t=0$. Therefore, it makes sense to expect the two dynamics to be ``close to each other'' in the limit $\omega\to\infty$ at fixed coupling $g$ and fixed detuning $\Delta$, where the oscillating phases of the counter-rotating terms become faster and faster. Our goal is to prove (Corollary~\ref{coroll:rwa} below) that, for all $t\in\mathbb{R}$ and $\psi\in\hfrak=\mathbb{C}^2\otimes\hilb$,
	\begin{equation}\label{eq:rwa-goal}
		\left\|\left(\e^{-\ii tH_{\rm QR}}-\e^{-\ii tH_{\rm JC}}\right)\psi\right\|\to0\qquad\text{as }\omega\to\infty,
	\end{equation}
	with $g$ and $\Delta$ fixed (and thus $\omega_0=\omega+\Delta\to\infty$ as well).
	
	\subsection{The integration-by-parts method}\index{integration-by-parts method}
	
	To explain the idea behind the method, in this subsection we consider bounded operators only. Let $U_1(t),U_2(t)$ be two one-parameter families of unitaries, with $U_1(0)=U_2(0)=I$, whose strong derivatives are given by two families of bounded operators $-\ii H_1(t)$ and $-\ii H_2(t)$, strongly continuous in $t$:
	\begin{equation}
		\frac{\mathrm{d}}{\mathrm{d}t}U_i(t)\psi=-\ii H_i(t)U_i(t)\psi,\qquad i=1,2,
	\end{equation}
	so that $\frac{\mathrm{d}}{\mathrm{d}t}U_i(t)^*\psi=\ii U_i(t)^*H_i(t)\psi$. By the product rule,
	\begin{equation}
		\frac{\mathrm{d}}{\mathrm{d}s}\left(U_1(s)^*U_2(s)\right)\psi=\ii U_1(s)^*\left(H_1(s)-H_2(s)\right)U_2(s)\psi,
	\end{equation}
	and therefore
	\begin{equation}\label{eq:rwa-ibp1}
		\ii\left(U_2(t)-U_1(t)\right)\psi=\ii U_1(t)\left[U_1(s)^*U_2(s)\psi\right]_{s=0}^{s=t}=U_1(t)\int_0^tU_1(s)^*\left(H_2(s)-H_1(s)\right)U_2(s)\psi\;\mathrm{d}s.
	\end{equation}
	Using the unitarity of $U_1(t)$ and $U_1(s)^*$, this equation gives
	\begin{equation}
		\left\|\left(U_2(t)-U_1(t)\right)\psi\right\|\leq\left|\int_0^t\left\|\left(H_2(s)-H_1(s)\right)U_2(s)\psi\right\|\mathrm{d}s\right|,
	\end{equation}
	so that we can bound the difference of the two unitary families by the difference of their generators. This bound, however, is only useful if $H_2(s)-H_1(s)$ is small; it is blind to the case we are interested in, in which the difference of the generators is not small, but oscillates rapidly in time, so that its effect averages out.
	
	To capture this averaging, we perform a second integration by parts. Define
	\begin{equation}\label{eq:rwa-S}
		S(s):=\int_0^s\left(H_2(r)-H_1(r)\right)\mathrm{d}r,\qquad\text{so that}\qquad H_2(s)-H_1(s)=\frac{\mathrm{d}}{\mathrm{d}s}S(s).
	\end{equation}
	By the product rule,
	\begin{align}
		U_1(s)^*\left(H_2(s)-H_1(s)\right)U_2(s)\psi&=\frac{\mathrm{d}}{\mathrm{d}s}\left(U_1(s)^*S(s)U_2(s)\psi\right)\nonumber\\
		&\quad+\ii U_1(s)^*\left(S(s)H_2(s)-H_1(s)S(s)\right)U_2(s)\psi.
	\end{align}
	Inserting this into Eq.~\eqref{eq:rwa-ibp1}, and using $S(0)=0$, we get
	\begin{equation}\label{eq:rwa-ibp2}
		\ii\left(U_2(t)-U_1(t)\right)\psi=S(t)U_2(t)\psi+\ii\int_0^tU_1(t)U_1(s)^*\left(S(s)H_2(s)-H_1(s)S(s)\right)U_2(s)\psi\;\mathrm{d}s.
	\end{equation}
	The advantage is that the right-hand side now contains the integral $S(s)$ of the difference of the generators, where averaging can happen. Indeed, by unitarity, Eq.~\eqref{eq:rwa-ibp2} yields
	\begin{equation}\label{eq:rwa-ibp2-bound}
		\left\|\left(U_2(t)-U_1(t)\right)\psi\right\|\leq\left\|S(t)\right\|\|\psi\|+\left|\int_0^t\left\|S(s)H_2(s)-H_1(s)S(s)\right\|\mathrm{d}s\right|\|\psi\|,
	\end{equation}
	which is small whenever $S(s)$ is small, uniformly for $s$ between $0$ and $t$, while $H_1(s)$ and $H_2(s)$ remain bounded---even if $H_2(s)-H_1(s)$ itself is not small. For instance, if $H_2(s)-H_1(s)=\e^{2\ii\omega s}B$, with $B$ and $\sup_s\|H_2(s)\|$ bounded independently of $\omega$, then $\|S(s)\|\leq\|B\|/\omega$, and the difference between the two evolutions is of order $1/\omega$, uniformly for $t$ in compact intervals---although $\|H_2(s)-H_1(s)\|=\|B\|$ does not decrease with $\omega$.
	
	\subsection{The integration-by-parts formula in the unbounded case}
	
	We now return to the quantum Rabi and Jaynes--Cummings models. From now on, we set
	\begin{equation}
		U_1(t):=\e^{\ii H_0t}\e^{-\ii H_{\rm QR}t},\qquad H_1(t):=\tilde{H}_{\rm QR}(t),\qquad U_2(t):=\e^{\ii H_0t}\e^{-\ii H_{\rm JC}t},\qquad H_2(t):=\tilde{H}_{\rm JC}(t),
	\end{equation}
	i.e. the evolutions in the interaction picture and their generators (Eqs.~\eqref{eq:rwa-hqr-int}--\eqref{eq:rwa-hjc-int}). In this case, the first integration by parts, Eq.~\eqref{eq:rwa-ibp1}, is of no use, since the counter-rotating terms are not small:
	
	\begin{exercise}\label{exerc:rwa-noaverage}
		Show that $\left\|\left(\tilde{H}_{\rm QR}(t)-\tilde{H}_{\rm JC}(t)\right)\psi\right\|$ is independent of $\omega$ (and $t$) for all $\psi\in\mathbb{C}^2\odot\Dom\N^{1/2}$. Hint: compute the difference via Eqs.~\eqref{eq:rwa-hqr-int}--\eqref{eq:rwa-hjc-int}, write $\psi=e_+\otimes\phi_++e_-\otimes\phi_-$, and use the orthogonality of $e_+$ and $e_-$ to show that the phases drop out of the norm.
		\seesol{rwa-noaverage}
	\end{exercise}
	
	We thus resort to the second formula~\eqref{eq:rwa-ibp2}. Its derivation, however, assumed $H_1(t)$ and $H_2(t)$ to be bounded, which is not the case for $\tilde{H}_{\rm QR}(t)$ and $\tilde{H}_{\rm JC}(t)$. To prove a similar equation in our case, the right-hand side of Eq.~\eqref{eq:rwa-ibp2} should at least be well-defined. This is the case, for example, if $\psi\in\mathcal{D}_0$, with $\mathcal{D}_0$ a subspace such that, for all $s$,
	\begin{equation}\label{eq:rwa-conditions}
		\mathcal{D}_0\subseteq\Dom H_1(s)\cap\Dom H_2(s)\cap\Dom S(s),\quad U_2(s)\mathcal{D}_0\subseteq\mathcal{D}_0,\quad H_i(s)\mathcal{D}_0\subseteq\mathcal{D}_0,\quad S(s)\mathcal{D}_0\subseteq\mathcal{D}_0.
	\end{equation}
	We choose
	\begin{equation}
		\mathcal{D}_0:=\mathbb{C}^2\odot\hilb_{\rm fin},\qquad\hilb_{\rm fin}=\operatorname{Span}\{\phi_n\}_{n\in\mathbb{N}},
	\end{equation}
	the space of states with a finite number of excitations. Equivalently, $\mathcal{D}_0=\bigcup_{N\in\mathbb{N}}\mathcal{K}_N$, where $\mathcal{K}_N:=\bigoplus_{n=0}^N\mathcal{E}_n$ is a finite sum of eigenspaces of the total number of excitations $\N_{\rm tot}$ (Definition~\ref{def:ntot} and Proposition~\ref{prop:hfrakn}). We will show that the integration-by-parts formula~\eqref{eq:rwa-ibp2} holds on $\mathcal{D}_0$.
	
	\begin{lemma}[Second integration-by-parts formula]\label{lem:rwa-ibp}
		For all $\psi\in\mathcal{D}_0$ and $t\in\mathbb{R}$,
		\begin{equation}\label{eq:rwa-ibp}
			\ii\left(U_2(t)-U_1(t)\right)\psi=S(t)U_2(t)\psi+\ii\int_0^tU_1(t)U_1(s)^*\left(S(s)H_2(s)-H_1(s)S(s)\right)U_2(s)\psi\;\mathrm{d}s.
		\end{equation}
	\end{lemma}
	
	We first check the conditions~\eqref{eq:rwa-conditions}. Directly from the definitions, $H_0\mathcal{D}_0\subseteq\mathcal{D}_0$, $H_{\rm QR}\mathcal{D}_0\subseteq\mathcal{D}_0$ and $H_{\rm JC}\mathcal{D}_0\subseteq\mathcal{D}_0$. Since $\mathcal{D}_0$ is spanned by eigenvectors $e_\pm\otimes\phi_n$ of $H_0$, we also have $\e^{\ii H_0t}\mathcal{D}_0=\mathcal{D}_0$, and in particular $H_i(t)\mathcal{D}_0\subseteq\mathcal{D}_0$ for $i=1,2$ and all $t\in\mathbb{R}$. Furthermore, $S(t)$ can be computed explicitly:
	
	\begin{lemma}\label{lem:rwa-S}
		For all $t\in\mathbb{R}$, $\mathcal{D}_0\subseteq\Dom S(t)$ and, for all $\psi\in\mathcal{D}_0$,
		\begin{equation}\label{eq:rwa-S-explicit}
			S(t)\psi=-\frac{g}{\omega}\sin(\omega t)\left(\e^{\ii\omega t}\sigma_+\otimes a^\dag+\e^{-\ii\omega t}\sigma_-\otimes a\right)\psi.
		\end{equation}
		In particular, $S(t)\mathcal{D}_0\subseteq\mathcal{D}_0$; more precisely, $S(t)\mathcal{K}_N\subseteq\mathcal{K}_{N+2}$ for all $N\in\mathbb{N}$.
	\end{lemma}
	
	\begin{exercise}\label{exerc:rwa-S}
		Prove Lemma~\ref{lem:rwa-S}. Hint: compute $H_2(r)-H_1(r)$ via Eqs.~\eqref{eq:rwa-hqr-int}--\eqref{eq:rwa-hjc-int}, and use $\int_0^t\e^{\pm2\ii\omega r}\,\mathrm{d}r=\e^{\pm\ii\omega t}\sin(\omega t)/\omega$.
		\seesol{rwa-S}
	\end{exercise}
	
	When diagonalizing $H_{\rm JC}$, we have seen that $H_{\rm JC}$ leaves the eigenspaces $\mathcal{E}_n$ of $\N_{\rm tot}=I\otimes\N+\sigma_+\sigma_-\otimes I$ invariant (Proposition~\ref{prop:solving_jc1}). We can use this fact to prove the following:
	
	\begin{lemma}\label{lem:rwa-U2}
		For all $t\in\mathbb{R}$ and $N\in\mathbb{N}$, $U_2(t)\mathcal{K}_N=\mathcal{K}_N$. In particular, $U_2(t)\mathcal{D}_0\subseteq\mathcal{D}_0$.
	\end{lemma}
	
	\begin{proof}
		Each $\mathcal{E}_n$ is a finite-dimensional invariant subspace of the self-adjoint operator $H_{\rm JC}$, and is therefore spanned by eigenvectors of $H_{\rm JC}$; hence $\e^{-\ii H_{\rm JC}t}\mathcal{E}_n=\mathcal{E}_n$. Since $H_0=\omega\N_{\rm tot}-\frac{\omega}{2}$, also $\e^{\ii H_0t}\mathcal{E}_n=\mathcal{E}_n$. Therefore $U_2(t)\mathcal{E}_n=\mathcal{E}_n$ for all $n$, and the claim follows by linearity.
	\end{proof}
	
	Note that the same is not true for $U_1(t)$, since $H_{\rm QR}$ does not preserve the number of excitations. We have now shown all the properties~\eqref{eq:rwa-conditions} needed for the right-hand side of Eq.~\eqref{eq:rwa-ibp} to be well-defined. To reproduce the argument of the bounded case, we still need the following:
	
	\begin{lemma}[Strong derivatives in the interaction picture]\label{lem:rwa-derivatives}
		For all $\psi\in\mathcal{D}_0$ and $t\in\mathbb{R}$,
		\begin{equation}\label{eq:rwa-der-U2}
			\frac{\mathrm{d}}{\mathrm{d}t}U_2(t)\psi=-\ii H_2(t)U_2(t)\psi
		\end{equation}
		and
		\begin{equation}\label{eq:rwa-der-Ustar}
			\frac{\mathrm{d}}{\mathrm{d}t}U_i(t)^*\psi=\ii U_i(t)^*H_i(t)\psi,\qquad i=1,2.
		\end{equation}
	\end{lemma}
	
	Note that $U_i(t)^*\neq U_i(-t)$, as $(U_i(t))_{t\in\mathbb{R}}$ is not a group; so Eq.~\eqref{eq:rwa-der-Ustar} does not trivially follow from Eq.~\eqref{eq:rwa-der-U2}.
	
	\begin{proof}
		The idea is to follow the proof of the usual product rule for differentiable functions, being careful with the fact that the individual strong derivatives are unbounded.
		
		\emph{Proof of Eq.~\eqref{eq:rwa-der-U2}.} Let $\psi_t:=\e^{-\ii tH_{\rm JC}}\psi$. By Lemma~\ref{lem:rwa-U2} (and $\e^{\ii H_0t}\mathcal{D}_0=\mathcal{D}_0$), $\psi_t\in\mathcal{D}_0\subseteq\Dom H_{\rm JC}\cap\Dom H_0$. By Stone's theorem (Theorem~\ref{thm:stone}),
		\begin{equation}
			\lim_{s\to0}\frac{\e^{-\ii sH_{\rm JC}}\psi_t-\psi_t}{s}=-\ii H_{\rm JC}\psi_t,\qquad\lim_{s\to0}\frac{\e^{\ii sH_0}\psi_t-\psi_t}{s}=\ii H_0\psi_t.
		\end{equation}
		Since $\e^{\ii sH_0}$ is unitary and $\e^{\ii sH_0}\phi\to\phi$ as $s\to0$ for all $\phi$,
		\begin{align}
			\left\|\e^{\ii sH_0}\frac{\e^{-\ii sH_{\rm JC}}\psi_t-\psi_t}{s}+\ii H_{\rm JC}\psi_t\right\|&\leq\left\|\left(\e^{\ii sH_0}-I\right)\ii H_{\rm JC}\psi_t\right\|+\left\|\frac{\e^{-\ii sH_{\rm JC}}\psi_t-\psi_t}{s}+\ii H_{\rm JC}\psi_t\right\|\to0.
		\end{align}
		Therefore, adding and subtracting $\e^{\ii H_0s}\psi_t$,
		\begin{align}
			\frac{U_2(t+s)\psi-U_2(t)\psi}{s}&=\e^{\ii H_0t}\left(\e^{\ii H_0s}\frac{\e^{-\ii sH_{\rm JC}}\psi_t-\psi_t}{s}+\frac{\e^{\ii H_0s}\psi_t-\psi_t}{s}\right)\to\e^{\ii H_0t}\left(\ii H_0-\ii H_{\rm JC}\right)\psi_t\nonumber\\
			&=-\ii\,\e^{\ii H_0t}(H_{\rm JC}-H_0)\e^{-\ii H_0t}\,\e^{\ii H_0t}\e^{-\ii H_{\rm JC}t}\psi=-\ii H_2(t)U_2(t)\psi.
		\end{align}
		
		\emph{Proof of Eq.~\eqref{eq:rwa-der-Ustar}.} Let $H$ be either $H_{\rm QR}$ or $H_{\rm JC}$, for $i=1$ or $i=2$ respectively, and $\psi^0_t:=\e^{-\ii H_0t}\psi\in\mathcal{D}_0\subseteq\Dom H\cap\Dom H_0$. By Stone's theorem (Theorem~\ref{thm:stone}),
		\begin{equation}
			\lim_{s\to0}\frac{\e^{\ii sH}\psi^0_t-\psi^0_t}{s}=\ii H\psi^0_t,\qquad\lim_{s\to0}\frac{\e^{-\ii sH_0}\psi^0_t-\psi^0_t}{s}=-\ii H_0\psi^0_t,
		\end{equation}
		and, exactly as before, $\e^{\ii sH}\left(\e^{-\ii sH_0}\psi^0_t-\psi^0_t\right)/s\to-\ii H_0\psi^0_t$. Therefore, adding and subtracting $\e^{\ii Hs}\psi^0_t$,
		\begin{align}
			\frac{U_i(t+s)^*\psi-U_i(t)^*\psi}{s}&=\e^{\ii Ht}\left(\e^{\ii Hs}\frac{\e^{-\ii H_0s}\psi^0_t-\psi^0_t}{s}+\frac{\e^{\ii Hs}\psi^0_t-\psi^0_t}{s}\right)\to\e^{\ii Ht}\left(\ii H-\ii H_0\right)\psi^0_t\nonumber\\
			&=\ii\,\e^{\ii Ht}\e^{-\ii H_0t}\,\e^{\ii H_0t}(H-H_0)\e^{-\ii H_0t}\psi=\ii U_i(t)^*H_i(t)\psi.\qedhere
		\end{align}
	\end{proof}
	
	We can finally prove the second integration-by-parts formula.
	
	\begin{proof}[Proof of Lemma~\ref{lem:rwa-ibp}]
		Let $\psi\in\mathcal{D}_0$. One has
		\begin{equation}\label{eq:rwa-prod1}
			\frac{\mathrm{d}}{\mathrm{d}s}\left(U_1(s)^*U_2(s)\right)\psi=\ii\,U_1(s)^*\left(H_1(s)-H_2(s)\right)U_2(s)\psi
		\end{equation}
		and
		\begin{align}
			\frac{\mathrm{d}}{\mathrm{d}s}\left(U_1(s)^*S(s)U_2(s)\right)\psi&=\ii\,U_1(s)^*\left(H_1(s)S(s)-S(s)H_2(s)\right)U_2(s)\psi\nonumber\\
			&\quad+U_1(s)^*\left(H_2(s)-H_1(s)\right)U_2(s)\psi,\label{eq:rwa-prod2}
		\end{align}
		both derivatives being continuous in $s$ (Exercise~\ref{exerc:rwa-derivatives}). Combining Eqs.~\eqref{eq:rwa-prod1}--\eqref{eq:rwa-prod2},
		\begin{equation}
			\ii\frac{\mathrm{d}}{\mathrm{d}s}\left(U_1(s)^*U_2(s)\right)\psi=\frac{\mathrm{d}}{\mathrm{d}s}\left(U_1(s)^*S(s)U_2(s)\right)\psi-\ii\,U_1(s)^*\left(H_1(s)S(s)-S(s)H_2(s)\right)U_2(s)\psi.
		\end{equation}
		Then, by the fundamental theorem of calculus, and since $U_1(0)=U_2(0)=I$ and $S(0)=0$,
		\begin{align}
			\ii\left(U_2(t)-U_1(t)\right)\psi&=\ii U_1(t)\int_0^t\frac{\mathrm{d}}{\mathrm{d}s}\left(U_1(s)^*U_2(s)\right)\psi\;\mathrm{d}s\nonumber\\
			&=U_1(t)\int_0^t\frac{\mathrm{d}}{\mathrm{d}s}\left(U_1(s)^*S(s)U_2(s)\right)\psi\;\mathrm{d}s\nonumber\\
			&\quad-\ii\int_0^tU_1(t)U_1(s)^*\left(H_1(s)S(s)-S(s)H_2(s)\right)U_2(s)\psi\;\mathrm{d}s\nonumber\\
			&=S(t)U_2(t)\psi+\ii\int_0^tU_1(t)U_1(s)^*\left(S(s)H_2(s)-H_1(s)S(s)\right)U_2(s)\psi\;\mathrm{d}s.\qedhere
		\end{align}
	\end{proof}
	
	\begin{exercise}\label{exerc:rwa-derivatives}
		Prove Eqs.~\eqref{eq:rwa-prod1}--\eqref{eq:rwa-prod2}, and the continuity in $s$ of both right-hand sides. Hint: recall that $U_2(s)\mathcal{K}_N=\mathcal{K}_N$ and that $H_1(s)$, $H_2(s)$, $S(s)$ map $\mathcal{K}_N$ into $\mathcal{K}_{N+2}$, all these spaces being finite-dimensional; then add and subtract suitable terms, as in the proof of Lemma~\ref{lem:rwa-derivatives}.
		\seesol{rwa-derivatives}
	\end{exercise}
	
	\subsection{An explicit error bound}
	
	We can now apply the integration-by-parts formula to our original problem. We prove the convergence~\eqref{eq:rwa-goal} by first establishing an explicit error bound, valid on the dense subspace $\mathcal{D}_0$. In the following, we write $\hat{\N}:=I\otimes\N$, and leave all trivial tensor products with the identity understood.
	
	\begin{theorem}[Upper error bound for the RWA]\label{thm:rwa-bound}\index{rotating-wave approximation!error bound}
		For all $\psi\in\mathcal{D}_0$ and $t\in\mathbb{R}$,
		\begin{multline}\label{eq:rwa-bound}
			\left\|\left(\e^{-\ii tH_{\rm QR}}-\e^{-\ii tH_{\rm JC}}\right)\psi\right\|\\\leq\frac{g}{\omega}\Bigl[|\sin\omega t|\left\|(\hat{\N}+2)^{1/2}\psi\right\|+|t|\left(|\Delta|\left\|(\hat{\N}+2)^{1/2}\psi\right\|+6g\left\|(\hat{\N}+3)\psi\right\|\right)\Bigr],
		\end{multline}
		where, without loss of generality, $g,\omega>0$.
	\end{theorem}
	
	We need a preliminary estimate. Recall that, for symmetric operators $A,B$, one writes $A\geq0$ if $\braket{\psi,A\psi}\geq0$ for all $\psi\in\Dom A$, and $A\leq B$ if $B-A\geq0$.
	
	\begin{lemma}\label{lem:rwa-number}
		Let $f:[0,\infty)\rightarrow[0,\infty)$ be nondecreasing. Then, for all $\psi\in\mathcal{D}_0$ and $t\in\mathbb{R}$,
		\begin{equation}
			\left\|f(\hat{\N})U_2(t)\psi\right\|\leq\left\|f(\hat{\N}+1)\psi\right\|.
		\end{equation}
		In particular, taking $f(x)=\sqrt{x}$, $U_2(t)^*\hat{\N}U_2(t)\leq\hat{\N}+I$ on $\mathcal{D}_0$.
	\end{lemma}
	
	\begin{proof}
		The operators $\hat{\N}$ and $\N_{\rm tot}=\hat{\N}+\sigma_+\sigma_-$ are diagonal in the orthonormal basis $(e_\pm\otimes\phi_m)_{m\in\mathbb{N}}$, with eigenvalues respectively $m$ and $m+\delta_{\pm,+}$. Since $0\leq\sigma_+\sigma_-\leq I$, the corresponding eigenvalues satisfy $m\leq m+\delta_{\pm,+}\leq m+1$, whence, $f$ being nondecreasing and nonnegative,
		\begin{equation}
			\left\|f(\hat{\N})\phi\right\|\leq\left\|f(\N_{\rm tot})\phi\right\|\leq\left\|f(\hat{\N}+1)\phi\right\|\qquad\text{for all }\phi\in\mathcal{D}_0.
		\end{equation}
		Besides, $f(\N_{\rm tot})$ acts as the multiple $f(n)$ of the identity on each $\mathcal{E}_n$, which is invariant under $U_2(t)$ (Lemma~\ref{lem:rwa-U2}); therefore $f(\N_{\rm tot})U_2(t)\psi=U_2(t)f(\N_{\rm tot})\psi$. Since $U_2(t)$ is unitary,
		\begin{equation}
			\left\|f(\hat{\N})U_2(t)\psi\right\|\leq\left\|f(\N_{\rm tot})U_2(t)\psi\right\|=\left\|f(\N_{\rm tot})\psi\right\|\leq\left\|f(\hat{\N}+1)\psi\right\|.\qedhere
		\end{equation}
	\end{proof}
	
	\begin{proof}[Proof of Theorem~\ref{thm:rwa-bound}]
		Let $\psi\in\mathcal{D}_0$, and recall that $\e^{-\ii tH_{\rm QR}}-\e^{-\ii tH_{\rm JC}}=\e^{-\ii H_0t}(U_1(t)-U_2(t))$. By Lemma~\ref{lem:rwa-ibp} and unitarity,
		\begin{equation}
			\left\|\left(\e^{-\ii tH_{\rm QR}}-\e^{-\ii tH_{\rm JC}}\right)\psi\right\|\leq\left\|S(t)U_2(t)\psi\right\|+\left|\int_0^t\left\|\left(S(s)H_2(s)-H_1(s)S(s)\right)U_2(s)\psi\right\|\mathrm{d}s\right|.
		\end{equation}
		We deal with both terms separately.
		
		\emph{First term.} Let $\phi:=U_2(t)\psi\in\mathcal{D}_0$. By Lemma~\ref{lem:rwa-S}, and using $\sigma_\pm^2=0$,
		\begin{align}
			\left\|S(t)\phi\right\|^2&=\frac{g^2}{\omega^2}\sin^2(\omega t)\Braket{\phi,\left(\e^{-\ii\omega t}\sigma_-\otimes a+\e^{\ii\omega t}\sigma_+\otimes a^\dag\right)\left(\e^{\ii\omega t}\sigma_+\otimes a^\dag+\e^{-\ii\omega t}\sigma_-\otimes a\right)\phi}\nonumber\\
			&=\frac{g^2}{\omega^2}\sin^2(\omega t)\Braket{\phi,\left(\sigma_-\sigma_+\otimes aa^\dag+\sigma_+\sigma_-\otimes a^\dag a\right)\phi}.
		\end{align}
		Using $a^\dag a=\N\leq\N+I$, $aa^\dag=\N+I$, and $\sigma_+\sigma_-+\sigma_-\sigma_+=I$, we get
		\begin{equation}
			\|S(t)\phi\|^2\leq\frac{g^2}{\omega^2}\sin^2(\omega t)\braket{\phi,(\hat{\N}+I)\phi},
		\end{equation}
		whence, by Lemma~\ref{lem:rwa-number},
		\begin{equation}
			\left\|S(t)U_2(t)\psi\right\|\leq\frac{g}{\omega}|\sin\omega t|\left\|(\hat{\N}+1)^{1/2}U_2(t)\psi\right\|\leq\frac{g}{\omega}|\sin\omega t|\left\|(\hat{\N}+2)^{1/2}\psi\right\|.
		\end{equation}
		
		\emph{Second term.} We rewrite the integrand to make use of possible cancellations. Using $[\sigma_\pm,\sigma_z]=\mp2\sigma_\pm$ and $\sigma_\pm^2=0$, a direct computation gives
		\begin{align}
			S(s)H_2(s)-H_1(s)S(s)&=[S(s),H_2(s)]+\left(H_2(s)-H_1(s)\right)S(s)\nonumber\\
			&=-\frac{g}{\omega}\sin(\omega s)\Bigl[\Delta\left(-\e^{\ii\omega s}\sigma_+\otimes a^\dag+\e^{-\ii\omega s}\sigma_-\otimes a\right)\nonumber\\
			&\qquad+g\,\sigma_+\sigma_-\otimes\left(\e^{\ii\omega s}(a^\dag)^2-\e^{-\ii\omega s}a^2-\e^{\ii\omega s}a^\dag a\right)\nonumber\\
			&\qquad+g\,\sigma_-\sigma_+\otimes\left(-\e^{\ii\omega s}(a^\dag)^2+\e^{-\ii\omega s}a^2-\e^{-\ii\omega s}aa^\dag\right)\Bigr]
		\end{align}
		on $\mathcal{D}_0$. We estimate the three terms on $\phi:=U_2(s)\psi\in\mathcal{D}_0$. For the first term, as in the estimate above,
		\begin{equation}
			\left\|\left(-\e^{\ii\omega s}\sigma_+\otimes a^\dag+\e^{-\ii\omega s}\sigma_-\otimes a\right)\phi\right\|^2=\Braket{\phi,\left(\sigma_-\sigma_+\otimes aa^\dag+\sigma_+\sigma_-\otimes a^\dag a\right)\phi}\leq\left\|(\hat{\N}+1)^{1/2}\phi\right\|^2.
		\end{equation}
		As for the other two terms, since $\|\sigma_\pm\sigma_\mp\|=1$, by the estimate on monomials of ladder operators (Proposition~\ref{prop:asharpvsn}) and $\sqrt{(n+1)(n+2)}\leq n+2$,
		\begin{equation}
			\left\|\sigma_+\sigma_-\otimes\left(\e^{\ii\omega s}(a^\dag)^2-\e^{-\ii\omega s}a^2-\e^{\ii\omega s}a^\dag a\right)\phi\right\|\leq3\left\|(\hat{\N}+1)^{1/2}(\hat{\N}+2)^{1/2}\phi\right\|\leq3\left\|(\hat{\N}+2)\phi\right\|,
		\end{equation}
		and likewise for the term with $\sigma_-\sigma_+$. Altogether, using Lemma~\ref{lem:rwa-number} once with $f(x)=(x+1)^{1/2}$ and once with $f(x)=x+2$, and $|\sin\omega s|\leq1$,
		\begin{align}
			\left\|\left(S(s)H_2(s)-H_1(s)S(s)\right)U_2(s)\psi\right\|&\leq\frac{g}{\omega}\left(|\Delta|\left\|(\hat{\N}+1)^{1/2}U_2(s)\psi\right\|+6g\left\|(\hat{\N}+2)U_2(s)\psi\right\|\right)\nonumber\\
			&\leq\frac{g}{\omega}\left(|\Delta|\left\|(\hat{\N}+2)^{1/2}\psi\right\|+6g\left\|(\hat{\N}+3)\psi\right\|\right).
		\end{align}
		Integrating over $s$ between $0$ and $t$ and combining with the estimate of the first term yields Eq.~\eqref{eq:rwa-bound}.
	\end{proof}
	
	The upper bound~\eqref{eq:rwa-bound} immediately implies the convergence~\eqref{eq:rwa-goal} on $\mathcal{D}_0$; by density, it extends to the whole Hilbert space:
	
	\begin{corollary}[Convergence of the rotating-wave approximation]\label{coroll:rwa}\index{rotating-wave approximation!convergence}
		For all $\psi\in\hfrak=\mathbb{C}^2\otimes\hilb$,
		\begin{equation}
			\lim_{\omega\to\infty}\left\|\left(\e^{-\ii tH_{\rm QR}}-\e^{-\ii tH_{\rm JC}}\right)\psi\right\|=0
		\end{equation}
		at fixed $g$ and $\Delta$, uniformly for $t$ in compact intervals.
	\end{corollary}
	
	\begin{proof}
		Fix $T>0$. For $\psi\in\mathcal{D}_0$, the claim follows from Eq.~\eqref{eq:rwa-bound}, whose right-hand side vanishes as $\omega\to\infty$, uniformly for $|t|\leq T$. Now let $\psi\in\hfrak$ and $\varepsilon>0$. As $\mathcal{D}_0$ is dense, there is $\psi_\varepsilon\in\mathcal{D}_0$ with $\|\psi-\psi_\varepsilon\|<\varepsilon/4$, and $\omega_\varepsilon>0$ such that $\|(\e^{-\ii tH_{\rm QR}}-\e^{-\ii tH_{\rm JC}})\psi_\varepsilon\|<\varepsilon/2$ for all $\omega\geq\omega_\varepsilon$ and $|t|\leq T$. Then, since $\e^{-\ii tH_{\rm QR}}-\e^{-\ii tH_{\rm JC}}$ has norm at most $2$, for all $\omega\geq\omega_\varepsilon$ and $|t|\leq T$,
		\begin{equation}
			\left\|\left(\e^{-\ii tH_{\rm QR}}-\e^{-\ii tH_{\rm JC}}\right)\psi\right\|\leq\left\|\left(\e^{-\ii tH_{\rm QR}}-\e^{-\ii tH_{\rm JC}}\right)\psi_\varepsilon\right\|+2\|\psi-\psi_\varepsilon\|<\varepsilon.\qedhere
		\end{equation}
	\end{proof}
	
	In other words, the Jaynes--Cummings model approximates the quantum Rabi model in the \emph{strong dynamical sense}, in the limit $\omega\to\infty$ of large frequencies at fixed $g$ and $\Delta$. Besides, for initial states $\psi\in\mathcal{D}_0$ with a finite number of excitations, Eq.~\eqref{eq:rwa-bound} provides an explicit upper bound on the error, which scales as $g/\omega$ and grows at most linearly in time. This bound essentially reproduces the one obtained in~\cite{burgarth}, whose main results this appendix summarizes; its linear growth in time is not sharp.
	
	\addtocontents{toc}{\protect\setcounter{tocdepth}{2}}
	
	\section{Solutions to the exercises}\label{app:solutions}
	
	The solutions are grouped according to the section in which the corresponding exercise appears.
	
	\subsection{Direct sums and tensor products \texorpdfstring{(Section~\ref{sec:directsum})}{}}\label{solsec:sec:directsum}
	
	\solsec{directsumbases}{Direct sum of separable Hilbert spaces}
	
	Let us write
	\begin{equation}
		\varepsilon_1^{(k)}:=(e_1^{(k)},0),\qquad\varepsilon_2^{(k)}:=(0,e_2^{(k)}),\qquad k\in\mathbb{N}.
	\end{equation}
	The countable family $(\varepsilon_j^{(k)})_{k\in\mathbb{N},\,j=1,2}$ is clearly orthonormal:
	\begin{equation}
		\Braket{\varepsilon_j^{(k)},\varepsilon_j^{(h)}}_{\hilb_1\oplus\hilb_2}=\Braket{e_j^{(k)},e_j^{(h)}}_{\hilb_j}=\delta_{kh}\quad(j=1,2),\qquad\Braket{\varepsilon_1^{(k)},\varepsilon_2^{(h)}}_{\hilb_1\oplus\hilb_2}=0.
	\end{equation}
	To prove that it is complete, consider an arbitrary $(u_1,u_2)\in\hilb_1\oplus\hilb_2$. Since $(e_j^{(k)})_{k\in\mathbb{N}}$ is an orthonormal basis of $\hilb_j$,
	\begin{equation}
		u_j=\sum_{k\in\mathbb{N}}\braket{e_j^{(k)},u_j}_{\hilb_j}e_j^{(k)},\qquad j=1,2;
	\end{equation}
	but clearly $\braket{e_j^{(k)},u_j}_{\hilb_j}=\braket{\varepsilon_j^{(k)},(u_1,u_2)}_{\hilb_1\oplus\hilb_2}$, whence
	\begin{equation}
		(u_1,u_2)=(u_1,0)+(0,u_2)=\sum_{k\in\mathbb{N}}\sum_{j=1,2}\braket{\varepsilon_j^{(k)},(u_1,u_2)}_{\hilb_1\oplus\hilb_2}\,\varepsilon_j^{(k)}.
	\end{equation}
	
	\solsec{hfin}{$\hilb_{\rm fin}$ is dense in a countable direct sum}
	
	Let $u=(u_n)_{n\in\mathbb{N}}\in\oplus_{n\in\mathbb{N}}\hilb_n$, and let $u^{(k)}\in\hilb_{\rm fin}$ be defined as in the hint, i.e. the sequence obtained from $u$ by replacing all entries with $n>k$ by zero. Then
	\begin{equation}
		\|u - u^{(k)}\|^2
		= \sum_{n\in\mathbb{N}} \|u_n - u^{(k)}_n\|^2
		= \sum_{n>k} \|u_n\|^2
		\to 0
	\end{equation}
	as $k\to \infty$, being the tail of the convergent series $\sum_{n\in\mathbb{N}}\|u_n\|^2=\|u\|^2$.
	
	\solsec{directsumesssa}{Direct sums of (essentially) self-adjoint operators}
	
	By Proposition~\ref{prop:adjoint_directsum}, $A^* = \oplus_{n\in\mathbb{N}}A_n^*$. Hence, if each $A_n$ is self-adjoint, then
	\begin{equation}
		A^* = \bigoplus_{n\in\mathbb{N}}A_n^* = \bigoplus_{n\in\mathbb{N}}A_n = A,
	\end{equation}
	i.e. $A$ is self-adjoint. In the general case, each $A_n$ is symmetric, $A_n\subseteq A_n^*$, and hence $A$ is symmetric as well, since $A=\oplus_nA_n\subseteq\oplus_nA_n^* = A^*$.
	
	Let $A_0$ be the restriction of $A$ to $\Dom A \cap \hilb_{\rm fin}$, which is symmetric. We will show that $A_0$ is essentially self-adjoint. This proves both remaining claims: $\Dom A\cap\hilb_{\rm fin}$ is a core for $A$, and $A$---being a symmetric extension of the essentially self-adjoint operator $A_0$---is essentially self-adjoint, with $\overline{A}=\overline{A_0}$ (indeed, $A_0\subseteq A\subseteq A^*\subseteq A_0^*=\overline{A_0}$).	
	By Theorem~\ref{thm:criterion-esssa}, it suffices to show that $\overline{\Ran(A_0\pm \ii)} = \hilb$. Let $v = (v_n)_{n\in\mathbb{N}}\in\hilb$, and $\epsilon>0$. Choose $N \in \mathbb{N}$ such that
	\begin{equation}
		\sum_{n>N} \|v_n\|^2 < \frac{\epsilon}{2}.
	\end{equation}
	As each $A_n$ is essentially self-adjoint, $\Ran(A_n\pm \ii)$ is dense in $\hilb_n$; thus, for each $n\leq N$, there is $u_n \in \Dom A_n$ such that
	\begin{equation}
		\|v_n - (A_n \pm \ii)u_n\|^2 < \frac{\epsilon}{2(N+1)}.
	\end{equation}
	Let $u=(u_0,\dots,u_N,0,0,\ldots)$. By construction, $u\in\Dom A_0$, and
	\begin{equation}
		\|v - (A_0\pm\ii)u\|^2
		= \sum_{n=0}^N \|v_n - (A_n\pm \ii)u_n\|^2 + \sum_{n>N} \|v_n\|^2
		< \sum_{n=0}^{N} \frac{\epsilon}{2(N+1)} + \frac{\epsilon}{2}
		= \epsilon.
	\end{equation}
	Hence $v \in \overline{\Ran(A_0 \pm \ii)}$. As $v$ was arbitrary, $\overline{\Ran(A_0\pm \ii)}=\hilb$, and $A_0$ is essentially self-adjoint.
	
	\solsec{uniformly}{Direct sum of uniformly bounded operators}
	
	Let $A = \oplus_{n\in\mathbb{N}}A_n$ and $v = (v_n)_{n\in\mathbb{N}} \in \hilb$. Since $\|A_n v_n\| \leq \|A_n\| \|v_n\| \leq M \|v_n\|$ for all $n$,
	\begin{equation}
		\sum_{n\in\mathbb{N}} \|A_n v_n\|^2 \leq M^2 \sum_{n\in\mathbb{N}} \|v_n\|^2 = M^2 \|v\|^2<\infty.
	\end{equation}
	This shows that $\Dom A=\hilb$ and $\|Av\|\leq M\|v\|$ for all $v\in\hilb$, i.e. $A$ is bounded with $\|A\|\leq M$.
	
	\solsec{tensorfinitedim}{The algebraic tensor product with a finite-dimensional space}
	
	Let $(e_j)_{j=1,\dots,d}$ be an orthonormal basis of $\hilb_1$. Any $\Psi\in\hilb_1\odot\hilb_2$ is of the form
	\begin{equation}
		\Psi=\sum_{r=1}^n v_r\otimes w_r,\qquad v_1,\dots,v_n\in\hilb_1,\;w_1,\dots,w_n\in\hilb_2.
	\end{equation}
	Expanding each $v_r=\sum_{j=1}^d\braket{e_j,v_r}e_j$ and using the bilinearity of the tensor product,
	\begin{equation}
		\Psi=\sum_{r=1}^n \sum_{j=1}^d \braket{e_j,v_r}e_j\otimes w_r=\sum_{j=1}^d e_j\otimes \underbrace{\left(\sum_{r=1}^n \braket{e_j,v_r}w_r\right)}_{=:u_j}.
	\end{equation}
	So any $\Psi\in\hilb_1\odot\hilb_2$ can be written as $\sum_{j=1}^de_j\otimes u_j$. This representation is unique: if $\sum_{j=1}^de_j\otimes u_j=\sum_{j=1}^de_j\otimes u'_j$, then $\Phi:=\sum_{j=1}^de_j\otimes(u_j-u'_j)=0$,
	i.e. $\Phi$ is the null form on $\hilb_1\times\hilb_2$. In particular, for every $w\in\hilb_2$ and every $k=1,\dots,d$,
	\begin{equation}
		0=\Phi(e_k,w)=\sum_{j=1}^d\braket{e_k,e_j}_{\hilb_1}\braket{w,u_j-u'_j}_{\hilb_2}=\braket{w,u_k-u'_k}_{\hilb_2},
	\end{equation}
	so that $u_k-u'_k$ is orthogonal to all of $\hilb_2$, i.e. $u_k=u'_k$.
	
	Because of uniqueness, the map
	\begin{equation}
		U:\hilb_1\odot\hilb_2\rightarrow(\hilb_2)^{\oplus d},\qquad U\left(\sum_{j=1}^de_j\otimes u_j\right):=(u_1,\dots,u_d)
	\end{equation}
	is well-defined, and it is clearly linear. It is an isometry:
	\begin{equation}
		\|\Psi\|^2=\sum_{j,k=1}^d\braket{e_j,e_k}_{\hilb_1}\braket{u_j,u_k}_{\hilb_2}=\sum_{j=1}^d\|u_j\|^2_{\hilb_2}=\|(u_1,\dots,u_d)\|^2_{(\hilb_2)^{\oplus d}}.
	\end{equation}
	Finally, it is surjective: given any $(u_1,\dots,u_d)\in(\hilb_2)^{\oplus d}$, the vector $\Psi:=\sum_{j=1}^de_j\otimes u_j$ satisfies $U\Psi=(u_1,\dots,u_d)$. Therefore $U$ is unitary. In particular, $\hilb_1\odot\hilb_2$ is complete, being isometrically isomorphic to the Hilbert space $(\hilb_2)^{\oplus d}$.
	
	\solsec{tensorbases}{Tensor product of separable spaces}
	
	The family $(e_{1h}\otimes e_{2k})_{h,k\in\mathbb{N}}$ is orthonormal:
	\begin{equation}
		\Braket{e_{1h}\otimes e_{2k},e_{1h'}\otimes e_{2k'}}=\braket{e_{1h},e_{1h'}}_{\hilb_1}\braket{e_{2k},e_{2k'}}_{\hilb_2}=\delta_{hh'}\delta_{kk'}.
	\end{equation}
	We are left with proving that its span is dense in $\hilb_1\otimes\hilb_2$. As $\hilb_1\odot\hilb_2$ is dense in $\hilb_1\otimes\hilb_2$ by construction, it suffices to show that every $\Psi\in\hilb_1\odot\hilb_2$ can be approximated arbitrarily well by elements of $\operatorname{Span}(e_{1h}\otimes e_{2k})_{h,k\in\mathbb{N}}$. By linearity, it suffices to do so for $\Psi=v\otimes w$. Let
	\begin{equation}
		v^{(N)}:=\sum_{h=0}^N\braket{e_{1h},v}e_{1h},\qquad w^{(N)}:=\sum_{k=0}^N\braket{e_{2k},w}e_{2k},
	\end{equation}
	so that $v^{(N)}\otimes w^{(N)}\in\operatorname{Span}(e_{1h}\otimes e_{2k})_{h,k\in\mathbb{N}}$, and $v^{(N)}\to v$, $w^{(N)}\to w$ as $N\to\infty$. Since $\|x\otimes y\|=\|x\|\|y\|$,
	\begin{equation}
		\|v\otimes w-v^{(N)}\otimes w^{(N)}\|\leq\|v-v^{(N)}\|\|w\|+\|v^{(N)}\|\|w-w^{(N)}\|\to0,
	\end{equation}
	which proves the claim.
	
	\solsec{denseindense}{The algebraic tensor product of dense subspaces is dense}
	
	We will show that $\mathcal{V}_1 \odot \mathcal{V}_2$ is dense in $\hilb_1 \odot \hilb_2$ (with respect to the norm of $\hilb_1\otimes\hilb_2$). As $\hilb_1 \odot \hilb_2$ is dense in $\hilb_1 \otimes \hilb_2$, the claim follows.
	
	Let $\psi \in \hilb_1 \odot \hilb_2$, i.e.
	\begin{equation}
		\psi = \sum_{j=1}^N u_j \otimes v_j
	\end{equation}
	for some $u_j \in \hilb_1$ and $v_j \in \hilb_2$. As $\mathcal{V}_1$ and $\mathcal{V}_2$ are dense, there are sequences $(u_j^{(n)})_{n\in \mathbb{N} } \subset \mathcal{V}_1$ and $(v_j^{(n)})_{n\in \mathbb{N}} \subset \mathcal{V}_2$ with $u_j^{(n)} \to u_j$ and $v_j^{(n)} \to v_j$. Let $\psi^{(n)} := \sum_{j=1}^N u_j^{(n)} \otimes v_j^{(n)}\in \mathcal{V}_1 \odot \mathcal{V}_2$. Writing $u_j \otimes v_j - u_j^{(n)} \otimes v_j^{(n)}=u_j\otimes(v_j-v_j^{(n)})+(u_j-u_j^{(n)})\otimes v_j^{(n)}$ and using $\|x\otimes y\|=\|x\|\|y\|$, we get
	\begin{align}
		\|\psi - \psi^{(n)}\|
		&\leq \sum_{j=1}^{N} \left( \|u_j\| \|v_j - v_j^{(n)} \| + \| u_j - u_j^{(n)} \| \|v_j^{(n)}\| \right)\nonumber
		\\
		&\leq \sum_{j=1}^{N} \left( \|u_j\| \|v_j - v_j^{(n)} \| + \| u_j - u_j^{(n)} \| \left(\|v_j^{(n)} - v_j\|+\|v_j\| \right) \right)\to 0,
	\end{align}
	which proves the claim.
	
	\solsec{finitedim2}{Tensor products of operators with a finite-dimensional factor}
	
	Let $V:\hilb_1\odot\hilb_2\rightarrow(\hilb_2)^{\oplus d}$ be the unitary operator $V(\sum_{j=1}^de_j\otimes\psi_j)=(\psi_1,\dots,\psi_d)$ of Proposition~\ref{prop:tensorproduct_finitedim} (cf. the solution of Exercise~\ref{exerc:tensorfinitedim}). Expanding the first factor of each pure tensor in the basis $(e_j)_{j=1,\dots,d}$, one sees that $\Dom(A\odot B)=\hilb_1\odot\Dom B$ consists precisely of the vectors $\sum_{j=1}^de_j\otimes\psi_j$ with $\psi_1,\dots,\psi_d\in\Dom B$, so that $V\Dom(A\odot B)=(\Dom B)^{\oplus d}$. On such vectors,
	\begin{equation}
		(A\odot B)\sum_{j=1}^de_j\otimes\psi_j=\sum_{j=1}^dAe_j\otimes B\psi_j=\sum_{j,k=1}^da_{kj}\,e_k\otimes B\psi_j=\sum_{k=1}^de_k\otimes\left(\sum_{j=1}^da_{kj}B\psi_j\right).
	\end{equation}
	Therefore $V(A\odot B)V^*=T$, where $T$ is the operator in Eq.~\eqref{eq:mapped}. Note that $T=MB^{\oplus d}$, where $B^{\oplus d}:=B\oplus\cdots\oplus B$ is defined on $(\Dom B)^{\oplus d}$, and $M\in\mathcal{B}((\hilb_2)^{\oplus d})$ is the bounded operator $M(\chi_1,\dots,\chi_d):=(\sum_ja_{1j}\chi_j,\dots,\sum_ja_{dj}\chi_j)$. Since unitary equivalence preserves closures, $A\otimes B=\overline{A\odot B}$ is unitarily equivalent to $\overline{T}$, while $A\odot\overline{B}$ is unitarily equivalent to $M\overline{B}^{\oplus d}$, i.e. the operator in Eq.~\eqref{eq:mapped2}.
	
	\emph{Inclusion.} Let $\psi=(\psi_1,\dots,\psi_d)\in(\Dom\overline{B})^{\oplus d}$. By definition of closure, for each $j$ there is a sequence $(\psi_j^{(n)})_{n\in\mathbb{N}}\subset\Dom B$ with $\psi_j^{(n)}\to\psi_j$ and $B\psi_j^{(n)}\to\overline{B}\psi_j$. Setting $\psi^{(n)}:=(\psi_1^{(n)},\dots,\psi_d^{(n)})$, we have $\psi^{(n)}\to\psi$ and, since $M$ is bounded,
	\begin{equation}
		T\psi^{(n)}=MB^{\oplus d}\psi^{(n)}\to M\overline{B}^{\oplus d}\psi.
	\end{equation}
	Hence $\psi\in\Dom\overline{T}$ and $\overline{T}\psi=M\overline{B}^{\oplus d}\psi$, i.e. $M\overline{B}^{\oplus d}\subseteq\overline{T}$, which translates into $A\odot\overline{B}\subseteq A\otimes B$.
	
	\emph{Equality for invertible $A$.} If $A$ is invertible, so is the matrix $(a_{jk})_{j,k}$, and thus $M$ has a bounded inverse. We claim that $M\overline{B}^{\oplus d}$ is then closed. Indeed, let $\psi^{(n)}\to\psi$ with $\psi^{(n)}\in(\Dom\overline{B})^{\oplus d}$ and $M\overline{B}^{\oplus d}\psi^{(n)}\to\chi$. Then $\overline{B}^{\oplus d}\psi^{(n)}\to M^{-1}\chi$, and since $\overline{B}$---and thus $\overline{B}^{\oplus d}$---is closed, $\psi\in(\Dom\overline{B})^{\oplus d}$ and $\overline{B}^{\oplus d}\psi=M^{-1}\chi$, i.e. $M\overline{B}^{\oplus d}\psi=\chi$. Thus $M\overline{B}^{\oplus d}$ is a closed extension of $T$ contained in $\overline{T}$; since $\overline{T}$ is the smallest closed extension of $T$, we conclude $\overline{T}=M\overline{B}^{\oplus d}$, i.e. $A\otimes B=A\odot\overline{B}$. If, in addition, $B$ is closed, then $A\otimes B=A\odot B$.
	
	The invertibility assumption cannot be dropped. For instance, if $A=0$ and $B$ is closed and unbounded, $A\odot B$ is the null operator on the dense subspace $\hilb_1\odot\Dom B$, whose closure is the null operator on the whole $\hilb_1\otimes\hilb_2$; this is a proper extension of $A\odot\overline{B}=A\odot B$.
	
	\subsection{The quantum harmonic oscillator and single-mode boson fields \texorpdfstring{(Section~\ref{sec:singlemode})}{}}\label{solsec:sec:singlemode}
	
	\solsec{commn}{A commutator formula for ladder operators}
	
	We proceed by induction on $k\geq1$. Let $\psi \in \mathcal{S}(\mathbb{R})$; recall that $\mathcal{S}(\mathbb{R})$ is invariant under $a_0$ and $a_0^\dag$. For $k=1$, the claim is the canonical commutation relation $[a_0, a_0^\dag]\psi = \psi$. Assume that the claim holds for some $k\geq1$. Then
	\begin{align}
		[a_0,(a_0^\dag)^{k+1}]\psi&= a_0 a_0^\dag (a_0^\dag)^{k} \psi - (a_0^\dag)^{k+1} a_0 \psi= \left(a_0^\dag a_0 + [a_0,a_0^\dag] \right)(a_0^\dag)^{k} \psi - (a_0^\dag)^{k+1} a_0 \psi\nonumber\\
		&= a_0^\dag [a_0, (a_0^\dag)^k]\psi + (a_0^\dag)^k \psi= k\,a_0^\dag (a_0^\dag)^{k-1}\psi + (a_0^\dag)^k \psi
		= (k+1) (a_0^\dag)^{k}\psi,
	\end{align}
	where we used $(a_0^\dag)^k\psi\in\mathcal{S}(\mathbb{R})$ and the induction hypothesis. This proves the claim for all $k\geq1$.
	
	\solsec{commn2}{Normalization of the number states}
	
	Fix $n\geq1$ (the case $n=0$ being trivial). We claim that
	\begin{equation}\label{eq:commn2-proof}
		a_0^m (a_0^\dag)^n \phi_0 = \frac{n!}{(n-m)!} (a_0^\dag)^{n-m} \phi_0\qquad\text{for all }m=1,\dots,n.
	\end{equation}
	For $m=1$, this follows from Exercise~\ref{exerc:commn} and $a_0\phi_0=0$:
	\begin{equation}
		a_0(a_0^\dag)^n\phi_0=(a_0^\dag)^na_0\phi_0+[a_0,(a_0^\dag)^n]\phi_0=n(a_0^\dag)^{n-1}\phi_0.
	\end{equation}
	Assume that Eq.~\eqref{eq:commn2-proof} holds for some $1\leq m<n$. Then, by the same argument,
	\begin{equation}
		a_0^{m+1} (a_0^\dag)^n \phi_0= \frac{n!}{(n-m)!} a_0 (a_0^\dag)^{n-m} \phi_0= \frac{n!}{(n-m-1)!} (a_0^\dag)^{n-m-1} \phi_0,
	\end{equation}
	where we used $a_0 (a_0^\dag)^{n-m} \phi_0=(n-m)(a_0^\dag)^{n-m-1} \phi_0$.
	By induction, Eq.~\eqref{eq:commn2-proof} holds for all $1\leq m\leq n$; for $m=n$, it reads $a_0^n (a_0^\dag)^n \phi_0= n! \phi_0$.
	
	\solsec{asharpvsn}{Estimating monomials of ladder operators}
	
	Let $\psi \in \mathcal{S}(\mathbb R)$, and write $\psi = \sum_{n\in\mathbb{N}} c_n \phi_n$, with $c_n=\braket{\phi_n,\psi}$. We first show that
	\begin{equation}\label{eq:expansion_a0}
		a_0\psi=\sum_{n\in\mathbb{N}} c_n\, a_0 \phi_n,\qquad a_0^\dag\psi=\sum_{n\in\mathbb{N}} c_n\, a_0^\dag \phi_n.
	\end{equation}
	Since $\phi_m\in\mathcal{S}(\mathbb{R})$, by Eq.~\eqref{eq:aadag1} we have, for all $m\in\mathbb{N}$,
	\begin{equation}
		\braket{\phi_m,a_0\psi}=\braket{a_0^\dag\phi_m,\psi}=\sqrt{m+1}\,c_{m+1},\qquad \braket{\phi_m,a_0^\dag\psi}=\braket{a_0\phi_m,\psi}=\sqrt{m}\,c_{m-1}
	\end{equation}
	(with $c_{-1}:=0$). Expanding $a_0\psi,a_0^\dag\psi\in L^2(\mathbb{R})$ in the basis $(\phi_m)_{m\in\mathbb{N}}$ gives
	\begin{align}
		a_0\psi&=\sum_{m\in\mathbb{N}}\sqrt{m+1}\,c_{m+1}\phi_m=\sum_{n\in\mathbb{N}}c_n\sqrt{n}\,\phi_{n-1},\nonumber\\
		a_0^\dag\psi&=\sum_{m\in\mathbb{N}}\sqrt{m}\,c_{m-1}\phi_m=\sum_{n\in\mathbb{N}}c_n\sqrt{n+1}\,\phi_{n+1},
	\end{align}
	which is Eq.~\eqref{eq:expansion_a0}. Since $\mathcal{S}(\mathbb{R})$ is invariant under $a_0,a_0^\dag$, applying Eq.~\eqref{eq:expansion_a0} repeatedly yields
	\begin{equation}
		A_1\cdots A_k\psi=\sum_{n\in\mathbb{N}}c_n\,A_1\cdots A_k\phi_n.
	\end{equation}
	
	Now fix $n\in\mathbb{N}$. Applying $A_k,A_{k-1},\dots,A_1$ in turn to $\phi_n$, at each step the current vector is a multiple of some $\phi_m$, and $a_0\phi_m=\sqrt{m}\,\phi_{m-1}$, $a_0^\dag\phi_m=\sqrt{m+1}\,\phi_{m+1}$. Before the $i$th step, $m\leq n+i-1$, so the factor picked up is at most $\sqrt{n+i}$. Hence
	\begin{equation}
		A_1\cdots A_k\phi_n=\mu_n\,\phi_{n+r},\qquad 0\leq\mu_n\leq\sqrt{(n+1)(n+2)\cdots(n+k)},
	\end{equation}
	where $r$ is the number of creation operators minus the number of annihilation operators in the monomial (with $\mu_n=0$ whenever $n+r<0$). Since $r$ does not depend on $n$, the vectors $(A_1\cdots A_k\phi_n)_{n\in\mathbb{N}}$ are mutually orthogonal, whence
	\begin{align}
		\|A_1\cdots A_k\psi\|^2&=\sum_{n\in\mathbb{N}}|c_n|^2\mu_n^2\leq\sum_{n\in\mathbb{N}}|c_n|^2(n+1)(n+2)\cdots(n+k)\nonumber\\
		&=\big\|\sqrt{(\N_0+1)(\N_0+2)\cdots(\N_0+k)}\,\psi\big\|^2,
	\end{align}
	the last equality following from $\N_0\phi_n=n\phi_n$.
	
	\solsec{adjointa}{The adjoint of the annihilation operator}
	
	Let $\psi\in\Dom a^*$, with $\tilde{\psi}=a^*\psi$ as in Eq.~\eqref{eq:adja}, and $d_n=\braket{\phi_n,\psi}$. Testing Eq.~\eqref{eq:adja} on $\varphi=\phi_m\in\Dom a$, $m\geq1$, we get
	\begin{equation}
		\braket{\tilde{\psi},\phi_m}=\braket{\psi,a\phi_m}=\sqrt{m}\,\braket{\psi,\phi_{m-1}}=\sqrt{m}\,\overline{d_{m-1}},
	\end{equation}
	that is, $\braket{\phi_m,\tilde{\psi}}=\sqrt{m}\,d_{m-1}$. By Parseval's identity,
	\begin{equation}
		\sum_{n\in\mathbb{N}}(n+1)|d_n|^2=\sum_{m\geq1}m|d_{m-1}|^2=\sum_{m\geq1}|\braket{\phi_m,\tilde{\psi}}|^2\leq\|\tilde{\psi}\|^2<\infty,
	\end{equation}
	and a fortiori $\sum_{n\in\mathbb{N}}n|d_n|^2<\infty$, i.e. $\psi\in\Dom\N^{1/2}=\Dom a^\dag$.
	
	\subsection{Single-mode spin--boson models \texorpdfstring{(Section~\ref{sec:singlemodesb})}{}}\label{solsec:sec:singlemodesb}
	
	\solsec{weyl}{Displacing the pure dephasing Hamiltonian}
	
	Write $c:=g/\omega$, so that $W_\pm=W(\pm c)$. By Proposition~\ref{prop:displacement},
	\begin{equation}
		W_\pm aW_\pm^*=a\mp c,\qquad W_\pm a^\dag W_\pm^*=a^\dag\mp c,
	\end{equation}
	both identities holding on $\Dom a=\Dom a^\dag=\Dom\N^{1/2}$. As $W_\pm$ is unitary, $W_\pm^*W_\pm=I$, and therefore, on $\Dom\N$,
	\begin{equation}
		W_\pm a^\dag aW_\pm^*=W_\pm a^\dag W_\pm^*\,W_\pm aW_\pm^*=\left(a^\dag\mp c\right)\left(a\mp c\right)=a^\dag a\mp c(a+a^\dag)+c^2,
	\end{equation}
	while $W_\pm(a+a^\dag)W_\pm^*=a+a^\dag\mp2c$. Hence
	\begin{align}
		W_\pm \left(\omega a^\dag a\pm\left(g(a+a^\dag)+\frac{\omega_0}{2}\right)\right) W_\pm^*&=\omega\left(a^\dag a\mp c(a+a^\dag)+c^2\right)\pm g\left(a+a^\dag\mp2c\right)\pm\frac{\omega_0}{2}\nonumber\\
		&=\omega a^\dag a+\frac{g^2}{\omega}-\frac{2g^2}{\omega}\pm\frac{\omega_0}{2}=\omega a^\dag a-\frac{g^2}{\omega}\pm\frac{\omega_0}{2}.
	\end{align}
	
	\solsec{dephasingdiag}{Diagonalizing the pure dephasing spin--boson model}
	
	By Proposition~\ref{prop:weyldisplaced}, $W_\pm H_\pm W_\pm^*=\omega\N-\frac{g^2}{\omega}\pm\frac{\omega_0}{2}$, and therefore, since $W_\pm$ is unitary,
	\begin{equation}
		H_\pm=W_\pm^*\left(\omega\N-\frac{g^2}{\omega}\pm\frac{\omega_0}{2}\right)W_\pm.
	\end{equation}
	Let $\phi_n^\pm:=W_\pm^*\phi_n$. The vectors $(\phi_n)_{n\in\mathbb{N}}$, the eigenvectors of $\N$, form an orthonormal basis of $\hilb$; since unitary operators map orthonormal bases into orthonormal bases (Lemma~\ref{lem:unitary}), so do the vectors $(\phi_n^\pm)_{n\in\mathbb{N}}$. Besides,
	\begin{equation}
		H_\pm\phi_n^\pm=W_\pm^*\left(\omega\N-\frac{g^2}{\omega}\pm\frac{\omega_0}{2}\right)W_\pm W_\pm^*\phi_n=W_\pm^*\left(\omega n-\frac{g^2}{\omega}\pm\frac{\omega_0}{2}\right)\phi_n=\left(\omega n-\frac{g^2}{\omega}\pm\frac{\omega_0}{2}\right)\phi_n^\pm.
	\end{equation}
	
	\solsec{jcinvariant}{Invariant subspaces of the Jaynes--Cummings Hamiltonian}
	
	Recall that $a\phi_n=\sqrt{n}\phi_{n-1}$, $a^\dag\phi_n=\sqrt{n+1}\phi_{n+1}$ and $a^\dag a\phi_n=n\phi_n$. The space $\mathcal{E}_0$ is spanned by $e_-\otimes\phi_0$, and
	\begin{equation}
		H\,e_-\otimes\phi_0\cong\begin{bmatrix}
			\frac{\omega_0}{2}+\omega a^\dag a&ga\\ga^\dag&-\frac{\omega_0}{2}+\omega a^\dag a
		\end{bmatrix}\begin{bmatrix}
			0\\\phi_0
		\end{bmatrix}=\begin{bmatrix}
			ga\phi_0\\\left(-\frac{\omega_0}{2}+\omega a^\dag a\right)\phi_0
		\end{bmatrix}=-\frac{\omega_0}{2}\begin{bmatrix}
			0\\\phi_0
		\end{bmatrix},
	\end{equation}
	since $a\phi_0=0$. That is, $e_-\otimes\phi_0$ is an eigenvector of $H$ with eigenvalue $-\frac{\omega_0}{2}$, and $H\mathcal{E}_0\subseteq\mathcal{E}_0$.
	
	For $n\geq1$, $\mathcal{E}_n$ is spanned by $e_+\otimes\phi_{n-1}$ and $e_-\otimes\phi_n$. We have
	\begin{equation}
		H\,e_+\otimes\phi_{n-1}\cong\begin{bmatrix}
			\left(\frac{\omega_0}{2}+\omega(n-1)\right)\phi_{n-1}\\g\sqrt{n}\,\phi_n
		\end{bmatrix},\qquad H\,e_-\otimes\phi_n\cong\begin{bmatrix}
			g\sqrt{n}\,\phi_{n-1}\\\left(-\frac{\omega_0}{2}+\omega n\right)\phi_n
		\end{bmatrix},
	\end{equation}
	that is,
	\begin{align}
		H\,e_+\otimes\phi_{n-1}&=\left(\frac{\omega_0}{2}+\omega(n-1)\right)e_+\otimes\phi_{n-1}+g\sqrt{n}\,e_-\otimes\phi_n,\label{eq:inside1}\\
		H\,e_-\otimes\phi_n&=g\sqrt{n}\,e_+\otimes\phi_{n-1}+\left(-\frac{\omega_0}{2}+\omega n\right)e_-\otimes\phi_n.\label{eq:inside2}
	\end{align}
	By linearity, $H\mathcal{E}_n\subseteq\mathcal{E}_n$. Moreover, for $\psi=c_+e_+\otimes\phi_{n-1}+c_-e_-\otimes\phi_n$, Eqs.~\eqref{eq:inside1}--\eqref{eq:inside2} give
	\begin{equation}
		H\psi=\left[\left(\frac{\omega_0}{2}+\omega(n-1)\right)c_++g\sqrt{n}\,c_-\right]e_+\otimes\phi_{n-1}+\left[g\sqrt{n}\,c_++\left(-\frac{\omega_0}{2}+\omega n\right)c_-\right]e_-\otimes\phi_n,
	\end{equation}
	so that, under the isomorphism $\mathcal{E}_n\cong\mathbb{C}^2$ mapping $e_+\otimes\phi_{n-1}$ and $e_-\otimes\phi_n$ into the canonical basis vectors,
	\begin{equation}
		H\psi\cong\begin{bmatrix}
			\frac{\omega_0}{2}+\omega(n-1)&g\sqrt{n}\\g\sqrt{n}&-\frac{\omega_0}{2}+\omega n
		\end{bmatrix}\begin{bmatrix}
			c_+\\c_-
		\end{bmatrix}.
	\end{equation}
	
	\solsec{ampdamp}{The Jaynes--Cummings model induces an amplitude-damping channel}
	
	Write $\rho=\sum_{s,s'=\pm}\rho_{ss'}\ket{e_s}\!\bra{e_{s'}}$, so that the initial state of the spin--boson system is
	\begin{equation}\label{eq:bra-ket}
		\rho\otimes\ket{\phi_0}\!\bra{\phi_0}=\sum_{s,s'=\pm}\rho_{ss'}\ket{e_s,\phi_0}\!\bra{e_{s'},\phi_0},
	\end{equation}
	where $\ket{e_s,\phi_n}\equiv e_s\otimes\phi_n$. We thus need the evolution of the two vectors $e_\pm\otimes\phi_0$.
	
	By Proposition~\ref{prop:solving_jc1}, $e_-\otimes\phi_0$ spans the invariant subspace $\mathcal{E}_0$, i.e. it is an eigenvector of $H$, with eigenvalue $-\frac{\omega_0}{2}$; thus
	\begin{equation}
		\e^{-\ii tH}e_-\otimes\phi_0=\e^{\ii\frac{\omega_0}{2}t}e_-\otimes\phi_0.
	\end{equation}
	Instead, $e_+\otimes\phi_0$ belongs to the two-dimensional invariant subspace $\mathcal{E}_1=\operatorname{Span}\{e_+\otimes\phi_0,e_-\otimes\phi_1\}$. Since $H\mathcal{E}_1\subseteq\mathcal{E}_1$ and $H$ acts on $\mathcal{E}_1$ as the Hermitian matrix $H_1$, the space $\mathcal{E}_1$ is spanned by two eigenvectors of $H$ (those of $H_1$), each of which only acquires a phase under $\e^{-\ii tH}$. Hence $\e^{-\ii tH}\mathcal{E}_1\subseteq\mathcal{E}_1$, and
	\begin{equation}
		\e^{-\ii tH}e_+\otimes\phi_0=\alpha_+(t)e_+\otimes\phi_0+\alpha_-(t)e_-\otimes\phi_1
	\end{equation}
	for some $\alpha_\pm(t)\in\mathbb{C}$. As $\e^{-\ii tH}$ is unitary, $|\alpha_+(t)|^2+|\alpha_-(t)|^2=1$. Consequently,
	\begin{align}
		\e^{-\ii tH}\ket{e_-,\phi_0}\!\bra{e_-,\phi_0}\e^{\ii tH}&=\ket{e_-,\phi_0}\!\bra{e_-,\phi_0};\\
		\e^{-\ii tH}\ket{e_+,\phi_0}\!\bra{e_-,\phi_0}\e^{\ii tH}&=\e^{-\ii\frac{\omega_0}{2}t}\left(\alpha_+(t)\ket{e_+,\phi_0}+\alpha_-(t)\ket{e_-,\phi_1}\right)\!\bra{e_-,\phi_0};\\
		\e^{-\ii tH}\ket{e_+,\phi_0}\!\bra{e_+,\phi_0}\e^{\ii tH}&=|\alpha_+(t)|^2\ket{e_+,\phi_0}\!\bra{e_+,\phi_0}+\alpha_+(t)\overline{\alpha_-(t)}\ket{e_+,\phi_0}\!\bra{e_-,\phi_1}\nonumber\\
		&\quad+\overline{\alpha_+(t)}\alpha_-(t)\ket{e_-,\phi_1}\!\bra{e_+,\phi_0}+|\alpha_-(t)|^2\ket{e_-,\phi_1}\!\bra{e_-,\phi_1},
	\end{align}
	the term with $\ket{e_-,\phi_0}\!\bra{e_+,\phi_0}$ being the adjoint of the second one. We now take the partial trace over $\hilb$. Since $(\phi_n)_{n\in\mathbb{N}}$ is orthonormal,
	\begin{equation}
		\operatorname{tr}_\hilb\ket{e_s,\phi_n}\!\bra{e_{s'},\phi_m}=\delta_{nm}\ket{e_s}\!\bra{e_{s'}}
	\end{equation}
	for all $s,s'=\pm$; hence the only surviving terms are those not coupling distinct eigenstates of the number operator:
	\begin{align}
		\operatorname{tr}_\hilb\left(\e^{-\ii tH}\ket{e_-,\phi_0}\!\bra{e_-,\phi_0}\e^{\ii tH}\right)&=\ket{e_-}\!\bra{e_-};\\
		\operatorname{tr}_\hilb\left(\e^{-\ii tH}\ket{e_+,\phi_0}\!\bra{e_-,\phi_0}\e^{\ii tH}\right)&=\alpha_+(t)\e^{-\ii\frac{\omega_0}{2}t}\ket{e_+}\!\bra{e_-};\\
		\operatorname{tr}_\hilb\left(\e^{-\ii tH}\ket{e_-,\phi_0}\!\bra{e_+,\phi_0}\e^{\ii tH}\right)&=\overline{\alpha_+(t)}\e^{\ii\frac{\omega_0}{2}t}\ket{e_-}\!\bra{e_+};\\
		\operatorname{tr}_\hilb\left(\e^{-\ii tH}\ket{e_+,\phi_0}\!\bra{e_+,\phi_0}\e^{\ii tH}\right)&=|\alpha_+(t)|^2\ket{e_+}\!\bra{e_+}+\left(1-|\alpha_+(t)|^2\right)\ket{e_-}\!\bra{e_-}.
	\end{align}
	Inserting these equalities into Eq.~\eqref{eq:bra-ket} and defining $\alpha(t):=\alpha_+(t)\e^{-\ii\frac{\omega_0}{2}t}$, so that $|\alpha(t)|=|\alpha_+(t)|$, one obtains the claimed expression for $\rho(t)$.
	
	\subsection{Fock spaces and second quantization \texorpdfstring{(Section~\ref{sec:sq})}{}}\label{solsec:sec:sq}
	
	\solsec{a0fprop}{Formal adjointness of $a_0(f)$ and $a_0^\dag(f)$}
	
	We start with the first claim. By sesquilinearity and linearity, it suffices to check it on pure tensors $\psi=\psi_1\otimes\cdots\otimes\psi_m$ and $\varphi=\varphi_1\otimes\cdots\otimes\varphi_n$. Since $a_0(f)\varphi\in\sph^{\otimes(n-1)}$ and $a_0^\dag(f)\psi\in\sph^{\otimes(m+1)}$, and vectors with different particle numbers are orthogonal, both sides vanish unless $m=n-1$. In the latter case,
	\begin{align}
		\Braket{\psi,a_0(f)\varphi}&=\sqrt{n}\Braket{\psi_1\otimes\cdots\otimes\psi_{n-1},\varphi_1\otimes\cdots\otimes\varphi_{n-1}}\braket{f,\varphi_n}\nonumber\\
		&=\sqrt{n}\Braket{\psi_1\otimes\cdots\otimes\psi_{n-1}\otimes f,\varphi_1\otimes\cdots\otimes\varphi_{n-1}\otimes\varphi_n}=\braket{a_0^\dag(f)\psi,\varphi}.
	\end{align}
	
	The second claim follows from the definition of the adjoint: $\psi\in\Dom a_0(f)^*$ if and only if there is $\tilde{\psi}\in\Gamma(\sph)$ such that $\braket{\tilde{\psi},\varphi}=\braket{\psi,a_0(f)\varphi}$ for all $\varphi\in\Dom a_0(f)=\Gamma_0(\sph)$. By the first claim, every $\psi\in\Gamma_0(\sph)$ satisfies this property with $\tilde{\psi}=a_0^\dag(f)\psi$, i.e. $a_0^\dag(f)\subseteq a_0(f)^*$; analogously, $a_0(f)\subseteq a_0^\dag(f)^*$. In particular, $\Dom a_0(f)^*\supseteq\Gamma_0(\sph)$ is dense, and thus $a_0(f)$ is closable, since an operator is closable if and only if its adjoint is densely defined. The same holds for $a_0^\dag(f)$.
	
	\solsec{afstep3}{Formal adjointness of $a(f)$ and $a^\dag(f)$}
	
	Let $\psi,\phi\in\Dom\N^{1/2}$. As $\Gamma_0(\sph)$ is a core for $\N^{1/2}$, there exist sequences $(\psi_k)_{k\in\mathbb{N}},(\phi_k)_{k\in\mathbb{N}}\subset\Gamma_0(\sph)$ converging to $\psi$ and $\phi$, respectively, in the graph norm of $\N^{1/2}$. For each $k$, by Proposition~\ref{prop:a(f)prop},
	\begin{equation}\label{eq:approxk}
		\Braket{\psi_k,a_0(f)\phi_k}=\Braket{a_0^\dag(f)\psi_k,\phi_k}.
	\end{equation}
	By Step 1, the graph norm of $\N^{1/2}$ dominates the graph norms of $a_0(f)$ and $a_0^\dag(f)$; hence $(\phi_k)_{k\in\mathbb{N}}$ and $(\psi_k)_{k\in\mathbb{N}}$ are Cauchy sequences in the graph norms of $a_0(f)$ and $a^\dag_0(f)$ respectively, and by the definition of closure,
	\begin{equation}
		a_0(f)\phi_k\to\a{f}\phi,\qquad a_0^\dag(f)\psi_k\to\adag{f}\psi.
	\end{equation}
	As the scalar product is jointly continuous, taking the limit $k\to\infty$ in Eq.~\eqref{eq:approxk} yields $\braket{\psi,\a{f}\phi}=\braket{\adag{f}\psi,\phi}$.
	
	\subsection{Bosonic and fermionic Fock spaces \texorpdfstring{(Section~\ref{sec:bf})}{}}\label{solsec:sec:bf}
	
	\solsec{usigma}{Permutation operators}
	
	Let $\phi_1,\dots,\phi_n\in\sph$. Clearly $U_{\mathrm{id}}\phi_1\otimes\cdots\otimes\phi_n=\phi_1\otimes\cdots\otimes\phi_n$. Given $\sigma,\tau\in\mathfrak{S}_n$, write $U_\tau\phi_1\otimes\cdots\otimes\phi_n=\chi_1\otimes\cdots\otimes\chi_n$, with $\chi_i:=\phi_{\tau^{-1}(i)}$. Then
	\begin{equation}
		U_\sigma U_\tau\phi_1\otimes\cdots\otimes\phi_n=\chi_{\sigma^{-1}(1)}\otimes\cdots\otimes\chi_{\sigma^{-1}(n)}=\phi_{\tau^{-1}\sigma^{-1}(1)}\otimes\cdots\otimes\phi_{\tau^{-1}\sigma^{-1}(n)}=U_{\sigma\tau}\phi_1\otimes\cdots\otimes\phi_n,
	\end{equation}
	since $\tau^{-1}\sigma^{-1}=(\sigma\tau)^{-1}$. Furthermore, for all $\phi_1,\dots,\phi_n,\chi_1,\dots,\chi_n\in\sph$,
	\begin{align}
		\Braket{U_\sigma\phi_1\otimes\cdots\otimes\phi_n,\chi_1\otimes\cdots\otimes\chi_n}&=\prod_{i=1}^n\braket{\phi_{\sigma^{-1}(i)},\chi_i}=\prod_{j=1}^n\braket{\phi_j,\chi_{\sigma(j)}}\nonumber\\
		&=\Braket{\phi_1\otimes\cdots\otimes\phi_n,U_{\sigma^{-1}}\chi_1\otimes\cdots\otimes\chi_n},
	\end{align}
	where we relabeled the factors via $j=\sigma^{-1}(i)$. All these identities are between bounded operators (resp. continuous sesquilinear expressions) and hold on pure tensors; by linearity and continuity, they extend to the whole $\sph^{\otimes n}$, since pure tensors span the dense subspace $\sph^{\odot n}$. This proves $U_{\mathrm{id}}=I$, $U_\sigma U_\tau=U_{\sigma\tau}$ and $U_\sigma^*=U_{\sigma^{-1}}$.
	
	\solsec{spsm}{Products of symmetrizers}
	
	By Definition~\ref{def:symmetrizers} and Lemma~\ref{lem:usigma},
	\begin{equation}
		S_n^+S_n^-=\frac{1}{(n!)^2}\sum_{\sigma,\tau\in\mathfrak{S}_n}\sgn(\tau)\,U_{\sigma\tau}.
	\end{equation}
	For fixed $\sigma$, the map $\tau\mapsto\pi:=\sigma\tau$ is a bijection of $\mathfrak{S}_n$, and $\sgn(\tau)=\sgn(\sigma^{-1}\pi)=\sgn(\sigma)\sgn(\pi)$ by Eq.~\eqref{eq:sgnmult}. Hence
	\begin{equation}
		S_n^+S_n^-=\frac{1}{(n!)^2}\sum_{\sigma\in\mathfrak{S}_n}\sgn(\sigma)\sum_{\pi\in\mathfrak{S}_n}\sgn(\pi)\,U_\pi=\frac{1}{n!}\left(\sum_{\sigma\in\mathfrak{S}_n}\sgn(\sigma)\right)S_n^-=0
	\end{equation}
	for $n\geq2$, by Eq.~\eqref{eq:sgnmult}. Analogously, writing $\pi:=\sigma\tau$ for fixed $\tau$, $S_n^-S_n^+=\frac{1}{(n!)^2}\sum_{\sigma,\tau}\sgn(\sigma)U_{\sigma\tau}=\frac{1}{n!}\bigl(\sum_\tau\sgn(\tau)\bigr)S_n^-=0$, where we used $\sgn(\sigma)=\sgn(\pi)\sgn(\tau)$, again by Eq.~\eqref{eq:sgnmult}.
	
	\solsec{basisantisym}{An orthonormal basis of $S_n^-\sph^{\otimes n}$}
	
	For a multi-index $J=(j_1,\dots,j_n)$, write $e_J:=e_{j_1}\otimes\cdots\otimes e_{j_n}$; the vectors $(e_J)_J$ form an orthonormal basis of $\sph^{\otimes n}$ (Proposition~\ref{prop:tensorproductbases}). By definition, $U_\sigma e_J=e_{J\circ\sigma^{-1}}$, where $J\circ\sigma^{-1}:=(j_{\sigma^{-1}(1)},\dots,j_{\sigma^{-1}(n)})$, so that
	\begin{equation}
		S_n^-e_J=\frac{1}{n!}\sum_{\sigma\in\mathfrak{S}_n}\sgn(\sigma)\,e_{J\circ\sigma^{-1}}.
	\end{equation}
	Recall also that $S_n^-U_\sigma=\sgn(\sigma)S_n^-$ (Lemma~\ref{lem:sunu}), and that $S_n^-e_J=0$ whenever $J$ contains a repeated index (Proposition~\ref{prop:repetitions}).
	
	\emph{Orthonormality.} If $j_1<\cdots<j_n$, the multi-indices $J\circ\sigma^{-1}$, $\sigma\in\mathfrak{S}_n$, are pairwise distinct, so $S_n^-e_J$ is a combination of $n!$ distinct orthonormal vectors with coefficients of modulus $1/n!$, and $\|S_n^-e_J\|^2=n!/(n!)^2=1/n!$. If $J\neq J'$ are both strictly increasing, they contain different sets of indices, so no rearrangement of $J$ coincides with a rearrangement of $J'$, and $S_n^-e_J\perp S_n^-e_{J'}$. Thus the vectors $\sqrt{n!}\,S_n^-e_J$, $J$ strictly increasing, are orthonormal.
	
	\emph{Completeness.} Since $S_n^-$ is a bounded projection and $(e_K)_K$ spans a dense subspace of $\sph^{\otimes n}$, the vectors $(S_n^-e_K)_K$ span a dense subspace of $S_n^-\sph^{\otimes n}$. For every multi-index $K$, either $K$ has a repeated index, whence $S_n^-e_K=0$, or $K=J\circ\sigma^{-1}$ for some strictly increasing $J$ and some $\sigma\in\mathfrak{S}_n$, whence $S_n^-e_K=S_n^-U_\sigma e_J=\sgn(\sigma)S_n^-e_J$. Hence the vectors $S_n^-e_J$, $J$ strictly increasing, span a dense subspace of $S_n^-\sph^{\otimes n}$.
	
	\emph{Dimension.} If $\dim\sph=d$, the strictly increasing multi-indices with entries in $\{1,\dots,d\}$ are in one-to-one correspondence with the $n$-element subsets of $\{1,\dots,d\}$, which are $\binom{d}{n}$ (none if $n>d$).
	
	\solsec{basissym}{An orthonormal basis of $S_n^+\sph^{\otimes n}$}
	
	We use the notation of the solution of Exercise~\ref{exerc:basisantisym} (Appendix~\ref{sol:basisantisym}). Now
	\begin{equation}
		S_n^+e_J=\frac{1}{n!}\sum_{\sigma\in\mathfrak{S}_n}e_{J\circ\sigma^{-1}},\qquad S_n^+U_\sigma=S_n^+,
	\end{equation}
	the latter by Lemma~\ref{lem:sunu}.
	Let $J$ have occupation numbers $m_1,m_2,\ldots$, i.e. the index $k$ appears $m_k$ times in $J$, so that $S_n^+e_J=\ket{m_1,m_2,\ldots}$. As $\sigma$ ranges over $\mathfrak{S}_n$, the multi-index $J\circ\sigma^{-1}$ ranges over the distinct rearrangements of $J$, which are $n!/(m_1!m_2!\cdots)$ in number, each of them being obtained exactly $m_1!m_2!\cdots$ times (the number of permutations exchanging equal indices among themselves). Hence
	\begin{align}
		\ket{m_1,m_2,\ldots}&=\frac{m_1!m_2!\cdots}{n!}\sum_{K\text{ rearrangement of }J}e_K,\\
		\big\|\ket{m_1,m_2,\ldots}\big\|^2&=\left(\frac{m_1!m_2!\cdots}{n!}\right)^2\frac{n!}{m_1!m_2!\cdots}=\frac{m_1!m_2!\cdots}{n!},
	\end{align}
	so the vectors $\sqrt{n!/(m_1!m_2!\cdots)}\ket{m_1,m_2,\ldots}$ are normalized. Vectors with different occupation numbers involve disjoint sets of basis vectors $e_K$, and are therefore orthogonal. Completeness follows as in the previous solution: for every multi-index $K$, $S_n^+e_K=S_n^+e_J$, where $J$ is the nondecreasing rearrangement of $K$, so the vectors $\ket{m_1,m_2,\ldots}$ span a dense subspace of $S_n^+\sph^{\otimes n}$.
	
	Finally, if $\dim\sph=d$, the basis is indexed by the families $(m_1,\dots,m_d)\in\mathbb{N}^d$ with $m_1+\cdots+m_d=n$. Representing each such family as a string of $n$ ``stars'' and $d-1$ ``bars'' (the bars separating the groups of $m_1,\dots,m_d$ stars), their number is $\binom{d+n-1}{n}$.
	
	\solsec{invariantlemma}{Restrictions of self-adjoint operators to reducing subspaces}
	
	Let $P$ be the orthogonal projection onto $\mathcal{K}$, so that $I-P$ is the orthogonal projection onto $\mathcal{K}^\perp$; note that $I-P$ also commutes with $T$. Let $T_1$ and $T_2$ be the restrictions of $T$ to $\Dom T\cap\mathcal{K}$ and $\Dom T\cap\mathcal{K}^\perp$.	
	These are operators on $\mathcal{K}$ and $\mathcal{K}^\perp$: if $\psi\in\Dom T\cap\mathcal{K}$, then $T\psi=TP\psi=PT\psi\in\mathcal{K}$, and analogously for $\mathcal{K}^\perp$. They are densely defined: $\Dom T_1=P\Dom T$, since $P\Dom T\subseteq\Dom T\cap\mathcal{K}$ and $\psi=P\psi$ for $\psi\in\Dom T\cap\mathcal{K}$; and $P\Dom T$ is dense in $\mathcal{K}=P\sph$, $P$ being continuous and $\Dom T$ dense. The same holds for $T_2$.
	
	Under the unitary identification $\sph\ni\psi\mapsto(P\psi,(I-P)\psi)\in\mathcal{K}\oplus\mathcal{K}^\perp$, the operator $T$ is mapped into $T_1\oplus T_2$. Indeed, if $\psi\in\Dom T$, then $P\psi\in\Dom T_1$, $(I-P)\psi\in\Dom T_2$, and $T\psi=TP\psi+T(I-P)\psi=T_1P\psi+T_2(I-P)\psi$; conversely, if $P\psi\in\Dom T_1$ and $(I-P)\psi\in\Dom T_2$, then $\psi=P\psi+(I-P)\psi\in\Dom T$. By Proposition~\ref{prop:adjoint_directsum}, $T_1\oplus T_2\cong T=T^*\cong T_1^*\oplus T_2^*$.
	Comparing the two direct sums on vectors of the form $(\psi_1,0)$ and $(0,\psi_2)$ yields $T_1=T_1^*$ and $T_2=T_2^*$.
	
	(Mere invariance of $\Dom T\cap\mathcal{K}$ under $T$ is not enough in general: this intersection could even be $\{0\}$.)
	
	\solsec{gammaU}{Second quantization of unitaries preserves (anti)symmetry}
	
	Recall that $\Gamma^\pm(\sph) = S^\pm \Gamma(\sph)$, with $S^\pm = \oplus_{n\in\mathbb{N}} S_n^\pm$. We prove that $\Gamma(U)$ commutes with $S^\pm$. Let $n\in \mathbb{N}$, $\sigma \in \mathfrak{S}_n$ and $\phi_1,\dots,\phi_n\in\sph$. Then
	$U^{\otimes n} U_\sigma \phi_1\otimes\cdots\otimes\phi_n= U \phi_{\sigma^{-1}(1)}\otimes \cdots \otimes U \phi_{\sigma^{-1}(n)}= U_\sigma U^{\otimes n} \phi_1\otimes\cdots\otimes \phi_n$ by Eq.~\eqref{eq:usigma}. By linearity and continuity, $U^{\otimes n} U_\sigma = U_\sigma U^{\otimes n}$ on the whole $\sph^{\otimes n}$; by Eq.~\eqref{eq:symmetrizers}, $U^{\otimes n}$ thus commutes with $S_n^\pm$, and taking direct sums, $\Gamma(U) S^\pm= S^\pm \Gamma(U)$. Hence
	\begin{equation}
		\Gamma(U)\Gamma^{\pm}(\sph)= \Gamma(U) S^\pm \Gamma(\sph)= S^\pm \Gamma(U) \Gamma(\sph)\subseteq S^\pm \Gamma(\sph)= \Gamma^{\pm}(\sph).
	\end{equation}
	
	\solsec{ccr/car}{The remaining canonical (anti)commutation relations}
	
	Let $f,g\in\sph=L^2(\mathcal{M},\Sigma,\mu)$, $n\geq2$, and $\psi_n\in S_n^\pm\sph^{\otimes n}$. By Proposition~\ref{prop:a(f)explicit}, applied twice (to $\psi_n$ and to the $(n-1)$-particle function $a_\pm(g)\psi_n$),
	\begin{align}
		\left(a_\pm(f)a_\pm(g)\psi_n\right)(x_1,\dots,x_{n-2})&=\sqrt{n-1}\int\overline{f(x_{n-1})}\left(a_\pm(g)\psi_n\right)(x_1,\dots,x_{n-1})\;\mathrm{d}\mu(x_{n-1})\nonumber\\
		&=\sqrt{n(n-1)}\iint\overline{f(x_{n-1})g(x_n)}\,\psi_n(x_1,\dots,x_{n-2},x_{n-1},x_n)\;\mathrm{d}^2\mu,
	\end{align}
	where $\mathrm{d}^2\mu:=\mathrm{d}\mu(x_{n-1})\,\mathrm{d}\mu(x_n)$; exchanging the roles of $f$ and $g$ and then renaming the integration variables $x_{n-1}\leftrightarrow x_n$,
	\begin{align}
		\left(a_\pm(g)a_\pm(f)\psi_n\right)(x_1,\dots,x_{n-2})&=\sqrt{n(n-1)}\iint\overline{g(x_{n-1})f(x_n)}\,\psi_n(x_1,\dots,x_{n-2},x_{n-1},x_n)\;\mathrm{d}^2\mu\nonumber\\
		&=\sqrt{n(n-1)}\iint\overline{f(x_{n-1})g(x_n)}\,\psi_n(x_1,\dots,x_{n-2},x_n,x_{n-1})\;\mathrm{d}^2\mu.
	\end{align}
	The two integrands differ by the transposition of the last two variables of $\psi_n$, which has sign $-1$ (Example~\ref{examp:sign}). As $\psi_n\in S_n^\pm\sph^{\otimes n}$,
	\begin{equation}
		\psi_n(x_1,\dots,x_{n-2},x_n,x_{n-1})=\pm\psi_n(x_1,\dots,x_{n-2},x_{n-1},x_n),
	\end{equation}
	whence $a_\pm(g)a_\pm(f)\psi_n=\pm a_\pm(f)a_\pm(g)\psi_n$, i.e. $[a_+(f),a_+(g)]\psi_n=0$ and $\{a_-(f),a_-(g)\}\psi_n=0$. For $n=0,1$, both sides vanish trivially. These relations extend to all $\psi\in\Dom\N\cap\Gamma^\pm(\sph)$ as in the proof of Proposition~\ref{prop:ccr/car}.
	
	\subsection{Multimode spin--boson models \texorpdfstring{(Section~\ref{sec:multimodesb})}{}}\label{solsec:sec:multimodesb}
	
	\solsec{expvectors}{Exponential vectors}
	
	Let $n\geq1$. By Definitions~\ref{def:a0(f)} and~\ref{def:apm}, $a_+(g)f^{\otimes n}=\sqrt{n}\braket{g,f}f^{\otimes(n-1)}$ and
	\begin{equation}\label{eq:adagfn}
		a_+^\dag(g)f^{\otimes(n-1)}=\sqrt{n}\,S_n^+\left(f^{\otimes(n-1)}\otimes g\right)=\frac{1}{\sqrt{n}}\sum_{j=1}^nf^{\otimes(j-1)}\otimes g\otimes f^{\otimes(n-j)},
	\end{equation}
	since the $n!$ permutations of the factors of $f^{\otimes(n-1)}\otimes g$ place $g$ in each of the $n$ slots exactly $(n-1)!$ times. As $\varepsilon(f)\in\Dom\N^{1/2}\cap\Gamma^+(\sph)$ (Lemma~\ref{lem:expvectors}(i)), its partial sums converge to it in the graph norm of $\N^{1/2}$, and thus, by Proposition~\ref{prop:apm(f)prop}(ii), $a_+^\sharp(g)$ can be applied to $\varepsilon(f)$ component by component.
	
	(ii) By the above, $a_+(g)\varepsilon(f)=\sum_{n\geq1}\sqrt{n}\braket{g,f}f^{\otimes(n-1)}/\sqrt{n!}=\braket{g,f}\varepsilon(f)$. As for the derivative, fix $s\in\mathbb{R}$ and let $k:=f+sg$. Given $0<|\delta|\leq1$, by multilinearity,
	\begin{equation}
		(k+\delta g)^{\otimes n}=k^{\otimes n}+\delta\sum_{j=1}^nk^{\otimes(j-1)}\otimes g\otimes k^{\otimes(n-j)}+R_n(\delta),
	\end{equation}
	where $R_n(\delta)$ is the sum of the terms containing $g$ at least twice. There are $\binom{n}{j}$ terms containing $g$ exactly $j$ times, each with norm $|\delta|^j\|g\|^j\|k\|^{n-j}$; hence, using $|\delta|^j\leq\delta^2$ and $\binom{n}{j}\leq\binom{n}{2}\binom{n-2}{j-2}$ for $j\geq2$, and the binomial theorem,
	\begin{equation}
		\|R_n(\delta)\|\leq\delta^2\sum_{j=2}^n\binom{n}{j}\|g\|^j\|k\|^{n-j}\leq\frac{n(n-1)}{2}\delta^2\|g\|^2\left(\|g\|+\|k\|\right)^{n-2}.
	\end{equation}
	By Eq.~\eqref{eq:adagfn}, the $n$-particle component of $a_+^\dag(g)\varepsilon(k)$ is $a_+^\dag(g)k^{\otimes(n-1)}/\sqrt{(n-1)!}=\sum_jk^{\otimes(j-1)}\otimes g\otimes k^{\otimes(n-j)}/\sqrt{n!}$. Therefore
	\begin{equation}
		\left\|\frac{\varepsilon(k+\delta g)-\varepsilon(k)}{\delta}-a_+^\dag(g)\varepsilon(k)\right\|^2=\sum_{n\geq2}\frac{\|R_n(\delta)\|^2}{\delta^2\,n!}\leq\frac{\delta^2\|g\|^4}{4}\sum_{n\geq2}\frac{n^4\left(\|g\|+\|k\|\right)^{2(n-2)}}{n!}\xrightarrow{\delta\to0}0.
	\end{equation}
	
	(iii) Since $U^{\otimes n}f^{\otimes n}=(Uf)^{\otimes n}$ for all $n$, we have $\Gamma(U)\varepsilon(f)=\varepsilon(Uf)$. Now let $f\in\Dom A$. Then $f^{\otimes n}\in\bigodot_{j=1}^n\Dom A$ and, by Definition~\ref{def:secondquantization} and Eq.~\eqref{eq:adagfn} with $g=Af$,
	\begin{equation}
		A_nf^{\otimes n}=\sum_{j=1}^nf^{\otimes(j-1)}\otimes Af\otimes f^{\otimes(n-j)}=\sqrt{n}\,a_+^\dag(Af)f^{\otimes(n-1)}.
	\end{equation}
	Hence the $n$-particle component of $a_+^\dag(Af)\varepsilon(f)$ is $A_nf^{\otimes n}/\sqrt{n!}$, and in particular
	\begin{equation}
		\sum_{n\in\mathbb{N}}\frac{\|A_nf^{\otimes n}\|^2}{n!}=\left\|a_+^\dag(Af)\varepsilon(f)\right\|^2<\infty.
	\end{equation} As $\mathrm{d}\Gamma(A)=\oplus_nA_n$, this means that $\varepsilon(f)\in\Dom\mathrm{d}\Gamma(A)$ and $\mathrm{d}\Gamma(A)\varepsilon(f)=a_+^\dag(Af)\varepsilon(f)$.
	
	\solsec{polarization}{The polarization identity}
	
	By multilinearity,
	\begin{equation}
		\left(s_1\phi_1+\ldots+s_n\phi_n\right)^{\otimes n}=\sum_{j_1,\dots,j_n=1}^ns_{j_1}\cdots s_{j_n}\,\phi_{j_1}\otimes\cdots\otimes\phi_{j_n}.
	\end{equation}
	Fix $(j_1,\dots,j_n)$, and let $m_i$ be the number of indices equal to $i$, so that $s_1\cdots s_n\,s_{j_1}\cdots s_{j_n}=\prod_{i=1}^ns_i^{1+m_i}$. Then
	\begin{equation}
		\sum_{s_1,\dots,s_n=\pm1}\prod_{i=1}^ns_i^{1+m_i}=\prod_{i=1}^n\sum_{s_i=\pm1}s_i^{1+m_i},
	\end{equation}
	which equals $2^n$ if all $m_i$ are odd, and $0$ otherwise. Since $m_1+\ldots+m_n=n$, all $m_i$ are odd if and only if $m_i=1$ for all $i$, i.e. if and only if $(j_1,\dots,j_n)=(\sigma(1),\dots,\sigma(n))$ for some permutation $\sigma\in\mathfrak{S}_n$. Therefore, the right-hand side of Eq.~\eqref{eq:polarization} equals
	\begin{equation}
		\frac{2^n}{2^nn!}\sum_{\sigma\in\mathfrak{S}_n}\phi_{\sigma(1)}\otimes\cdots\otimes\phi_{\sigma(n)}=S_n^+(\phi_1\otimes\cdots\otimes\phi_n),
	\end{equation}
	by Eq.~\eqref{eq:symmetrizers} (as $\sigma$ ranges over $\mathfrak{S}_n$, so does $\sigma^{-1}$).
	
	\solsec{weylcov}{Covariance of Weyl operators}
	
	Recall that $\Gamma(U)$ maps $\Gamma^+(\sph)$ onto itself (Lemma~\ref{lem:gammacov2}), and that $\Gamma(U)^*=\Gamma(U^*)$, since $(U^{\otimes n})^*=(U^*)^{\otimes n}$. Let $f\in\sph$. By Lemma~\ref{lem:expvectors}(iii) and Eq.~\eqref{eq:weylexp}, using $\|Uh\|=\|h\|$ and $\braket{h,U^*f}=\braket{Uh,f}$,
	\begin{align}
		\Gamma(U)W(h)\Gamma(U)^*\varepsilon(f)&=\Gamma(U)W(h)\varepsilon(U^*f)=\e^{-\|h\|^2/2-\braket{h,U^*f}}\,\Gamma(U)\varepsilon(U^*f+h)\nonumber\\
		&=\e^{-\|Uh\|^2/2-\braket{Uh,f}}\,\varepsilon(f+Uh)=W(Uh)\varepsilon(f).
	\end{align}
	Since both sides are bounded operators and $\operatorname{Exp}(\sph)$ is dense (Lemma~\ref{lem:expvectors}(iv)), we conclude that $\Gamma(U)W(h)\Gamma(U)^*=W(Uh)$. Finally, taking $U=\e^{-\ii t\omega}$, by Proposition~\ref{prop:gammaexp} we have $\Gamma(\e^{-\ii t\omega})=\e^{-\ii t\mathrm{d}\Gamma(\omega)}$, and $\Gamma(\e^{-\ii t\omega})^*=\e^{\ii t\mathrm{d}\Gamma(\omega)}$, which proves the last claim.
	
	\solsec{multijc}{Excited-state amplitude in the multimode Jaynes--Cummings model}
	
	Let $P_1$ be the orthogonal projection onto $\mathcal{E}_1$. Since $H$ commutes with $P_1$, so does $\e^{-\ii tH}$ (by the spectral theorem); hence $\Psi(t):=\e^{-\ii tH}e_+\otimes\varepsilon_0=\e^{-\ii tH}P_1e_+\otimes\varepsilon_0\in\mathcal{E}_1$, and we can write $\Psi(t)=c(t)\,e_+\otimes\varepsilon_0+e_-\otimes g_t$, with $c(0)=1$ and $g_0=0$. Besides, as $e_+\otimes\varepsilon_0\in\Dom H$, we have $\Psi(t)\in\Dom H$, i.e. $g_t\in\Dom\omega$, and $\Psi(t)$ is differentiable with $\ii\dot\Psi(t)=H\Psi(t)$ (Proposition~\ref{prop:unitarygroup}). Taking the components along $e_+\otimes\varepsilon_0$ and $e_-\otimes\sph$, which are bounded operations, and using the action of $H$ on $\mathcal{E}_1$, we get
	\begin{equation}
		\ii\dot c(t)=\frac{\omega_0}{2}c(t)+\braket{f,g_t},\qquad\ii\dot g_t=c(t)f+\omega'g_t,\qquad\omega':=\omega-\frac{\omega_0}{2}.
	\end{equation}
	As $g_s\in\Dom\omega'$ is differentiable in $s$, so is $\e^{\ii s\omega'}g_s$, with derivative $\e^{\ii s\omega'}(\ii\omega'g_s+\dot g_s)=-\ii c(s)\e^{\ii s\omega'}f$; this follows by writing the difference quotient as
	\begin{equation}
		\frac{\e^{\ii(s+\delta)\omega'}g_{s+\delta}-\e^{\ii s\omega'}g_s}{\delta}=\e^{\ii(s+\delta)\omega'}\frac{g_{s+\delta}-g_s}{\delta}+\frac{\e^{\ii(s+\delta)\omega'}-\e^{\ii s\omega'}}{\delta}g_s,
	\end{equation}
	and using the uniform boundedness and strong continuity of $\e^{\ii s\omega'}$, together with Proposition~\ref{prop:unitarygroup}. Integrating between $0$ and $t$ (variation of constants),
	\begin{equation}
		g_t=-\ii\int_0^tc(s)\,\e^{-\ii(t-s)\omega'}f\;\mathrm{d}s,\qquad\text{whence}\qquad\braket{f,g_t}=-\ii\int_0^tc(s)\,\e^{\ii\frac{\omega_0}{2}(t-s)}\int\e^{-\ii\lambda(t-s)}\;\mathrm{d}J(\lambda)\;\mathrm{d}s,
	\end{equation}
	where we used $\braket{f,\e^{-\ii\tau\omega}f}=\int\e^{-\ii\tau\omega(k)}|f(k)|^2\,\mathrm{d}\mu(k)=\int\e^{-\ii\tau\lambda}\,\mathrm{d}J(\lambda)$. Finally, $\alpha(t)=\e^{-\ii\frac{\omega_0}{2}t}c(t)$, so that $\beta(t)=\e^{\ii\omega_0t}\alpha(t)=\e^{\ii\frac{\omega_0}{2}t}c(t)$, $\beta(0)=1$, and, by the equation for $c(t)$,
	\begin{equation}
		\dot\beta(t)=\e^{\ii\frac{\omega_0}{2}t}\left(\ii\frac{\omega_0}{2}c(t)+\dot c(t)\right)=-\ii\,\e^{\ii\frac{\omega_0}{2}t}\braket{f,g_t}=-\int_0^t\int\e^{-\ii(\lambda-\omega_0)(t-s)}\;\mathrm{d}J(\lambda)\;\beta(s)\;\mathrm{d}s,
	\end{equation}
	since $c(s)=\e^{-\ii\frac{\omega_0}{2}s}\beta(s)$ and $\frac{\omega_0}{2}t+\frac{\omega_0}{2}(t-s)-\frac{\omega_0}{2}s=\omega_0(t-s)$. This is Eq.~\eqref{eq:multijc}.
	
	\subsection{The rotating-wave approximation \texorpdfstring{(Appendix~\ref{app:rwa})}{}}\label{solsec:app:rwa}
	
	\solsec{rwa-interaction}{The Rabi and Jaynes--Cummings models in the interaction picture}
	
	Let $\phi=\sum_{n\in\mathbb{N}}c_n\phi_n\in\Dom\N^{1/2}$. Recall that $a$ and $a^\dag$ act on such vectors coefficientwise, $a\phi=\sum_nc_n\sqrt{n}\,\phi_{n-1}$ and $a^\dag\phi=\sum_nc_n\sqrt{n+1}\,\phi_{n+1}$: the partial sums of $\phi$ converge to $\phi$ in the graph norm of $\N^{1/2}$, which coincides with the graph norm of $a$ and is equivalent to that of $a^\dag$ (cf. the proof of Proposition~\ref{prop:aprop2}). Since $\e^{-\ii\omega t\N}\phi=\sum_nc_n\e^{-\ii\omega nt}\phi_n\in\Dom\N^{1/2}$,
	\begin{equation}
		\e^{\ii\omega t\N}a^\dag\e^{-\ii\omega t\N}\phi=\sum_{n\in\mathbb{N}}c_n\e^{\ii\omega(n+1)t}\sqrt{n+1}\,\e^{-\ii\omega nt}\phi_{n+1}=\e^{\ii\omega t}a^\dag\phi,
	\end{equation}
	and analogously $\e^{\ii\omega t\N}a\,\e^{-\ii\omega t\N}\phi=\sum_nc_n\e^{\ii\omega(n-1)t}\sqrt{n}\,\e^{-\ii\omega nt}\phi_{n-1}=\e^{-\ii\omega t}a\phi$. As for the spin, $\e^{\ii\frac{\omega t}{2}\sigma_z}e_\pm=\e^{\pm\ii\frac{\omega t}{2}}e_\pm$, while $\sigma_+e_-=e_+$, $\sigma_-e_+=e_-$ and $\sigma_\pm e_\pm=0$. Hence
	\begin{equation}
		\e^{\ii\frac{\omega t}{2}\sigma_z}\sigma_+\e^{-\ii\frac{\omega t}{2}\sigma_z}e_-=\e^{\ii\frac{\omega t}{2}}\e^{\ii\frac{\omega t}{2}}e_+=\e^{\ii\omega t}\sigma_+e_-,\qquad \e^{\ii\frac{\omega t}{2}\sigma_z}\sigma_-\e^{-\ii\frac{\omega t}{2}\sigma_z}e_+=\e^{-\ii\omega t}\sigma_-e_+,
	\end{equation}
	both sides vanishing on the other basis vector; by linearity, $\e^{\ii\frac{\omega t}{2}\sigma_z}\sigma_\pm\e^{-\ii\frac{\omega t}{2}\sigma_z}=\e^{\pm\ii\omega t}\sigma_\pm$.
	
	Since $\frac{\omega}{2}\sigma_z\otimes I$ and $I\otimes\omega\N$ commute, $\e^{\pm\ii H_0t}=\e^{\pm\ii\frac{\omega t}{2}\sigma_z}\otimes\e^{\pm\ii\omega t\N}$, which leaves $\mathbb{C}^2\odot\Dom\N$ invariant. We have $H_{\rm QR}-H_0=\frac{\Delta}{2}\sigma_z\otimes I+g(\sigma_++\sigma_-)\otimes(a+a^\dag)$ and $H_{\rm JC}-H_0=\frac{\Delta}{2}\sigma_z\otimes I+g(\sigma_+\otimes a+\sigma_-\otimes a^\dag)$. On pure tensors $v\otimes\phi$ with $\phi\in\Dom\N$, the term $\frac{\Delta}{2}\sigma_z\otimes I$ commutes with $\e^{\pm\ii H_0t}$, while, by the identities above,
	\begin{align}
		\e^{\ii H_0t}(\sigma_+\otimes a)\e^{-\ii H_0t}&=\e^{\ii\omega t}\e^{-\ii\omega t}\sigma_+\otimes a=\sigma_+\otimes a,&\e^{\ii H_0t}(\sigma_-\otimes a^\dag)\e^{-\ii H_0t}&=\sigma_-\otimes a^\dag,\\
		\e^{\ii H_0t}(\sigma_+\otimes a^\dag)\e^{-\ii H_0t}&=\e^{2\ii\omega t}\sigma_+\otimes a^\dag,&\e^{\ii H_0t}(\sigma_-\otimes a)\e^{-\ii H_0t}&=\e^{-2\ii\omega t}\sigma_-\otimes a.
	\end{align}
	By linearity, these identities hold on $\mathbb{C}^2\odot\Dom\N$, and Eqs.~\eqref{eq:rwa-hqr-int}--\eqref{eq:rwa-hjc-int} follow.
	
	\solsec{rwa-noaverage}{The first integration by parts does not suffice}
	
	By Eqs.~\eqref{eq:rwa-hqr-int}--\eqref{eq:rwa-hjc-int},
	\begin{equation}
		\tilde{H}_{\rm QR}(t)-\tilde{H}_{\rm JC}(t)=g\left(\e^{2\ii\omega t}\sigma_+\otimes a^\dag+\e^{-2\ii\omega t}\sigma_-\otimes a\right).
	\end{equation}
	Every $\psi\in\mathbb{C}^2\odot\Dom\N^{1/2}$ can be written as $\psi=e_+\otimes\phi_++e_-\otimes\phi_-$ with $\phi_\pm\in\Dom\N^{1/2}$. Since $\sigma_+e_-=e_+$, $\sigma_-e_+=e_-$ and $\sigma_\pm e_\pm=0$,
	\begin{equation}
		\left(\tilde{H}_{\rm QR}(t)-\tilde{H}_{\rm JC}(t)\right)\psi=g\left(\e^{2\ii\omega t}e_+\otimes a^\dag\phi_-+\e^{-2\ii\omega t}e_-\otimes a\phi_+\right),
	\end{equation}
	and, as $e_+\perp e_-$,
	\begin{equation}
		\left\|\left(\tilde{H}_{\rm QR}(t)-\tilde{H}_{\rm JC}(t)\right)\psi\right\|^2=g^2\left(\|a^\dag\phi_-\|^2+\|a\phi_+\|^2\right)=g^2\left(\left\|(\N+1)^{1/2}\phi_-\right\|^2+\left\|\N^{1/2}\phi_+\right\|^2\right),
	\end{equation}
	which is independent of $\omega$ and $t$. In particular, the bound obtained from the first integration by parts grows linearly in $|t|$, and does not vanish as $\omega\to\infty$.
	
	\solsec{rwa-S}{Computing $S(t)$}
	
	Let $\psi\in\mathcal{D}_0$. By Eqs.~\eqref{eq:rwa-hqr-int}--\eqref{eq:rwa-hjc-int}, the function
	\begin{equation}
		r\mapsto(H_2(r)-H_1(r))\psi=-g\left(\e^{2\ii\omega r}\sigma_+\otimes a^\dag+\e^{-2\ii\omega r}\sigma_-\otimes a\right)\psi
	\end{equation}
	is continuous with values in $\mathcal{D}_0$, so that $S(t)\psi=\int_0^t(H_2(r)-H_1(r))\psi\,\mathrm{d}r$ is well-defined. Since
	\begin{equation}
		\int_0^t\e^{\pm2\ii\omega r}\;\mathrm{d}r=\frac{\e^{\pm2\ii\omega t}-1}{\pm2\ii\omega}=\e^{\pm\ii\omega t}\,\frac{\e^{\ii\omega t}-\e^{-\ii\omega t}}{2\ii\omega}=\e^{\pm\ii\omega t}\,\frac{\sin(\omega t)}{\omega},
	\end{equation}
	we obtain
	\begin{equation}
		S(t)\psi=-g\left(\e^{\ii\omega t}\frac{\sin(\omega t)}{\omega}\sigma_+\otimes a^\dag+\e^{-\ii\omega t}\frac{\sin(\omega t)}{\omega}\sigma_-\otimes a\right)\psi,
	\end{equation}
	which is Eq.~\eqref{eq:rwa-S-explicit}. Finally, $\sigma_+\otimes a^\dag$ maps $e_-\otimes\phi_n\in\mathcal{E}_n$ into $\sqrt{n+1}\,e_+\otimes\phi_{n+1}\in\mathcal{E}_{n+2}$ and annihilates $e_+\otimes\phi_{n-1}$, while $\sigma_-\otimes a$ maps $e_+\otimes\phi_{n-1}\in\mathcal{E}_n$ into $\sqrt{n-1}\,e_-\otimes\phi_{n-2}\in\mathcal{E}_{n-2}$ (or into zero) and annihilates $e_-\otimes\phi_n$. Hence $S(t)\mathcal{K}_N\subseteq\mathcal{K}_{N+2}$, and in particular $S(t)\mathcal{D}_0\subseteq\mathcal{D}_0$.
	
	\solsec{rwa-derivatives}{Product rules in the interaction picture}
	
	Let $\psi\in\mathcal{K}_N$. Note that $H_2(s)=\tilde{H}_{\rm JC}$ does not depend on $s$ and leaves each $\mathcal{E}_n$ invariant, while $H_1(s)$, $S(s)$ and $H_2(s)-H_1(s)$ map $\mathcal{K}_M$ into $\mathcal{K}_{M+2}$ for every $M$, depending continuously on $s$ as linear maps between these finite-dimensional spaces (the counter-rotating terms change the number of excitations by $\pm2$; cf. the solution of Exercise~\ref{exerc:rwa-S}).
	
	Let $\phi(s):=U_2(s)\psi$. By Lemma~\ref{lem:rwa-U2}, $\phi(s)\in\mathcal{K}_N$ for all $s$, and by Lemma~\ref{lem:rwa-derivatives}, $\phi'(s)=-\ii H_2\phi(s)$; being a solution of a linear differential equation in the finite-dimensional space $\mathcal{K}_N$, $\phi$ is continuously differentiable.
	
	We first prove a general statement: if $\chi:\mathbb{R}\rightarrow\mathcal{K}_M$ is continuously differentiable for some $M$, then
	\begin{equation}\label{eq:rwa-sol-general}
		\frac{\mathrm{d}}{\mathrm{d}s}\left(U_1(s)^*\chi(s)\right)=U_1(s)^*\chi'(s)+\ii U_1(s)^*H_1(s)\chi(s),
	\end{equation}
	and the right-hand side is continuous in $s$. Indeed, adding and subtracting $U_1(s+h)^*\chi(s)$,
	\begin{equation}
		\frac{U_1(s+h)^*\chi(s+h)-U_1(s)^*\chi(s)}{h}=U_1(s+h)^*\frac{\chi(s+h)-\chi(s)}{h}+\frac{U_1(s+h)^*-U_1(s)^*}{h}\chi(s).
	\end{equation}
	The second term converges to $\ii U_1(s)^*H_1(s)\chi(s)$ by Lemma~\ref{lem:rwa-derivatives}, applied to the fixed vector $\chi(s)\in\mathcal{D}_0$. As for the first term,
	\begin{align}
		\left\|U_1(s+h)^*\frac{\chi(s+h)-\chi(s)}{h}-U_1(s)^*\chi'(s)\right\|&\leq\left\|\frac{\chi(s+h)-\chi(s)}{h}-\chi'(s)\right\|\nonumber\\
		&\quad+\left\|\left(U_1(s+h)^*-U_1(s)^*\right)\chi'(s)\right\|,
	\end{align}
	which vanishes as $h\to0$, using $\|U_1(s+h)^*\|=1$ and the strong continuity of $s\mapsto U_1(s)^*=\e^{\ii H_{\rm QR}s}\e^{-\ii H_0s}$. The continuity of the right-hand side of Eq.~\eqref{eq:rwa-sol-general} follows likewise: $\chi'(s)$ and $H_1(s)\chi(s)$ are continuous with values in finite-dimensional spaces, and $U_1(s)^*$ is uniformly bounded and strongly continuous.
	
	\emph{Eq.~\eqref{eq:rwa-prod1}.} Apply Eq.~\eqref{eq:rwa-sol-general} with $\chi=\phi$:
	\begin{align}
		\frac{\mathrm{d}}{\mathrm{d}s}\left(U_1(s)^*U_2(s)\right)\psi&=U_1(s)^*\left(-\ii H_2(s)\right)U_2(s)\psi+\ii U_1(s)^*H_1(s)U_2(s)\psi\nonumber\\
		&=\ii U_1(s)^*\left(H_1(s)-H_2(s)\right)U_2(s)\psi.
	\end{align}
	
	\emph{Eq.~\eqref{eq:rwa-prod2}.} Let $\chi(s):=S(s)\phi(s)\in\mathcal{K}_{N+2}$. By the product rule for differentiable matrix- and vector-valued functions in finite dimensions, and since $S'(s)=H_2(s)-H_1(s)$ on $\mathcal{K}_N$, $\chi$ is continuously differentiable with
	\begin{equation}
		\chi'(s)=\left(H_2(s)-H_1(s)\right)\phi(s)-\ii S(s)H_2(s)\phi(s).
	\end{equation}
	Applying Eq.~\eqref{eq:rwa-sol-general},
	\begin{align}
		\frac{\mathrm{d}}{\mathrm{d}s}\left(U_1(s)^*S(s)U_2(s)\right)\psi&=U_1(s)^*\left(H_2(s)-H_1(s)\right)U_2(s)\psi-\ii U_1(s)^*S(s)H_2(s)U_2(s)\psi\nonumber\\
		&\quad+\ii U_1(s)^*H_1(s)S(s)U_2(s)\psi\nonumber\\
		&=\ii\,U_1(s)^*\left(H_1(s)S(s)-S(s)H_2(s)\right)U_2(s)\psi\nonumber\\
		&\quad+U_1(s)^*\left(H_2(s)-H_1(s)\right)U_2(s)\psi.
	\end{align}
	Since every $\psi\in\mathcal{D}_0$ belongs to some $\mathcal{K}_N$, this proves the claims.

	\clearpage
	\phantomsection
	\addcontentsline{toc}{section}{List of symbols}
	\section*{List of symbols}\label{sec:symbols}
	
	The following table lists the main symbols used throughout the notes, together with the place where they are introduced. Symbols used only locally (e.g. within a single proof) are omitted.
	
	{\small
		\renewcommand{\arraystretch}{1.25}
		\begin{longtable}{@{}>{\raggedright\arraybackslash}p{0.25\textwidth}>{\raggedright\arraybackslash}p{0.5\textwidth}>{\raggedright\arraybackslash}p{0.19\textwidth}@{}}
			\toprule
			\textbf{Symbol} & \textbf{Meaning} & \textbf{See}\\
			\midrule
			\endfirsthead
			\toprule
			\textbf{Symbol} & \textbf{Meaning} & \textbf{See}\\
			\midrule
			\endhead
			\bottomrule
			\endfoot
			\multicolumn{3}{@{}l}{\emph{General notation}}\\
			$\mathbb{N}$ & natural numbers, including zero & Sec.~\ref{subsec:notation}\\
			$\ii$, $\e$, $\mathrm{d}x$ & imaginary unit, Euler's number, differential & Sec.~\ref{subsec:notation}\\
			$\overline{z}$, $\overline{f}$ & complex conjugate of a number or (pointwise) of a function & Sec.~\ref{subsec:notation}\\
			$\braket{\cdot,\cdot}$, $\|\cdot\|$ & scalar product (antilinear in the first entry) and norm & Def.~\ref{def:scalarproduct}\\
			$\cong$ & unitary equivalence & Sec.~\ref{subsec:notation}\\
			$\bm{x}$, $\bm{E}$, $\bm{\sigma}$, \dots & vectors and vector fields in $\mathbb{R}^3$, vector of Pauli matrices & Sec.~\ref{subsec:notation}\\
			$[A,B]$, $\{A,B\}$ & commutator and anticommutator & Sec.~\ref{subsec:notation}\\
			$\ket{\cdot}$, $\bra{\cdot}$ & bra-ket notation & Sec.~\ref{subsec:notation}\\
			\midrule
			\multicolumn{3}{@{}l}{\emph{Measure theory}}\\
			$\mathfrak{B}(\mathbb{R}^d)$ & Borel $\sigma$-algebra & Def.~\ref{def:pvm}, App.~\ref{app:measure}\\
			$\chi_E$ & characteristic function of a set $E$ & App.~\ref{app:measure}\\
			$\delta_x$ & Dirac measure (point mass) at $x$ & App.~\ref{app:measure}\\
			$\Sigma_1\otimes\Sigma_2$, $\mu_1\otimes\mu_2$ & product $\sigma$-algebra and product measure & App.~\ref{app:measure}\\
			$\omega_*\mu$ & image measure & App.~\ref{app:measure}\\
			\midrule
			\multicolumn{3}{@{}l}{\emph{Hilbert spaces and function spaces}}\\
			$\hilb$ & complex separable Hilbert space & Def.~\ref{def:scalarproduct}\\
			$\ell^2$ & square-summable sequences & Ex.~\ref{examp:ell2}\\
			$L^2(\mathcal{M},\Sigma,\mu)$ & square-integrable functions on a $\sigma$-finite measure space & Ex.~\ref{examp:l2}\\
			$L^2(\mathcal{M},\Sigma,\mu;\hfrak)$ & square-integrable functions with values in a Hilbert space $\hfrak$ & Ex.~\ref{examp:l2vec}\\
			$L^p$, $L^\infty$ & $p$-integrable and essentially bounded functions & App.~\ref{app:measure}\\
			$\mathrm{H}^1$, $\mathrm{H}^2$ & first and second Sobolev spaces & Ex.~\ref{examp:momentum},~\ref{examp:laplace}\\
			$\mathrm{C}^k_{\rm c}(\mathbb{R})$, $\mathcal{S}(\mathbb{R})$ & compactly supported $\mathrm{C}^k$ functions, Schwartz space & Ex.~\ref{examp:momentumlaplacian}\\
			\midrule
			\multicolumn{3}{@{}l}{\emph{Operators}}\\
			$\Dom A$, $\Ran A$, $\Ker A$ & domain, range and kernel of $A$ & Def.~\ref{def:linop}\\
			$\mathcal{B}(\hilb)$ & bounded operators on $\hilb$ & Def.~\ref{def:linop}\\
			$I$ & identity operator & Sec.~\ref{subsec:notation}\\
			$A\subseteq B$, $A\restriction\mathcal{D}$ & extension, restriction & Def.~\ref{def:extension}\\
			$A^*$ & adjoint & Def.~\ref{def:adjoint}\\
			$\overline{A}$ & closure & Def.~\ref{def:closure}\\
			$\operatorname{Graph}A$ & graph of $A$ & Def.~\ref{def:closure}\\
			$A\geq0$, $A\geq-c$ & nonnegative, bounded from below & Sec.~\ref{subsec:notation}\\
			$M_v$ & multiplication operator by $v$ & Ex.~\ref{examp:multiplication}\\
			$p$, $q$ & momentum and position operators & Ex.~\ref{examp:multiplication},~\ref{examp:momentum}\\
			$p_\alpha$ & self-adjoint momenta on the interval & Ex.~\ref{examp:momentuminterval}\\
			$-\Delta_U$ & self-adjoint Laplacians on the segment & Ex.~\ref{examp:laplacesegment}\\
			$\Delta$ & Laplace operator (in Section~\ref{sec:singlemodesb} and App.~\ref{app:rwa}: detuning) & Ex.~\ref{examp:laplace}\\
			$\rho(A)$, $\sigma(A)$, $R_A(z)$ & resolvent set, spectrum, resolvent operator & Def.~\ref{def:resolvent}\\
			$P$, $P_A$ & projection-valued measure, the one associated with $A$ & Def.~\ref{def:pvm}, Thm.~\ref{thm:spectraltheorem1}\\
			$\mu_{\psi,A}$ & spectral measure of $\psi$ with respect to $A$ & Thm.~\ref{thm:spectraltheorem1}\\
			$f(A)$, $\e^{-\ii tA}$ & functional calculus, unitary group generated by $A$ & Sec.~\ref{subsec:spectrum}\\
			$a_\infty$ & relative bound & Def.~\ref{def:relbound}\\
			$\operatorname{tr}T$ & trace of a trace-class operator & Sec.~\ref{subsec:mixed}\\
			$\mathcal{B}_2(\hilb)$, $\mathcal{L}$ & Hilbert--Schmidt operators; Liouvillian & Sec.~\ref{subsec:mixed}\\
			$\rho$, $\ket{\psi}\!\bra{\psi}$ & density operator, pure state & Def.~\ref{def:density}\\
			$\operatorname{tr}_{\hilb_2}T$ & partial trace & Prop.~\ref{prop:partialtrace}\\
			\midrule
			\multicolumn{3}{@{}l}{\emph{Sesquilinear forms}}\\
			$q(u,v)$, $q(u)$ & sesquilinear form and associated quadratic form & Def.~\ref{def:forms}\\
			$\Dom q$ & form domain & Def.~\ref{def:forms}\\
			$q_A$, $\mathcal{Q}(A)$ & form associated with $A$, and its domain & Ex.~\ref{examp:operatorform}\\
			$\|\cdot\|_q$ & form norm & Def.~\ref{def:closedform}\\
			$A\dotplus B$ & form sum & Thm.~\ref{thm:klmn}\\
			\midrule
			\multicolumn{3}{@{}l}{\emph{Direct sums and tensor products}}\\
			$\hilb_1\oplus\hilb_2$, $\bigoplus_n\hilb_n$ & orthogonal direct sum of Hilbert spaces & Def.~\ref{def:directsum_two},~\ref{def:directsum_countable}\\
			$\hilb_{\rm fin}$ & finite sequences in $\bigoplus_n\hilb_n$; span of the number states & Def.~\ref{def:hfin}, Sec.~\ref{subsec:maximal}\\
			$(\hilb)^{\oplus d}$ & direct sum of $d$ copies of $\hilb$ & Prop.~\ref{prop:tensorproduct_finitedim}\\
			$\bigoplus_nA_n$ & direct sum of operators & Def.~\ref{def:directsumoperators}\\
			$\hilb_1\odot\hilb_2$, $\hilb^{\odot n}$ & algebraic tensor product & Def.~\ref{def:algtensorhs}\\
			$\hilb_1\otimes\hilb_2$, $\hilb^{\otimes n}$ & Hilbert space tensor product & Def.~\ref{def:hs_tensorproduct}\\
			$A_1\odot A_2$ & algebraic tensor product of operators & Def.~\ref{def:tensor_operators}\\
			$A_1\otimes A_2$ & Hilbert space tensor product of operators & Def.~\ref{def:hs_tensorproduct_op}\\
			$p(A_1,\dots,A_n)$ & polynomial in self-adjoint operators on a tensor product & Prop.~\ref{prop:polytensor}\\
			\midrule
			\multicolumn{3}{@{}l}{\emph{Single-mode boson fields and spin--boson models}}\\
			$\phi_n$ & Hermite functions; number states & Ex.~\ref{examp:qho}, Def.~\ref{def:abstractladder}\\
			$a_0$, $a_0^\dag$ & ladder operators on $\mathcal{S}(\mathbb{R})$ & Def.~\ref{def:a}\\
			$a$, $a^\dag$, $\N$ & annihilation, creation and number operators & Def.~\ref{defin:aadagnclose},~\ref{def:abstractladder}\\
			$\sigma_x,\sigma_y,\sigma_z$, $\sigma_\pm$ & Pauli matrices, raising and lowering matrices & Def.~\ref{def:singlemodesb}, Sec.~\ref{subsec:jc}\\
			$e_\pm$ & eigenvectors of $\sigma_z$ & Sec.~\ref{subsec:dephasing}\\
			$H_{\rm free}$ & free spin--boson Hamiltonian & Sec.~\ref{subsec:singlemodesb}\\
			$W(\lambda)$ & Weyl displacement operator & Def.~\ref{def:weyl}\\
			$\N_{\rm tot}$, $\mathcal{E}_n$ & number of excitations and its eigenspaces & Def.~\ref{def:ntot}, Prop.~\ref{prop:hfrakn}\\
			$H_{\rm JC}$, $H_{\rm QR}$ & Jaynes--Cummings and quantum Rabi Hamiltonians & Sec.~\ref{subsec:jc},~\ref{subsec:rabi}\\
			\midrule
			\multicolumn{3}{@{}l}{\emph{Fock spaces and second quantization}}\\
			$\sph$ & single-particle Hilbert space & Sec.~\ref{subsec:fock}\\
			$\Gamma(\sph)$, $\varepsilon_0$ & Fock space and vacuum & Def.~\ref{def:fock}\\
			$\Gamma_{\rm fin}(\sph)$, $\Gamma_0(\sph)$ & dense subspaces of finite-particle vectors & Prop.~\ref{prop:finitefock_dense},~\ref{prop:finitefock_dense2}\\
			$\mathrm{d}\Gamma(A)$ & second quantization of $A$ & Def.~\ref{def:secondquantization}\\
			$\N$ & number operator on a direct sum, and on the Fock space & Ex.~\ref{examp:protonumber},~\ref{examp:number_fock}\\
			$\Gamma(U)$ & second quantization of a unitary $U$ & Def.~\ref{defin:gammau}\\
			$a_0(f)$, $a_0^\dag(f)$; $a(f)$, $a^\dag(f)$ & annihilation and creation operators on $\Gamma(\sph)$ & Def.~\ref{def:a0(f)},~\ref{def:a(f)}\\
			$\mathfrak{S}_n$, $\sgn(\sigma)$ & permutation group and sign of a permutation & Def.~\ref{def:permutations},~\ref{def:sign}\\
			$U_\sigma$ & permutation operator on $\sph^{\otimes n}$ & Def.~\ref{def:usigma}\\
			$S_n^\pm$, $S^\pm$ & (anti)symmetrizers & Def.~\ref{def:symmetrizers}\\
			$\Gamma^\pm(\sph)$ & bosonic and fermionic Fock spaces & Def.~\ref{def:fockpm}\\
			$a_\pm(f)$, $a_\pm^\dag(f)$ & bosonic and fermionic annihilation and creation operators & Def.~\ref{def:apm}\\
			$\omega$, $f$ & dispersion relation and form factor & Def.~\ref{def:spinboson}\\
			$J$ & spectral density & Sec.~\ref{subsec:multidephasing}\\
			$W(h)$ & Weyl operator on the Fock space & Def.~\ref{def:weylmulti}\\
			$\varepsilon(f)$, $\psi_h$ & exponential vector and coherent state & Def.~\ref{def:coherentmulti}\\
			$\operatorname{Exp}(\mathcal{V})$ & span of the exponential vectors $\varepsilon(f)$, $f\in\mathcal{V}$ & Def.~\ref{def:coherentmulti}\\
			$\sph_{-s}$ & form factors with $\|\omega^{-s/2}f\|<\infty$ & Sec.~\ref{subsec:creationforms}\\
			$f_\Lambda$ & form factor with ultraviolet cutoff $\Lambda$ & Sec.~\ref{subsec:creationforms}\\
			$v_f$, $H_f$ & interaction form and spin--boson model as a form sum & Thm.~\ref{thm:singularsb}\\
		\end{longtable}
	}
	
	\clearpage
	\phantomsection
	\addcontentsline{toc}{section}{Index}

	\begin{theindex}
		
		\item adjoint, \hyperpage{9}
		\item amplitude-damping channel, \hyperpage{56}
		\item annihilation and creation operators, \hyperpage{65}
		\subitem bosonic, \hyperpage{75}
		\subitem fermionic, \hyperpage{75}
		\item axioms of quantum mechanics, \hyperpage{17}
		
		\indexspace
		
		\item Bloch ball, \hyperpage{18}
		\item Borel $\sigma$-algebra, \hyperpage{98}
		\item boundary conditions, \hyperpage{11}
		
		\indexspace
		
		\item canonical anticommutation relations, \hyperpage{79}
		\item canonical commutation relations, \hyperpage{79}
		\item Cauchy--Schwarz inequality, \hyperpage{7}
		\item closed form, \hyperpage{22}
		\item closed graph theorem, \hyperpage{14}
		\item closure, \hyperpage{11}
		\item coherent states, \hyperpage{53}, \hyperpage{89}
		\item compact operator, \hyperpage{17}
		\item continuous linear extension theorem, \hyperpage{8}
		\item core, \hyperpage{12}
		\item counter-rotating terms, \hyperpage{100}
		\item criterion of essential self-adjointness, \hyperpage{13}
		\item criterion of self-adjointness, \hyperpage{13}
		
		\indexspace
		
		\item decoherence, \hyperpage{92}
		\item delta potential, \hyperpage{24}
		\item dense subspace, \hyperpage{7}
		\item density operator, \hyperpage{18}
		\item dephasing channel, \hyperpage{51}
		\item detuning, \hyperpage{100}
		\item Dicke model, \hyperpage{100}
		\item direct sum
		\subitem countable, \hyperpage{26}
		\subitem of Hilbert spaces, \hyperpage{25}
		\subitem of operators, \hyperpage{27}
		\item dispersion relation, \hyperpage{84}
		\item dominated convergence theorem, \hyperpage{99}
		\item dressed states, \hyperpage{56}
		
		\indexspace
		
		\item electromagnetic field, \hyperpage{41}
		\item entanglement, \hyperpage{33}
		\subitem entropy, \hyperpage{34}
		\item exponential decay, \hyperpage{94}
		\item exponential vectors, \hyperpage{89}
		
		\indexspace
		
		\item Fermi's golden rule, \hyperpage{94}
		\item finite-particle vectors, \hyperpage{60}
		\item Fock space, \hyperpage{59}
		\subitem bosonic, \hyperpage{73}
		\subitem fermionic, \hyperpage{73}
		\item form core, \hyperpage{22}
		\item form domain, \hyperpage{22}
		\item form factor, \hyperpage{84}
		\item form sum, \hyperpage{23}
		\item form-bounded operator, \hyperpage{23}
		\item Friedrichs--Lee model, \hyperpage{93}
		\item Fubini's theorem, \hyperpage{99}
		\item functional calculus, \hyperpage{16}
		
		\indexspace
		
		\item Gelfand triple, \hyperpage{83}
		\item graph norm, \hyperpage{12}, \hyperpage{45}
		\item graph of an operator, \hyperpage{11}
		
		\indexspace
		
		\item Hamiltonian, \hyperpage{17}
		\item harmonic oscillator, \hyperpage{13}
		\subitem ladder representation, \hyperpage{41}
		\item Hermite functions, \hyperpage{13}
		\item Hilbert space, \hyperpage{6}
		\subitem separable, \hyperpage{7}
		\item Hilbert--Schmidt operator, \hyperpage{18}
		
		\indexspace
		
		\item indistinguishable particles, \hyperpage{68}
		\item infrared condition, \hyperpage{87}
		\item integration-by-parts method, \hyperpage{101}
		\item interaction picture, \hyperpage{58}
		\item interior boundary conditions, \hyperpage{98}
		
		\indexspace
		
		\item Jaynes--Cummings model, \hyperpage{54}
		\subitem multimode, \hyperpage{93}
		
		\indexspace
		
		\item Kato--Rellich theorem, \hyperpage{20}
		\item KLMN theorem, \hyperpage{23}
		
		\indexspace
		
		\item $L^2$ space, \hyperpage{8}
		\item $\ell^2$ space, \hyperpage{7}
		\item ladder operators, \hyperpage{42}
		\subitem abstract definition, \hyperpage{47}
		\item Laplace operator, \hyperpage{10}
		\subitem Dirichlet, \hyperpage{20}
		\item Lebesgue integral, \hyperpage{99}
		\item Lebesgue measure, \hyperpage{99}
		\item Liouville--von Neumann equation, \hyperpage{18}
		\item Liouvillian, \hyperpage{18}
		
		\indexspace
		
		\item massive field, \hyperpage{84}
		\item massless field, \hyperpage{84}
		\item measurable function, \hyperpage{99}
		\item mixed state, \hyperpage{18}
		\item momentum operator, \hyperpage{10}
		\item monotone convergence theorem, \hyperpage{99}
		\item multiplication operator, \hyperpage{9}
		
		\indexspace
		
		\item $\N$-representation theorem, \hyperpage{44}
		\item Nelson model, \hyperpage{98}
		\item Nelson's analytic vector theorem, \hyperpage{58}
		\item normal modes, \hyperpage{41}
		\item number of excitations, \hyperpage{54}
		\item number operator, \hyperpage{44}
		\subitem on Fock space, \hyperpage{63}
		
		\indexspace
		
		\item occupation number, \hyperpage{72}
		\item operator
		\subitem bounded, \hyperpage{8}
		\subitem closable, \hyperpage{11}
		\subitem closed, \hyperpage{11}
		\subitem densely defined, \hyperpage{8}
		\subitem essentially self-adjoint, \hyperpage{12}
		\subitem extension, \hyperpage{9}
		\subitem linear, \hyperpage{8}
		\subitem restriction, \hyperpage{9}
		\subitem self-adjoint, \hyperpage{9}
		\subitem symmetric, \hyperpage{9}
		\item orthogonal projection, \hyperpage{7}
		\item orthonormal basis, \hyperpage{7}
		
		\indexspace
		
		\item partial trace, \hyperpage{32}
		\item Pauli exclusion principle, \hyperpage{72}
		\item Pauli matrices, \hyperpage{48}
		\item Pauli--Fierz model, \hyperpage{4}
		\item permutation, \hyperpage{68}
		\subitem operator, \hyperpage{70}
		\subitem sign, \hyperpage{69}
		\subitem transposition, \hyperpage{69}
		\item polarization identity, \hyperpage{89}
		\item position operator, \hyperpage{9}
		\item pre-Hilbert space, \hyperpage{7}
		\item product measure, \hyperpage{99}
		\item projection theorem, \hyperpage{7}
		\item projection-valued measure, \hyperpage{15}
		\item pure state, \hyperpage{17}
		
		\indexspace
		
		\item quadratic form, \hyperpage{22}
		\item quantum channel, \hyperpage{51}
		\item quantum Rabi model, \hyperpage{57}
		\item quantum regression, \hyperpage{93}
		\item qubit
		\subitem populations and coherences, \hyperpage{18}
		
		\indexspace
		
		\item Rabi oscillations, \hyperpage{56}
		\subitem vacuum, \hyperpage{57}
		\item reduced state, \hyperpage{33}
		\item relative boundedness, \hyperpage{19}
		\subitem infinitesimal, \hyperpage{19}
		\subitem relative bound, \hyperpage{19}
		\item relative form bound, \hyperpage{23}
		\item renormalization, \hyperpage{98}
		\item representation theorem, \hyperpage{22}
		\item resolvent, \hyperpage{14}
		\item resolvent convergence, \hyperpage{16}
		\item Riemann--Lebesgue lemma, \hyperpage{100}
		\item Riesz representation theorem, \hyperpage{29}
		\item rotating-wave approximation, \hyperpage{58}
		\subitem convergence, \hyperpage{106}
		\subitem error bound, \hyperpage{105}
		\item rotating-wave spin--boson model, \hyperpage{93}
		
		\indexspace
		
		\item scalar product, \hyperpage{6}
		\item Schmidt decomposition, \hyperpage{33}
		\item Schr\"odinger equation, \hyperpage{17}
		\item Schr\"odinger operator, \hyperpage{20}
		\item Schwartz space, \hyperpage{12}
		\item second quantization, \hyperpage{61}
		\subitem of unitary operators, \hyperpage{63}
		\subitem unitary covariance, \hyperpage{64}
		\item sesquilinear form, \hyperpage{22}
		\item $\sigma$-algebra, \hyperpage{98}
		\item Sobolev space, \hyperpage{12}
		\item spectral density, \hyperpage{88}
		\subitem flat, \hyperpage{93, 94}
		\subitem ohmic, \hyperpage{93}
		\item spectral measure, \hyperpage{15}
		\item spectral theorem, \hyperpage{15}
		\subitem multiplicative form, \hyperpage{15}
		\item spectrum, \hyperpage{14}
		\item spin--boson model, \hyperpage{84}
		\subitem pure dephasing, \hyperpage{51}, \hyperpage{88}
		\subitem single-mode, \hyperpage{48}
		\item spin-$\frac12$ particle, \hyperpage{25}
		\item Stone's theorem, \hyperpage{16}
		\item symmetric group, \hyperpage{68}
		\item symmetrizer, \hyperpage{70}
		
		\indexspace
		
		\item tensor product
		\subitem of Hilbert spaces, \hyperpage{31}
		\subitem of Hilbert spaces, algebraic, \hyperpage{29}
		\subitem of $L^2$ spaces, \hyperpage{34}
		\subitem of operators, \hyperpage{38}
		\subitem of operators, algebraic, \hyperpage{35}
		\item trace class, \hyperpage{17}
		\item Trotter's theorem, \hyperpage{16}
		
		\indexspace
		
		\item ultraviolet cutoff, \hyperpage{96}
		\item unitary equivalence, \hyperpage{5}
		\item unitary group
		\subitem generator, \hyperpage{16}
		\item unitary group, strongly continuous, \hyperpage{16}
		\item unitary operator, \hyperpage{8}
		\item UV divergence, \hyperpage{87}
		
		\indexspace
		
		\item vacuum, \hyperpage{59}
		\item van Hove model, \hyperpage{91}, \hyperpage{98}
		
		\indexspace
		
		\item W\"ust theorem, \hyperpage{20}
		\item Weyl displacement operator, \hyperpage{52}
		\subitem on the Fock space, \hyperpage{88}
		\subitem Weyl relations, \hyperpage{90}
		\item Wigner--Weisskopf theory, \hyperpage{94}
		
		\indexspace
		
		\item Young inequality, \hyperpage{19}
		
	\end{theindex}
	

\begin{thebibliography}{99}
		
		
		
		
		
		
		
		
		
		
		
		\bibitem{fetterwalecka} A. L. Fetter and J. D. Walecka, \emph{Quantum Theory of Many-Particle Systems}, McGraw-Hill, New York (1971).
		\bibitem{negeleorland} J. W. Negele and H. Orland, \emph{Quantum Many-Particle Systems}, Frontiers in Physics \textbf{68}, Addison-Wesley, Redwood City (1988).
		\bibitem{altlandsimons} A. Altland and B. D. Simons, \emph{Condensed Matter Field Theory}, 2nd ed., Cambridge University Press, Cambridge (2010).
		\bibitem{gerryknight} C. C. Gerry and P. L. Knight, \emph{Introductory Quantum Optics}, Cambridge University Press, Cambridge (2005).
		\bibitem{reedsimon2} M. Reed and B. Simon, \emph{Methods of Modern Mathematical Physics II: Fourier Analysis, Self-Adjointness}, Academic Press, New York (1975).
		\bibitem{brattelirobinson2} O. Bratteli and D. W. Robinson, \emph{Operator Algebras and Quantum Statistical Mechanics 2: Equilibrium States. Models in Quantum Statistical Mechanics}, 2nd ed., Texts and Monographs in Physics, Springer, Berlin (1997).
		\bibitem{spohn} H. Spohn, \emph{Dynamics of Charged Particles and their Radiation Field}, Cambridge University Press, Cambridge (2004).
		\bibitem{derezinskigerard} J. Derezi\'nski and C. G\'erard, \emph{Mathematics of Quantization and Quantum Fields}, Cambridge Monographs on Mathematical Physics, Cambridge University Press, Cambridge (2013).
		\bibitem{arai} A. Arai, \emph{Analysis on Fock Spaces and Mathematical Theory of Quantum Fields: An Introduction to Mathematical Analysis of Quantum Fields}, World Scientific, Singapore (2018).
		\bibitem{derezinskiccr} J. Derezi\'nski, ``Introduction to representations of the canonical commutation and anticommutation relations'', in J. Derezi\'nski and H. Siedentop (eds.), \emph{Large Coulomb Systems: Lecture Notes on Mathematical Aspects of QED}, Lecture Notes in Physics \textbf{695}, 63--143 (2006), Springer, Berlin, \doi{10.1007/3-540-32579-4_3}.
		\bibitem{gustafsonsigal} S. J. Gustafson and I. M. Sigal, \emph{Mathematical Concepts of Quantum Mechanics}, 2nd ed., Universitext, Springer, Berlin (2011).
		\bibitem{teschl} G. Teschl, \emph{Mathematical Methods in Quantum Mechanics: With Applications to Schr\"odinger Operators}, 2nd ed., Graduate Studies in Mathematics \textbf{157}, American Mathematical Society, Providence (2014).
		\bibitem{reedsimon1} M. Reed and B. Simon, \emph{Methods of Modern Mathematical Physics I: Functional Analysis}, revised and enlarged ed., Academic Press, New York (1980).
		\bibitem{hall} B. C. Hall, \emph{Quantum Theory for Mathematicians}, Graduate Texts in Mathematics \textbf{267}, Springer, New York (2013).
		\bibitem{deoliveira} C. R. de Oliveira, \emph{Intermediate Spectral Theory and Quantum Dynamics}, Progress in Mathematical Physics \textbf{54}, Birkh\"auser, Basel (2009).
		\bibitem{schmuedgen} K. Schm\"udgen, \emph{Unbounded Self-adjoint Operators on Hilbert Space}, Graduate Texts in Mathematics \textbf{265}, Springer, Dordrecht (2012).
		\bibitem{katobook} T. Kato, \emph{Perturbation Theory for Linear Operators}, reprint of the 1980 ed., Classics in Mathematics, Springer, Berlin (1995).
		\bibitem{bonneau2001} G. Bonneau, J. Faraut and G. Valent, ``Self-adjoint extensions of operators and the teaching of quantum mechanics'', Am. J. Phys. \textbf{69}, 322--331 (2001), \doi{10.1119/1.1328351}.
		\bibitem{asorey2005} M. Asorey, A. Ibort and G. Marmo, ``Global theory of quantum boundary conditions and topology change'', Int. J. Mod. Phys. A \textbf{20}, 1001--1025 (2005), \doi{10.1142/S0217751X05019798}.
		\bibitem{benderboettcher1998} C. M. Bender and S. Boettcher, ``Real spectra in non-Hermitian Hamiltonians having $\mathcal{PT}$ symmetry'', Phys. Rev. Lett. \textbf{80}, 5243 (1998), \doi{10.1103/PhysRevLett.80.5243}.
		\bibitem{nielsenchuang} M. A. Nielsen and I. L. Chuang, \emph{Quantum Computation and Quantum Information}, 10th anniversary ed., Cambridge University Press, Cambridge (2010).
		\bibitem{lonigro2024lvn} D. Lonigro, A. Hahn and D. Burgarth, ``On the Liouville--von Neumann equation for unbounded Hamiltonians'', Open Syst. Inf. Dyn. \textbf{31}, 2450018 (2024), \doi{10.1142/S1230161224500185}.
		
		\bibitem{rellich1939} F. Rellich, ``St\"orungstheorie der Spektralzerlegung. III'', Math. Ann. \textbf{116}, 555--570 (1939), \doi{10.1007/BF01597374}.
		\bibitem{kato1951} T. Kato, ``Fundamental properties of Hamiltonian operators of Schr\"odinger type'', Trans. Amer. Math. Soc. \textbf{70}, 195--211 (1951), \doi{10.1090/S0002-9947-1951-0041010-X}.
		\bibitem{wust1971} R. W\"ust, ``Generalisations of Rellich's theorem on perturbation of (essentially) selfadjoint operators'', Math. Z. \textbf{119}, 276--280 (1971), \doi{10.1007/BF01113402}. 
		\bibitem{kato1955} T. Kato, \emph{Quadratic forms in Hilbert spaces and asymptotic perturbation series}, Technical Report No.~7, Department of Mathematics, University of California, Berkeley (1955).
		\bibitem{lions1961} J.-L. Lions, \emph{\'Equations Diff\'erentielles Op\'erationnelles et Probl\`emes aux Limites}, Springer, Berlin (1961).
		\bibitem{laxmilgram1954} P. D. Lax and A. N. Milgram, ``Parabolic equations'', in \emph{Contributions to the Theory of Partial Differential Equations}, Annals of Mathematics Studies \textbf{33}, Princeton University Press, Princeton (1954), pp.~167--190, \doi{10.1515/9781400882182-010}.
		\bibitem{nelson1964} E. Nelson, ``Interaction of nonrelativistic particles with a quantized scalar field'', J. Math. Phys. \textbf{5}, 1190--1197 (1964), \doi{10.1063/1.1704225}.
		\bibitem{albeverio2005} S. Albeverio, F. Gesztesy, R. H\o egh-Krohn and H. Holden, \emph{Solvable Models in Quantum Mechanics}, 2nd ed., AMS Chelsea Publishing, Providence (2005).
		\bibitem{cohentannoudji1989} C. Cohen-Tannoudji, J. Dupont-Roc and G. Grynberg, \emph{Photons and Atoms: Introduction to Quantum Electrodynamics}, Wiley, New York (1989).
		\bibitem{vonneumann1932} J. von Neumann, ``\"Uber adjungierte Funktionaloperatoren'', Ann. of Math. \textbf{33}, 294--310 (1932), \doi{10.2307/1968331}.
		\bibitem{jaynescummings1963} E. T. Jaynes and F. W. Cummings, ``Comparison of quantum and semiclassical radiation theories with application to the beam maser'', Proc. IEEE \textbf{51}, 89--109 (1963), \doi{10.1109/PROC.1963.1664}.
		\bibitem{shoreknight1993} B. W. Shore and P. L. Knight, ``The Jaynes--Cummings model'', J. Mod. Opt. \textbf{40}, 1195--1238 (1993), \doi{10.1080/09500349314551321}.
		\bibitem{rabi1937} I. I. Rabi, ``Space quantization in a gyrating magnetic field'', Phys. Rev. \textbf{51}, 652--654 (1937), \doi{10.1103/PhysRev.51.652}.
		\bibitem{larsonmavrogordatos} J. Larson and T. Mavrogordatos, \emph{The Jaynes--Cummings Model and its Descendants: Modern Research Directions}, 2nd ed., IOP Series in Quantum Technology, IOP Publishing, Bristol (2024).
		\bibitem{palma1996} G. M. Palma, K.-A. Suominen and A. K. Ekert, ``Quantum computers and dissipation'', Proc. R. Soc. Lond. A \textbf{452}, 567--584 (1996), \doi{10.1098/rspa.1996.0029}.
		\bibitem{breuerpetruccione} H.-P. Breuer and F. Petruccione, \emph{The Theory of Open Quantum Systems}, Oxford University Press, Oxford (2002).
		\bibitem{cahillglauber1969} K. E. Cahill and R. J. Glauber, ``Ordered expansions in boson amplitude operators'', Phys. Rev. \textbf{177}, 1857--1881 (1969), \doi{10.1103/PhysRev.177.1857}.
		\bibitem{braak2011} D. Braak, ``Integrability of the Rabi model'', Phys. Rev. Lett. \textbf{107}, 100401 (2011), \doi{10.1103/PhysRevLett.107.100401}.
		\bibitem{burgarth} D. Burgarth, P. Facchi, R. Hillier and M. Ligab\`o, ``Taming the rotating wave approximation'', Quantum \textbf{8}, 1262 (2024), \doi{10.22331/q-2024-02-21-1262}.
		\bibitem{weedbrook2012} C. Weedbrook, S. Pirandola, R. Garc\'ia-Patr\'on, N. J. Cerf, T. C. Ralph, J. H. Shapiro and S. Lloyd, ``Gaussian quantum information'', Rev. Mod. Phys. \textbf{84}, 621--669 (2012), \doi{10.1103/RevModPhys.84.621}.
		\bibitem{lloydbraunstein1999} S. Lloyd and S. L. Braunstein, ``Quantum computation over continuous variables'', Phys. Rev. Lett. \textbf{82}, 1784--1787 (1999), \doi{10.1103/PhysRevLett.82.1784}.
		\bibitem{fischer2026} F. Fischer, D. Burgarth and D. Lonigro, ``Self-adjoint realizations of higher-order squeezing operators'', J. Phys. A: Math. Theor. \textbf{59}, 255203 (2026), \doi{10.1088/1751-8121/ae7acb}.
		\bibitem{fischerknapp2026} F. Fischer, F. Knapp, D. Burgarth and D. Lonigro, ``Self-adjoint extensions of $k$-photon light--matter Hamiltonians'', arXiv:2607.22378 (2026), \doi{10.48550/arXiv.2607.22378}.
		\bibitem{fischer2026hamiltonian} F. Fischer, H. Schulz-Baldes, D. Burgarth and D. Lonigro, ``When is a bosonic operator a Hamiltonian?'' arXiv:2610.33545 (2026), \doi{10.48550/arXiv.2610.33545}.
		\bibitem{ashhab2026} S. Ashhab, F. Fischer, D. Lonigro, D. Braak and D. Burgarth, ``Finite-dimensional approximations of generalized squeezing'', Phys. Rev. A \textbf{113}, 013703 (2026), \doi{10.1103/9vwp-f35c}.
		\bibitem{fock1932} V. Fock, ``Konfigurationsraum und zweite Quantelung'', Z. Phys. \textbf{75}, 622--647 (1932), \doi{10.1007/BF01344458}.
		\bibitem{gelfandvilenkin4} I. M. Gel'fand and N. Ya. Vilenkin, \emph{Generalized Functions, Vol.~4: Applications of Harmonic Analysis}, Academic Press, New York (1964).
		\bibitem{delamadrid2005} R. de la Madrid, ``The role of the rigged Hilbert space in quantum mechanics'', Eur. J. Phys. \textbf{26}, 287--312 (2005), \doi{10.1088/0143-0807/26/2/008}.
		\bibitem{leggett1987} A. J. Leggett, S. Chakravarty, A. T. Dorsey, M. P. A. Fisher, A. Garg and W. Zwerger, ``Dynamics of the dissipative two-state system'', Rev. Mod. Phys. \textbf{59}, 1--85 (1987), \doi{10.1103/RevModPhys.59.1}.
		\bibitem{spohn1989} H. Spohn, ``Ground state(s) of the spin-boson Hamiltonian'', Commun. Math. Phys. \textbf{123}, 277--304 (1989), \doi{10.1007/BF01238859}.
		\bibitem{hubnerspohn1995} M. H\"ubner and H. Spohn, ``Spectral properties of the spin-boson Hamiltonian'', Ann. Inst. Henri Poincar\'e Phys. Th\'eor. \textbf{62}, 289--323 (1995).
		\bibitem{araihirokawa1997} A. Arai and M. Hirokawa, ``On the existence and uniqueness of ground states of a generalized spin-boson model'', J. Funct. Anal. \textbf{151}, 455--503 (1997), \doi{10.1006/jfan.1997.3140}.
		\bibitem{haslerherbst2011} D. Hasler and I. Herbst, ``Ground states in the spin boson model'', Ann. Henri Poincar\'e \textbf{12}, 621--677 (2011), \doi{10.1007/s00023-011-0091-6}.
		\bibitem{lonigro2022} D. Lonigro, ``Generalized spin--boson models with non-normalizable form factors'', J. Math. Phys. \textbf{63}, 072105 (2022), \doi{10.1063/5.0085576}.
		\bibitem{derezinski2003} J. Derezi\'nski, ``Van Hove Hamiltonians -- Exactly solvable models of the infrared and ultraviolet problem'', Ann. Henri Poincar\'e \textbf{4}, 713--738 (2003), \doi{10.1007/s00023-003-0145-5}.
		\bibitem{lonigrochruscinski2022dephasing} D. Lonigro and D. Chru\'sci\'nski, ``Quantum regression in dephasing phenomena'', J. Phys. A: Math. Theor. \textbf{55}, 225308 (2022), \doi{10.1088/1751-8121/ac6a2d}.
		\bibitem{friedrichs1948} K. O. Friedrichs, ``On the perturbation of continuous spectra'', Commun. Pure Appl. Math. \textbf{1}, 361--406 (1948), \doi{10.1002/cpa.3160010404}.
		\bibitem{lee1954} T. D. Lee, ``Some special examples in renormalizable field theory'', Phys. Rev. \textbf{95}, 1329--1334 (1954), \doi{10.1103/PhysRev.95.1329}.
		\bibitem{khalfin1958} L. A. Khalfin, ``Contribution to the decay theory of a quasi-stationary state'', Sov. Phys. JETP \textbf{6}, 1053--1063 (1958).
		\bibitem{lonigrochruscinski2022regression} D. Lonigro and D. Chru\'sci\'nski, ``Quantum regression beyond the Born--Markov approximation for generalized spin--boson models'', Phys. Rev. A \textbf{105}, 052435 (2022), \doi{10.1103/PhysRevA.105.052435}.
		\bibitem{lonigro2023mpag} D. Lonigro, ``Self-adjointness of a class of multi-spin--boson models with ultraviolet divergences'', Math. Phys. Anal. Geom. \textbf{26}, 15 (2023), \doi{10.1007/s11040-023-09457-6}.
		\bibitem{lilllonigro2025} S. Lill and D. Lonigro, ``Self-adjointness and domain of generalized spin--boson models with mild ultraviolet divergences'', in A. Michelangeli, B. Cassano, M. Ligab\`o, M. Gallone and F. Deelan Cunden (eds.), \emph{Singularities, Asymptotics, and Limiting Models}, Springer INdAM Series \textbf{64}, Springer (2025), pp.~159--185, \doi{10.1007/978-981-96-3584-9_7}.
		\bibitem{schmidt2018} J. Schmidt, J. Lampart, S. Teufel and R. Tumulka, ``Particle creation at a point source by means of interior-boundary conditions'', Math. Phys. Anal. Geom. \textbf{21}, 12 (2018), \doi{10.1007/s11040-018-9270-8}.
		\bibitem{lampartschmidt2019} J. Lampart and J. Schmidt, ``On Nelson-type Hamiltonians and abstract boundary conditions'', Commun. Math. Phys. \textbf{367}, 629--663 (2019), \doi{10.1007/s00220-019-03294-x}.
		\bibitem{lonigro2023} D. Lonigro, ``Renormalization of spin--boson interactions mediated by singular form factors'', in M. Correggi and M. Falconi (eds.), \emph{Quantum Mathematics II}, Springer INdAM Series \textbf{58}, Springer, Singapore (2023), pp.~103--122, \doi{10.1007/978-981-99-5884-9_3}.
		\bibitem{dammoller2020} T. N. Dam and J. S. M\o ller, ``Asymptotics in spin-boson type models'', Commun. Math. Phys. \textbf{374}, 1389--1415 (2020), \doi{10.1007/s00220-020-03685-5}.
		\bibitem{hinrichs2025} B. Hinrichs, J. Lampart and J. Valent\'in Mart\'in, ``Ultraviolet renormalization of spin boson models I. Normal and 2-nilpotent interactions'', arXiv:2502.04876 (2025), \doi{10.48550/arXiv.2502.04876}.
		\bibitem{alvarez2025} B. Alvarez, S. Lill, D. Lonigro and J. Valent\'in Mart\'in, ``Renormalization of generalized spin--boson models with critical ultraviolet divergences'', arXiv:2508.00803 (2025), \doi{10.48550/arXiv.2508.00803}.
		\bibitem{lillthesis} S. Lill, \emph{Time Dynamics in Quantum Field Theory Systems}, PhD thesis, Eberhard Karls Universit\"at T\"ubingen (2022), \url{http://hdl.handle.net/10900/129763}.
		\bibitem{richterdicke} L. Richter, D. Burgarth and D. Lonigro, ``Quantifying the rotating-wave approximation of the Dicke model'', J. Phys. A: Math. Theor. \textbf{59}, 075203 (2026), \doi{10.1088/1751-8121/ae42a3}.
		\bibitem{rudin} W. Rudin, \emph{Real and Complex Analysis}, 3rd ed., McGraw-Hill, New York (1987).
		\bibitem{folland} G. B. Folland, \emph{Real Analysis: Modern Techniques and Their Applications}, 2nd ed., Wiley, New York (1999).
		\bibitem{bogachev} V. I. Bogachev, \emph{Measure Theory}, Vols.~I--II, Springer, Berlin (2007).
		
	\end{thebibliography}
\end{document}